\DocumentMetadata{  
    lang = en,
    pdfstandard = ua-2,
}

\documentclass[11pt]{report}
\usepackage[utf8]{inputenc}
\usepackage{graphicx}
\graphicspath{ {images/} }
\usepackage{caption}
\usepackage{subfigure}
\usepackage{float}
\usepackage[width=150mm,top=35mm,bottom=25mm,bindingoffset=6mm]{geometry}
\usepackage[twoside]{fancyhdr}
\usepackage{setspace}
\usepackage{hyperref}
\usepackage{amsmath, amssymb}
\usepackage{longtable}
\usepackage[referable]{threeparttablex}
\usepackage{deluxetable} %apparently a load-bearing package for the "landscape" environment, I ended up using longtable and threeparttablex
\usepackage{titlesec}
\usepackage[strict]{changepage}
\newcommand\chaptitlefont{
  \fontfamily{libertine}\fontseries{b}
  \libertine
  \fontshape{n}\fontsize{25}{35}\selectfont\raggedright}

\titleformat{\chapter}[display]{\chaptitlefont\huge\bfseries}{\chaptertitlename\ \thechapter}{20pt}{\Huge}
\titleformat{\section}{\chaptitlefont\Large\bfseries}{\thesection}{1em}{}
\titleformat{\subsection}{\normalfont\large\itshape\raggedright}{\thesubsection}{1em}{}
\titleformat{\subsubsection}
  {\normalfont\normalsize\bfseries\raggedright}{\thesubsubsection}{1em}{}
\titleformat{\paragraph}[runin]{\normalfont\normalsize\sc}{\theparagraph}{1em}{}

\titlespacing{\paragraph}
  {\parindent}{*2}{\wordsep}

\usepackage{libertine}\chaptitlefont

\usepackage[T1]{fontenc}\normalfont

\newcommand\mosaicwidth{0.24\textwidth}
\newcommand\mostopclip{41pt}
\newcommand\mosleftclip{101pt}
\newcommand\mosrightclip{91pt}
\newcommand\mosbotclip{37pt}

\fancypagestyle{plain}{%
  \fancyhf{}%
  \fancyhf[roh,reh]{\thepage}%
}

\usepackage{natbib}
\usepackage[nottoc]{tocbibind}

\newcommand\thetitle{Taking Inventory of the Most Promising Lensed Radio Sources for Constraining Fundamental Properties of Dark Matter}

\defcitealias{martinez2025}{M25}

\title{\thetitle} 
\author{Michael N. Martinez}
\date{1 May 2026}

\newcommand\ion[2]{#1$\;${%
\ifx\@currsize\normalsize\small \else
\ifx\@currsize\small\footnotesize \else
\ifx\@currsize\footnotesize\scriptsize \else
\ifx\@currsize\scriptsize\tiny \else
\ifx\@currsize\large\normalsize \else
\ifx\@currsize\Large\large
\fi\fi\fi\fi\fi\fi
\rmfamily\@Roman{#2}}\relax}% 

{\catcode`\$=\active
}% 

 \newcommand\HA{\textsc{ha}}

\newcommand\sbond{\chem@bnd{\@sbnd}}%
\newcommand\dbond{\chem@bnd{\@dbnd}}%
\newcommand\tbond{\chem@bnd{\@tbnd}}%

\begin{document}
\pagenumbering{roman}

\begin{titlepage}
    \begin{center}
        \vspace*{1cm}
        
        \Huge
        \libertine\textbf{Taking Inventory of the Most Promising Lensed Radio Sources\\ for Constraining Fundamental Properties of Dark Matter} \\

        \vspace{0.5cm}
        \LARGE
        \normalfont
        %subtitle\\
        by\\
        \vspace{0.5cm}
        \libertine
        \textbf{Michael N. Martinez}
        
        \vfill
        \normalfont
        A dissertation submitted in partial fulfillment\\
        of the requirements for the degree of\\
        Doctor of Philosophy\\
        (Physics)\\

        \vspace{1.8cm}

        \Large
        at the\\University of Wisconsin-Madison\\
        2026\\
        \vspace{1.0cm}
        \begin{flushleft}
        \large
        Date of Final Oral Exam:  5/1/2026\\
        The dissertation is approved by the following members of the Final Oral Committee: \\
        \setlength{\parindent}{10ex}
        Keith Bechtol, Professor, Physics\\Sebastian Heinz, Professor, Astronomy\\Peter Timbie, Professor, Physics\\ Christy Tremonti, Professor, Astronomy\\
        \end{flushleft}
        
    \end{center}
    
\end{titlepage}

%remove for arxiv
%\newpage 
%\
%\newpage

\begin{center}
    \libertine
    \Large
    \textbf{\thetitle}
    
    %\vspace{0.4cm}
    %\large
    %Thesis Subtitle
    
    \vspace{0.4cm}
    \textbf{Michael N. Martinez}
    
    \vspace{0.9cm}
    \textbf{Abstract}
    \normalfont
\end{center}

While dark matter (DM) makes up roughly 80\% of the total matter in the Universe, its microscopic properties remain one of the biggest questions in Cosmology today. Fortunately, those properties dictate the distribution and form of macro-scale gravitational structures in the universe, allowing for indirect studies which can distinguish between competing particle models. One such avenue for this research is via strong gravitational lensing systems, where dark halos in the lens substructure and along the line of sight perturb image positions and flux. However, the current population of sources suitable for this analysis is limited, especially at radio wavelengths where astrometric perturbations are observable. I will first discuss which properties of lens systems make them especially useful for DM constraints and examine the minimum amount of information necessary for such an experiment. Then, I present the results of two successful searches for new radio lenses in existing radio and optical surveys, utilizing a new method to expand the potential follow-up population for dark matter studies in the future. I conclude with a discussion of the completeness of this population.

\newpage 
%\
%\newpage

\textit{To my family.}

%\chapter*{Declaration}

%I declare that..

\chapter*{Acknowledgements}

It is difficult to collect five years of education and research into 150-odd pages, but summing up everything else about those years in just a few may be impossible. Nevertheless I will do my best to briefly capture as much as I can. To anyone who I've neglected here, please treat it as an omission borne of forgetfulness and stress rather than malice.

To Keith Bechtol: Over the years there have been many times where I was concerned with a research problem, a career choice, or any other thing however small, and I always felt able to come to you for advice. Just as importantly, I always left with a clear picture of what to do next. Thank you for being an incredible advisor.

To Yjan Gordon and Peter Ferguson: Thank you for the mentorship over the years as postdocs in the Observational Cosmology group and beyond. Be it presentation feedback, job applications, or general life advice, your contributions to my growth as a scientist can't be overstated, and I'm glad to call you my friends. 

To all my other coworkers over the years in the Bechtol Group: Rob Morgan, Megan Tabbutt, Mitch McNanna, Jimena Gonz\'{a}lez, Miranda Gorsuch, Juli\'{a}n Beas-Gonz\'{a}lez, Kayleigh Excell, Jaemyoung Lee, Gillian Cartwright, Kyle Boone, Alex Tellez, and Anna Castello. I wish you all the best in your efforts, be that finishing your own Ph.D's, starting graduate school, working in industry, or anything in between. I will miss our meetings, happy hours, and general hanging out in the office throughout the years. To Yurii Kvasiuk, Faizah Siddique, Sophia Nowack, Anderson Lai, and everyone else in Observational Cosmology: The same goes for you -- thank you for letting me show up unannounced at your desks and talk about whatever for a bit. To Carrie Laber-Smith, Stephen McKay, Zain Abhari, and the rest of my UW Physics cohort: I can't believe it's been five years. While we've all gone our separate ways, I'll never forget struggling through core classes with you all.

To everyone working in the UW Astronomy department, particularly Eric Hooper, Marsha Wolf, Melinda Soares-Furtado, Sebastian Heinz, Christy Tremonti, Rachel McClure, Melissa Morris, Leon Trapman, Michael Nicandro Rosenthal, Jennifer Stafford, Yiting Wang, Talia O'Shea, and Francisco Sequeira Murillo: Thank you for making me feel welcome in your department, be it at star parties, sherry hours, or jets group meetings.

To Peter Timbie and Moritz Munchmeyer: thank you for your advice and feedback over the years, and for making the UW Cosmology what it is today.
To Jim Reardon, Ben Spike and Abdollah Mohammadi: thank you for the mentorship in teaching each of you have provided me. I can only hope to emulate you in my future teaching endeavors.
To my fellow TAs and, of course, all my students over the years, you also contributed greatly to my growth as a teacher and deserve just as much gratitude.
To Michael Gladders, Richard Kron, Gourav Khullar, Binhua Lin, and the memory of Stuart A. Rice: your mentorship during my undergraduate and post-graduate research career led me to Madison, and it was a pleasure working with and learning from you all.

To Fritz Hofmann, thank you for being the best flatmate I could ask for the last five years. I'll miss you dearly and wish you the best, especially when you finally decide to go to grad school.

To everyone who attended my dissertation defense: Thank you for the support! I was astounded so many people were there both in person and online, and it means a lot to me.

Special thanks also goes to the developers of the innumerable programs, software packages, and science platforms I have used in the course of my research. While I am certain I have failed in my attempts to credit you all, know your dedication does not go unappreciated.
Likewise, the engineers and telescope operators at NRAO, the system administrators and help-desk staff at CHTC, and the employees of \texttt{arxiv.org} and NASA ADS have a claim to this work as well. 
Also deserving of thanks is the staff of every bar, coffeeshop, and restaurant I had the pleasure of working at during my time at UW, especially those of Leopold's Books Bar Caff\`{e}, where I wrote at least half of this thesis.

I would like to give a special acknowledgment to the memory of my grandfather, Arturo Lauro Martinez. In the past five years, my exposure to radio astronomy has only magnified my regret that I cannot discuss the subject with you.

Finally, to my family: Danny Martinez, Nettie Martinez, Andrew Martinez, and Mary Martinez. It's easy to say ``I couldn't have done this without you'' to everyone listed here, but the sentiment is only really true of you. I love you all so much.

\tableofcontents

\listoffigures

\listoftables

%\newpage
%\
%\newpage

\doublespacing
\chapter{Introduction}
\pagenumbering{arabic}

\paragraph{Dark Matter is all around us,}
 surrounding our galaxy and dictating the formation of structures from that size to the largest we can observe in the Universe.
Despite this, however, the identity of the particle (or particles) that make up dark matter remains a mystery, and has resisted discovery for decades in direct detection experiments on Earth.
Fortunately, the larger-scale distribution of dark matter clumps, or ``halos'', gravitationally affects things we can observe, and those effects provide indirect probes of the particle's microscopic properties.
Gravitational lenses provide one such laboratory for these studies, and ultra-high resolution observations at radio wavelengths have the potential to put very competitive limits on both the overall statistical properties of dark matter and those of individual halos.
 This introduction is structured as follows: in \ref{sec:LCDM} we will introduce $\Lambda$CDM, the current picture of the universe on its largest scales.
 We also will review structure formation driven by Cold Dark Matter and briefly introduce alternative dark matter models, before \ref{sec:lensing} covers the basics of strong gravitational lensing.
Then, we will discuss techniques of Radio Astronomy (\ref{sec:radio}) and briefly review the structure of Active Galactic Nuclei (AGN, \ref{sec:AGN}).
Finally, \ref{sec:introutline} will bring everything together by reviewing the ways gravitational lensing can probe Dark Matter properties, and outline the rest of this document. 

\section{Dark Matter and Cosmology} \label{sec:LCDM}

Cosmologists typically assume the universe to be spatially ``homogeneous and isotropic,'' meaning that on the largest scales, space largely looks the same no matter where you look.
That assumption implies a structure of spacetime called Friedman-Lema\^{i}tre-Robertson-Walker (FLRW) space, in which spatial and time dimensions are linked in the following way:
\begin{equation}
ds^2 = -c^2\text{d}t^2 + a(t)^2\left(\frac{\text{d}r^2}{1-k r^2}+r^2d\Omega^2\right); \quad k\in \{-1, 0, 1\} .
\end{equation}
The parameter $a(t)$, which is normalized to be 1 today (notated $t_0$), is the \textbf{scale factor}, which defines the relation between physical distance and ``comoving'' distance, defined below.
We use the scale factor to define the familiar observational notion of \textbf{cosmological redshift} $z$ of an object at cosmological distance emitting light with wavelength $\lambda$ and frequency $\nu$:
\begin{equation}
1 + z = \frac{\lambda_{obs}}{\lambda_{emit}} = \frac{\nu_{emit}}{\nu_{obs}} = \frac{1}{a(t_{emit})}.
\end{equation}
We also define the \textbf{Hubble parameter} $H = \dot{a}/a$, which describes how the scale factor changes over time.
Related is the \textbf{Hubble constant} $H_0 = H(t_0)$, its present-day value and the proportionality constant between the physical distance $d$ to a $(z\ll1)$ galaxy\footnote{with no peculiar velocity} and the velocity $v$ with which it moves away from us:
\begin{equation}
v = H_0d
\end{equation}
Observation by multiple astronomers in the 1920's of this phenomenon, now known as the Hubble-Lema\^{i}tre law, was the first evidence of an expanding universe and kicked off development of cosmology as we know it today \citep{hubblelaw}.
The law gives an approximation of the \textbf{comoving distance} $\chi = d/a \approx zc/H_0$, a measure that ignores the bulk expansion of the universe and is used for the distances in the gravitational lens equation:
\begin{equation}\label{eqn:comoving}
\chi (t) = c\int^{t_0}_t \frac{dt'}{a(t')} = c\int^z_0\frac{dz'}{H(z')}.
\end{equation}
The closely-related \textbf{comoving horizon} $\eta$ gives the distance a particle could have traveled between the Big Bang and now:
\begin{equation}\label{eqn:conformal}
\eta = c\int^{t}_0\frac{dt'}{a(t')}.
\end{equation}
Regions of space can only causally interact with each other when they are less than $\eta$ apart, and the entry and exit of certain scales into the horizon dictates how perturbations evolve in the early universe.

The evolution of $H$ with time is governed by the \textbf{First Friedman Equation}, which is the $00$ (time-time) component of Einstein's equations for the FLRW metric:
\begin{equation}\label{eqn:intfried}
H^2 = \frac{H_0^2\rho}{\rho_{cr}} - \frac{c^2 k}{a^2},
\end{equation}
where the \textbf{critical density} $\rho_{cr} = {3H_0^2}/{8\pi G} \simeq 10^{-26} $ kg/m$^3$, is the density today corresponding to a flat, $k=0$ universe.
Since the universe is not a single substance, we can re-write this in terms of different dimensionless ``density parameters'' $\Omega_i = \rho_i/\rho_{cr}$ that correspond to its different components.
These pieces behave differently as the universe expands and dominate at different times.
When considering a universe with radiation, a cosmological constant $\Lambda$, curvature, and matter (including dark matter), we can rewrite \ref{eqn:intfried} as:
\begin{equation}
H^2 = H_0^2(\Omega_ra^{-4} + \Omega_ma^{-3} + \Omega_ka^{-2}+\Omega_{\Lambda}).
\end{equation}
For a more detailed derivation of these quantities see Appendix \ref{sec:flatflrw}.
As $H = H_0$ and $a = 1$ today, we can see that in the past, the radiation and matter components were much more important in the earlier universe.
While $\Lambda$ dominates today, the universe was matter-dominated before about redshift 0.3, and radiation-dominated before about redshift 1100 \citep{particlephysicsreview}.

\subsection{\texorpdfstring{$\Lambda$}{Lambda-}CDM: The Concordance Cosmology} 
Pronounced ``Lambda-CDM,'' we call $\Lambda$CDM the ``Concordance'' cosmology because it fits nearly all of the observations we have of the universe.
Built on top of a flat $(k=0)$ FLRW cosmology with baryonic matter, neutrino, and radiation components, it is named for two of its three big unknowns:
 a Cosmological Constant $(\Lambda)$ (a.k.a. dark energy) and a nonrelativistic, nonbaryonic component called Cold Dark Matter (CDM).
While quantum field theory predicts a constant, negative energy component from the vacuum that one would hope corresponds to $\Lambda$, the observed acceleration of the universe disagrees with this prediction by over 100 orders of magnitude \citep{dodelson}, so the nature of dark energy remains a mystery.
Furthermore, $\Lambda$CDM sets the dark energy equation of state to a constant $w=-1$ but this is not set in stone -- basic extensions such as $w$CDM and $w_0w_a$CDM use for a different or time-varying value, respectively.
The most recent (at press time) results of the Dark Energy Spectroscopic Instrument (DESI) has indicated a preference for a dark energy $w$ that is not only time-varying but was also $< -1$ in the past \citep{DESI24}%make sure its actually desi 24
, raising many very exciting questions that are far from the scope of this dissertation.

The third mystery component of $\Lambda$CDM is inflation, proposed as a solution to the seemingly unlikely flatness of the universe, as well as homogeneity in across regions of space which ought to be causally disconnected \citep{guth}.
The original idea of inflation was a phase transition in the early universe where space supercooled 28 orders of magnitude, undergoing de Sitter (exponential) expansion in the process, and blowing up inhomogeneities to a scale larger than today's observations can reach.
Today, inflation is thought to be the result of an unobserved ``inflaton'' particle decoupling from its associated scalar field and dumping immense amounts of negative pressure into spacetime before decaying into Standard Model particles we know today, ``reheating'' the universe and causing the Big Bang as we commonly think of it \citep{baumann}.
However, this reheating imparts tensor and scalar perturbations in the curvature of space, the letter of which are ultimately responsible for the growth of structure in the universe.

Now, we return to the unknown quantity in $\Lambda$CDM we skipped earlier. 
The idea of a ``dark'' component of the universe has been proposed to explain various phenomena since at least the 1800s \citep{hooperDM}, but this usually corresponded to faint stars or ``nebulous matter''.
The idea of dark matter as a source of ``missing mass'' that dominated luminous matter dates to the 1930s with \citet{zwickycoma} and \citet{smithcoma}, whose early applications of Newtonian dynamics to the Coma cluster produced anomalously large masses for individual galaxies.
The measurements remained a topic of debate in the study of galaxy clusters for decades \citep{swart}, but meanwhile galactic astronomers were uncovering their own `missing mass' in the rotation curves of nearby galaxies.
Thanks to neutral hydrogen observations by radio astronomers \citep[e.g.][]{21cmDM} and advancing optical techniques \citep[e.g.][]{rubinford}, by the late 1970s it was generally accepted that the mass distribution in galaxies extended much farther than what was visible from starlight.
However, it was cosmology that motivated the idea that rather than unobserved dust or small compact objects, the bulk of the mass was in some new particle. 
Hierarchical structure formation consisting solely of baryons would lead to greater-than-observed amounts of anisotropy in the Cosmic Microwave Background, but driving it with a weakly- or non-interacting dark matter could bring those fluctuations down to the level that was eventually observed by COBE \citep{trimbleDM}.
While active research continues into alternative theories (e.g., modified gravity) to explain the various observational dark matter phenomena, it is this ``Cold,'' structure-driving form that is most accepted today.

\subsection{Structure Formation}

%The reheating of the universe after inflation left spacetime with a spectrum of intrinsic curvature fluctuations $\mathcal{R}$ at all spatial scales $k$, with a power spectrum commonly parameterized by Amplitude $A_s$ and ``pivot scale'' $k_p$:
%\begin{equation}
%P_\mathcal{R}(k) = \frac{2\pi^2A_s}{k^3}\left(\frac{k}{k_p}\right)^{n_s-1};
%\end{equation} 
%where $n_s$ is the power spectrum of primordial fluctuations and is close to 1, which means perturbations are nearly scale-invariant.

The reheating of the universe after inflation left spacetime with a spectrum of intrinsic curvature fluctuations at all spatial scales $k$.
These initial perturbations have been measured to have spectral index of close to 1, meaning the perturbations are nearly scale-invariant.
While the generation of the perturbations is not strongly dependent on $k$, their evolution depends on when they enter the horizon $\eta$ -- modes that are larger than the horizon have a constant curvature perturbation until they enter it.
Before matter-radiation equality, modes within the horizon oscillate and decay from interactions with the photon-baryon plasma, while afterward, the potentials remain constant along with modes that entered after equality (though all modes' potentials decay after $\Lambda$-matter equality).
Dark matter overdensities associated with these perturbations, however, generally grow monotonically with time as soon as their modes enter the horizon, meaning that the largest scales (lowest $k$) are suppressed by entering late \citep{dodelson}.

\begin{figure}
	\centering %left bot right top
	\subfigure{\includegraphics[trim={8pt, 10pt, 10pt, 5pt}, clip, height=2.4in]{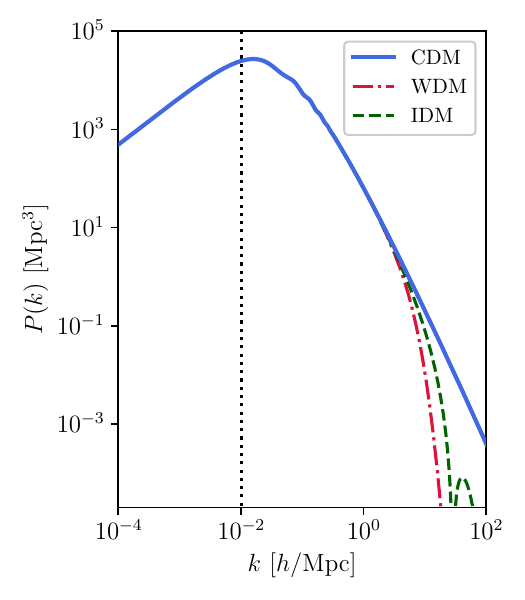}}
	\subfigure{\includegraphics[trim={8pt, 10pt, 10pt, 5pt}, clip, height=2.4in]{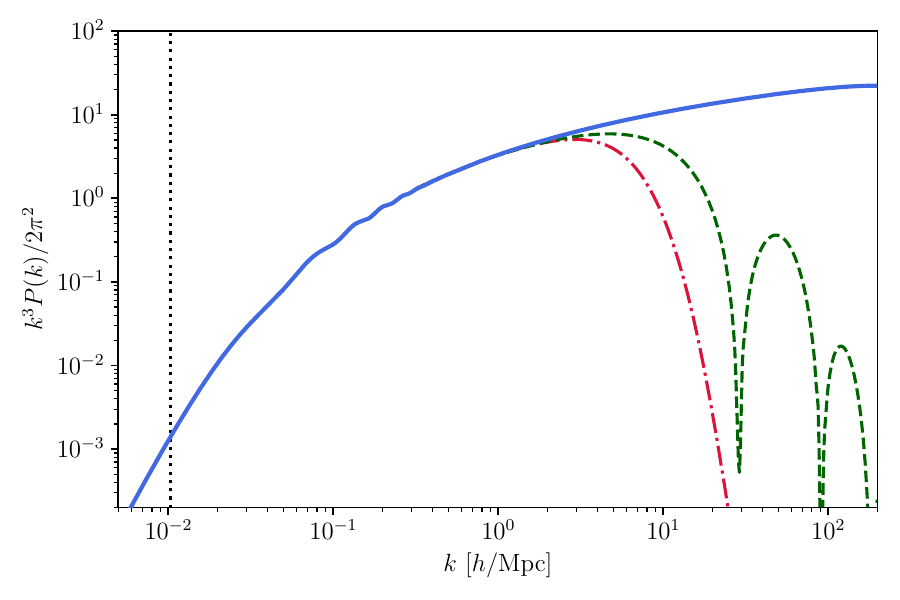}}
	\caption[Linearized Matter Power Spectrum]{Left: The linear matter power spectrum at $z\!=\!0$ for three different dark matter models. Right: The dimensionless linear matter power spectrum for the same. The dotted vertical line in both panels represents $k_eq=0.0104$ Mpc$^{-1}$. Blue solid line: Cold Dark Matter, following \citep{planck18} parameters for Flat $\Lambda$CDM. Red dash-dotted line: Warm Dark Matter, represented by a thermal relic with particle mass 3 keV, corresponding to a $M_{hm}=3.26\times10^8M_\odot$. Green dashed line: Interacting Dark Matter, represented by a 40 GeV particle with baryonic scattering cross-section $10^{-25}$cm$^2$. Calculated using \texttt{CLASS} \citep{class2, class4, classidm}.}
	\label{fig:powerspectrum}
\end{figure}

The left panel of Figure \ref{fig:powerspectrum} shows the power spectrum of those matter overdensities today, for several different dark matter models.
The spectrum turns over near the dotted line, which represents $k_{eq}=0.0104$ Mpc$^{-1}$ \citep{planck18}, the scale that entered the horizon at matter-radiation equality.
To the right of this are modes that decayed before then -- including most cluster and galaxy-size scales.
The figure also exhibits ``wiggles'' caused by Baryon Acoustic Oscillations, corresponding to the scale of standing waves in the photon-baryon plasma before recombination.
These wiggles are a baryonic effect and wouldn't show up if only the dark matter contribution was plotted.

\begin{figure}
	\centering %left bot right top
	\includegraphics[trim={8pt, 10pt, 10pt, 5pt}, clip, width=\textwidth]{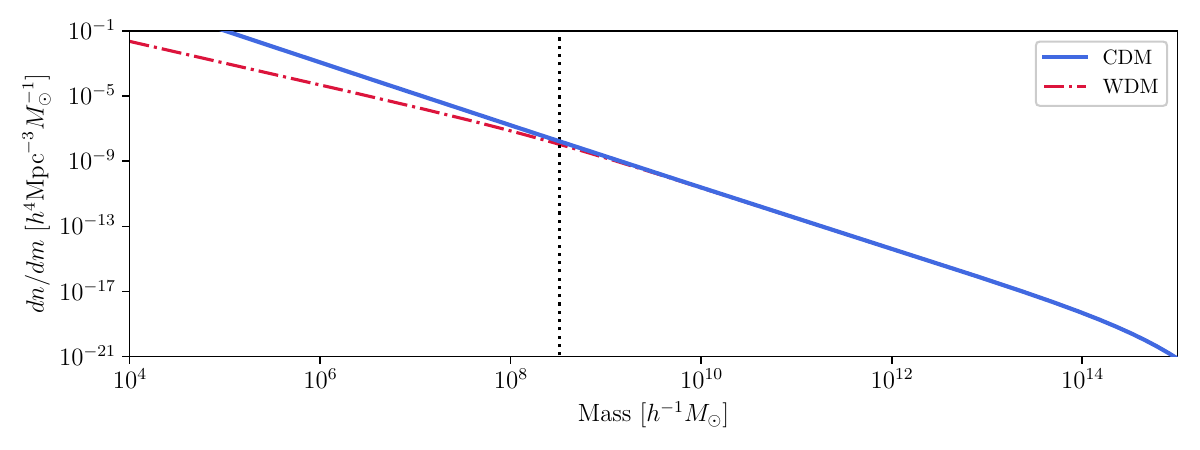}
	\caption[Halo Mass Function]{Halo mass function at $z\!=\!0$ for CDM and the same WDM model from Figure \ref{fig:powerspectrum}. The dotted vertical line is at $M_{hm} = 3.26\times 10^8 M_\odot$, the half-mode mass of the WDM model. Mass functions were calculated with Python package \texttt{hmf} \citep{hmf}.}
	\label{fig:halomass}
\end{figure}

As matter overdensities grow with time, they eventually can collapse and virialize into ``halos'' of dark matter.
However, while Figure \ref{fig:powerspectrum} shows the most power near $k_{eq}$, halos of corresponding mass are quite rare and correspond to only the largest galaxy clusters.
However, in real space, an actual overdensity consists of a superposition of these modes, and so smaller perturbations are more likely to reach the threshold for collapse, especially at early times \citep{dodelson}.
The resulting exponential suppression of large halos produces the halo mass function shown in Figure \ref{fig:halomass}, where the smallest halos are the most common and form first.
The interactions between these small halos ultimately lead to nonlinear perturbations to the power spectrum, and cause a ``bottom-up'' version of structure formation -- large halos containing galaxies are surrounded by smaller ``subhalos'' that contain satellites.

\subsection{Cold Dark Matter}
The above outline of structure formation is largely the same in all competitive models of dark matter, as the astrophysical probes have verified the linear matter power spectrum down to galactic and subgalactic scales.
Currently, nearly all observations have been consistent with ``Cold Dark Matter'' (CDM), in which the cosmological fluid making up the majority of $\Omega_m$ is nonrelativistic, collisionless and pressureless.
This leads to linear perturbations that follow a smooth power law on small scales, and a halo mass function that also follows a power law with decreasing mass. 
However, this concordance model is purely phenomenological, and doesn't point to a specific particle or particles.
In fact it is impossible for a particle theory to exactly match CDM at every scale -- things must break down at some point \citep{bechtol22}.
Nevertheless, particle theories can be constructed that do agree down to an arbitrarily small scale.
One such family of theories is the Weakly Interacting Massive Particle (WIMP), some Beyond-Standard-Model (BSM) particle or particles that decays via the weak force into more familiar particles and high-energy photons.
WIMP theories, though consistent with observations, have been thwarted by lack of detection of that electromagnetic signal, as well as a non-detection on Earth in particle accelerators \citep{particlephysicsreview}, leading theorists to propose myriad other DM models.

\subsection{Alternative Dark Matter Models}
Alternative DM models predict deviations from the CDM predictions at larger scales than something like WIMPs.
There are innumerable theories and families of theories for non-CDM, and we will only cover a few of the more popular groups.

\subsubsection{Warm Dark Matter}
Warm Dark Matter (WDM) presents one of the simpler modifications to CDM, which must be non-relativistic when it decouples from the early universe.
Instead, WDM particles are still relatively energetic when decoupling, which suppresses the growth of small overdensities as particles can ``free stream'' out of them until they slow down enough (via expansion) to be captured.
This leads to a steeper cutoff of the matter power spectrum as well as a turnover in the halo mass function as smaller halos can only form later, effects visible in Figures \ref{fig:powerspectrum} and \ref{fig:halomass}.
Like CDM, WDM is a phenomenological model that could correspond to a number of candidate BSM particles\footnote{Even ``CDM particles'' like WIMPs end up tapering off at very low halo masses.}.
WDM candidate particles are often some ``thermal relic'' particle, such as a new species of ``sterile'' neutrino, which could also emit (still unobserved) EM radiation upon decaying \citep{particlephysicsreview}.
WDM models are often described by particle mass or the ``half-mode mass'', the halo mass corresponding to the scale $k_{hm}$, where $\sqrt{P_{WDM}(k_{hm})/P_{CDM}(k_{hm})} = 0.5$ \citep{Lovell_2020}. 

\subsubsection{Fuzzy/Axion Dark Matter}
``Fuzzy'' models are inspired by the QCD axion, a hypothetical BSM particle to proposed to explain the strong CP problem \citep{particlephysicsreview}.
Such a particle (or an ``axion-like particle'') is ``ultra-light'' with mass in the area of $10^{-22}$ eV, giving it a de broglie wavelength the size of galaxies.
In fuzzy models, DM halos are strongly suppressed at low masses, and exhibit ``granule'' effects in halo mass density profiles due to wave interference \citep{powell23}.

\subsubsection{(Self-) Interacting Dark Matter}
%https://arxiv.org/abs/2512.01998v1
Interacting Dark Matter (IDM) scenarios are those in which the DM particle has some form of interaction with itself, normal matter, or with other unknown particle species in the ``dark sector''.
Figure \ref{fig:powerspectrum} exhibits a feature common in these sorts of models where rather than a smooth cutoff, the power spectrum exhibits enveloped oscillating behavior due to ``Dark Acoustic Oscillations'' -- analogues to BAOs but in the dark sector \citep{DAO}.
IDM models can provide different mass concentrations from CDM and WDM, especially given their ability to thermalize \citep{bechtol22}.
This has recently been invoked to explain the discovery of a $10^6M_\odot$ dark halo by \citet{vegettiSIDM}.

\section{Strong Gravitational Lensing} \label{sec:lensing}

Of all the effects predicted by General Relativity, the bending of light by a massive object may be the most beautiful to look at.
It was certainly one of the first observed, with the detection of a $\sim 1.75$ arcsecond deflection of stars around the sun during the 1919 total solar eclipse by \citeauthor*{eddington} providing an early success for the theory.
The Newtonian\footnote{A quote from Newton's \textit{Opticks} is often cited as the first consideration of gravitational light bending, but this line actually referred to more ordinary effects of refraction and diffraction \textendash{} an unpublished early 1800s notebook by Cavendish seems to contain the first Newtonian calculation of the effect \citep{trimblelens, gabaud}.} calculation, which takes the massless limit of a particle's trajectory, predicts only half the actual deflection.
After the 1919 discovery, several predictions were made regarding multiple imaging in star-star lensing cases, most famously by \citet{einsteinlens}, who concluded the magnification effect we now know of as microlensing was too small to be observed.
Of special note, however, is the anticipation by \citet{zwickylens} of galaxy-galaxy lenses as ``cosmic telescopes'' and their usefulness in mass modeling of galaxies -- predictions motivated in part by the author's measurement of a dark matter-dominated Coma cluster.
Lensing attracted more attention after World War II, gaining a theoretical framework closely resembling its present form by the time the first strongly lensed object, B0957$+$561, was serendipitously discovered during follow-up of quasar candidates \citep{walsh79}.
The first strong lenses were bright radio sources, but the launch of the Hubble Space Telescope (HST) and highly successful lens searches in the Sloan Digital Sky Survey (SDSS) set off an exponential growth of optical lenses, with thousands known today \citep{SLED}.

This section will outline basic lensing terminology and observable quantities, broadly following \citet{blandfordnarayan}.

\subsection{Lensing Observables} \label{sec:observables}

\begin{figure} 
	\centering
	\includegraphics[trim={0pt, 0pt, 0pt, 0pt}, clip, width=\textwidth]{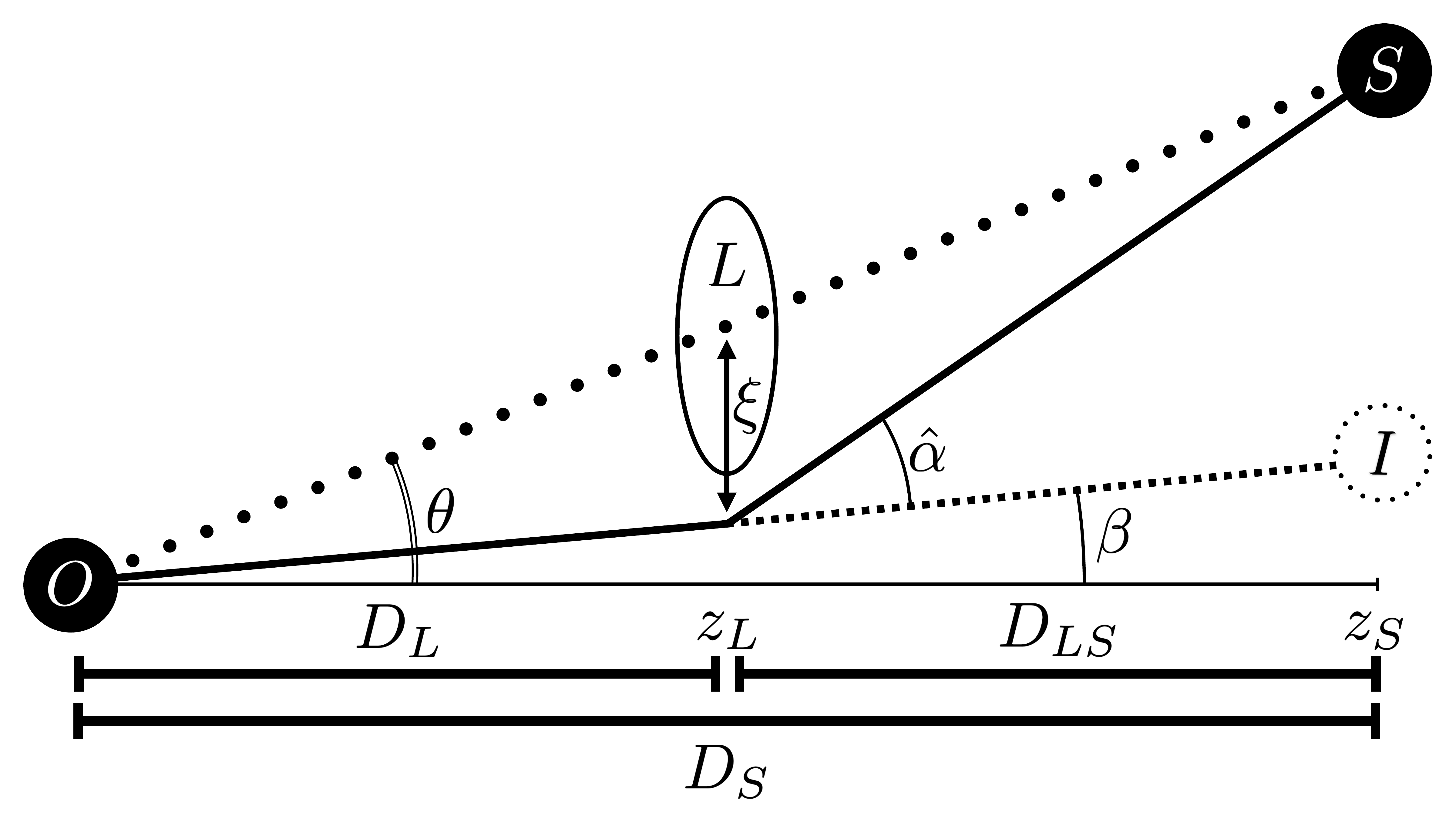}
	\caption[The Thin Lens Approximation]{Diagram of a thin lens. Light travels from the source ($S$) and is deflected at the lens ($L$) redshift, appearing to the observer ($O$) at the image location ($I$). The x-axis in the diagram is arbitrary, and the angles are two-dimensional vectors in practice.}
	\label{fig:thinlens}
\end{figure}

The basic schematic of a (thin) gravitational lens is shown in \ref{fig:thinlens}.
We define two related 2-D potentials on the sky that encode all the lensing information: the \textbf{lensing potential} $\psi$ and the \textbf{Fermat potential} $\tau$:
\begin{equation}
\tau(\vec{\beta}, \vec{\theta}) = \frac{1}{2}|\vec{\theta} - \vec{\beta}|^2 - \psi(\vec{\theta}); \quad \psi(\vec{\theta}) = \frac{1}{\pi}\int \kappa (\vec{\theta}') \ln (\vec{\theta} - \vec{\theta}') \text{d}^2\vec{\theta}'.
\end{equation}
The $\kappa$ in the lensing potential is the \textbf{convergence}, a dimensionless measure of projected mass density along the line of sight.
The lensing potential encodes the deflection by light rays at different source positions, while the Fermat potential is a scaled version of the time delay along the perturbed light path.
As light takes the path that extremizes arrival time (the relativistic version of Fermat's principle), images of a source at unlensed sky position $\vec{\theta}$ form at $\vec{\beta}$ where $\nabla \tau(\vec{\beta}, \vec{\theta}) = 0$.
Derivations of these potentials are found in Appendix \ref{sec:thinlens}.

We also cannot directly measure the source's position, only the positions of the lensed images on the sky, but can recover the source position using the \textbf{lens equation}, which relies on the thin lens approximation illustrated in Figure \ref{fig:thinlens}:
 \begin{equation}\label{eqn:lenseqn}
	\vec{\beta} = \vec{\theta} - \frac{D_{LS}}{D_S} \hat{\alpha},
\end{equation}
where $\hat{\alpha}$ is the deflection angle and depends on the mass distribution of the lens (derived in detail in \ref{sec:appdxlens}).

Gravitationally lensed sources have three main classes of observables: arrival time, sky position, and magnification.
These can only be defined in relative terms -- rather than measure an arrival time, we measure time delays, and instead of measuring magnifications, we must measure their ratios (via ratios of flux).
The observables also depend on the derivatives of the Fermat potential $\tau$, with the time delay probed by the potential itself.
As previously mentioned, images form when $\nabla \tau = 0$, and their magnifications depend on the second derivatives of $\tau$. 
For a thin lens, the Hessian of the Fermat potential forms an \textbf{amplification matrix}\footnote{The term ``magnification matrix'' refers to $\mathcal{A}$ in some authors and to its inverse in others, so I have chosen a more neutral (and acronymic) name from \citet{kovner87}.}$\mathcal{A}$:
\begin{equation} \label{eqn:ampmatrix}
\mathcal{A}_{ij} = \frac{\partial^2\tau(\vec{\theta},\vec{\beta})}{\partial\theta_i\partial\theta_j} \Rightarrow \mathcal{A} =  \begin{pmatrix}
1 - \kappa - \gamma_1 & -\gamma_2 \\ 
-\gamma_2 & 1 - \kappa + \gamma_1\\
\end{pmatrix}.
\end{equation}
As shown on the right side of the above equation, the amplification matrix can be expressed using three components: the convergence $\kappa$ (as seen in the lensing potential) and the two shears $\gamma_1$ and $\gamma_2$.
In a real lensing system, the convergence is not measurable in absolute terms, an phenomenon known as the ``mass sheet theorem'' \citep[MST,][]{mst}.
These parameters are especially useful in weak lensing, but we will examine them more in Chapter \ref{sec:chapter2}.
Due to the thin-lens approximation, $\mathcal{A}$ is symmetric, meaning that images can be sheared, reflected, and scaled, but not rotated.
Additionally, its two eigenvalues correspond to distortions in the tangential and radial directions:
\begin{equation}
\lambda_r = 1 - \kappa - \sqrt{\gamma_1^2 + \gamma_2^2}; \quad \lambda_t = 1 - \kappa - \sqrt{\gamma_1^2 - \gamma_2^2}.
\end{equation}
Finally, the \textbf{magnification} $\mu$ of an image encodes both the scaling factor and whether a given image changes parity:
\begin{equation}
\mu = \frac{1}{|\mathcal{A}|} = \frac{1}{(1-\kappa)^2 - \gamma_1^2 - \gamma_2^2} = \frac{1}{\lambda_r\lambda_t}.
\end{equation}
For a given lens, the surface(s) over which magnification is infinite in the source plane is called the ``caustic'', which can be projected forward to the image plane to form the ``critical curve''. 
Multiple images are formed when a source ``crosses'' the caustic -- a source just within a caustic will have two more images (with opposite parities to each other) than a source just outside it.
This implies a lens system can only ever have an odd number of images, a rule that is difficult to verify as one of these images is typically both de-magnified and obscured by the lens galaxy \citep{1984ComAp..10...75B}.

\section{Radio Astronomy} \label{sec:radio}
Radio waves were the first electromagnetic radiation to be discovered after Maxwell's Equations, and the first to be produced in a lab setting before being observed in nature.
Initial attempts in the late 1800s to detect radio activity from the Sun failed due to sensitivity or lack of solar activity, and the field of radio astronomy didn't really kick off until after Jansky's serendipitous discovery of the galactic center in 1932 and Reber's early surveys into the 1940s \citep{radiohist}.
In the meantime, radio telecommunications had become ubiquitous and played an increasing role in society as transmission and reception technologies improved.
Newly unemployed wartime engineers became the first generation of radio astronomers in the late 1940s, with teams worldwide (and especially in the commonwealth) pioneering new techniques for greater sensitivity and resolution.
In the more than 70 years since, radio astronomy methods have only become more sophisticated, and this trend is expected to continue with the next generation of facilities.
In this section we will review the basics of interferometry, which forms the backbone of radio astronomy at any appreciable resolution. 

%it is interesting though that nobody seems to have tried between Nordman in 1900 and Jansky. Perhaps the Blackbody radiation/Heaviside layer thing is correct, even if it is speculation on F. Ghigo's part

\subsection{Basics of Radio Interferometry}

The main difference between astronomical observations at short wavelengths (optical, ultraviolet, x-ray, etc.) and at longer (radio, submillimeter) wavelengths is the increased effect of diffraction.
Suppose we can model a telescope as a simple circular aperture with diameter $D$, which observes (far-field) electromagnetic radiation at some wavelength $\lambda$.
The point-spread function (PSF) for such a telescope has the form of an \textbf{Airy disk}, with the following normalized intensity profile:
\begin{equation}\label{eqn:airy}
I(\theta) = \left(\frac{2J_1(x)}{x}\right)^2; \quad x = \frac{D\pi}{\lambda} \sin(\theta),
\end{equation}
with $J_1$ the Bessel function of the first kind with parameter 1.
This is in fact the two-dimensional Fourier Transform of the aperture's circular shape (in optical astronomy, the ``pupil''), a relation which holds true for more complex shapes of telescope.
The Airy disk appears on the sky as alternating bright and dark rings, and we consider two sources to be ``resolved'' if they are separated by at least the angular distance to the first dark ring, which gives us the \textbf{Rayleigh criterion}, or angular resolution equation:
\begin{equation}\label{eqn:rayleigh}
\theta = \sin^{-1}(1.22\frac{\lambda}{D}) \simeq 1.22\frac{\lambda}{D},
\end{equation}
assuming $\theta$, in radians, is small (the constant in the equation is related to $J_1$'s first positive zero).
For optical light, even a one-meter telescope is more than enough for the diffraction effect to be subdominant to atmospheric seeing and ray-optical effects, but radio observations are a different story, and even the largest dishes struggle to achieve better than arcminute resolution at centimeter wavelengths.

\begin{figure}
    \includegraphics[trim={0pt, 20pt, 0pt, 0pt}, clip, width=\textwidth,
    alt = {test alt text for signal path figure}
    ]{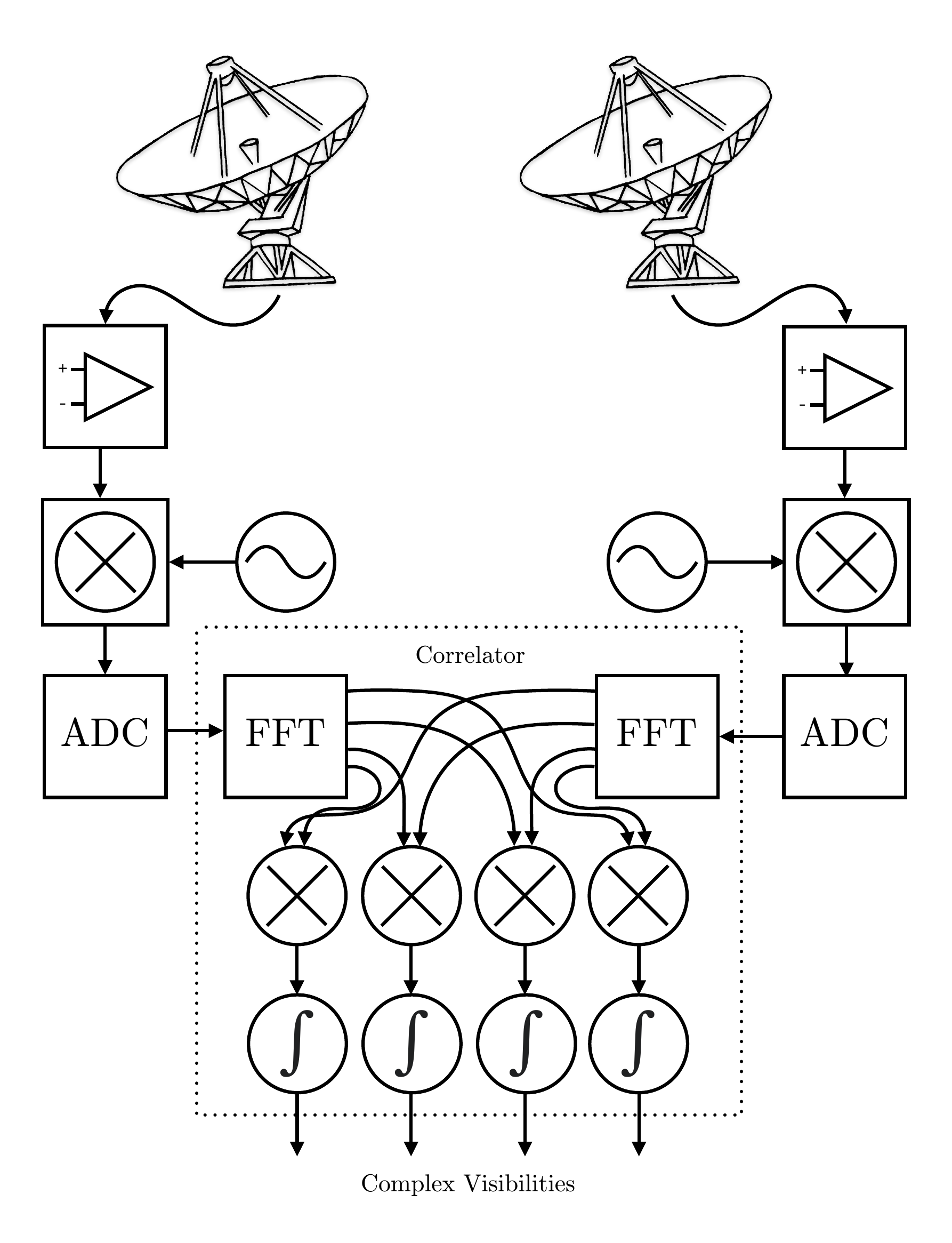}
    \caption[Interferometer Signal Path]{Two-element interferometer signal path. Each array element converts incoming radiation into a voltage, which is amplified and downconverted before being digitized. The digital signals enter the (FX) correlator, where it is split into frequency bins via Fourier Transform before being cross-correlated and integrated to produce visibilities. In a lag or XF correlator the cross-correlation and FT steps would be reversed. In this diagram separate polarization channels are omitted, and in an interferometer with more than two elements, the correlation is done on each pair of signals.}
        \label{fig:signalpath}
\end{figure}

We can get around this limitation by considering a telescope that is kilometers wide, but with the majority of its collecting area missing.
This is in essence what the individual dishes of an interferometer represent, though they each have their own optical paths and amplifier systems.
However, the individual antenna signals cannot simply be stacked \textendash{} the phase and amplitude information from each must be combined in a specific way to get the full picture\footnote{this is also the difference between multi-lens array telescopes such as Dragonfly \citep{dragonfly} and optical interferometers like the VLTI \citep{VLTI}}. 
This is accomplished in a radio array by using a delay line to account for the light travel time between antennas, followed by cross-correlation between every pair of interferometer elements.
Figure \ref{fig:signalpath} gives a simplified schematic of the path taken by the radio signals from antenna to correlator.
For each frequency $\nu$ considered, the correlator outputs two signals $R_C$ and $R_S$, which correspond to the even and odd-parity (aka sine and cosine) structure of the signal on the sky $I_\nu(\vec{s})$, and from these we construct the \textbf{visibility}: 

\begin{equation}\label{eqn:vis}
V_\nu = R_C - iR_S = \iint I_\nu(\vec{s})e^{-2\pi i \vec{b}\cdot \vec{s}}d\Omega,
\end{equation}
where $\vec{b}$ is known as the ``baseline vector''.
In addition to the physical coordinate distances between the two, the baseline depends on the direction the array is pointed, and is expressed in the following $u,v,w$ coordinate system:
\begin{equation}
\vec{b} = 
\lambda
\begin{pmatrix}
u \\
v \\
w \\ 
\end{pmatrix}
 = 
\frac{1}{\lambda}
 \begin{pmatrix}
 \sin(\HA) & \cos(\HA)  & 0 \\
 -\sin(\delta)\cos(\HA) & \sin(\delta)\sin(\HA)  & \cos(\delta) \\
 \cos(\delta)\cos(\HA) & -\cos(\delta)\sin(\HA) & \sin(\delta) \\
 \end{pmatrix}
 \begin{pmatrix}
L_x \\
L_y \\
L_z \\ 
\end{pmatrix},
\end{equation}
with $\delta$ and $\HA$ the declination and local Hour Angle of the antenna pointing, respectively, the rightmost vector representing the $xyz$ distance on Earth between the antennas in meters, and $\lambda$ the wavelength.
While ``interferometry'' can refer to any analysis of the visibilities (or any other technique utilizing wave interference in a different context), the use of visibilities to construct an image is typically referred to as ``Aperture Synthesis'' or \textbf{Synthesis Imaging}.
The $u$ and $v$ components of $\vec{b}$ are often considered together, and graphed on a Cartesian grid known as the ``$uv$ plane''.
The dependence on hour angle means that the baseline vector changes with time, and each baseline traces an ellipse on the plane over 24 hours.
This means that an interferometer with relatively few antennas can take on a large number of $uv$ values given a long observation, a technique known as ``Earth Rotation Synthesis''.
An array's $uv$ coverage can be further improved with ``Multi-Frequency Synthesis'', which uses the $\lambda$-dependence of $\vec{b}$ to scale a single physical baseline to multiple $uv$ points.

The third component of $\vec{b}$, $w$, is the path length difference between antennas, which causes a time delay that must be corrected for at the correlator \citep{thompsonbook}.
$w$ is only zero for a source at zenith but can be ignored via coordinate change if the interferometer is ``coplanar'' and form a 2D plane in $uvw$-space.
This is possible for East-West arrays and 2D arrays (e.g., VLA) when taking short observations.
For longer observations, $w$ changes as Earth rotates, causing a phase error that increases from the phase center\footnote{The phase center is the zero point of an imaging field and can be changed, it is not necessarily the pointing of the telescope.} and creating coma-like aberrations towards the edge of the field \citep{facet}.
When the $w$ term doesn't change appreciably, the visibility \ref{eqn:vis} can be re-written as a 2D Fourier Transform:
\begin{equation}\label{eqn:fouriervis}
V_\nu = \iint I_\nu(l,m)e^{-2\pi i ul+vm}dldm,
\end{equation}
where $l$ and $m$ are ``direction cosines,'' projections of $\vec{s}$ onto the $x$ and $y$ axes.
The Fourier Transform property of the above equation makes imaging of radio data exceedingly simple, and it can be used even when the $w$ term isn't negligible.
These wide-field effects can be countered via algorithms such as mosaicking, faceting \citep{facet} and $w$-projection \citep{wproj}, which can increase image fidelity at the cost of increased computational complexity.

\subsubsection{Very Long Baseline Interferometry}
The theory of interferometry is agnostic to the length of a baseline, and one could conceivably achieve arbitrarily high resolution given arbitrarily separated array elements.
This is the principle behind Very Long Baseline Interferometry (VLBI), which traditionally achieves these large (1000s of kilometers) separations by disconnecting individual antennas from the signal path and recording voltage data for later, offline correlation\footnote{however, the recent emergence of ``e-VLBI'' arrays that transfer data over optical fibers blurs the line between connected-element arrays and standard VLBI.}.
VLBI facilities regularly achieve resolutions upwards of 1 milliarcsecond (mas), and are the undisputed champions of precision astrometry \textendash{} the National Radio Astronomical Observatory (NRAO)'s Very Long Baseline Array (VLBA) is responsible for measuring continental drift and changes in the Earth's orientation through regular quasar observations.
More recently, the Event Horizon Telescope, a millimeter-wave consortium of dishes spanning the entire earth, has successfully observed the shadows of supermassive black holes M87* and Sagittarius A$\!$* \citep{EHTM87, EHTSAGA}.
The technique is also not limited to Earth's surface -- in the past, space VLBI missions have achieved a resolution of up to 12 microarcseconds \citep{radioastron}, and proposed missions like the Black Hole Explorer \citep{bhex} hope to break that record.

Unfortunately, with this increased resolution comes increased challenges in calibration and processing, and data volumes for VLBI observations are typically much larger than those of traditional interferometric arrays.
Additionally, the distance between VLBI stations drastically reduces the field of view possible due to time and frequency-averaging effects, and a single pointing can capture less than a square arcminute -- typically fine for observing single sources but not conducive to a wide-area survey.
This effect can be ameliorated by correlating at multiple phase centers, which allows for theoretically maximal coverage of the primary beam, at the cost of massively increased data volume \citep{2024MNRAS.529.2428D}.
Despite these limitations, VLBI remains the primary tool for high-resolution science, and is expected to see even more widespread adoption as computational techniques improve and new, advanced facilities are brought online.

\subsection{Radio Calibration}

Synthesis imaging observations are typically characterized in the literature by the \textbf{Radio Interferometer Measurement Equation} (RIME), which relates the visibilities measured by an interferometer to the properties of the observed source, as modified by various effects along the signal path.
In general the RIME takes the form:
\begin{equation}\label{eqn:RIME_gen}
V_{ij} = \mathbf{J}_i B \mathbf{J}_j^\dagger,
\end{equation}
where the dagger symbol represents the Hermitian conjugate \citep{RIME}. 
In the equation, the $V_{ij}$ are the interferometer visibilities, and $B$ is a $2\times 2$ matrix built from the Stokes polarization parameters of the observed source.
The $\mathbf{J}_i$ are known as ``Jones Matrices'' and encode propagation effects along the path from the source to the $i$-th antenna.
As there are always multiple effects, the $\mathbf{J}_i$ are decomposed into a set of matrices which are applied in the order of propagation.
These matrices usually represent simple 2D linear transformations \textendash{} for example time-variable complex gain from an antenna is a non-uniform scaling, and Faraday rotation is (perhaps unsurprisingly) a rotation.
In most cases, it is desirable (and sometimes mandatory) to correct for these effects via \textbf{calibration}, by observing a source with known properties.

For radio continuum observations at $3-10$GHz, which this thesis is concerned with, the main propagation effects are: absolute flux calibration, delay errors (caused by errors in antenna position), phase errors (caused by atmospheric turbulence), and instrumental response, which causes an uneven and time-varying frequency response \citep{thompsonbook}.
To correct for the absolute flux scale, a bright source with known brightness is observed, usually one of a handful of slowly or non-varying quasars which the NRAO moniters.
This source is typically also bright enough to serve as a bandpass calibrator and delay calibrator, and is typically observed first during an observation (though this is not a requirement).
Complex gain (amplitude and phase) errors are the trickiest and most crucial effects to correct for, and require observation of a phase calibrator, which must be a pointlike source within a few degrees of the target.
As phase errors are related to atmospheric turbulence, this source must be observed every few minutes (higher frequencies require more frequent switches) to account for the time-based component.
The NRAO keeps a list of gain calibrators, but on occasion these calibrators are found to be slightly extended and not true point sources.
When this happens, the gain calibrator must be self-calibrated (see \ref{sec:imaging}) to account for the extra structure before the calibration can proceed to the target source.
Observations incorporating polarization are exposed to additional effects, as are VLBI, low-frequency ($\lesssim 2$ GHz), and high-frequency ($\gtrsim 20$ GHz) observations, and these may require additional calibrator sources and procedures \citep{thompsonbook}.
Finally, as Radio Frequency Interference (RFI) is near-ubiquitous in modern radio astronomy, temporal and spectral sections of data affected by it must be excised either automatically or by hand, a step that happens alongside the calibration process.

\subsection{Imaging and Self-Calibration} \label{sec:imaging}
An interferometer measures visibilities, or points on the Fourier Transform (FT) of the source's brightness distribution.
However, recovering the image from the visibilities is significantly more complicated than FTing back due to the typical sparseness of the uv-plane.
Imaging must first include a \textbf{gridding} step, essentially a 2D Kernel Density Estimator that makes the non-uniform sampling into a uniform one.
In this step the visibilities are also weighted, and choice of weighting schemes affect the sensitivity and resolution of the resulting image.
The FT of the gridded visibilities then produces a sky image, but this is rarely the final product of a radio observation.

Recall that the PSF of a telescope is the FT of its pupil \textendash{} the ``spider'' holding up the secondary mirror of a Newtonian reflector causes spikes in the star images it produces.
The $uv$ coverage of an interferometer is in essence its pupil, and its PSF is found the same way, but while an optical telescope pupil is mostly uncovered, radio $uv$ coverage is usually sparse, which leads to many more artifacts that can dominate an image simply made by FTing gridded visibilities.
This ``dirty'' image must be made usable through deconvolution, more commonly known as ``cleaning''.
The deconvolution process involves construction of a flux model for the source, which is convolved with the instrumental PSF and subtracted from the dirty image, with the subtracted flux replaced by a ``synthesized'' or ``clean'' beam (typically a Gaussian ellipse fit to the center of the PSF or ``dirty beam'').
Cleaning takes advantage of the undersampling of the $uv$ plane, and can be accomplished through several methods, many more sophisticated than the original \citet{hogbom} algorithm.

The construction of a flux model by the cleaning algorithm can serve an additional purpose, that of \textbf{self-calibration} \citep{selfcalcornwell}.
Instead of using only the complex gain calibrator to correct for amplitude and phase variations, the observer may use the model instead, theoretically allowing for better corrections.
However, as the self-calibration model relies on the clean process, errors in that process can propagate and potentially introduce structure that does not actually exist into the sky model.
When self-calibrating it is vital to be cautious and not over-clean for this reason!
Additionally, self-calibrating may introduce absolute positional errors (which are equivalent to a global phase shift), but it will leave relative distances alone, and this effect is not a big concern when not conducting high-precision astrometric observations.

The radio data presented in this thesis was calibrated using the Common Astronomy Software Applications \citep[CASA,][]{casa}, a suite of software packages for calibration, manipulation, and imaging of radio data developed by the NRAO.

\section{Active Galactic Nuclei} \label{sec:AGN} %shorter section

Active galactic nuclei have long been drivers of both cosmological studies and radio astronomical techniques \textendash{} spectroscopic follow-up of early radio catalogues revealed a population of pointlike objects which implied the sources were incredibly energetic yet very small and possessing some of the highest redshifts known at the time \citep{kellerman}.
In addition, the relatively fast optical variability of these ``Quasi-Stellar Radio Sources,'' or ``quasars,'' implied an object light-months across, far smaller than anything else observable at cosmological redshift.
Quasars, along with other extragalactic curiosities such as blazars and radio galaxies, were eventually understood to all be the same thing \textendash{} emissions from the region surrounding a galaxy's supermassive black hole \textendash{} only viewed at different angles through obscuring material \citep[e.g.][]{urry}.
This ``unification'' of AGN, however, is still an ongoing process in the field \citep{2025NewAR.10101733A}, and a simple geometric argument is insufficient to describe the myriad flavors of active galaxy that have been observed.

\subsection{AGN Spectra and Regions}

To first order, AGN get colder and redder as you go outward from the central SMBH.
Just outside the event horizon is a hot corona of low-mass charged particles facilitating x-ray emission via inverse Compton scattering.
Further out, gas in a thermalized accretion disk (extending down to the SMBH's minimal stable orbit) converts its gravitational energy into heat as it travels inward, emitting blackbody radiation from the UV into the near-infrared \citep{2017NatAs...1..679R}.
Near the outer part of the disk (<.01 parsec), before the dust sublimation temperature, fast and dense clouds of gas in the broad-line region create the signature wide emission lines familiar in Type 1 AGN \citep{2011A&A...525L...8C}.
Outside the disk is a multiphase dusty ``torus'' with polar and radial structure where matter can flow in and out of the core.
The torus emits blackbody radiation at a temperature of hundreds to tens of Kelvin as radius increases, and can also contain astrophysical masers and molecular clouds in the outer regions (10s of parsecs) \citep{honigagn}.
Farthest out, extending potentially up to kiloparsecs at all azimuthal angles from the disk axis, is the narrow-line region -- slow-moving clouds of ionized interstellar media with low enough density for ``forbidden'' emission lines \citep{2011ApJ...739...69M}.
This scheme should be taken as a very broad overview, as every AGN is different and level of activity can influence the presence and size of these regions.

\subsection{Lobes, Jets and AGN Radio Emission}

One of the many divisions drawn when characterizing different types of AGN is the presence or lack of radio emission, although many so-called ``radio quiet'' objects have well-measured radio flux.
Generally, the radio emitted by an AGN is in the form of synchrotron radiation, produced by charged particles rapidly rotating in a magnetic field.
Synchrotron radiation is polarized and smooth, without spectral lines, and a jet's overall spectral slope depends on its highest energy electrons, with ``steeper'' spectra caused by depletion of those electrons over time.
The morphology of this emission in AGN strongly depends on the environment, evolutionary stage, and viewing angle of a given object, but the classic ``radio-loud'' picture is somewhat consistent: a collimated jet extending outward in both directions from the central SMBH.
The jet is powered by matter infalling from the accretion disk, but the exact mechanism of jet launching is still unknown, and work is ongoing to connect the immediate environment of the black hole to the innermost portions of the jet \citep[e.g.,][]{2025A&A...696A.169K, ehtcena}.
Jet emission is often relativistically beamed, and the innermost regions are characterized by helical filaments and hotspots whose (observed superluminal) motion can be measured over time \citep{walker2018, 2023NatAs...7.1359F}
As the jet extends outward, its spectral index steepens, with the steepest and ``oldest'' emission far from the flat-spectrum core \citep{jets}.
Many jets end in large ``lobes'' that can extend hundreds of kpc from the core.

%\section{Astronomical Surveys} \label{sec:surveys}%shorter section
%The recent [explosion] in precision of cosmological results has largely ben due to the proliferation of wide and deep astronomical surveys.
%While the idea of a sky survey and associated catalog of objects has its roots in the star catalogues of the ancient world and the early deep sky catalogues of Messier and the Herschels, the first \textit{imaging} surveys took shape with the dawn of plate astrophotography in the late 1800s.
%At optical wavelengths, CCD technology paved the way for the age of surveys we live in today [??? come on dude]

\section{Radio Lenses as a Probe of Dark Matter} \label{sec:introutline}

\begin{figure}
	\centering
	\includegraphics[trim={67pt, 56pt, 81pt, 105pt}, clip, width=\textwidth]{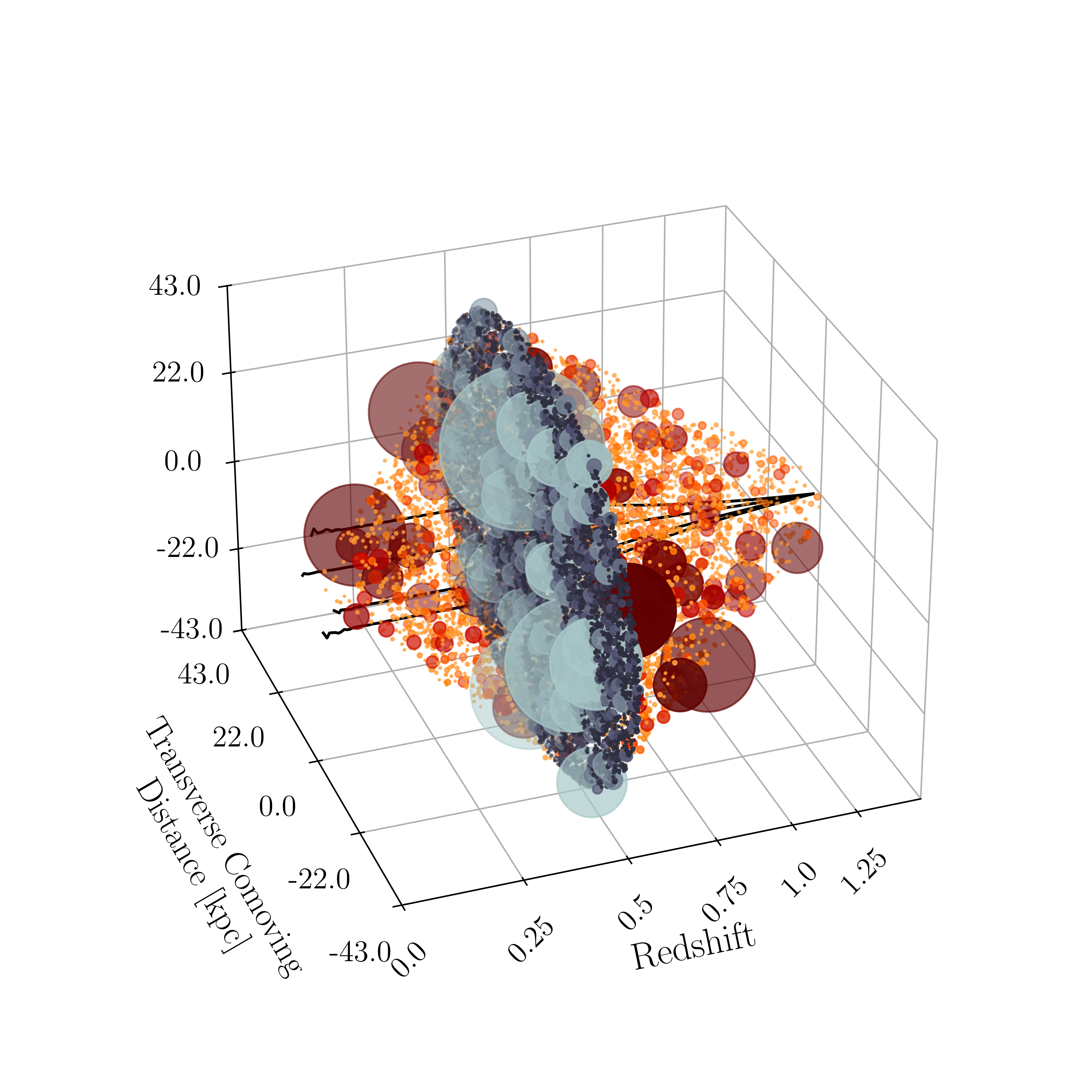}
	\caption[Lens with Subhalos and Field Halos]{A strong gravitational lens system illustrating the presence of subhalos (gray/blue) and field halos orange/red). Size and color of each circle corresponds with the logarithm of halo mass, with the most massive halos shown reaching about $10^{10} M_{\odot}$. The main deflector is not shown and would be at least the size of the figure at this scale. Black lines indicate the paths taken by a lensed source through the halos, but have been magnified by a factor of 2000 for the multiple images to be visible.}
	\label{fig:subhalos}
\end{figure}

Strong gravitational lenses have long been proposed as a probe of dark matter \citep{metcalfmadau}, and used to good effect in multiple cases.
As the lensing effect is sensitive to all mass along the light path, the perturbations of the gravitational potential from dark matter halos will cause perturbations in lensing observables.
An illustration of this scenario is shown in Figure \ref{fig:subhalos}, where the light path (magnified for legibility) is additionally deflected by low-mass halos on its way from source to observer.
To exploit this effect, several methods have been developed, each utilizing different source populations and wavelengths.

Currently, the most successful of these studies probe anomalies in the magnification by measuring the ratios of fluxes in lensed quasars.
This approach has several advantages, as magnification, being a second derivative of the Fermat potential, is more sensitive to anomalies than lower derivatives like position or time-delay anomalies \citep{vegetti23}.
However, only specific lensed systems are suitable for such an analysis.
Beyond the constraint that the source be an AGN (about 1/10th of lenses), it also must be at least quadruply lensed \citep{2003ApJ...598..138K}, and those four images must be in specific configurations to be most sensitive to perturbations.
Additionally, the flux ratio analysis cannot simply be carried out on optical measurements of each image, as the optical emission from an AGN is subject to variability (both intrinsic and due to microlensing in the immediate environment of the quasar core) -- these variations are delayed by different amounts for each image, making flux ratios vary with time as well \citep{2003astro.ph..4480S}.
To get around this, a flux ratio analysis must use a region of the AGN that is sufficiently large that microlensing effects are averaged out, and which doesn't intrinsically vary on month- or shorter timescales.
Common choices for these regions are the warm portions of the dust torus and the narrow-line region, which require mid-infrared, space-based (JWST) imaging and high-resolution spectroscopy, respectively \citep{vegetti23}.
These studies are especially effective when including additional constraints from other lensed components (e.g. arcs formed by the quasar host galaxy), and have established the most stringent constraints on a WDM half-mode mass to date \citep{gilman25}
However, concerns have been raised about possible degeneracies due to kinetic and angular structures within the lens galaxy itself \citep{2024MNRAS.531.3431C}, motivating independent checks of the flux ratio method.

The other family of methods, ``Gravitational Imaging,'' uses aggregate lensed image properties, especially small perturbations caused by low-mass halos, to constrain some property of a DM particle.
Several methods exist that leverage different techniques at different wavelengths, and these will be expanded on in the introduction to the next Chapter. 
Statistics-focused methods in this family, such as \citet{wagnercarena2}, utilize recent developments in Simulation-Based Inference (SBI) to directly fit quantities like the halo mass function from a population of observed lenses.
The most successful of these so far is the technique used in \citet{powell23, powell25}, utilizing situations where a bright and extended source is lensed into a giant arc.
Using a Bayesian forward modeling approach in visibility-space, a map of sensitivity to low-mass halos is constructed, and individual halos are not only identified but measured precisely.
However, the resolution required to conduct these measurements is currently only achievable by VLBI at radio wavelengths, so the lenses in question must be radio loud.
This sets very specific limits on how many sources, and therefore DM halos could be directly measured in this way.

The rest of this thesis will address this problem in two ways. 
First, in Chapter \ref{sec:chapter2} we attempt to find the lower limit of information necessary to make a dark matter constraint, for use in a hybrid astrometry/flux ratio method. 
Then, Chapter \ref{sec:chapter3} will present the results of a pilot survey to locate more radio lenses for potential follow-up with VLBI for gravitational imaging.
Following this, Chapter \ref{sec:chapter4} expands on the successful method in the pilot to exhibit the findings from a more expansive survey for even more radio lenses.
We summarize and examine next steps for the analysis, as well as general prospects for radio lenses, in \ref{sec:conclusion}.

\chapter{Astrometry-Based Studies of Lens Properties} %rework title
\label{sec:chapter2}

\paragraph{This chapter collects the results}
 of calculations conducted between 2024 and 2026 investigating the possibility of a simple statistical dark matter constraint based only on high-precision astrometry of sources with simple (pointlike) morphologies.

%ch2 notes
%anomalies are very detectable by *current* VLBI, assuming two astrometrically localizable features
%?even higher mass halos do produce anomalies? - makes looking at lower mass halos harder, kind of like a nuisance parameter
%a real physics statement
%?is there a minimum number of localizable features??
%limited by sensitivity, not angular resolution
%one would localize feaures with data -> some ML thing -> results
%?point cloud?
%ML tells you if there is a statistic that can distinguish, does the information exist that you can use in a more interpretable way
%limited by number of lens systems, that number grows with sensitivity

\section{Introduction}

Current efforts to constrain the properties of dark matter via gravitational lenses can be divided roughly into two approaches: ``direct gravitational imaging'' and ``indirect statistical constraints''.
The direct approach identifies specific dark halos within an individual lens system, measuring their properties and limiting corresponding particle properties.
This method is best exemplified by VLBI measurements of radio-loud lenses, such as \citep{powell25}, but similar methods have also been employed in clusters \citep{2023A&A...679A..31D}, and measurements of specifically anomalous flux ratios such as \citet{2004ApJ...604L...5M} also fall under this category (though the latter method is fairly out of fashion these days).
The direct method is highly effective for measuring the mass and even mass profile of an individual perturber, but is limited by a need for very high-resolution observations and a bright, extended source, limiting its usefulness to a handful of radio-loud lensed quasars.

On the other hand, the indirect method forgoes modeling of individual, anomalous lenses and uses a population of targets to constrain a statistical property of a dark matter model.
These methods rely heavily on forward modeling or simulation-based inference (SBI) methods like Approximate Bayesian Computation \citep[e.g.][]{pyhalo} or Neural Posterior Estimation \citep[e.g.][]{wagnercarena1, wagnercarena2}, and rely on a large number ($10^6-10^8$) of simulations to sample the parameter space of the DM population.
Recent flux ratio-based constraints such as \citep{keeley24} and \citep{gilman25}, which have established the most stringent limits on WDM particle mass to date, also fall into this category.
The indirect method is less starved for a source population, but does suffer from a high level of computing power to simulate millions of lenses with DM populations, making it infeasible for high-resolution data.

In this Chapter, we consider a family of methods that combines elements of both approaches by using high-resolution observations of many ``ordinary'' lenses to detect lens properties and indirectly probe DM ones.
We land on an astrometry-based approach which appears to be as effective as current flux ratio analyses and could be used as a complementary approach, provided the source morphology is suitable.

%state of the field - direct vs indirect (maybe better names for these?)
%and motivation

\section{Transfer Matrices} \label{sec:transfermatrices}
In \ref{sec:observables} we introduced the amplification matrix $\mathcal{A}$ (\ref{eqn:ampmatrix}), which encodes the image magnification as well as other lens information like convergence and shear.
As previously mentioned, the magnification is not directly accessible, but the ratio of magnifications of two distinct images can be probed by measuring a flux ratio.
However, we can also measure magnification ratios via astrometry, a technique first proposed by \citet{1984ApJ...287..538G} and \citet{blandfordnarayan}, and given a full mathematical treatment in a recent paper series \citep[][and references therin]{wagner19uni, wagner22}.
In the following we will follow the ``point matching'' technique of \citet{wagnertessore}, which is valid for sources relatively far from the caustic where the convergence and shear are roughly constant.

\begin{figure}
    \centering % left bot right top
    \includegraphics[trim={35pt, 37pt, 35pt, 36pt}, clip, width=0.9\textwidth]{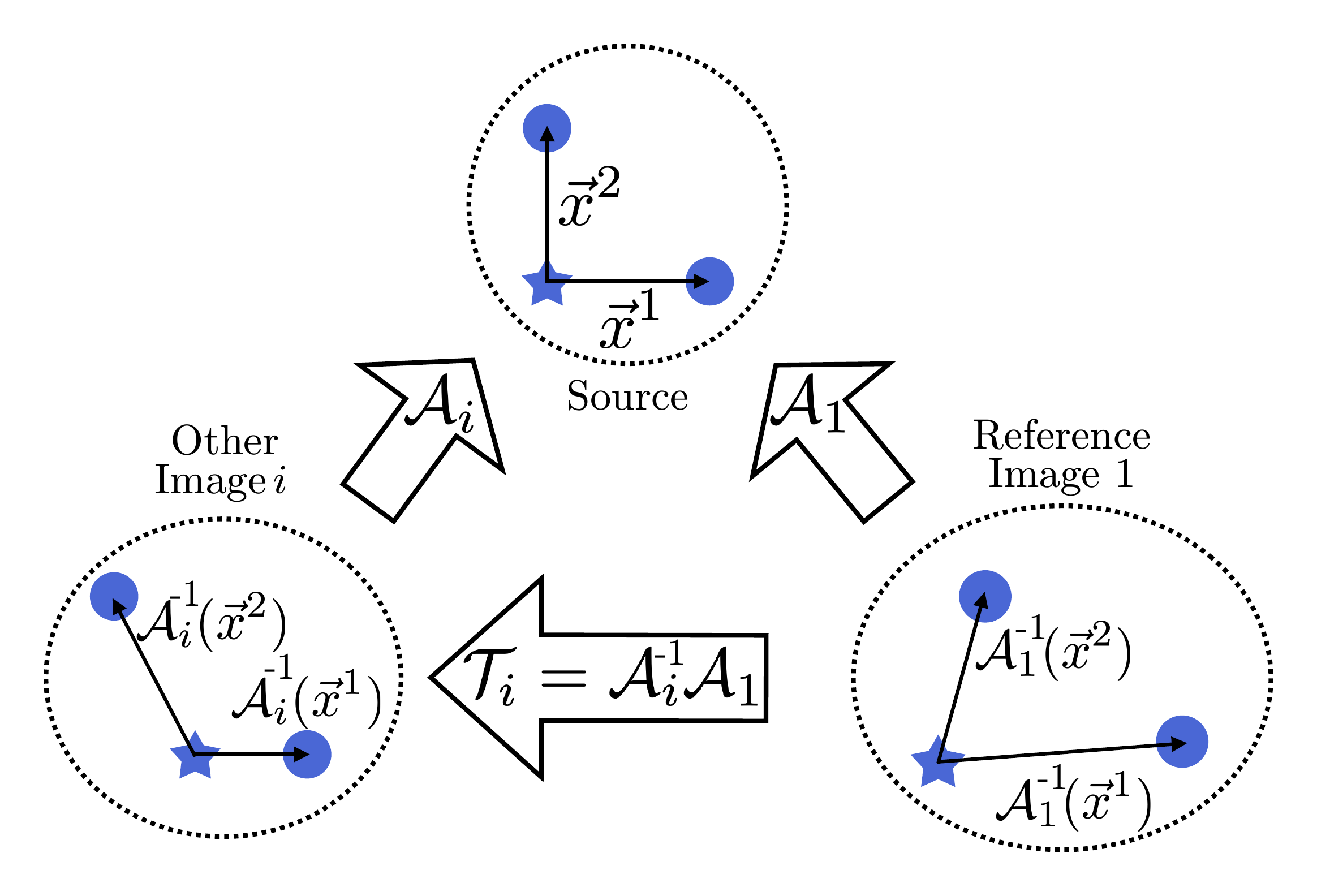}
    \caption[Illustration of the Point-Matching Scheme]{Illustration of the point-matching scheme of \citet{wagnertessore}. Relative vectors $\vec{x}^1$ and $\vec{x}^2$ are mapped from the source plane to two images, the reference image 1 and another image denoted $i$. The transfer matrix $\mathcal{T}_i$ maps the observed vectors in image 1 to image $i$.}
    \label{fig:transmatrix}
\end{figure}

We first reparameterize $\mathcal{A}$ using the \textbf{reduced shears} $g_1$ and $g_2$:
\begin{equation}
\mathcal{A} = (1 - \kappa)\begin{pmatrix}
1 - g_1 & -g_2 \\ 
-g_2 & 1 + g_1\\
\end{pmatrix}; \quad g_i = \frac{\gamma_i}{1-\kappa}.
\end{equation}
The typical, ``gamma'' shears are unobservable\footnote{if the $\gamma_i$ were absolutely measurable we could measure $\kappa$ and violate the MST.}, but the reduced shears are often estimated in the context of weak gravitational lensing.
Next, as we can't observe the source, we consider the ``Transformation matrix'' $\mathcal{T}_i = \mathcal{A}_i^{-1}\mathcal{A}_1$, which transforms features in a given image $i$ to corresponding features in some reference image $1$.
This correspondence is illustrated in Figure \ref{fig:transmatrix}, and it should be clear from the definition of $\mathcal{T}$ that its determinant gives the magnification ratio between the image in question and the reference; $|\mathcal{T}_i| = \mu_i/\mu_1$.

The elements of $\mathcal{T}$ can be found in \citet{wagnertessore} and in Appendix \ref{sec:transformation}, but it should be noted that each entry has the prefactor $f_i = \frac{1-\kappa_1}{1-\kappa_i}$, and that this is the only place the convergence appears.
Thus we can solve for the convergence ``ratios'' $f_i$, but not the absolute convergence, as expected from the MST.
Additionally, by solving equations formed by the $\mathcal{T}_i$, the reduced shears of all images are recoverable, when there are enough images in the lens system.
For $n$ observed images, there are $4(n-1)$ observable transformation matrix components, and $3n - 1$ shears and convergence ratios to solve for, and so two (or one) image lens systems are underdetermined.
Three-image systems are exactly solvable, provided the three images are not collinear \citep{wagnertessore} -- for instance, if the lens is axisymmetric.
A ``naked cusp'' configuration 3-image lens, for instance, would work, and a double lens with its demagnified third image may be solvable provided the third image is observed at high enough significance.
In this Chapter we will discuss four-image systems, for which the system of equations is overdetermined with 11 unknowns and 12 transformation matrix elements.

\section{Simulation Procedure}
The suite of simulations prepared for this analysis are based around a single lens and source configuration.
This base is then inserted into a double-ended cone of low-mass halos following one of 4 number density distributions: CDM, and WDM with half-mode mass cutoffs $M_{hm}$ of $10^7$, $10^8$, and $10^9 M_{\odot}$, and we generated 2000 instances for each of these scenarios to create a mock lens sample.
Unfortunately, in the real world, not every gravitational lens has the same configuration, nor does every lensed source have identical morphology, and so this population ought to have much more uniform properties than any comparable sample of actual lenses.
However, this also means that if an analysis fails on this highly simplified model, it is sure to fail in practice, allowing us to focus on methods that are more likely to work in reality.

\subsubsection{Base Model}
For the lens macromodel, we use a singular isothermal ellipsoid (SIE) at redshift $z=0.5$, with circularized Einstein Radius $\theta_E = 1$ arcsecond and eccentricity 0.77.
The source model, also the same for each dark matter configuration, consists of three points at $z=1.5$ arranged in a right triangular configuration.
One of these points is designated the ``core'', and the other two points (the ``hotposts'') are 1 milliarcsecond to the East and North.
This arrangement, while highly unphysical, lends itself to a more natural interpretation of the output image configurations, as the relative positions of the hotspots will simply give the elements of the amplification matrix.
In a more realistic scenario with three non-collinear points, one would simply work in the basis formed by the hotspots' positions relative to the core\footnote{Additionally, in a real lens, the source morphology is unknown}.
While the terminology used in this chapter suggests specific radio AGN regions and three discrete points, any three measurable points could be used to the same effect, such a as a core and two ends of a small jet, or even the center and major axes of an elliptical patch of emission -- as long as they are sufficiently high signal-to-noise.

The source is placed in an ``Einstein Cross''-like configuration, offset from directly behind the lens but relatively far from the caustics, an arrangement similar to known lenses B1413+117 \citep{1988Natur.334..325M} and B2237+0305 \citep{1985AJ.....90..691H}.
\begin{figure}
	\centering
	\includegraphics[trim={25pt, 18pt, 30pt, 17pt}, clip, width=\textwidth]{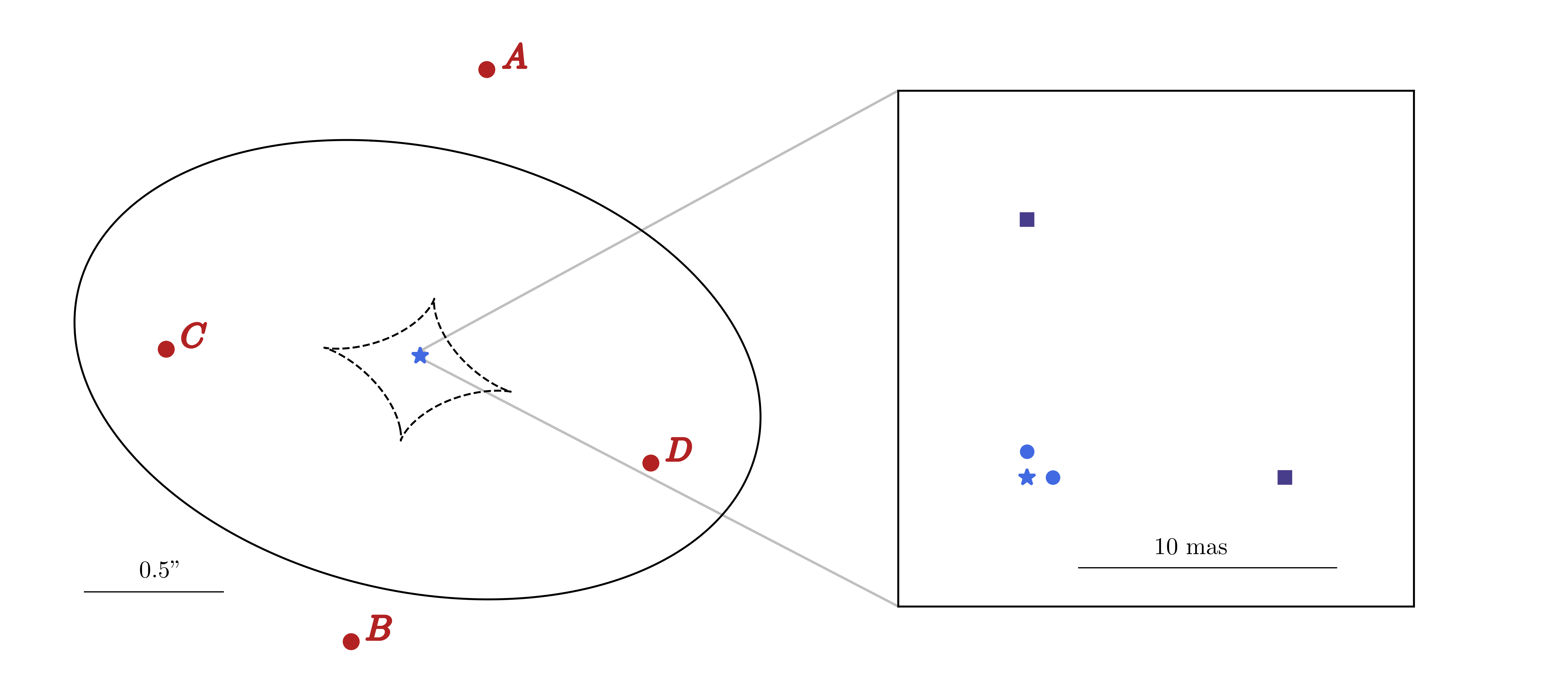}
	\caption[Simulation Image Configuration]{The basic image configuration for the simulations. The source position is indicated with a blue star, and images are shown in red. The tangential critical curve and caustic are shown by the solid and dotted lines, respectively. Inset: the scale of the source is shown. The `core' is represented by the star, and the `hotspots' by circles. For the 10 milliarcsecond source size, the hotspots are shown as purple squares.}
\end{figure}

\subsubsection{Low-Mass Halos}
To add extra halos to the base model, we use the \texttt{pyhalo} package \citep{pyhalo} to create a population of low-mass dark matter halos.
We use the default \texttt{pyhalo} parameters, with low-mass halos only being generated in a 10 arcsecond cone -- initial tests showed that increasing this radius didn't change lensing quantities much.
We used the \citet{2016MNRAS.460.1214L} mass concentration model for the CDM runs, and followed \citet{2016MNRAS.455..318B} for the WDM ones.
Additionally, only low-mass halos with masses between $10^6$ and $10^{10} M_{\odot}$ were generated, meaning that the CDM model is somewhat ``warmer'' than true CDM would be. 
\texttt{pyHalo} automatically incorporates sheets of negative mass when adding line-of-sight halos in order to keep the total mass considered constant, an operation that thanks to the MST doesn't change any lensing observables.

\subsubsection{Implementation Details}
The core and hotspots are ray-traced through the macromodel and halos using \texttt{lenstronomy} \citep{lenstronomy}, and their magnifications and relative arrival times are also reported. 
This method is relatively quick as we are only lensing two points and not a whole image, but the DM models with more halos still took an appreciable amount of time to both generate and raytrace through the whole population.
To speed things along, we turned to UW Madison's Center for High-Throughput Computing (CHTC)\footnote{\url{https://chtc.cs.wisc.edu}} and ran these more intense simulations on multiple machines in parallel, resulting in all simulations finishing in under five hours of wall time.
Results were pickled and downloaded from the CHTC for analysis.

\section{Deflection and Magnification Anomalies}

\begin{figure}
	\centering
	\includegraphics[trim={10pt, 18pt, 10pt, 10pt}, clip, width=\textwidth]{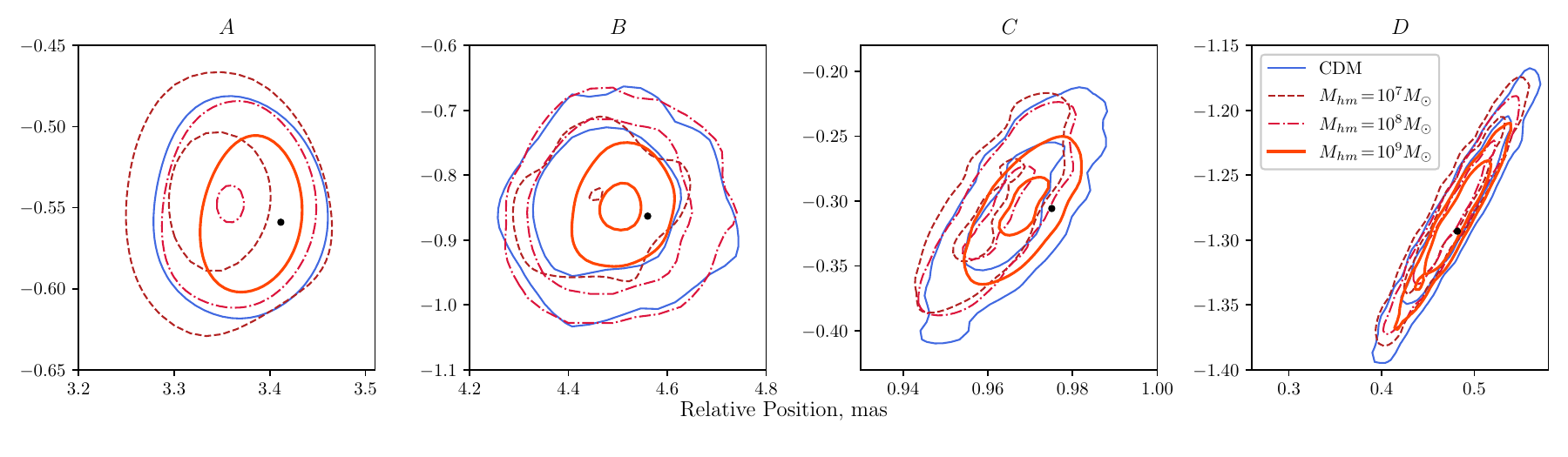}
	\caption[Position Anomaly Distribution]{Distribution of the separation vector between core and hotspot for each DM model. The position of the lensed core image is locked to (0,0) with the relative positions of the hotspot in each iteration shown. That distribution was convolved with a KDE and 3- and 6-$\sigma$ contours are plotted. The black dot represents the relative position for the macromodel only.}
	\label{fig:astrometric}
\end{figure}

Figure \ref{fig:astrometric} illustrates the distribution of a lensed hotspot position relative to the core for the milliarcsecond simulations.
The figure was constructed using a Kernel Density Estimator (KDE), with contours for each dark matter model corresponding to 3- and 6-$\sigma$ levels of the scored distribution samples.
This gives a good picture of the typical extent of the DM realizations, but ignores outliers. 
From the figure, it is clear that the separation sharply peaks near the unperturbed value, but the general shape and extent of this distribution isn't significantly different between CDM and the $10^8$ and $10^7$ WDM models.
The $10^9$ model (thick orange line) does appear more concentrated than the other three, however, which does suggest a trend of greater spread in clumpier mass distributions.

\begin{figure}
	\centering % left bot right top
	\subfigure{\includegraphics[trim={18pt 22pt 12pt 11pt}, clip, width=\textwidth]{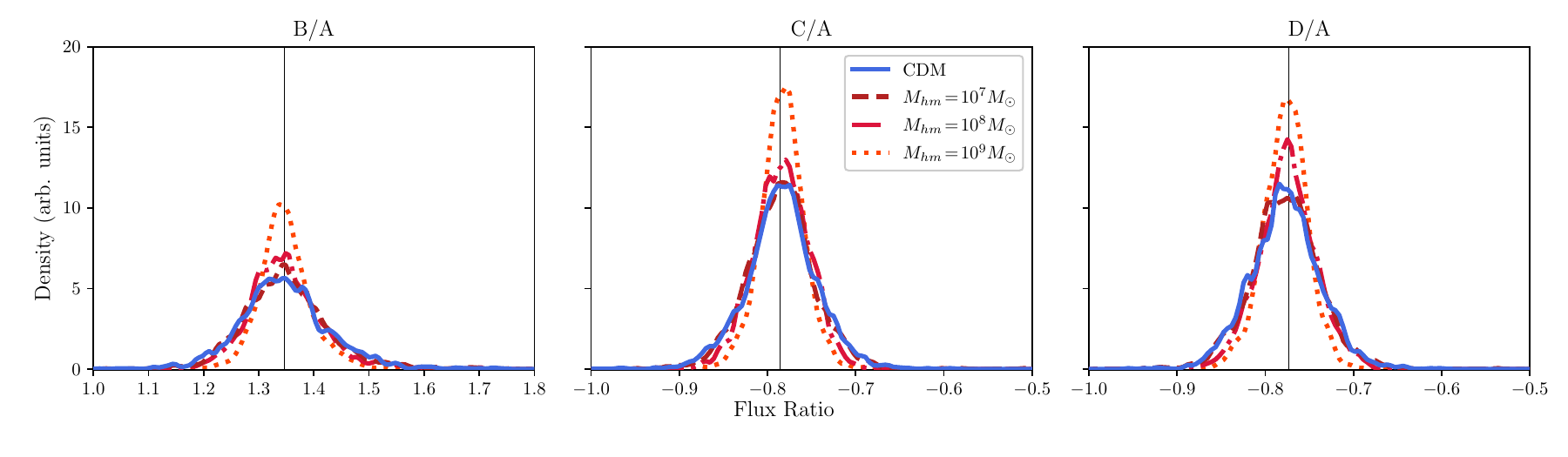}}
	\subfigure{\includegraphics[trim={18pt 22pt 12pt 11pt}, clip, width=\textwidth]{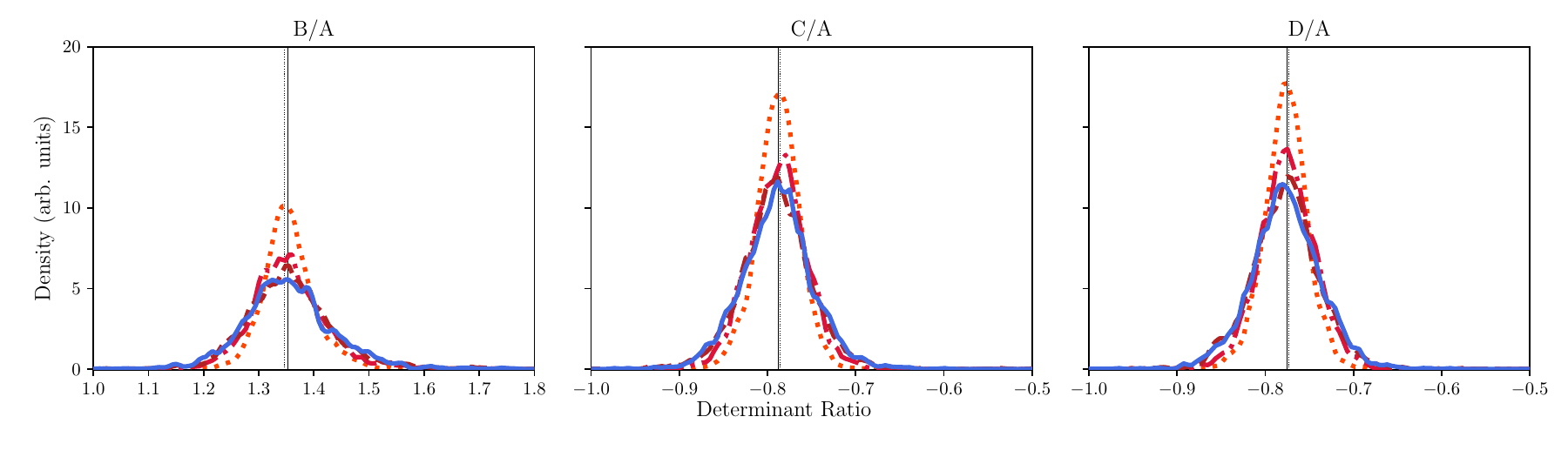}}
	\subfigure{\includegraphics[trim={18pt 20pt 11pt 11pt}, clip, width=\textwidth]{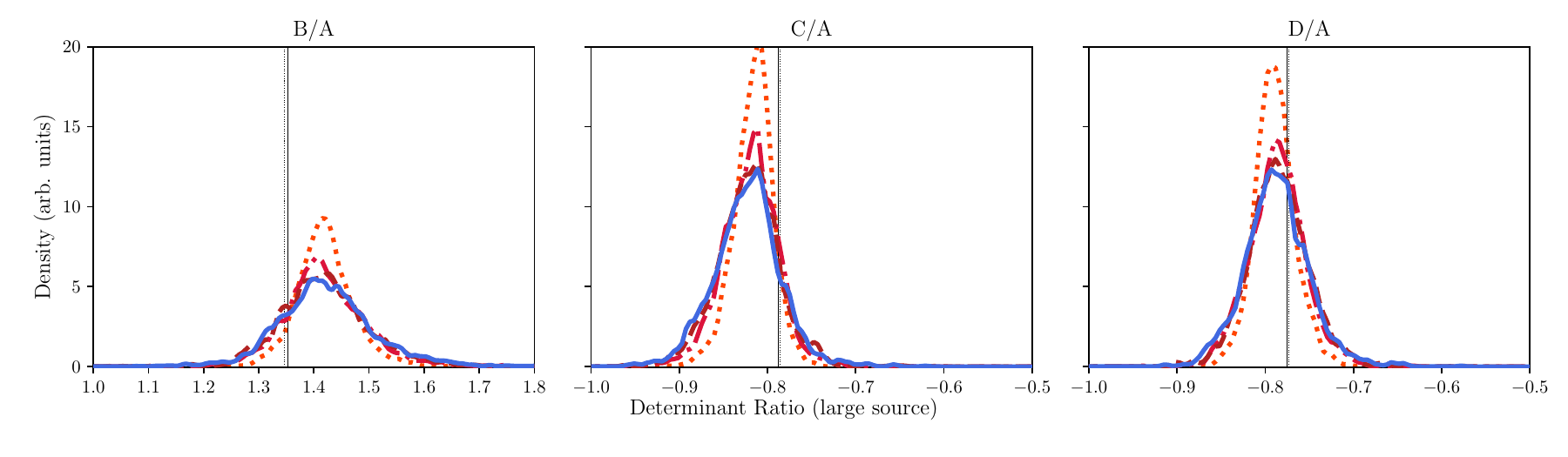}}
	\caption[Flux and Determinant Ratios]{Top: Flux ratio distribution of the simulations for each DM model. While the $M_{hm} = 10^7 M\odot$ model is fairly indistinguishable from CDM, the other two WDM models exhibit a wider distribution of the observable quantity. The flux ratio for the macromodel is shown by the solid vertical line. Middle: Determinant ratio distributions for the different models, which follow the flux ratios nearly exactly. The solid vertical line again shows the macromodel flux ratio, and the additional dotted vertical line shows the determinant ratio for the macromodel. Bottom: Determinant ratio distributions for the larger souce model (hotspot distance 10 mas instead of 1). While the overall shapes are consistent with the other two probes, the means are shifted in the direction of a larger ratio.}
	\label{fig:ratios}
\end{figure}

We can do a bit better by incorporating the other hotspot as well.
Recalling that the magnification ratio is the determinant of transfer matrix $\mathcal{T}$, we construct these matrices for each image and DM model, and show the distribution of ratios in Figure \ref{fig:ratios}.
The figure was constructed with another KDE to better compare the distributions, with the bandwidth kept constant between DM models per image.
The top row of the figure shows the magnification ratio of the core image (as would be measured by a flux ratio analysis), while the middle row gives the ratio by the determinant-of-transfer-matrix.
As the top two rows are nearly identical, we also include determinant ratios for a second run of simulations, where the hotspots have been moved ten times farther from the core, an easier to obtain observing prospect.
The distributions from these simulations are shown on the bottom row of the figure.

%% 	REDO FIGURES TO UNIFY AXES, CLEAN UP GENERALLY

{\footnotesize
\onehalfspacing
\begin{ThreePartTable}
\begin{longtable}[c]{lccc}
\caption[$p$-value of KS Test for Each WDM Model]{$p$-value of the KS test for each WDM model versus CDM \label{tab:KS_flux}} \\
\hline \hline
{WDM Model $M_{hm}$} & {$B/A$} & {$C/A$} & {$D/A$}\\
\hline 
\endhead
Flux Ratio & & & \\
\hline \\[-2ex]
$10^7 M_{\odot}$ & 0.349 & 0.413 & 0.692 \\
$10^8 M_{\odot}$ & 0.00168 & 0.00377 & 0.000627 \\
$10^9 M_{\odot}$ & $2.18 \times 10^{-19}$ & $1.76 \times 10^{-10}$ & $3.80 \times 10^{-11}$  \\
\hline
Determinant Ratio & & & \\
\hline \\[-2ex]
$10^7 M_{\odot}$& 0.329 & 0.413 & 0.560 \\
$10^8 M_{\odot}$ & 0.00117 & 0.0149 & 0.00168 \\
$10^9 M_{\odot}$ & $2.18 \times 10^{-19}$ & $6.40 \times 10^{-14}$ & $9.44 \times 10^{-13}$  \\
\hline
Determinant Ratio (Large Source) & & & \\
\hline \\[-2ex]
$10^7 M_{\odot}$& 0.212 & 0.370 & 0.139 \\
$10^8 M_{\odot}$ & 0.0135 & 0.000251 & 0.0693 \\
$10^9 M_{\odot}$ & $1.82 \times 10^{-13}$ & $1.14 \times 10^{-29}$ & $7.43\times 10^{-13}$  \\
\hline
\end{longtable}
\end{ThreePartTable}
}

\subsection{Statistical Tests}
While the distributions in Figure \ref{fig:ratios} do appear somewhat different, especially in the $D/A$ case, it is necessary to statistically quantify this.
We use the 2-sample Kolmogorov-Smirnov (KS) test, which quantifies the difference between two empirical cumulative distribution functions \citep{KSTEST}
The $p$-value of the KS test can be interpreted as the probability the two samples are drawn from the same distribution.
The KS test is most sensitive near the center of the distribution, and so for each pair of models we also ran the Epps-Singleton \citep{ESTEST} and Anderson-Darling \citep{ADTEST} tests, which are more sensitive to the tails of the distribution.
In every case, these other two tests found drastically lower $p$-values than the KS test, and so we report the latter to err on the side of caution.
Table \ref{tab:KS_flux} shows the test results for the true image magnification ratios, the determinant-derived magnification ratios, and the larger source simulation determinant ratios.
Due to the symmetric image configuration, no one ratio stands out as especially constraining, but the table as a whole suggests that the $M_{hm}=10^7$ model is largely indistinguishable from CDM, even using flux ratios.
However, all three situations, the $M_{hm}=10^8$ does appear different at a statistically significant level, although the standard flux ratio gives generally higher significance.
The $M_{hm}=10^9$ distribution is, as expected, very different from CDM in all cases, and we discourage reading into the specific values of the large negative exponents in those rows of the table.

%%INCLUDE TABLES

\section{Shear, Convergence, and Rotation}

While the magnification ratios are one avenue to probing dark matter models, the point matching technique with three points can provide more information which could be potentially useful. 
Following the \citet{wagnertessore} method outlined in \ref{sec:transfermatrices}, we measured the convergence ratios and reduced shears for the four images in each DM realization.
For four images, the point-matching method is overconstrained - to measure all four shears we only need to rely on three of the images, and can leave one out.
Therefore, we ran the point matching four times, using each image as the reference. 
For the unperturbed lens, where the Fermat potential is roughly constant between core and hotspot, the choice of reference image should be unimportant, but a clumpy lens breaks this assumption of smoothness.

\begin{figure}
	\centering %LBRT
	\includegraphics[trim={18pt 7pt 8pt 10pt}, clip, width=\textwidth]{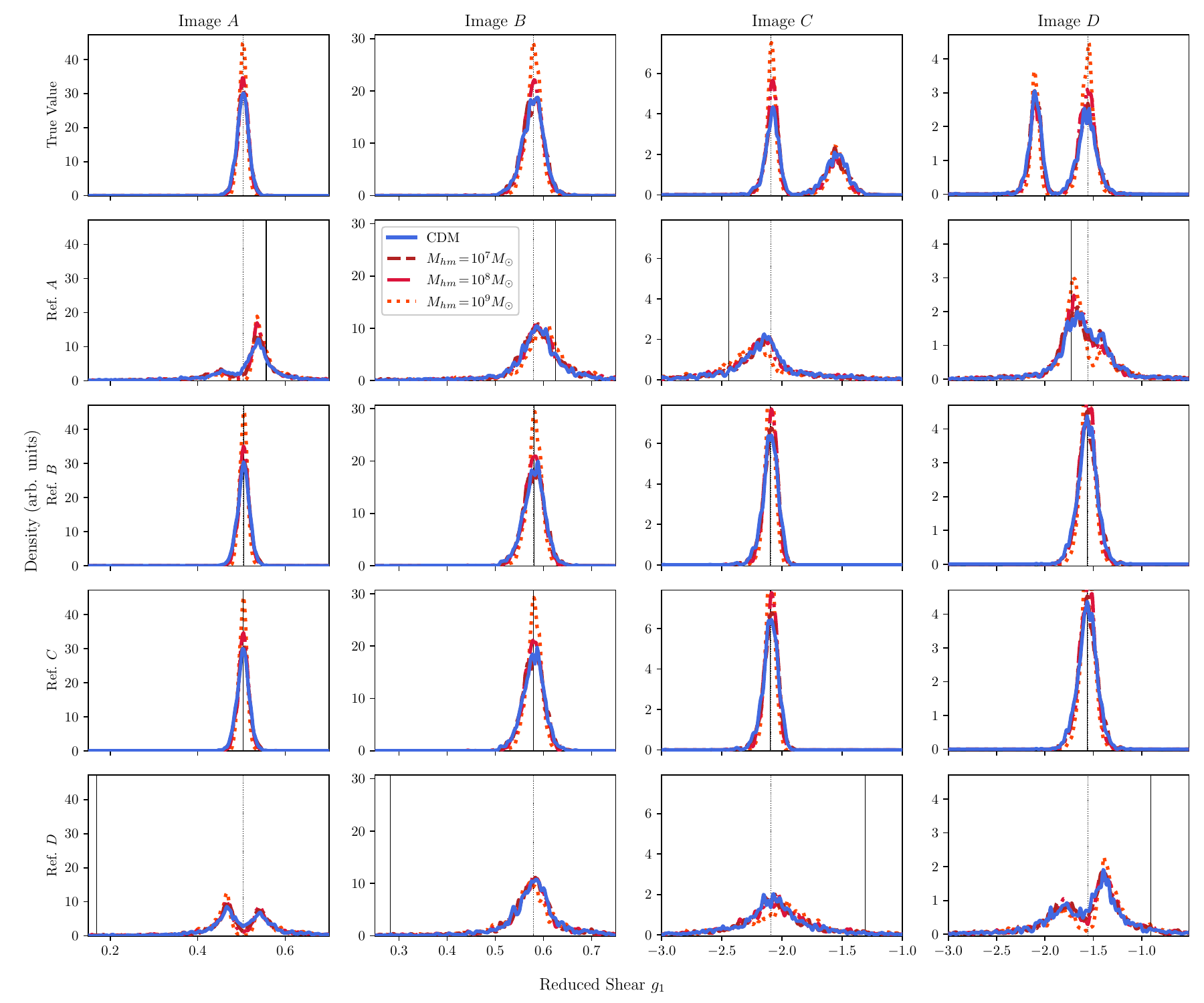}
	\caption[$g_1$ Reduced Shears]{Distribution of reduced shear $g_1$ for each image and DM model. Columns correspond to each lensed image, while rows correspond to the image used as a reference, with the top row the true value of the shear at the core location. The dotted vertical line shows the shear of the macromodel, while the solid line gives the shear predicted by the using the point-matching method on the macromodel. }
	\label{fig:g1}
\end{figure}

\begin{figure}
	\centering
	\includegraphics[trim={18pt 7pt 8pt 10pt}, clip, width=\textwidth]{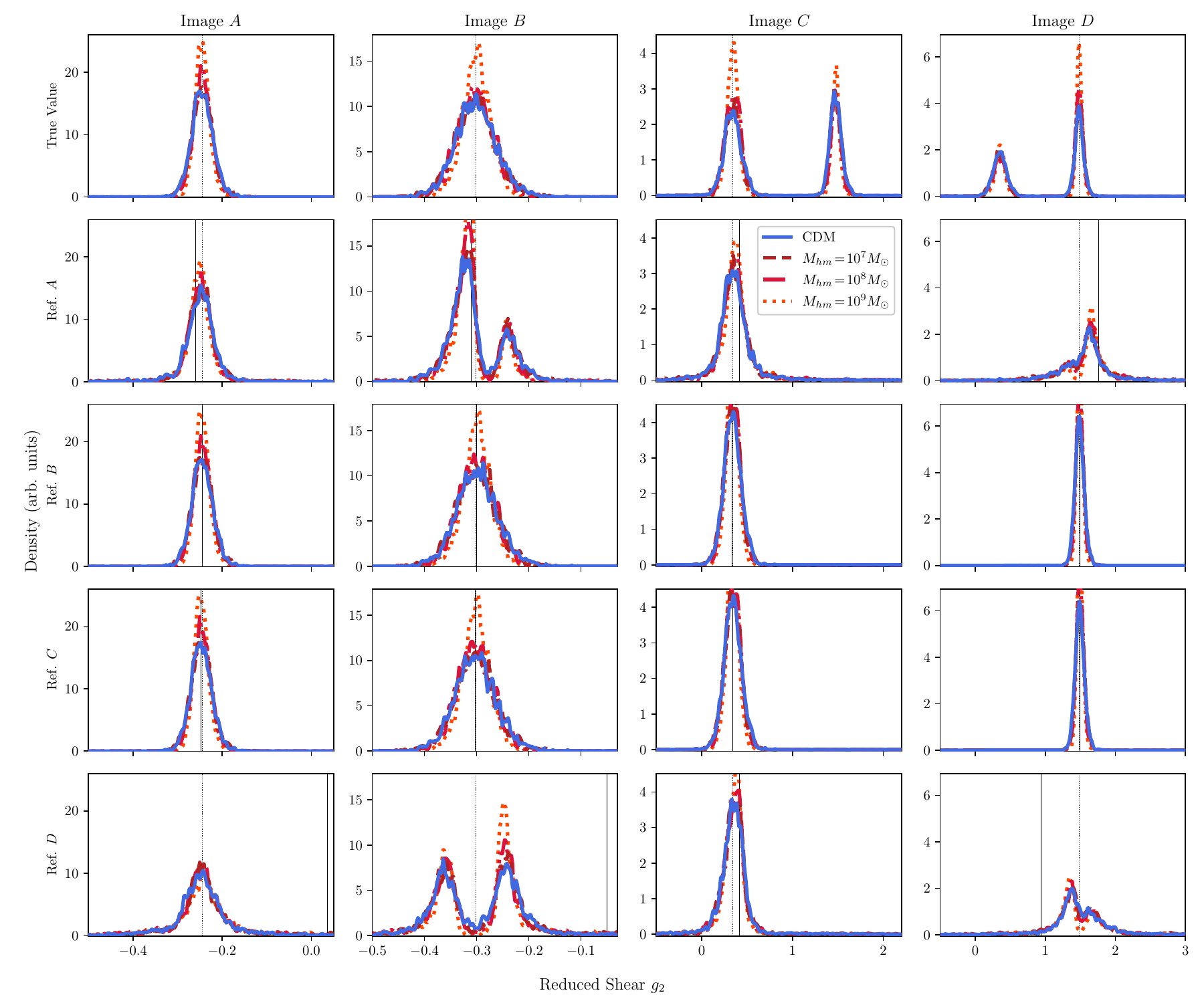}
	\caption[$g_2$ Reduced Shears]{Distribution of $g_2$ reduced shears, following the same scheme as Figure \ref{fig:g1}.}
	\label{fig:g2}
\end{figure}

\begin{figure}
	\centering
	\includegraphics[trim={18pt 6pt 8pt 10pt}, clip, width=\textwidth]{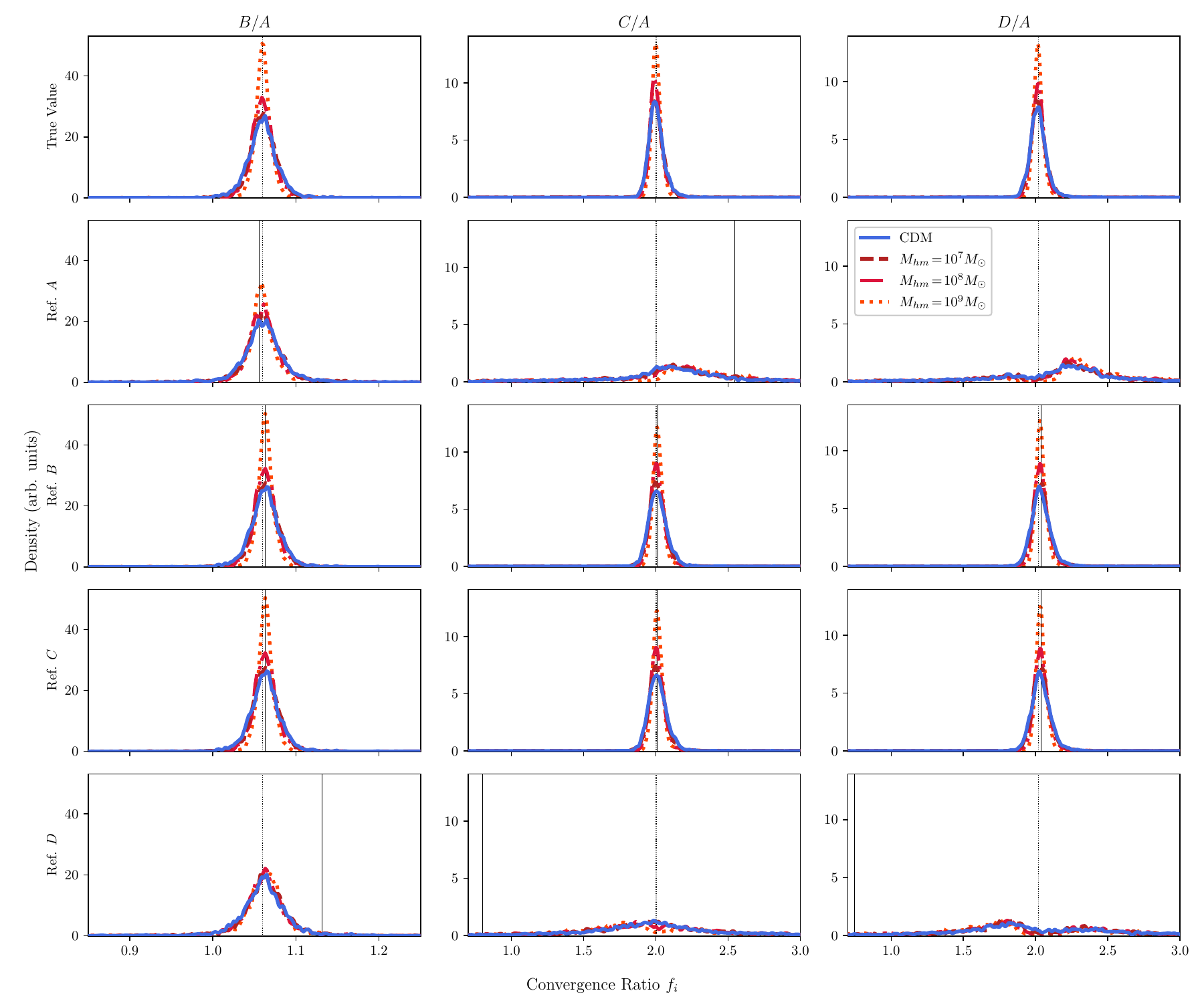}
	\caption[Convergence Ratios $f_i$]{Distribution of $f_i$ the ratio of (1-convergence) between each image and $A$. Follows the same scheme as Figure \ref{fig:g1}.}
	\label{fig:fi}
\end{figure}

In Figures \ref{fig:g1}, \ref{fig:g2}, and \ref{fig:fi}, we show the distributions of convergence ratios and reduced shears for each image and reference image, as well as their true values (taken from \texttt{lenstronomy}).
As with the magnification ratios, the other lensing quantities seem to be more spread out for CDM than the WDM models, but this trend is less clear than in the former case.
Additionally, depending on the choice of reference image, the measured shears follow significantly different distributions, even becoming bimodal in some cases.
Also of note is the bimodality of the true shears for the $C$ and $D$ images, a curious effect which is nonetheless unobservable in practice.

\subsection{Image Rotations}
Until this point we have only employed the thin-lens approximation, which ensures that the amplification matrix $\mathcal{A}$ is symmetric.
However, every Dark Matter model we've considered includes Line-of-Sight halos (see Figure \ref{fig:subhalos} for a representative example for CDM).
The analytical formalism for a ``thick'' gravitational lens is much less developed and much more computationally expensive to calculate than for a thin lens, as each lens plane couples to all the previous ones \citep{blandfordnarayan}.
Although each lens can be treated as thin, the products of all amplification matrices is not in general symmetric, meaning that in addition to a scaling, reflection and shear, a thick lens will also produce a differential rotation between images \citep{kovner87}.
The amplification matrix then takes the following form:
\begin{equation}
\mathcal{A} = \begin{pmatrix}
1 - \kappa - \gamma_1 & -\gamma_2 + \omega \\
-\gamma_2 - \omega & 1 - \kappa + \gamma_1\\
\end{pmatrix}.
\end{equation}
The rotation $\omega$ is in general not easily calculable from lensing quantities, but can be obtained by conducting a Singular Value Decomposition (SVD) of $\mathcal{A}$.
However, \citet{fleury21} provide an analytical form for the rotation given the special case of a ``Dominant Lens'' that dictates the bulk of the image deflection -- exactly the case we consider here -- and show it arises entirely from coupling terms between the perturbing halos and the macro lens.

\begin{figure}
	\centering
	\includegraphics[trim={10pt 20pt 11pt 10pt}, clip, width=\textwidth]{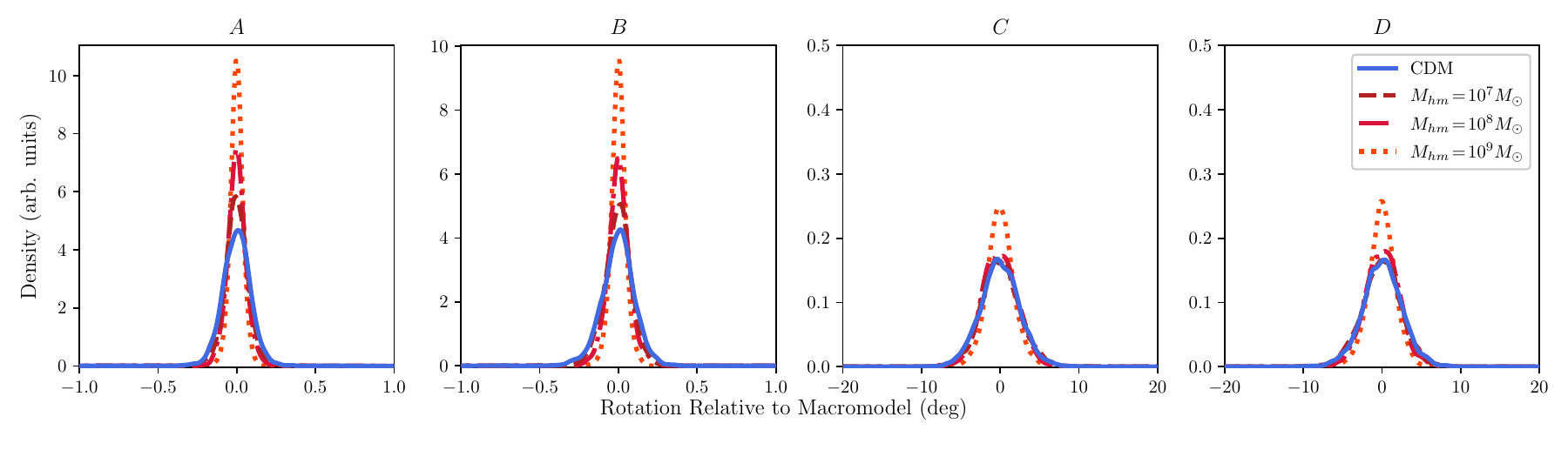}
	\caption[Relative Image Rotations]{Distribution of image rotations relative to the macromodel.}
	\label{fig:rotation}
\end{figure}

Given our simulation data, we can measure $\omega$ using an SVD on our measured amplification matrices.
We show the distribution of relative rotations from LOS halos in Figure \ref{fig:rotation}.
Like the magnification ratios, the WDM models show a wider distribution than CDM, but we note that the quantity graphed is not an observable.
However, using the point-matching technique it may be possible to constrain a single image rotation, as proposed by \citet{penmao}.
We explore this possibility in Appendix \ref{sec:transformation}.

\section{Discussion}
The results of Table \ref{tab:KS_flux} show some promise for an astrometric constraint with radio-emitting lensed quasars.
While the performance of the method for this lensing configuration doesn't lend itself to an especially constraining result, the broad agreement between the flux ratio result and the determinant ratio result suggests the latter could be competitive as a way to detect magnification anomalies in the same way the former is today.
This Chapter has mostly focused on the broad distribution of lensing properties obtainable by each method, but the tails of the distributions are arguably just as important.
One way to extend this analysis to incorporate those tails would be to construct more realistically sized subsets (100-200 lenses) in each DM model and analyze the occurrence of magnification anomalies there.
However, a better choice would be to instead investigate a more realistic sample of simulated lenses with multiple different configurations, especially fold and cusp configuration quads.
If the determinant ratio method also performs similarly to flux ratios in these situations, the astrometric data could provide a check for flux ratio data at other wavelengths.

A clear advantage of the method presented here is its achievability by already-existing radio telescopes. 
Examining the trends in Figure \ref{fig:astrometric}, the distances between core and hotspot in the lensed images are of order 1 milliarcsecond, well within the capabilities of the VLBA.
The method's robustness to slightly larger sources is also promising, as VLBI observations could go to lower frequencies where jet hotspots are more bright and therefore easier to measure\footnote{A caveat here is that when a source is sufficiently close to a caustic the assumption of a constant $\kappa$ is broken, but the method may still be feasible using the higher-order expansion presented in \citet{wagner22}.}.
The primary roadblock, therefore, appears to be a lack of existing sources with hotspots, which is primarily an issue of sensitivity -- while the VLBA may be able to resolve a core and hotspot it isn't guaranteed to detect them both in the first place.
Higher-SNR observations (e.g. with the High Sensitivity Array) would help to observe greater structure in a lens.
A particularly visceral example of this is the giant radio arc that appears in \citet{b1938vlbi} which went unobserved in \citet{1997MNRAS.289..450K}.
The Next-Generation VLA (ngVLA) will ultimately revolutionize these sorts of observations with its unprecedented resolution and sensitivity at centimeter wavelengths. 

A major concern with flux ratio analysis is the effect of microlensing by stars, which can affect magnification measurements if measured source size is too small.
The effect of microlensing on the analysis in this Chapter is not known but may be non-negligible, a possibility that certainly warrants further exploration.
Such an effect would depend on source size, although this dependence may be less than for flux ratios, as the deflection arises from the first derivative of $\tau$ rather than the second.

In this analysis we have focused on constructing quantities that are directly related to lensing quantities, such as the reduced shears.
With an additional number of identifiable points in each lensed image, it may be possible to construct some other quantity.
An investigation of this, probably employing some machine learning method to construct a viable statistic, would also be a natural extension to the work presented here.
Regardless of the analysis method, however, the population of radio-emitting lensed quasars would need to be significantly extended for a pure astrometric constraint to be viable.

%-look at VLBA observations of various CLASS+ objects (look at their data, if possible)

%-some existing radio lenses (CLASS?) have features like this, probably. That should be studied. maybe list them?

\chapter{Finding Lensed Radio Sources in VLASS}
\label{sec:chapter3}
\paragraph{This chapter originally appeared as}
\citet{martinez2025} in The Astrophysical Journal.
It appears here in slightly modified form, with changes made to better fit the formatting of the larger document, but the scientific results have not changed.
Most notably, Figure \ref{fig:fluxdist} originally appeared as two separate figures, which have been combined to better fit the single-column style of this document.
Additionally, while Table \ref{tab:allknown} has been updated to include more radio lenses discovered since publication,  we have not updated the results of Section \ref{sec:discussion} to reflect these additions.

\section{Introduction}  \label{sec:intro}

Strong gravitational lensing, the phenomenon by which multiple images of a background source are created by a foreground lens, has been an active and growing field of study since the discovery of the first lensed object by \citet{walsh79}.
Since then, the advent of high-resolution space-based optical imaging from the Hubble Space Telescope and large ground-based optical surveys have increased the number of known lenses today to many hundreds \citep[e.g.,][]{Jacobs2019, Huang2020, zaborowski23, Lemon2024}.

Gravitational lensing is achromatic, and strong lenses can be observed in any wavelength of light, though relative abundances vary across the electromagnetic spectrum.
At radio frequencies, under 100 lensed sources are known, as opposed to the thousands of optical ones.
This is due in part to the relative scarcity of radio sources. For example, the Faint Images of the Radio Sky at Twenty-centimeters \citep[FIRST,][]{FIRST, FIRSTfinal} and the imaging portion of the optical Sloan Digital Sky Survey \citep[SDSS][]{SDSS, SDSSDR7}, which covered the same area and were roughly contemporaneous, had source densities of $\approx90\,\text{deg}^{-2}$ and $\approx30,000\,\text{deg}^{-2}$, respectively.
Furthermore, the angular resolution needed to identify strong lensing, typically on the scale of $\sim1\,$arcsecond for galaxy-galaxy lenses \citep{McKean2015, collett15}, also presents a large barrier to finding lensed radio sources as the angular resolution of wide area surveys has historically been on the order of a few tens of arcseconds \citep[e.g,][]{Condon1998, Bock1999, Intema2017}.
This has historically resulted in samples of rare candidate lensed radio sources being overwhelmingly contaminated by non-lensed objects \citep[e.g.][]{JB07}.
Successful radio searches for lensing, such as the Jodrell Bank Astrometric Survey \citep[JVAS, ][]{1999MNRAS.307..225K} and the Cosmic Lens All-Sky Survey \citep[CLASS, ][]{Myers2003, 2003MNRAS.341...13B}, began with a flux-limited, flat-spectrum sample to limit the amount of necessary high-resolution follow-up to confirm lensing.
More recently, radio lens searches have taken advantage of the abundance of optical lensed quasars by conducting deep observations of these to try to detect radio emission \citep{2015MNRAS.454..287J, dobie23}.
In the future, facilities such as the Square Kilometer Array \citep[SKA, ][]{SKA} and next generation Very Large Array \citep[ngVLA,][]{Carilli2015} will provide depth and sub-arcsecond resolution in combination with high survey-speeds, making them efficient lens-finding tools.
Currently however, only limited sky areas (of order a few square degrees) have been observed with the requisite combination of depth and angular resolution to readily identify strong lensing at radio wavelengths \citep{Morabito2022}. 

The Very Large Array Sky Survey \citep[VLASS,][]{VLASS} provides $\approx 2''.5$ angular resolution across $\approx 34,000\,\text{deg}^2$ of sky at 3 GHz.
By 2025 VLASS will have observed its entire footprint over three distinct epochs, and at the time of writing, VLASS has already completed two of these epochs with the third epoch already underway\footnote{While this statement was accurate at the original publication's press time, as of thesis press time VLASS has completed all three full epochs and the additional fourth half-epoch.}.
While VLASS does not posses the resolution necessary to separate the images of most lensed quasars \citep{Lemon19}, the $2''.5$ beam of survey allows for high confidence associations with optical sources and is less subject to contamination from interloping sources than other near-all-sky radio surveys.

The scientific applications of radio lenses are numerous, and range from probing the structure of AGN jets at high redshift \citep{spignola2019} to studying the magnetic fields of lens galaxies \citep{mao17}.
Radio lenses have also been used to constrain the Hubble Constant via lensing time delays in lens systems such as CLASS B0218$+$357 \citep{biggs99, biggs18}, CLASS 1600$+$434 \citep{koopmans2000}, and CLASS 1608$+$656 \citep{class1608A, class1608B}.

One particularly exciting possibility lies in the characterization of low-mass dark matter halos to constrain the microphysics of dark matter.
Due to the sensitivity of image magnifications and deflections to all mass along the line of sight between source and observer, lensing observations are sensitive to the lower end of the dark matter halo mass function, especially the ``completely dark halos'' not massive enough to form stars \citep{vegetti23, bechtol22}.
The milliarcsecond-scale astrometric perturbations caused by these halos \citep{metcalfmadau} currently can only be accessed using the resolution of radio Very Long Baseline Interferometry (VLBI).
Such gravitational imaging analyses can potentially differentiate between different models of dark matter phenomenology \citep[e.g.,][]{2019MNRAS.483.2125S, powell23}.
Next-generation telescopes such as SKA and ngVLA will be able to perform observations of lens systems quickly and robustly -- larger samples of candidate systems are important to inform both the theory and technical development of those dark matter analyses.

In this paper, we present the results of a VLASS-based search for strong lensed radio sources, and report the detection of radio emission from five previously known optically lensed quasars.
In Section \ref{sec:search} we describe our candidate selection process, 
Section \ref{sec:obs} provides a summary of our observations, and Section \ref{sec:p1results} presents the results of each candidate observed in detail.
In Section \ref{sec:discussion} we discuss the population of known lensed radio sources and the potential for future survey-based radio lens searches.
We summarize this work in Section \ref{sec:summary}.

%%%%%%%%%%%%%%%%%%%%%%%%%%%%%%%%%%%%%%%%%%%%%%%%%%%%%%%%%%%%
\section{Candidate Identification} \label{sec:search}

In selecting sources for the VLA observations, we took a two-pronged approach based on both known lens systems and catalog-based optical-radio cross-matching.
We identify radio sources using the VLASS Epoch 1 quick-look catalog from \citet{Gordon2021}, which contains $\sim 1.8\times10^{6}$ reliable detections with $S_{3\text{GHz}}\gtrsim 1\,$mJy at $\delta > -40^{\circ}$.
To account for the known $\sim 0''.25$ astrometric errors in the quick-look data, we have corrected the source positions based on the method of \citet{Bruzewski2021}\footnote{Since the identification of these targets in 2022, a version of the epoch 1 VLASS Quick Look catalog with corrected astrometry has been made available (B. Sebastian et al., in prep.).}.

\subsection{Known Lensed Optical Sources} \label{existing}

As lensed radio sources are rare, knowing a priori that a system is a gravitational lens maximises the efficiency in searching for these objects.
To this end, we cross match the VLASS catalog with two catalogs of known optical lenses using data from Gaia \citep{Gaia2016} and the Dark Energy Survey \citep[DES,][]{DES2016}. 

We first used the catalog of lensed quasars in Gaia \citep{Lemon2017, Lemon2018, Lemon19}, finding 43 matches with VLASS.
Of these, 31 were previously known lensed radio sources, and a further 7 had existing archival observations at sufficient resolution and depth to confirm or reject the radio lensing hypothesis without the need for additional telescope time.
An additional candidate was also rejected after visual inspection of the VLASS data showed the lens galaxy to be an FR I radio galaxy, implying the radio emission in the system came from the lens rather than the lensed source.
After these cuts we were left with 5 candidate new radio lenses.

We also cross-matched VLASS with strongly lensed systems in DES \citep{Jacobs2019}.
Here we found 17 matches, none of which had archival high-resolution VLA data.
Visually inspecting these 17 objects showed that in most of these cases, the radio emission was more likely due to the lens galaxy being a radio galaxy. While in theory it is possible to observe radio emission from both the lens and source, we did not prioritize these targets.
In two cases we found the radio emission to be consistent with being from the \textit{lensed images} and require higher resolution follow up to confirm their nature.
However, due to limited observing time we only observed one of these with the VLA for this paper.

\subsection{Blind Search for Lensed Sources From Optical/Radio Associations} \label{catalog}

\begin{figure*}
    \centering  
    \subfigure{\includegraphics[trim={10mm 10mm 0pt 0pt}, clip, width=0.23\textwidth]{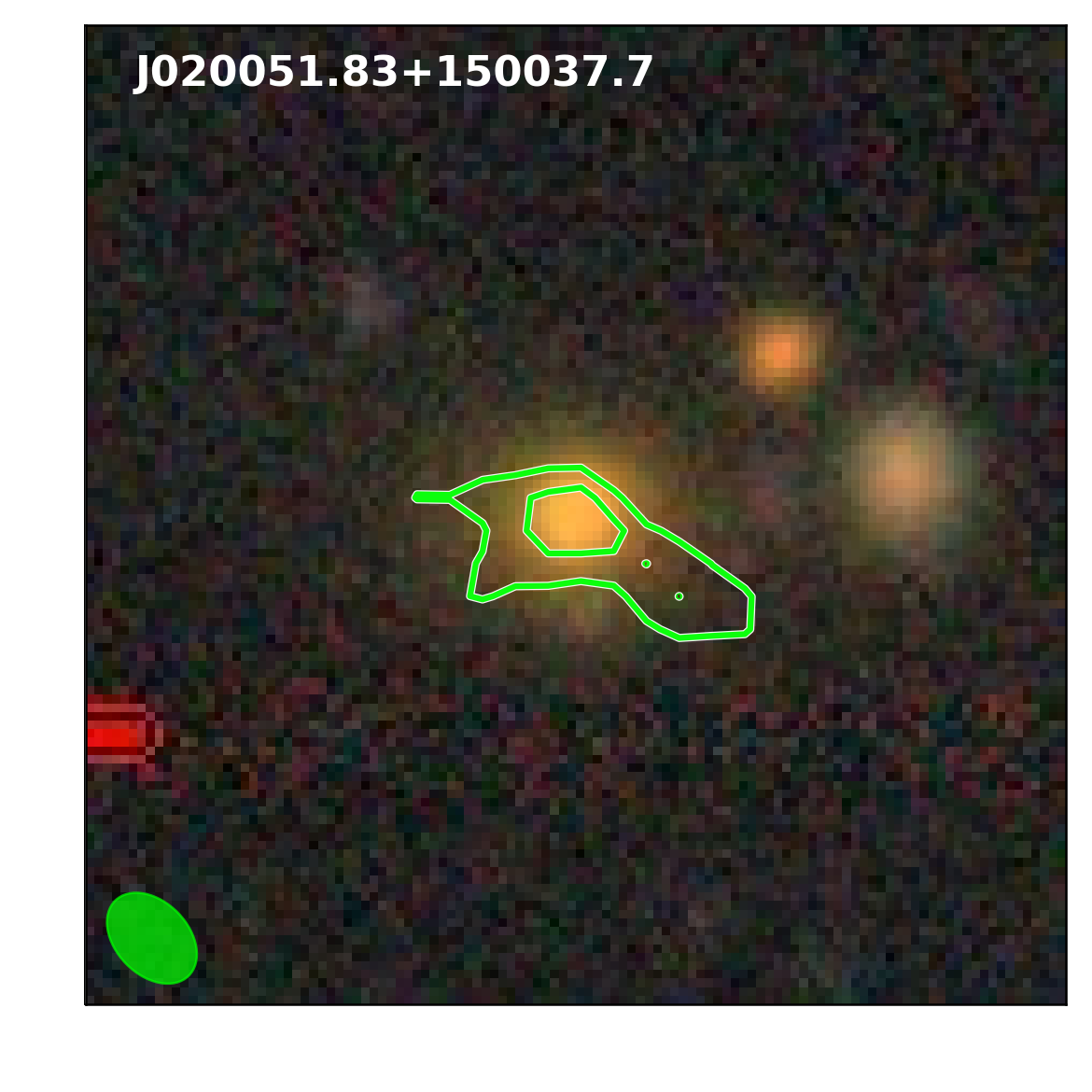}}
    \hspace{1mm}
    \subfigure{\includegraphics[trim={10mm 10mm 0pt 0pt}, clip, width=0.23\textwidth]{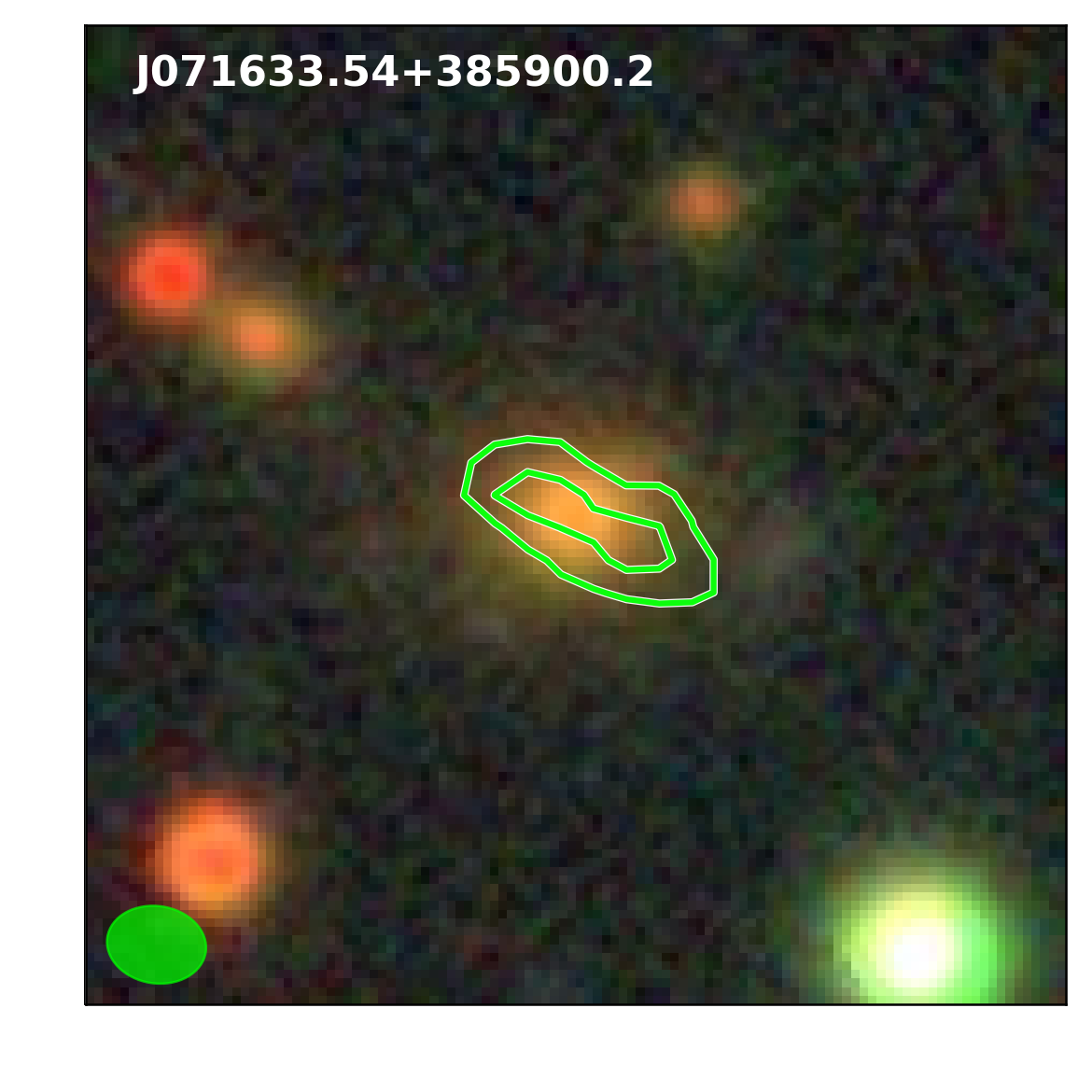}}
    \hspace{1mm}
    \subfigure{\includegraphics[trim={10mm 10mm 0pt 0pt}, clip, width=0.23\textwidth]{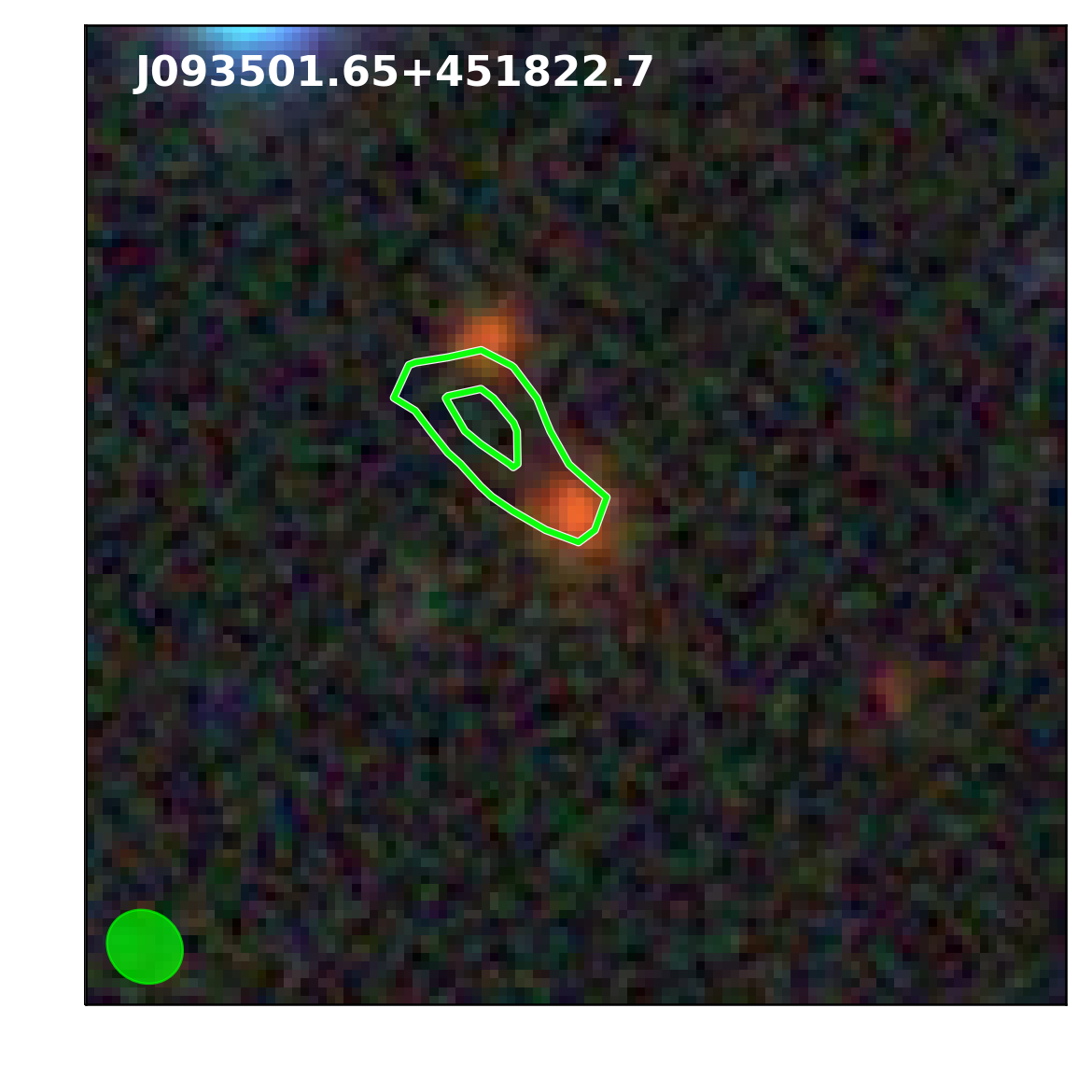}}
    \hspace{1mm}
    \subfigure{\includegraphics[trim={10mm 10mm 0pt 0pt}, clip, width=0.23\textwidth]{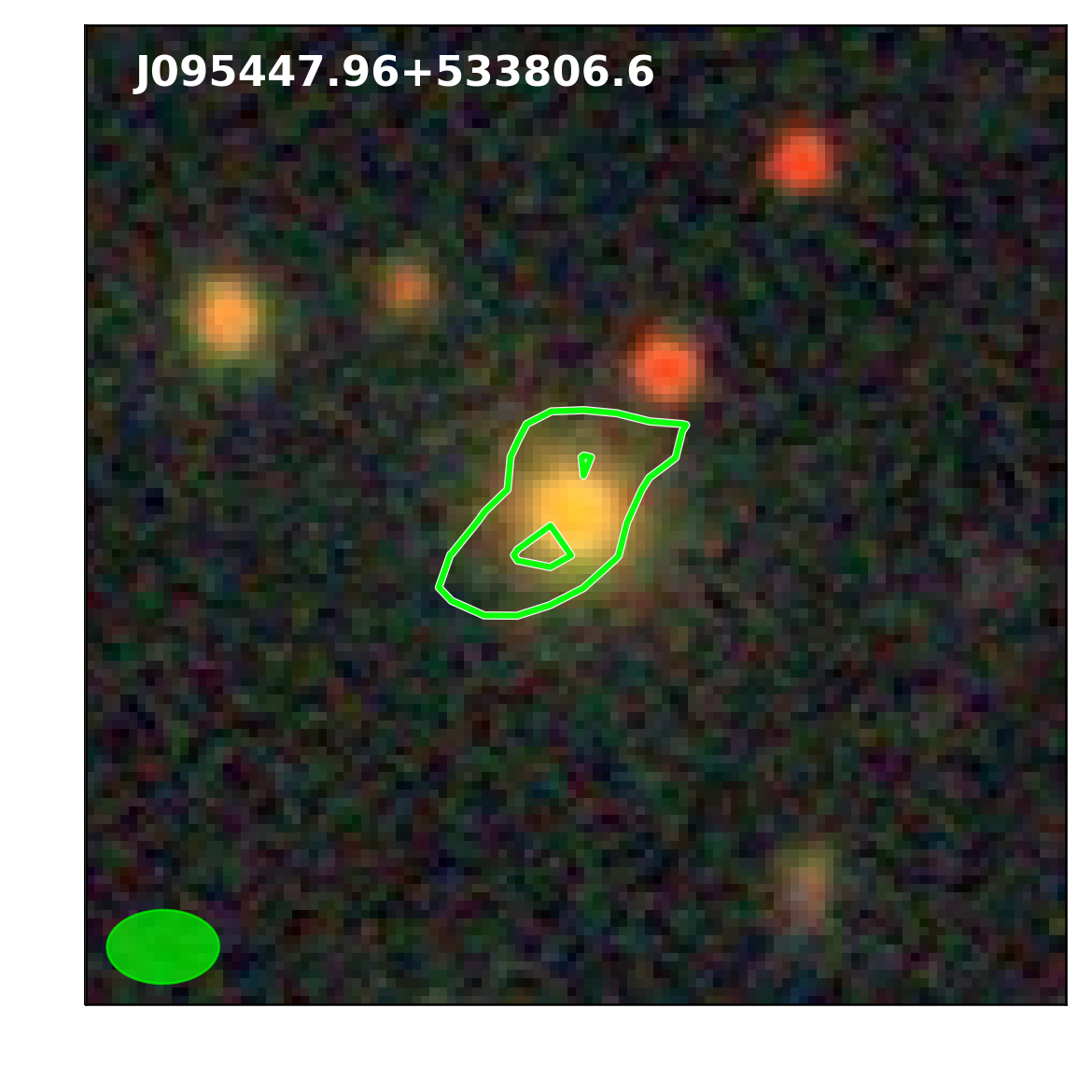}}
    
    \caption[Cutouts of Non-Candidate Radio Galaxies]{Postage stamp optical cutouts ($30''\times30''$) of sources where the radio emission is attributed to the lens galaxy on visual inspection and thus rejected as candidate lensed radio sources.
    VLASS contour levels (green) start at $0.5\,\text{mJy}\,\text{beam}^{-1}$ and increase in linear increments of $0.5\,\text{mJy}\,\text{beam}^{-1}$.}
    \label{fig:eg_rejects}
\end{figure*}

In addition to combining VLASS with catalogs of known optical lenses, we adopted the approach of \citet{JB07} (hereafter JB07) to conduct a blind search for lensed systems in the radio catalog data.
The JB07 method assumes the lensed source flux is blended together into a single detection at the survey resolution, and predicts an offset from the lens galaxy due to the unequal magnifications inherent in lensed images.
Additionally, {as illustrated in Figure 1 of \citet{JB07}}, these blended components should have position angles either close to or perpendicular to that of the lens galaxy's optical position angle for 2 and 4-image systems, respectively.
JB07 matched the SDSS and FIRST surveys, identifying $\sim 70$ candidates, none of which were lenses.
However, the wealth of additional candidates afforded by increased depth and sky coverage since JB07 has led us to use their method with VLASS and DES to attempt to identify further candidate lensed radio systems.

We begin by narrowing our optical selection to luminous red galaxies (LRGs), which due to their high mass are the most common type of lens galaxy, and are often embedded in larger structures which can increase lensing probability.
This restriction was adopted by both JB07 and \citet{2001ApJ...547...60L}, the latter being a joint optical-radio search for lensed radio lobes.
We used the Dark Energy Spectroscopic Instrument Legacy Survey 9th data release \citep[LS-DR9,][]{Dey2019} as the optical survey. 
LS-DR9 covers nearly the entire sky at $|b| > 18^{\circ}$ in the $g,r,z$ bands down to a point source depth of $r\lesssim23\,$mag in the Legacy Survey northern fields ($\delta > +32^{\circ}, b>+18^{\circ}$) and $r\lesssim 23.5\,$mag in the southern sky.
Additionally LS-DR9 provides mid-infrared forced photometry from the unblended Wide-field Infrared Survey Experiment \cite[unWISE,][]{WISE, unWISE} bands. 
We follow the selection criteria of \citet{Zhou2020},  to select LRGs with high purity.
We then cross matched these with VLASS sources that were marginally resolved ($0 < \Psi < 0''.5$; where $\Psi$ is the major axis of the source after deconvolution from the beam), selecting only those matches that satisfy the angular separation and misalignment criteria used in JB07.

\begin{figure*}
    \centering  
    \subfigure{\includegraphics[trim={10mm 10mm 0 0}, clip, width=0.23\textwidth]{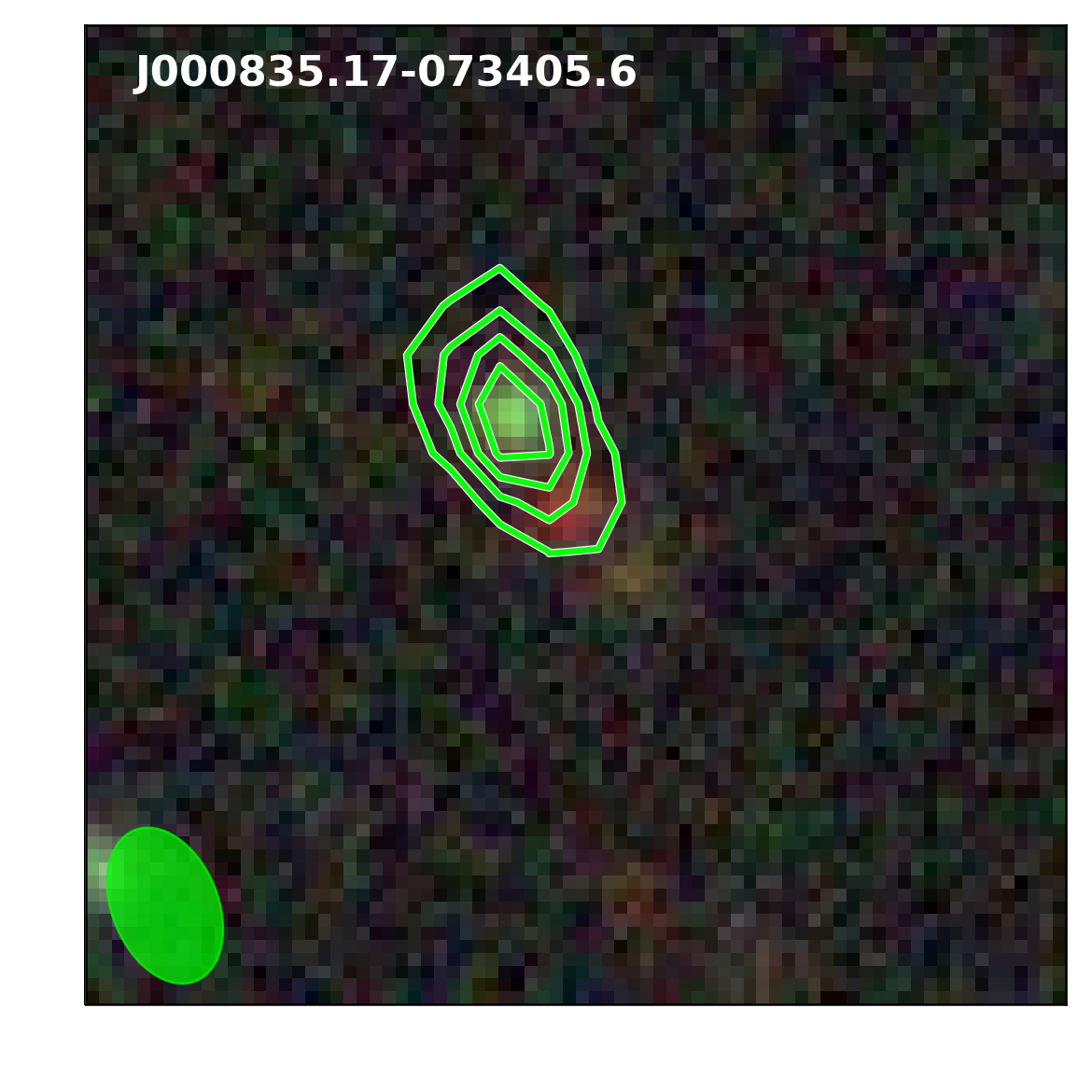}}
    \hspace{1mm}
    \subfigure{\includegraphics[trim={10mm 10mm 0 0}, clip, width=0.23\textwidth]{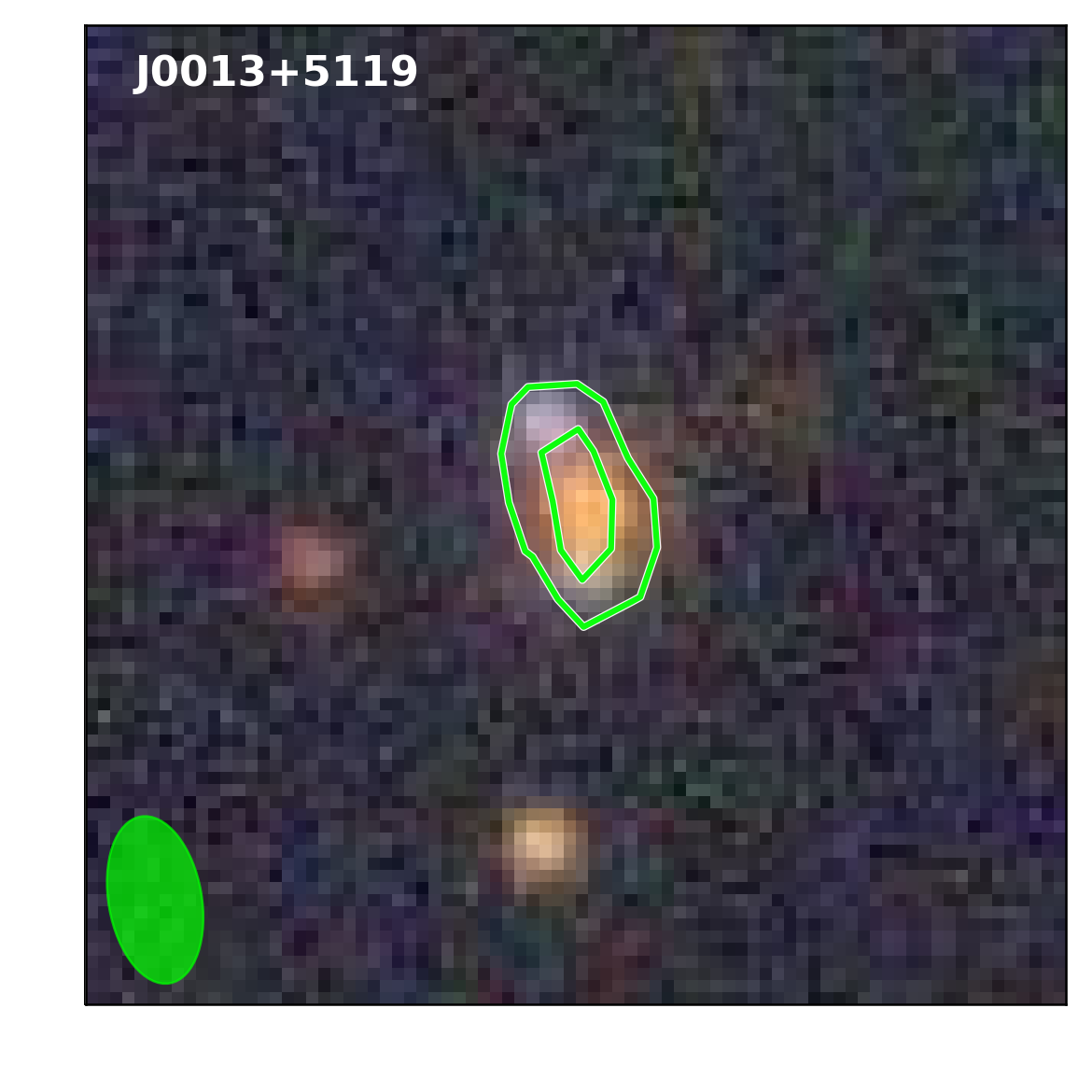}}
    \hspace{1mm}
    \subfigure{\includegraphics[trim={10mm 10mm 0 0}, clip, width=0.23\textwidth]{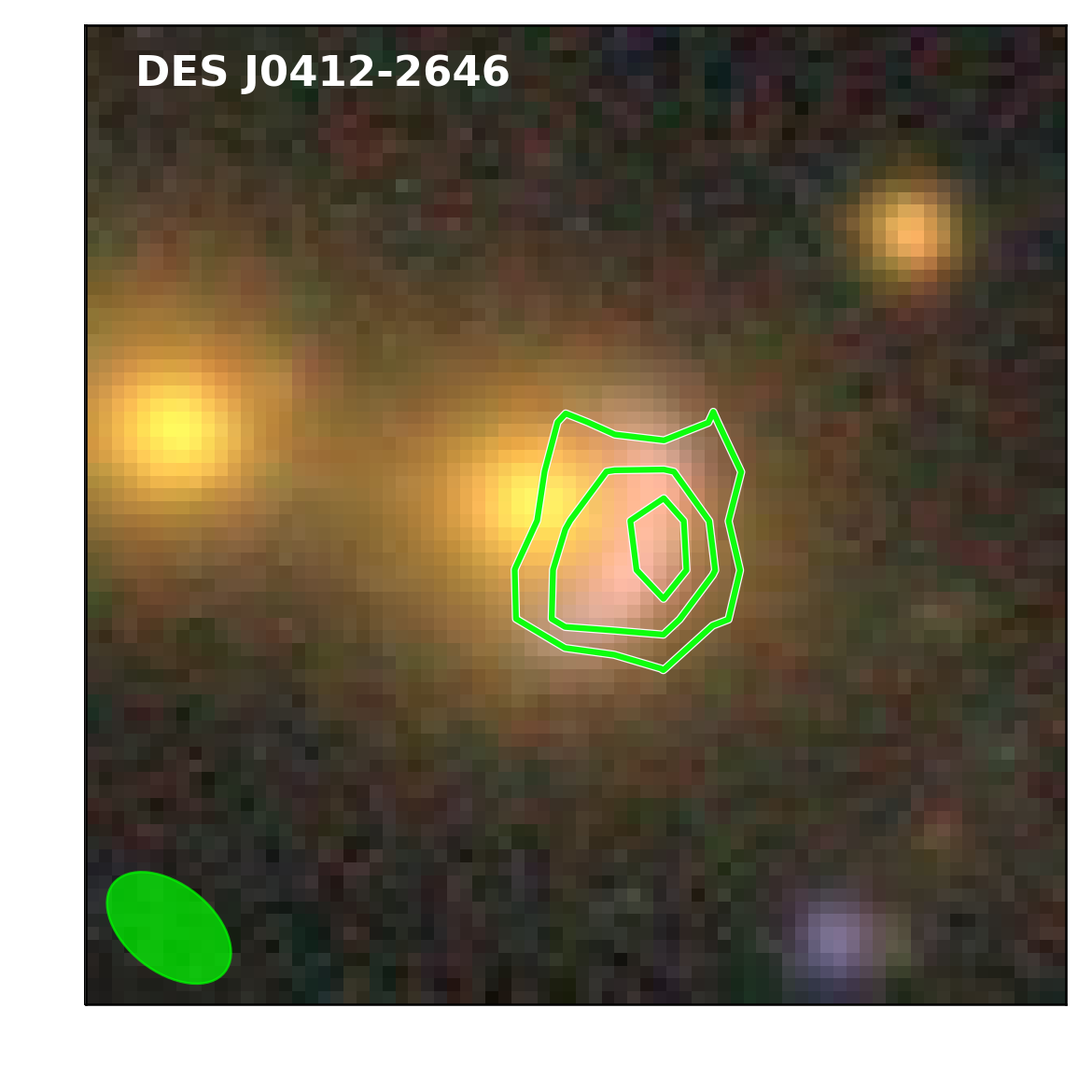}}
    \hspace{1mm}
    \subfigure{\includegraphics[trim={10mm 10mm 0 0}, clip, width=0.23\textwidth]{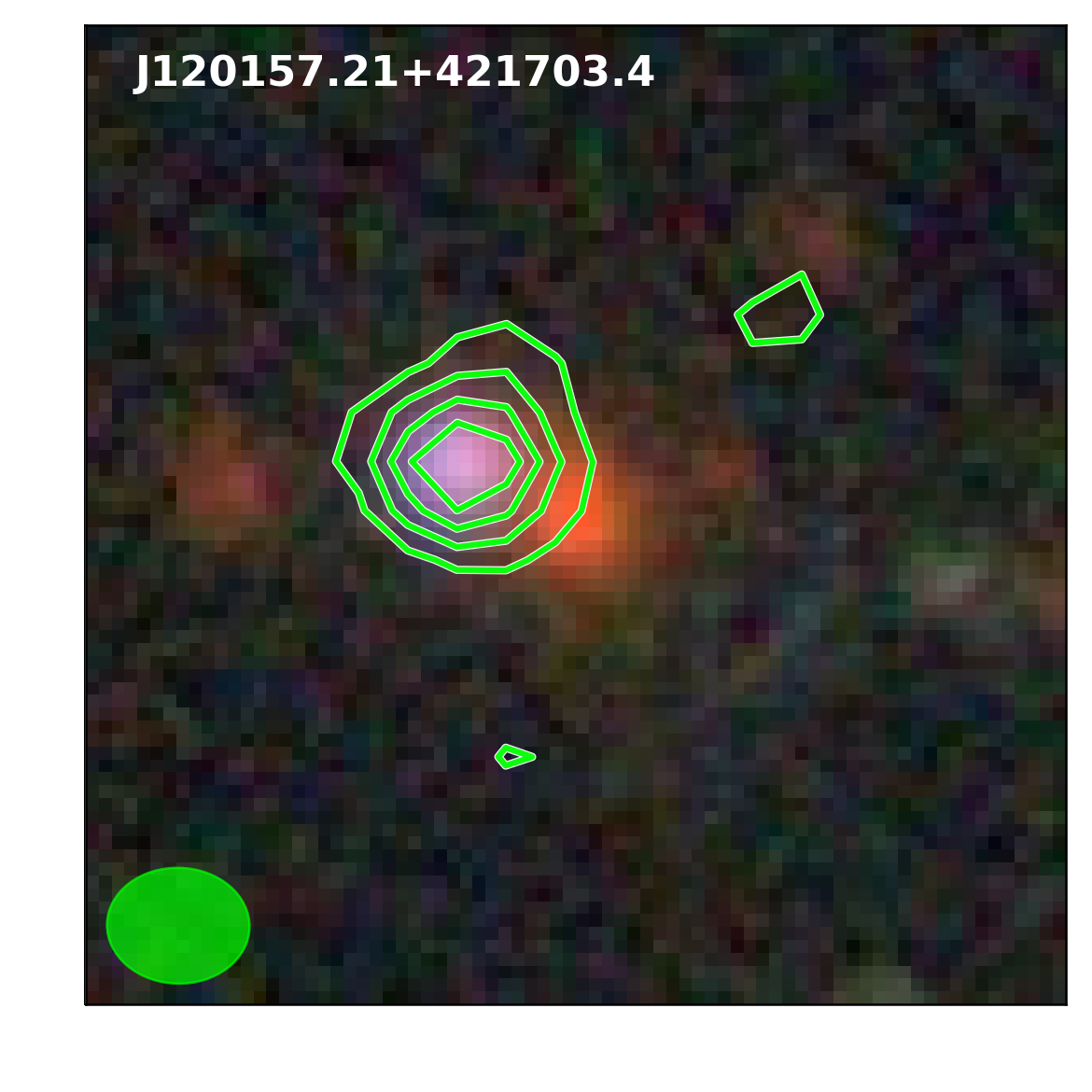}}
    \subfigure{\includegraphics[trim={10mm 10mm 0 0}, clip, width=0.23\textwidth]{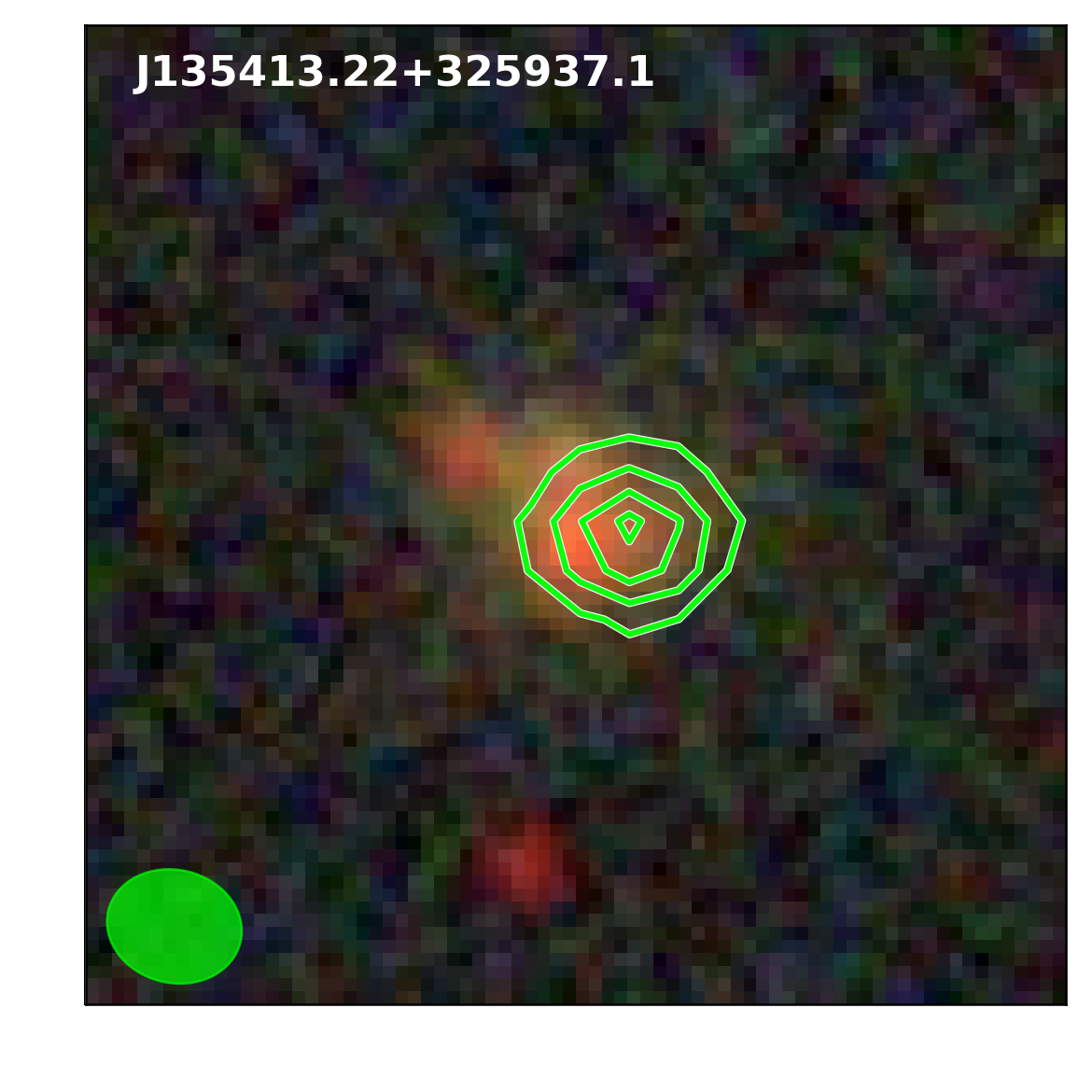}}
    \hspace{1mm}
    \subfigure{\includegraphics[trim={10mm 10mm 0 0}, clip, width=0.23\textwidth]{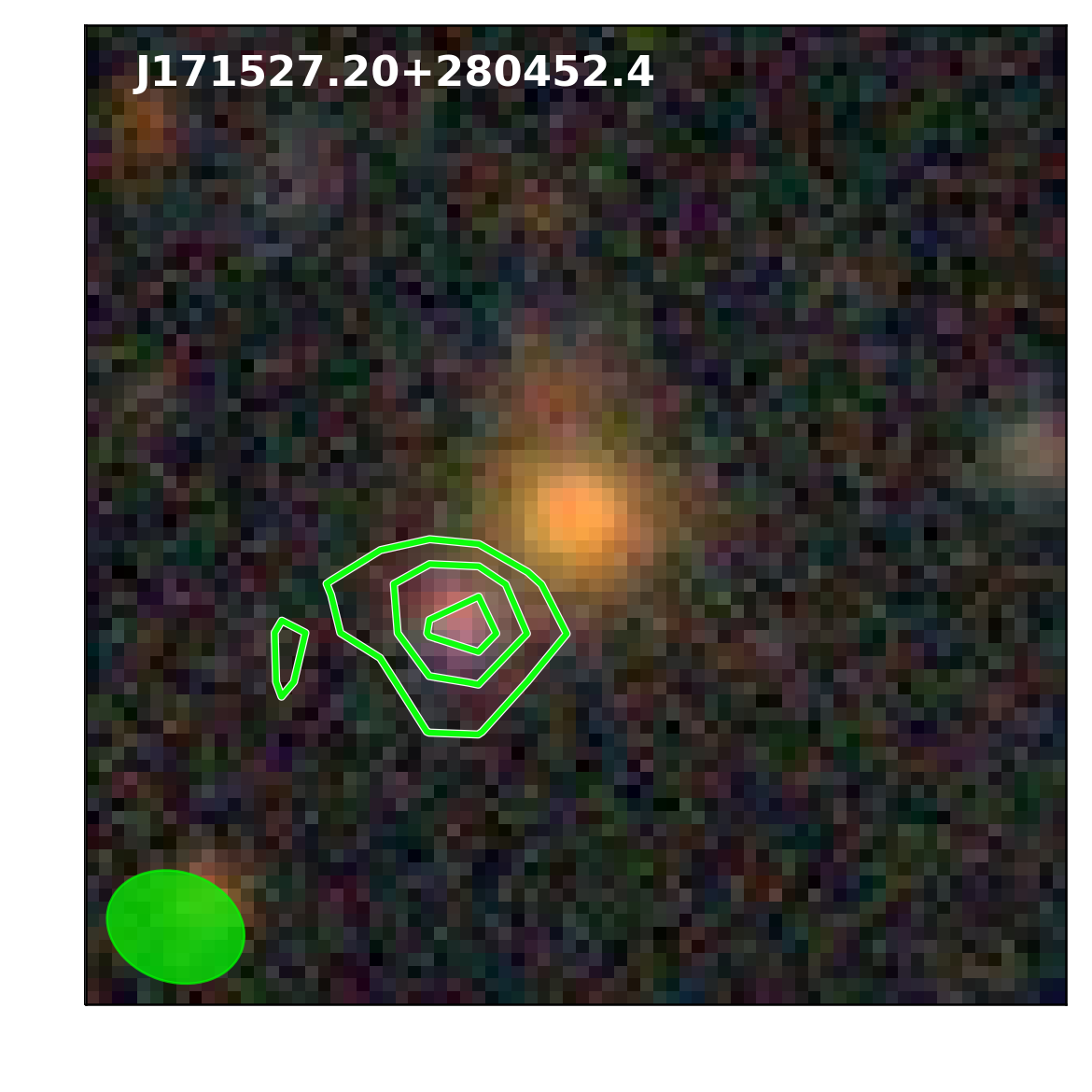}}
    \hspace{1mm}
    \subfigure{\includegraphics[trim={10mm 10mm 0 0}, clip, width=0.23\textwidth]{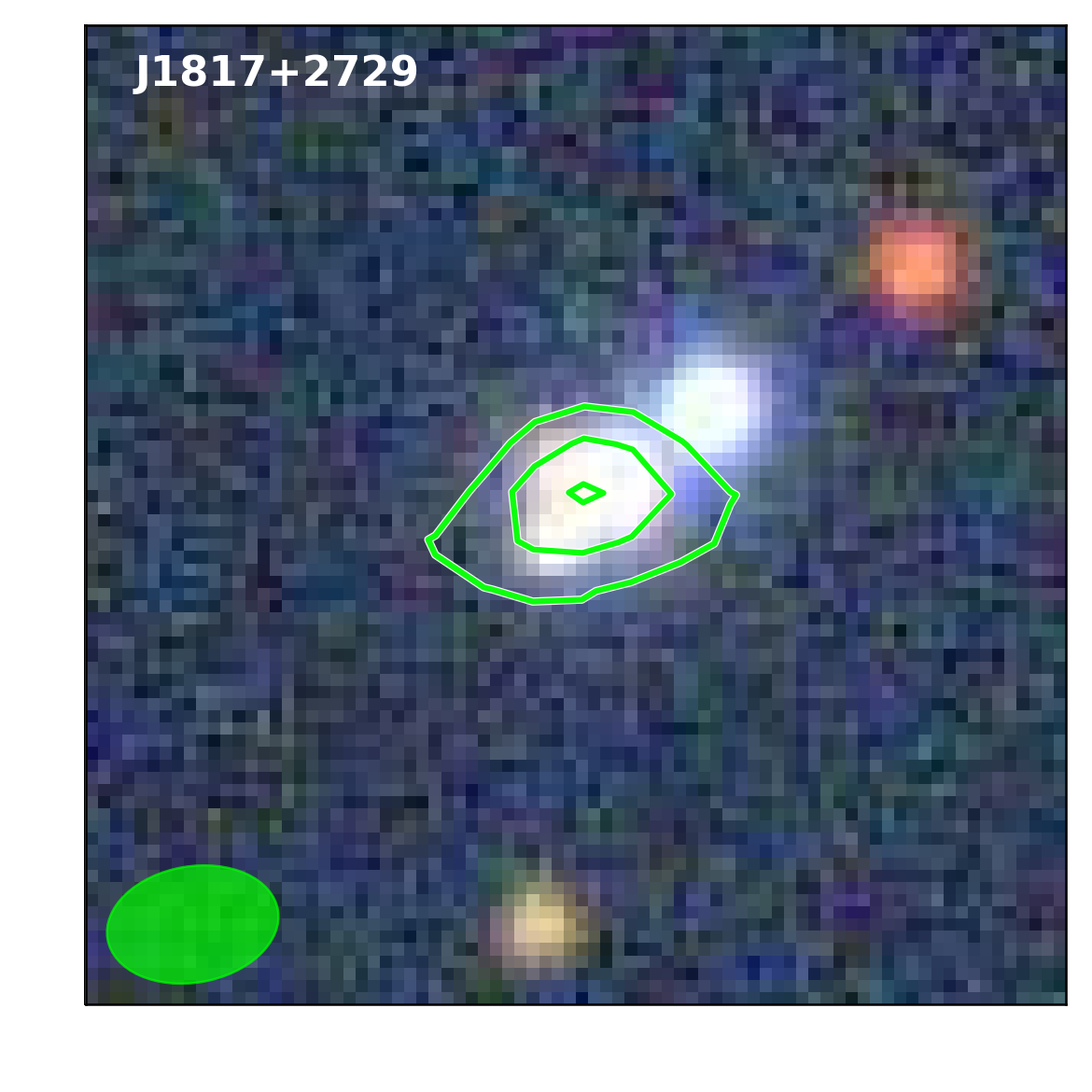}}
    \hspace{1mm}
    \subfigure{\includegraphics[trim={10mm 10mm 0 0}, clip, width=0.23\textwidth]{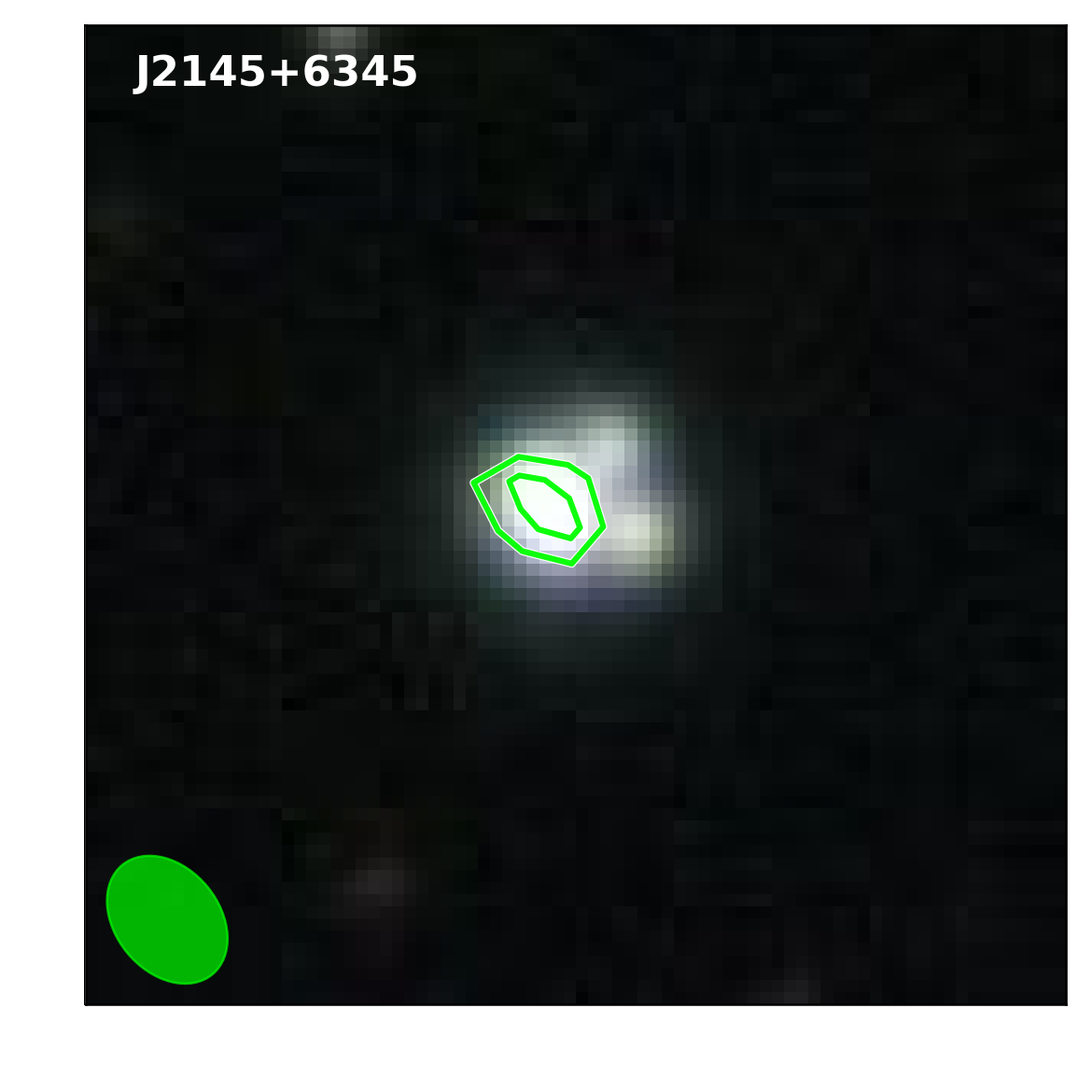}}
    \subfigure{\includegraphics[trim={10mm 10mm 0 0}, clip, width=0.23\textwidth]{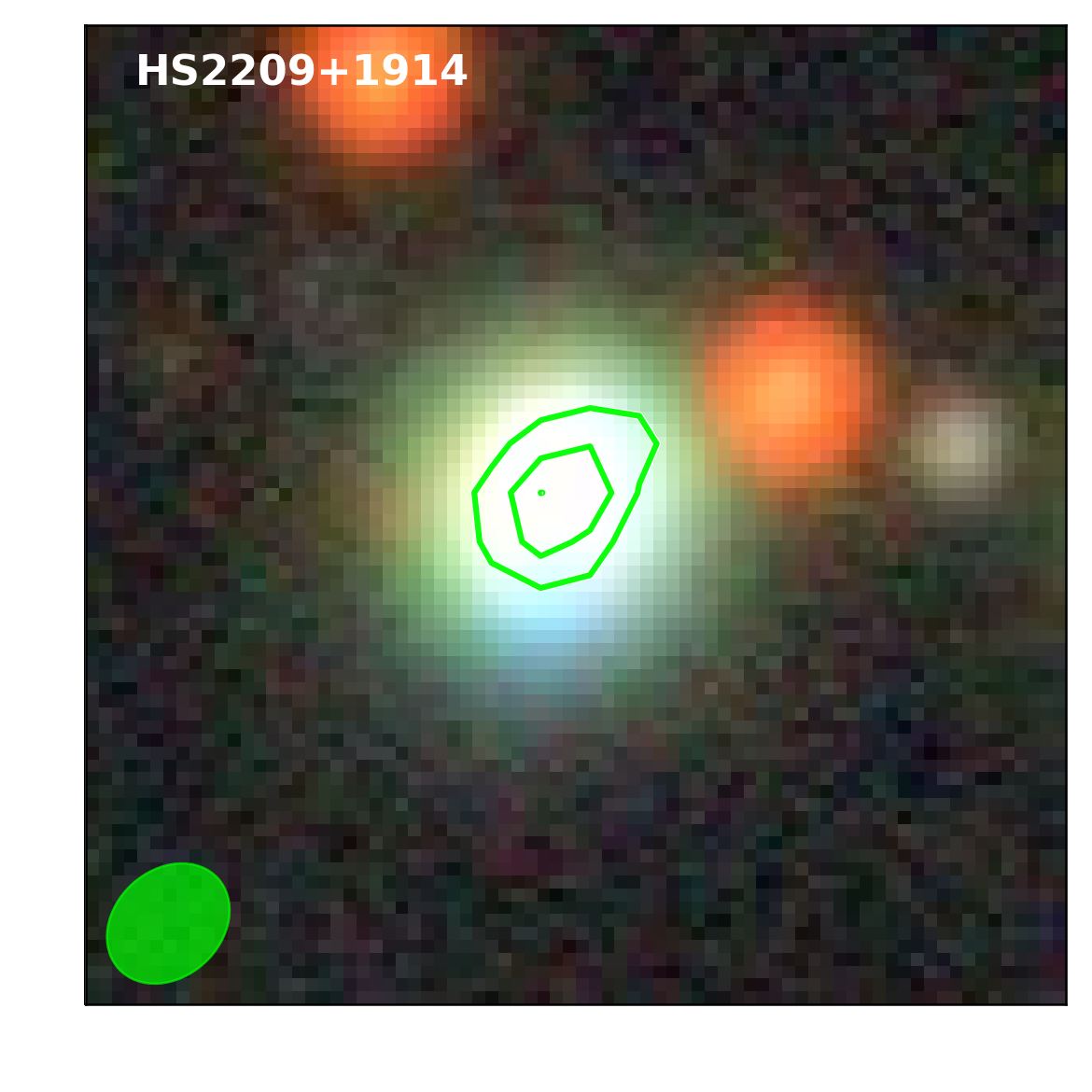}}
    \hspace{1mm}
    \subfigure{\includegraphics[trim={10mm 10mm 0 0}, clip, width=0.23\textwidth]{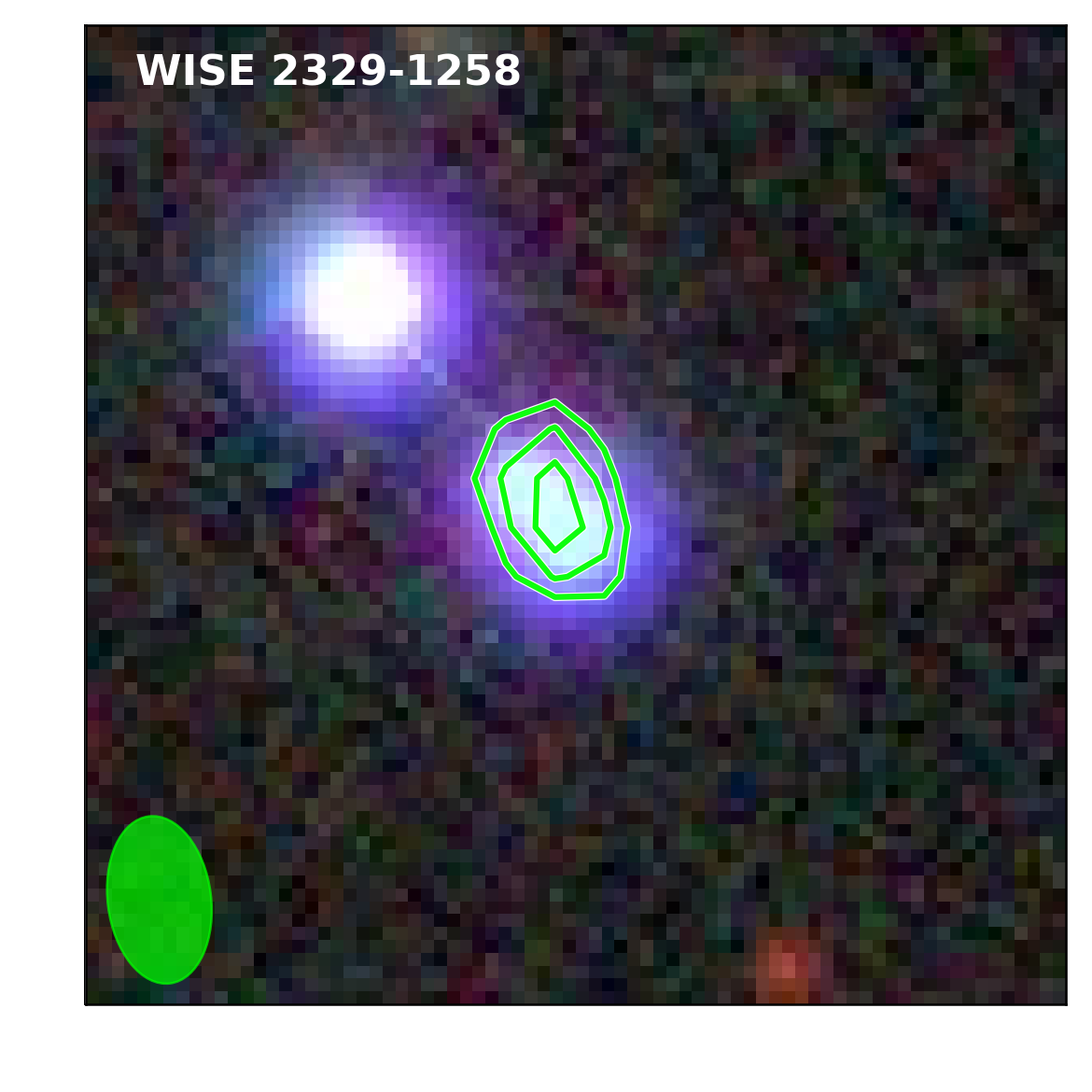}}
    \hspace{1mm}
    \subfigure{\includegraphics[trim={10mm 10mm 0 0}, clip, width=0.23\textwidth]{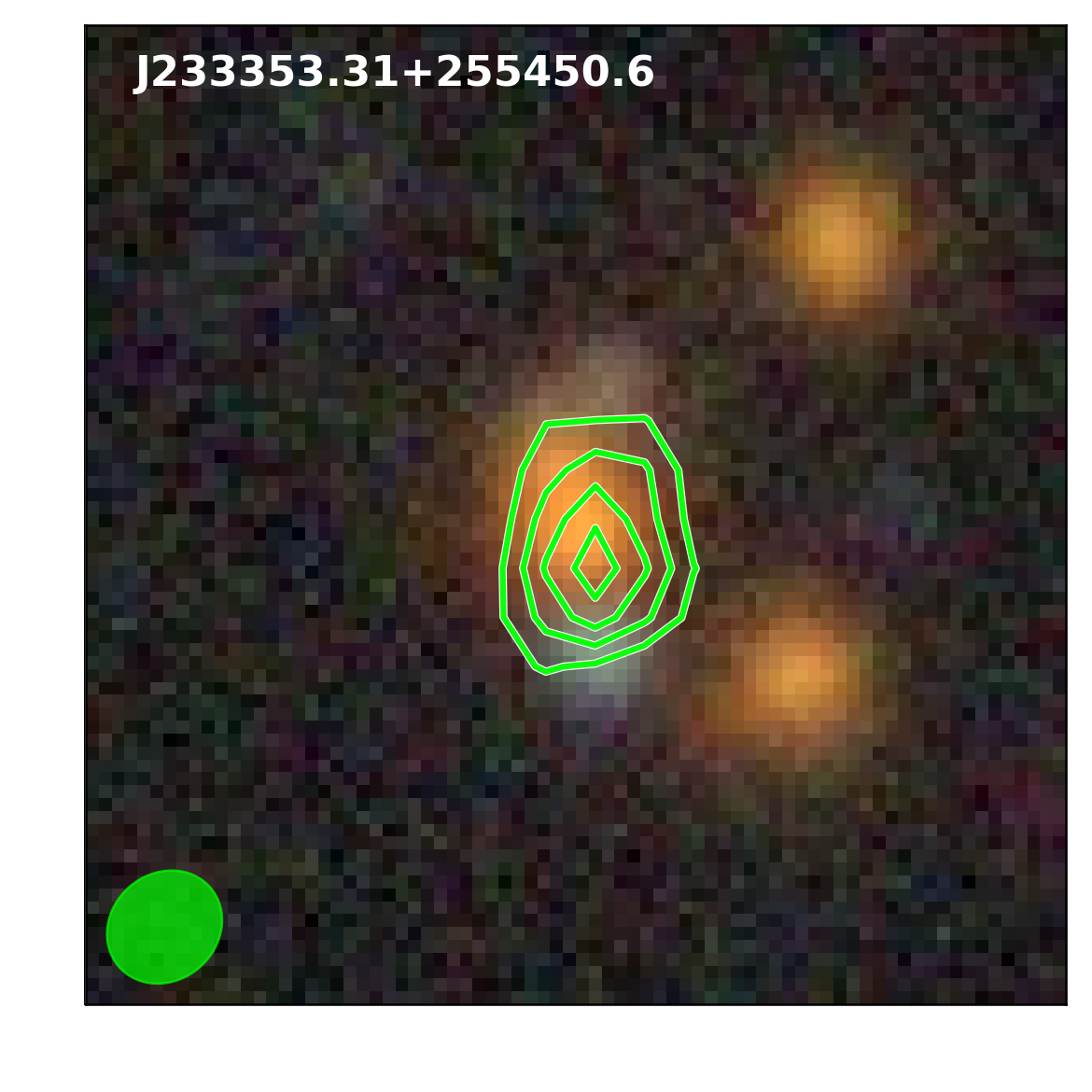}}
    
    \caption[Optical Cutouts of Candidates]{Postage stamp optical cutouts ($20'' \times 20''$) of candidate lensed radio sources with VLASS contours overlaid (green).
    The optical images are three color (grz) images from LS-DR9 except for J0013$+$5119, J1817$+$2729 and J2145$+$6345 where PanSTARRS gri images are used instead.
    Contour levels are the same as in Figure \ref{fig:eg_rejects} except for the fainter radio sources DES J0412$-$2646, J2145$+$6345and WISE 2329$-$1258 where contours increase by $0.3\,\text{mJy}\,\text{beam}^{-1}$.
    The green ellipse in the lower left of each panel shows the VLASS beam.}
    \label{fig:eg_targets}
\end{figure*}

\section{VLA Observations} \label{sec:obs}

Using our two selection approaches and removing those previously known lensed radio sources and those targets for which there are high resolution observations in the VLA archive, we are left with 11 targets.
We show optical images overlaid with VLASS contours in Figure \ref{fig:eg_targets}.
We observe these targets with the Karl G. Jansky Very Large Array (VLA) in A-configuration (VLA Proposal: 23A-249, P.I. Gordon).
Observations were conducted in the X-band using NRAO default correlator setup \texttt{X32f2A}, corresponding to 3-bit sampling and 2 second integration times, and basebands centered at 9 and 11 GHz with 2GHz bandwidth. 
The primary calibrator used was 3C 48 for all targets except J120157.21+421703.4, 135413.22+325937.1, 171527.20+280452.4, and J1817+2729, which used 3C 286.
Table \ref{tab:observations} shows our target list with the VLA integration times and complex gain calibrators used, alongside some of their selection criteria. 
The raw data were calibrated by the NRAO as part of the Science Ready Data Products (SRDP) initiative, which creates calibrated measurement sets optimized for continuum (Stokes I) imaging.

% 

%%%%%%%%%%%%%%%%%%%%%%%%%%%%%%%%%%%%%%%%%%%%%%%%%%%%%%%%%%%%
{\footnotesize
 \onehalfspacing
\begin{ThreePartTable}
\begin{TableNotes}[flushleft]
    \setlength{\labelsep}{0pt}
    \item[a] \label{p1obs-a}{J2145+6345 is not detected in NVSS and is outside the footprint of FIRST. As such we estimate a spectral index limit based on the $2\,$mJy detection limit of NVSS.}
    \item[b] \label{p1obs-b}{The image RMS for J233353+255450 is given for the uv-tapered image, see Section \ref{2333}.}
    \item[] Note. --- This table lists (1) the name of the candidate lensed radio source; (2) the flux density in VLASS epoch 1; (3) the estimated spectral index between $1.4\,$GHz and $3\,$GHz based on measurements from VLASS and either FIRST or NVSS depending on sky location; (4) the time for which we observed the target; (5) the RMS noise of our cleaned image; and (6) the complex gain calibrator for that source. Column (7) notes whether the target was identified from known lensed quasars in Gaia \citep{Lemon19}, lensed galaxies in DES \citep{Jacobs2019} or by applying the method of \citet{JB07} to the VLASS and LS DR9 catalogs.
  \end{TableNotes}
\begin{longtable}[c]{lcccccr}
    \caption[Targets and VLA Observation Details]{Targets and VLA observation details. \label{tab:observations}} \\
\hline \hline
    {Name} & {$S_{\text{VLASS}}$} & {$\alpha_{1.4}^{3}$} & {Time} & {Image RMS} & {Calibrator} & {Method}\\
    {} & {[mJy]} & {} & {[s]} & {[$\mu\text{Jy}/\text{beam}$]} & {} & {} \\
    \hline 
    \endfirsthead
    \caption[]{\textit{(continued)}} \\
    \hline \hline
    {Name} & {$S_{\text{VLASS}}$} & {$\alpha_{1.4}^{3}$} & {Time} & {Image RMS} & {Calibrator} & {Method}\\
    {} & {[mJy]} & {} & {[s]} & {[$\mu\text{Jy}/\text{beam}$]} & {} & {} \\
    \hline
    \endhead
    \hline
    \endfoot
    \hline
    \insertTableNotes
    \endlastfoot
    %\tablecolumns
    J000835.17$-$073405.6 & 3.5 & +0.13 & 80 & 25 & J0006$-$0623 & JB07 method \\
    J0013+5119 & 2.4 & -0.53 & 130 & 21 & J2355+4950 & Gaia lensed QSO \\
    DES J0412$-$2646 & 2.8 & -0.30 & 90 & 30 & J0416$-$1851 & DES lensed galaxy \\
    J120157.21+421703.4 & 4.2 & +0.16 & 90 & 119 & J1146+3958 &  JB07 method \\
    J135413.22+325937.1 & 3.0 & -0.11 & 90 & 107 & J1416+3444 & JB07 method \\
    J171527.20+280452.4 & 2.6 & -0.31 & 100 & 107 & J1753+2848 & JB07 method \\
    J1817+2729 & 3.0 & +0.09 & 90 & 115 & J1753+2848 & Gaia lensed QSO \\
    J2145+6345 & 1.3 & $>$-0.60\tnotex{p1obs-a} & 299 & 15 & J2022+6136 & Gaia lensed QSO \\
    HS B2209+1914 & 2.3 & -0.88 & 219 & 17 & J2212+2355 & Gaia lensed QSO \\
    WISE J2329$-$1258 & 1.2 & -0.96 & 448 & 11 & J2331$-$1556 & Gaia lensed QSO \\
    J233353.31+255450.6 & 4.9 & -0.22 & 75 & 52\tnotex{p1obs-b} & J2340+2641 & JB07 method\\
\end{longtable}
\end{ThreePartTable}
}

We imaged our visibilities using the \texttt{tclean} task in the NRAO's Common Astronomy Software Applications (CASA) suite of processing tools \citep{casa}.
As lensed radio quasar systems tend to be composed mainly of point sources, we used the \citet{hogbom} deconvolution method with single-term multi-frequency synthesis \citep{mfs}.
Given our targets are never more than a few arcseconds across, we did not use any wide-field imaging procedures.
Cleaning was done using an interactive mask, with a stopping threshold of 0.1mJy, which was usually between two and five times the noise floor.
After imaging, model visibilities were examined in order to assess the efficacy of increasing dynamic range via self-calibration \citep{selfcal}, but in each case our snapshot observation signal-to-noise was too low for a useful gain solution.
In a few cases this general imaging procedure was augmented with extra steps as required by the situation, these will be discussed individually in the following section.

%%%%%%%%%%%%%%%%%%%%%%%%%%%%%%%%%%%%%%%%%%%%%%%%%%%%%%%%%%%%
\section{Results} \label{sec:p1results}

Table \ref{tab:photometry} summarizes the targeted VLA observation results, including the position and flux of each detected radio component.
These were calculated with the CASA task \texttt{imfit}, which fits elliptical gaussians to image-plane radio maps.
We opted for image-plane modeling rather than fitting visibilities directly due to the \texttt{imfit} task's ability to simultaneously fit components, which is not possible in CASA's corresponding visibility space modelling algorithm.
As expected for AGN cores, most observed components were fit as point sources with some exceptions noted in the table and discussed below.
Of the 11 targets observed, we found evidence of lensed radio emission in 4 previously known lensed quasar systems. The fifth previously known system studied was not detected but was found to be a lensed radio source by \citet{dobie23}. In 4 other sources, we found unlensed radio emission; we attribute the VLASS emission to either the putative lens galaxy or an unlensed quasar. Another source had no significant detection whatsoever, and the final source is an ambiguous case discussed further in Section \ref{0412}.

\newpage

{\footnotesize
 \onehalfspacing
\begin{ThreePartTable}
\begin{TableNotes}[flushleft]
    \setlength{\labelsep}{0pt}
    \item[a] \label{p1comps-a}This component was fit as an extended source by \texttt{imfit} rather than a point source. 
    \item[] Note. --- Lensed quasar images are indicated by capital letters.
  \end{TableNotes}
\begin{longtable}[c]{lcccccc}
\caption[Position and Flux Measurements of Detected Radio Components.]{Position and flux measurements of detected radio components.} \label{tab:photometry} \\
\hline \hline
 {Target} &  {Component} &  \multicolumn{1}{c}{RA} & {$\sigma$RA} & \multicolumn{1}{c}{Dec} & {$\sigma$Dec} & {Flux Density} \\
 {} &  {} & \multicolumn{1}{c}{[deg]} &  {[mas]} &  \multicolumn{1}{c}{[deg]} &  {[mas]} &  {[$\mu$Jy]} \\
\hline \endfirsthead
\caption[]{\textit{(continued)}} \\
\hline \hline
 {Target} &  {Component} &  \multicolumn{1}{c}{RA} & {$\sigma$RA} & \multicolumn{1}{c}{Dec} & {$\sigma$Dec} & {Flux Density} \\
 {} &  {} & \multicolumn{1}{c}{[deg]} &  {[mas]} &  \multicolumn{1}{c}{[deg]} &  {[mas]} &  {[$\mu$Jy]} \\
\hline
\endhead
\hline
\endfoot
\hline
\insertTableNotes
\endlastfoot
J000835.17-073405.6 & Single Quasar & 2.1465188 & 2 & -7.5683175 & 2 & $2370 \pm 40$ \\
J0013+5119 & A & 3.348415 & 14 & 51.318736 & 10 & $240 \pm 40$ \\
 & B & 3.348112 & 5 & 51.3179497 & 3 & $250 \pm 30$ \\
 & Lens Galaxy & 3.348073 & 4 & 51.3182923 & 3 & $490 \pm 30$ \\
DES J0412-2646 & North & 63.179016 & 11 & -26.775585 & 23 & $160 \pm 40$ \\
 & South\tnotex{p1comps-a} & 63.179 & 26 & -26.77575 & 66 & $260 \pm 90$ \\
J120157.21+421703.4 & Single Quasar & 180.4883875 & 1 & 42.2842668 & 12 & $7000 \pm 200$ \\
J135413.22+325937.1 & Single Quasar & 208.5550522 & 2 & 32.993648 & 4 & $2800 \pm 200$ \\
J171527.20+280452.4 & Not Detected &  &  &  &  &  \\
J1817+2729 & Not Detected &  &  &  &  &  \\
J2145+6345 & A & 326.2717218 & 7 & 63.7613599 & 2 & $430 \pm 30$ \\
 & B & 326.27193 & 10 & 63.76152 & 4 & $250 \pm 30$ \\
 & C & 326.270737 & 25 & 63.761261 & 7 & $130 \pm 30$ \\
HS B2209+1914 & A & 332.876315 & 13 & 19.487111 & 7 & $290 \pm 30$ \\
 & B & 332.876415 & 21 & 19.48684 & 12 & $270 \pm 40$ \\
WISE J2329-1258 & A & 352.49105 & 47 & -12.98315 & 96 & $160 \pm 50$ \\
 & Northeast\tnotex{p1comps-a} & 352.491016 & 30 & -12.98286 & 63 & $80 \pm 20$ \\
 & Southwest\tnotex{p1comps-a} & 352.49132 & 99 & -12.98298 & 223 & $150 \pm 70$ \\
J233353.31+255450.6 & Extended\tnotex{p1comps-a} & 353.47219 & 26 & 25.91395 & 146 & $1700 \pm 200$ \\
\end{longtable}
\end{ThreePartTable}
}

\subsection{Statistical Considerations} \label{statistics}
For our observations, especially those which we claim are indeed radio-loud lenses, we wish to reject the possibility that the radio emission is indeed from the quasar and lens separately, rather than a chance alignment of radio sources and optical ones.
We adopt a frequentist approach based on \citet{SKA} to give the probability each radio detection is associated with its corresponding optical detection.
As it detects all the lensed images of the existing optical lensed quasars, \textit{Gaia} was used as the optical survey in this analysis.
Let $\rho_0$ be the density of optically detected sources, which in the case of \textit{Gaia} DR3 \citep{Gaiadr3} is approximately $45,000\,\text{deg}^{-2}$.
Assuming no correlation between radio and optical, the number of expected optical detections within $r$ seconds of arc of a given radio detection is given by
$\int_0^r \rho_0(2\pi r') dr'$, or $\rho_0(\pi r^2)$. 
As \textit{Gaia}'s astrometric precision is typically less than one milliarcsecond, and our VLA precision (in A-config X-band) is on the order of $20\sim40$ milliarcseconds, a typical value of $r$ is expected to be tens of milliarcseconds for a real match, corresponding to an individual source random probability of between $10^{-5}$ and $10^{-7}$.
By contrast, two unrelated sources separated by $1''$ would give a random probability of closer to 1/100.
The alignment between \textit{Gaia}'s astrometric frame and the International Celestial Reference Frame (as used by the VLA) is consistent to 0.01 mas \citep{Gaiadr3}, and thus our radio-optical alignments are dominated by the peculiar errors of each source in the optical and radio measurements.
We expect for a real radio lens to observe emission from each quasar image, and thus will measure a random probability for each of them.
Multiplying these probabilities together gives an estimated total probability that the radio sources are chance alignments with the lensed optical images, and we will report this number for each claimed radio-loud gravitational lens in the next section.

\begin{figure}
    \centering
    \subfigure{\includegraphics[trim={36mm, 12mm, 32mm, 12mm}, clip, width=0.45\textwidth]{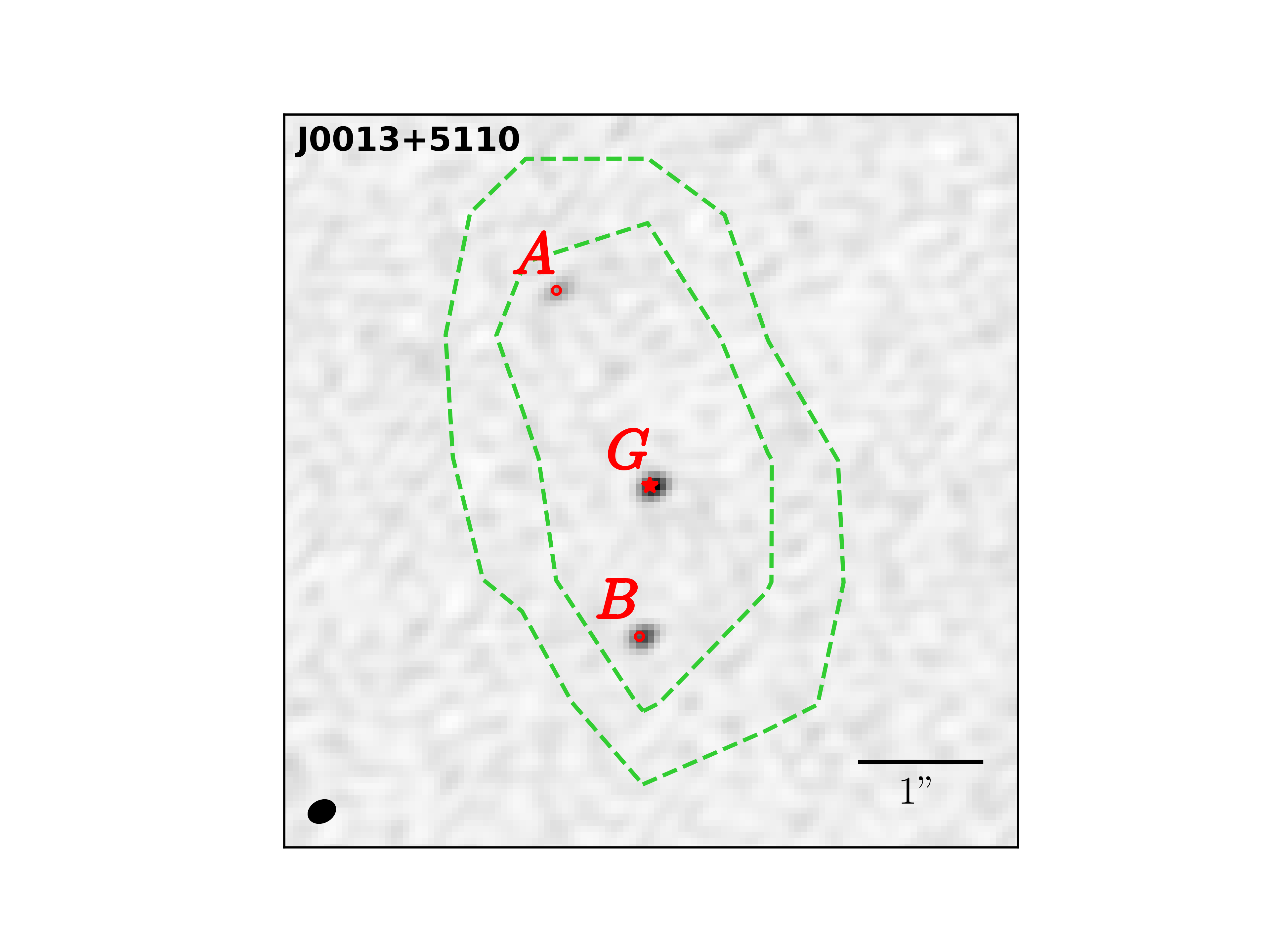}}
    \hspace{4mm}
    \subfigure{\includegraphics[trim={36mm, 12mm, 32mm, 12mm}, clip, width=0.45\textwidth]{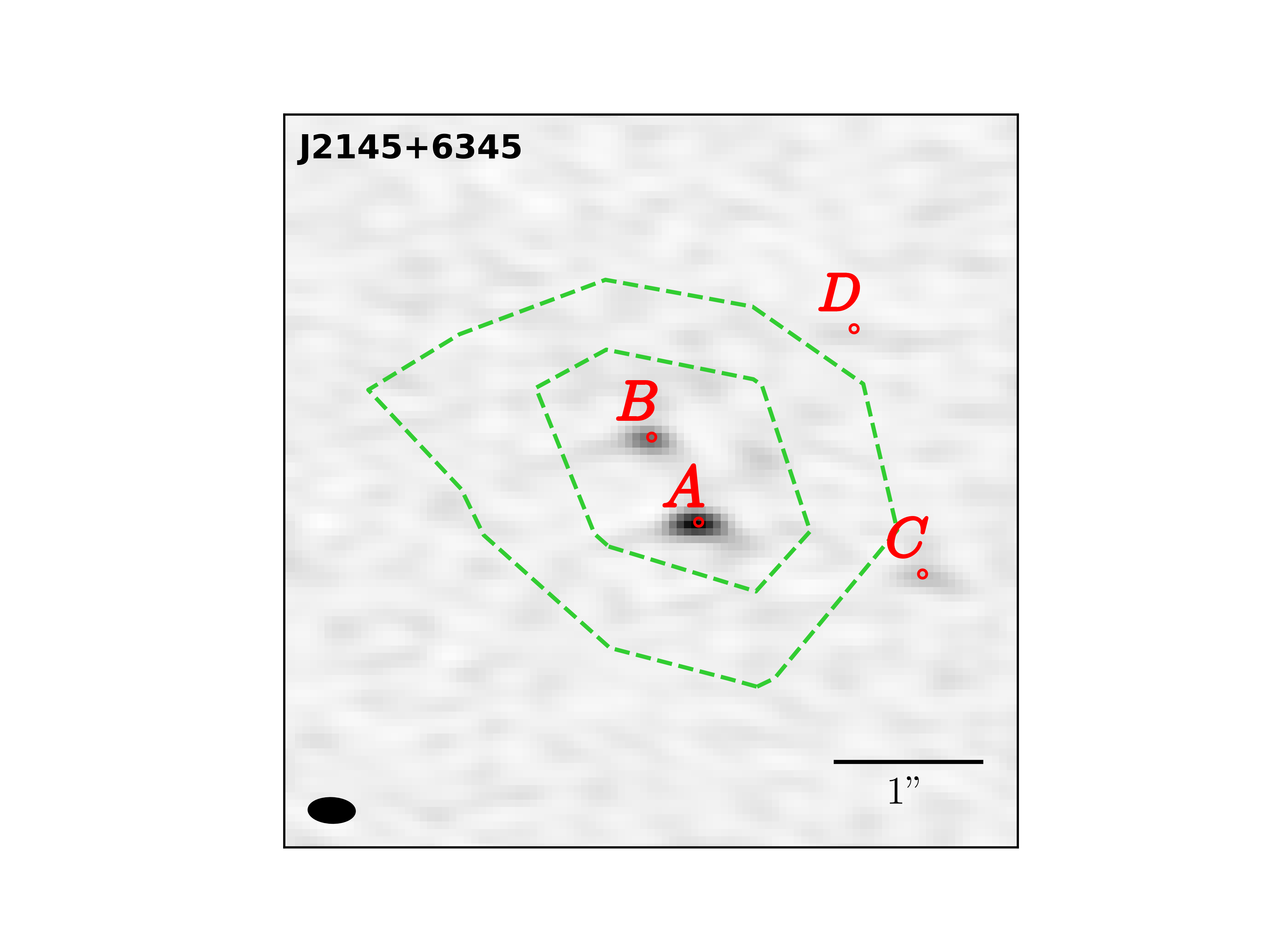}}
    % \hspace{1mm}
    \subfigure{\includegraphics[trim={36mm, 12mm, 32mm, 12mm}, clip, width=0.45\textwidth]{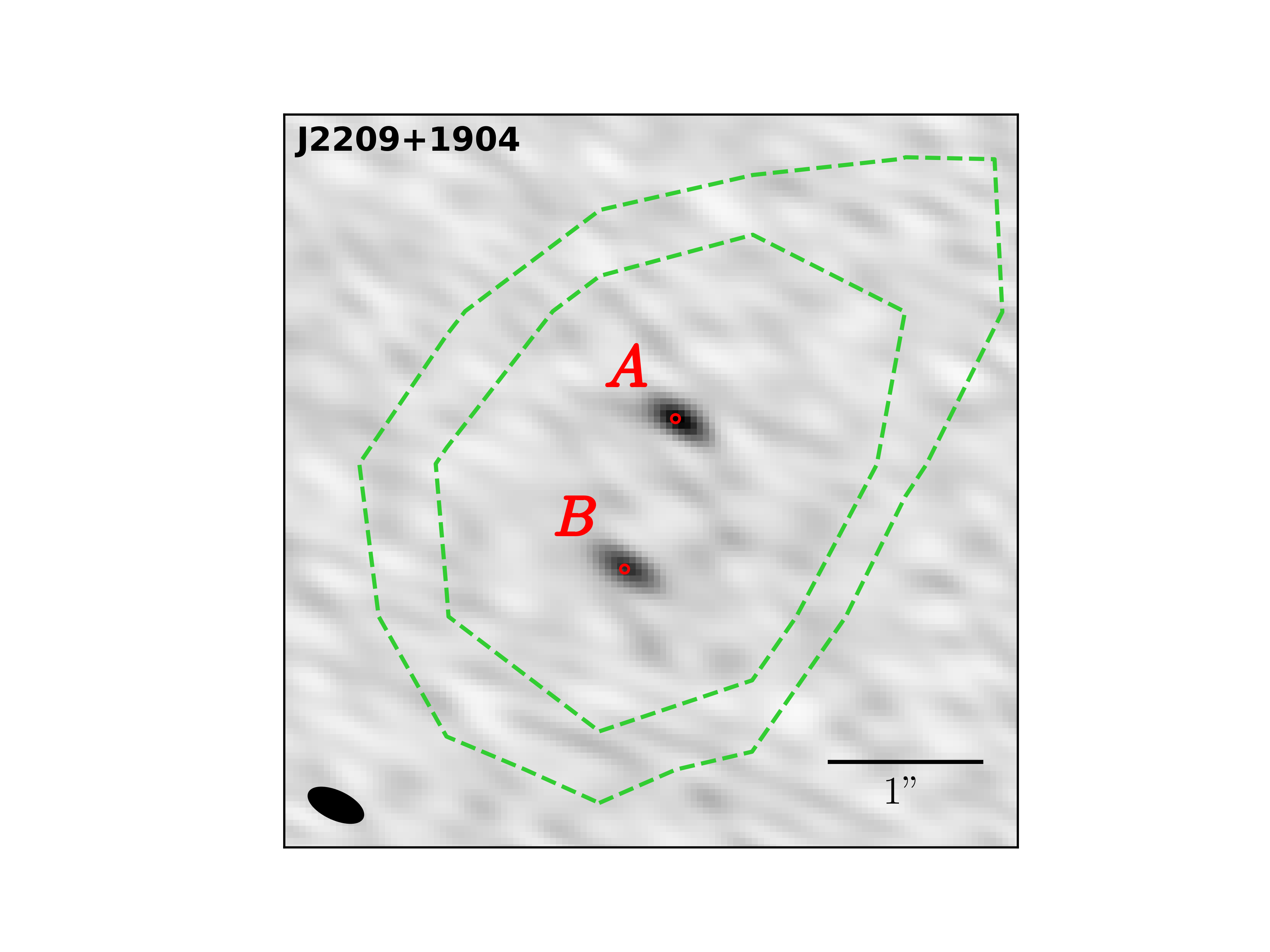}}
    \hspace{4mm}
    \subfigure{\includegraphics[trim={36mm, 12mm, 32mm, 12mm}, clip, width=0.45\textwidth]{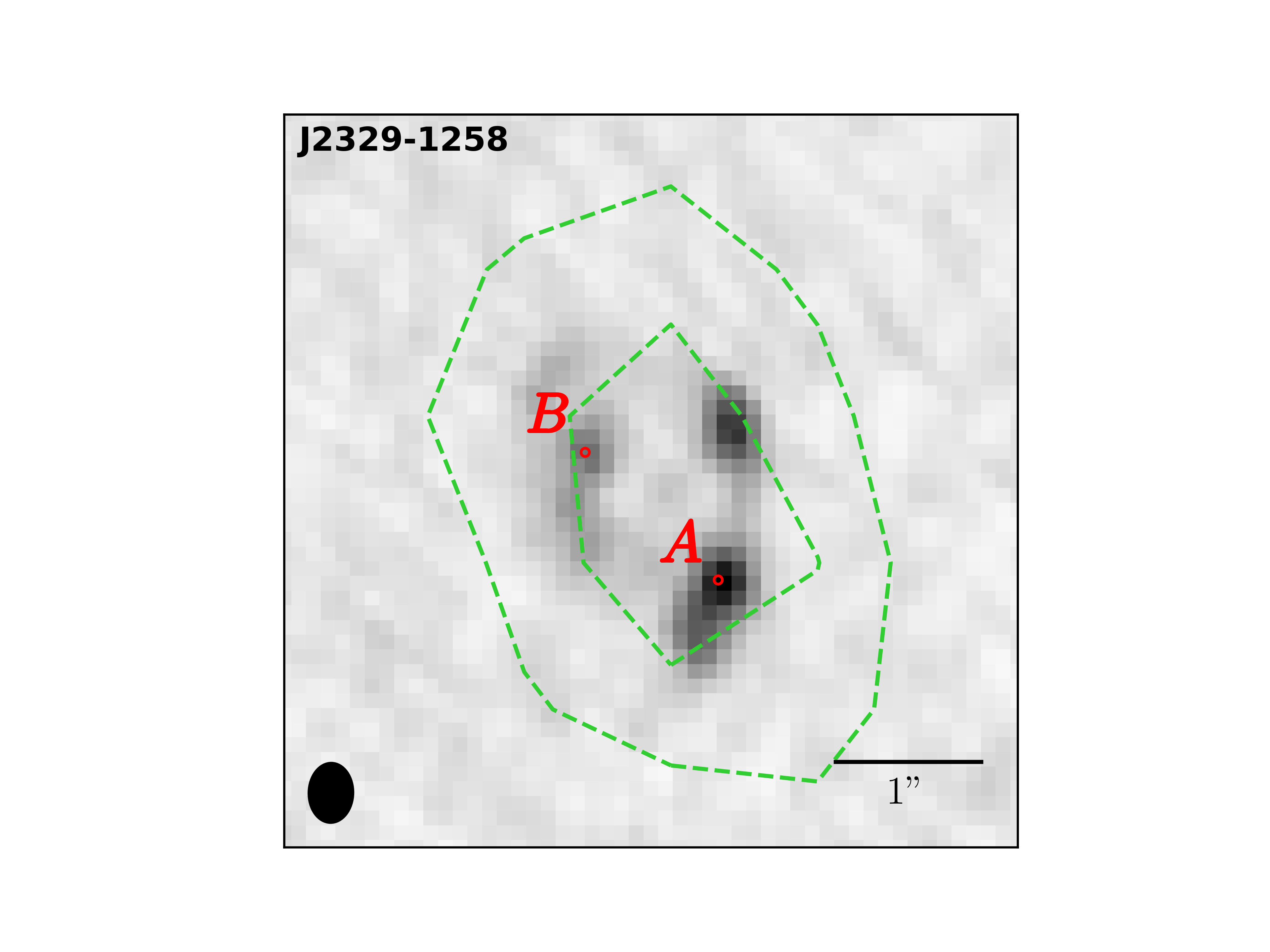}}
    \caption[New Radio Lenses]{Radio images of the four lenses discussed in this paper. Top Left: J0013+5119 (see Section \ref{0013}); Top Right: J2145+6345 (Section \ref{2145}); Bottom Left: HS B2209+1904 (Section \ref{2209}); Bottom Right: J2329$-$1258 (Section \ref{2329}). The final image was constructed using additional archival data as described in the text. These systems are not resolved in the VLASS epoch one quick-look images, as indicated by the green contours (3- and 5-$\sigma$ levels shown). The targeted VLA observations show multiple source images coinciding with \textit{Gaia} positions, shown in red. For the source J0013+5119, the PanSTARRS position of the lens galaxy is also shown.}
    \label{fig:lensfig}
\end{figure}

\subsection{New Radio-Loud Lenses}

The radio loud lenses presented below are displayed in Figure \ref{fig:lensfig}.

\subsubsection{J0013+5119} \label{0013}
J0013+5119 was discovered as a doubly imaged quasar by \citet{Lemon19} using the Wide-field Infrared Survey Explorer (WISE) and \textit{Gaia} DR2 catalogues. The quasar source is at redshift $z=2.63$ and the two images are separated by $2''.92$.
Our VLA A-config observations revealed radio emission from both the lens and the quasar images, which are all fit to point sources in X-band using the CASA task \texttt{imfit} but appear as one source in VLASS.
The flux densities for the lensed images $A$ and $B$ taken from \texttt{imfit} give a flux ratio $ A/B = 1.0 \pm 0.2$, similar to the optical flux ratio of $1.2$ based on the \textit{Gaia} \text{$g$-band} measurements.
The lens galaxy was not detected in \textit{Gaia} and was too blended with quasar light to get a precise position measurement in any other available optical survey, so our probability consideration only includes the two measured quasar positions.
The system probability of random coincidence is then $3 \times 10^{-11}$.

As this is the only one of our new lenses to not have a published lens model, we made an effort to provide one in this paper. 
Using the Lenstronomy \citep{lenstronomy, lenstronomy2} software suite, we fit a simple Singular Isothermal Ellipsoid \citep{SIE} model with external shear to both VLA data and data from the PanSTARRS 1 survey  \citep[PS1,][]{Chambers2016}.
However, when testing our best-fit results from this method, we found the source plane positions of images $A$ and $B$ did not match, i.e. the model was not accurately reproducing observations.
We suspect this is due to the environment of the lens, and examining wider-field survey images of the J0013+5119 system show other galaxies of similar redshift in the vicinity of the lens, which could lead to a more complex lens model.
Modeling such a lens system would require deeper and sharper optical data and is beyond the scope of this paper.

\subsubsection{J2145+6345} \label{2145}
Quad lens J2145+6345 was also discovered by \citet{Lemon19} using the same method as J0013+5100, and was singled out by the authors as being ideal for time-delay studies given its reasonably large image separation (a max of $2".07$) and bright images. The quasar is located at $z=1.56$, and \citet{Lemon19} report no detection of a lens galaxy in the PanSTARRS survey.

We significantly detected the three brightest images of J2145+6345 in X-band as point sources, and also detected a noise bump coincident with the \textit{Gaia} position of the fourth quasar image. In VLASS, the system is blurred together into one component.
Excluding the faintest image, which was not significantly detected, we obtained a system chance of random of $3\times10^{-15}$.
We calculated the flux ratios between our significantly detected images as $A/B = 1.7 \pm 0.2$ and $A/C = 3.3 \pm 0.5$.
These radio flux ratios do not differ significantly from the \textit{Gaia} \text{$g$-band} flux ratios of $A/B = 1.4$ and $A/C = 3.9$.

\subsubsection{HS B2209+1904} \label{2209}

B2209+1904 (aka J2211+1929), a doubly-imaged quasar at $z=1.07$, was catalogued, along with its lens galaxy, in the Hamburg Quasar Survey \citep{hamburg}.
Our X-band observations detected both quasar images as point sources with a flux ratio of $A/B = 1.1 \pm 0.2$.
This is slightly, but not significantly, lower than the optical flux ratio observed by \textit{Gaia} in the \text{$g$-band} of $1.5$.
The chance of two random radio sources being in these positions is $7\times 10^{-12}$.

\subsubsection{J1817+2729} \label{1817}

J1817+2729, a quadruply imaged source at $z=3.07$ \citep{Lemon19}, was discovered by \citet{1817} using a blind catalog search in \textit{Gaia} DR2. 
Despite a strong detection in VLASS, our X-band observations report no significant emission at $10\,$GHz, and a manual re-reduction of the data showed the same.
Our radio map of this source is shown in the bottom panel of Figure \ref{fig:nondetection}.
Fortunately, the target was also observed by \citet{dobie23} in C band (6 GHz), and was confirmed as a lensed radio source therein.
J1817+2729 shows no variability between epochs 1 (May 2019) and 2 (Sept 2021) of VLASS, so we assume no significant variability for the source. From VLASS epoch 1 and the summed flux densities of all images in the C-band by \citet{dobie23}, we estimate a spectral index between 3 GHz and 6 GHz of $\alpha_{3\,\text{GHz}}^{6\,\text{GHz}} = -1.6\pm0.2$, substantially steeper than the relatively flat spectrum estimated from NVSS and VLASS.
This may be the result of genuine spectral curvature\textemdash for instance the spectral index between $1.4\,$GHz and $3\,$GHz might be capturing the spectral turnover of a peaked spectrum radio source \citep[e.g.,][]{ODea2021}. 
Extrapolating the C-band flux density to X-band using $\alpha_{3\,\text{GHz}}^{6\,\text{GHz}}$, we would expect the sum of the lensed images to have $S_{10\,\text{GHz}} \approx 440\,\mu$Jy. With the distribution of image brightness reported in \citet{dobie23}, we would expect the brightest lensed image to have a 10 GHz flux density of $\approx 230\,\mu$Jy, corresponding to a $<2\sigma$ detection in our image.
We conclude that our observations were simply not sensitive enough to detect the lensed images in X-band, a consequence of estimating the required integration time based on a lower-frequency spectral index and assuming no spectral curvature.

\begin{figure*}
    \centering
    \subfigure{\includegraphics[trim = {36mm, 12mm, 32mm, 12mm}, clip, width=0.30\textwidth]{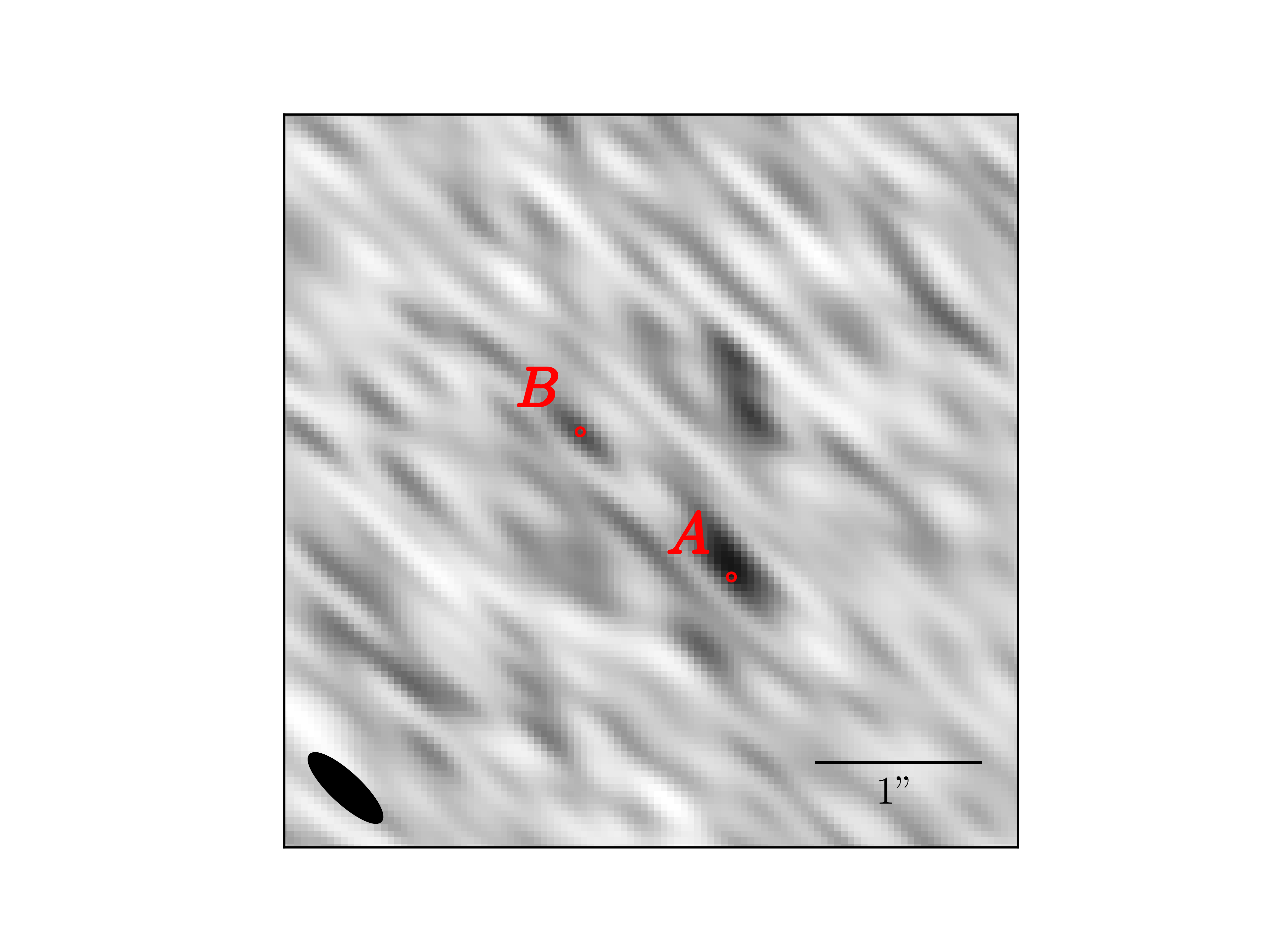}}
    \hspace{1mm}
    \subfigure{\includegraphics[trim = {36mm, 12mm, 32mm, 12mm}, clip, width=0.30\textwidth]{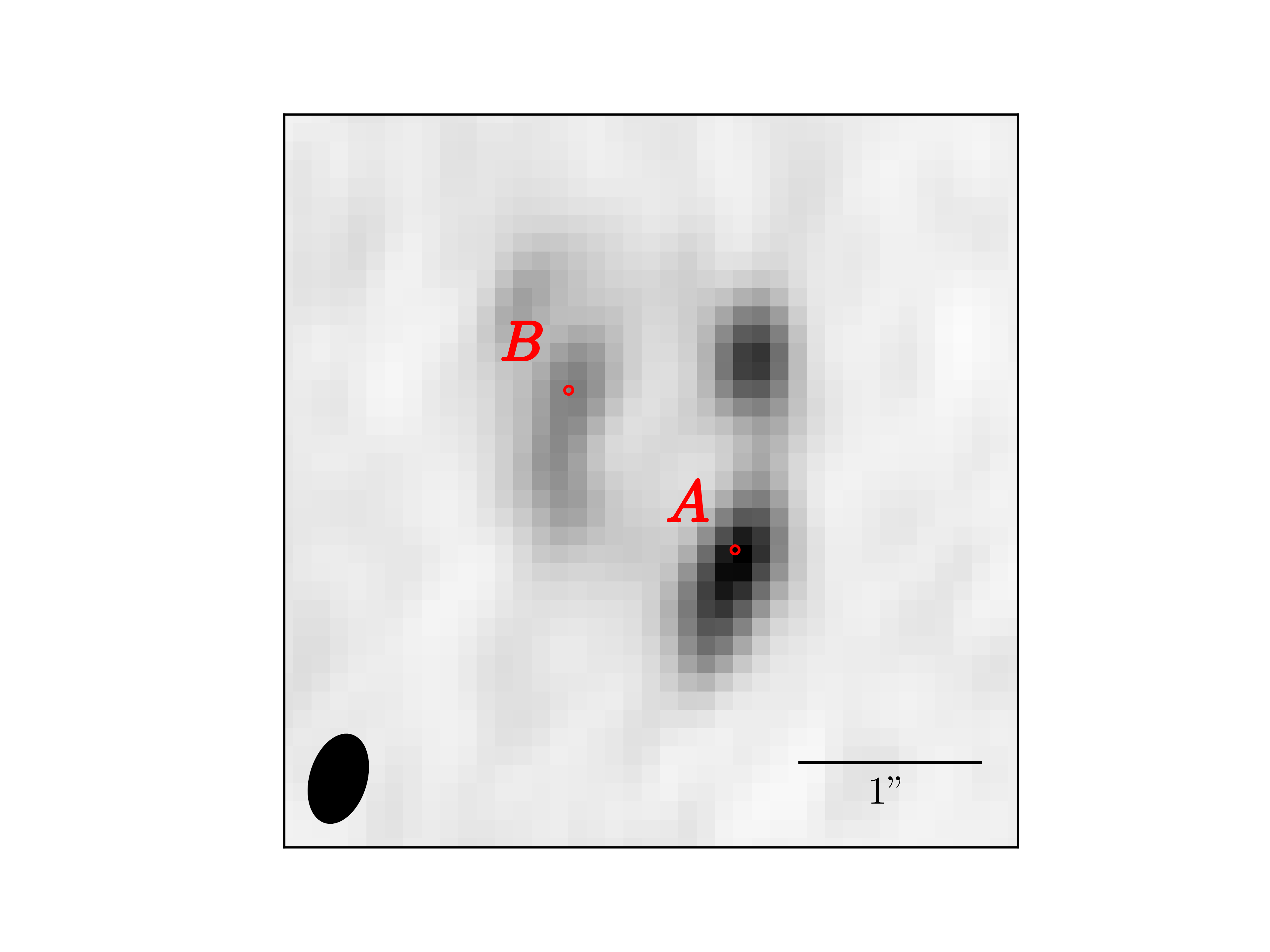}}
    \hspace{1mm}    
    \subfigure{\includegraphics[trim = {29mm, 12.5mm, 25mm, 13mm}, clip, width=0.30\textwidth]{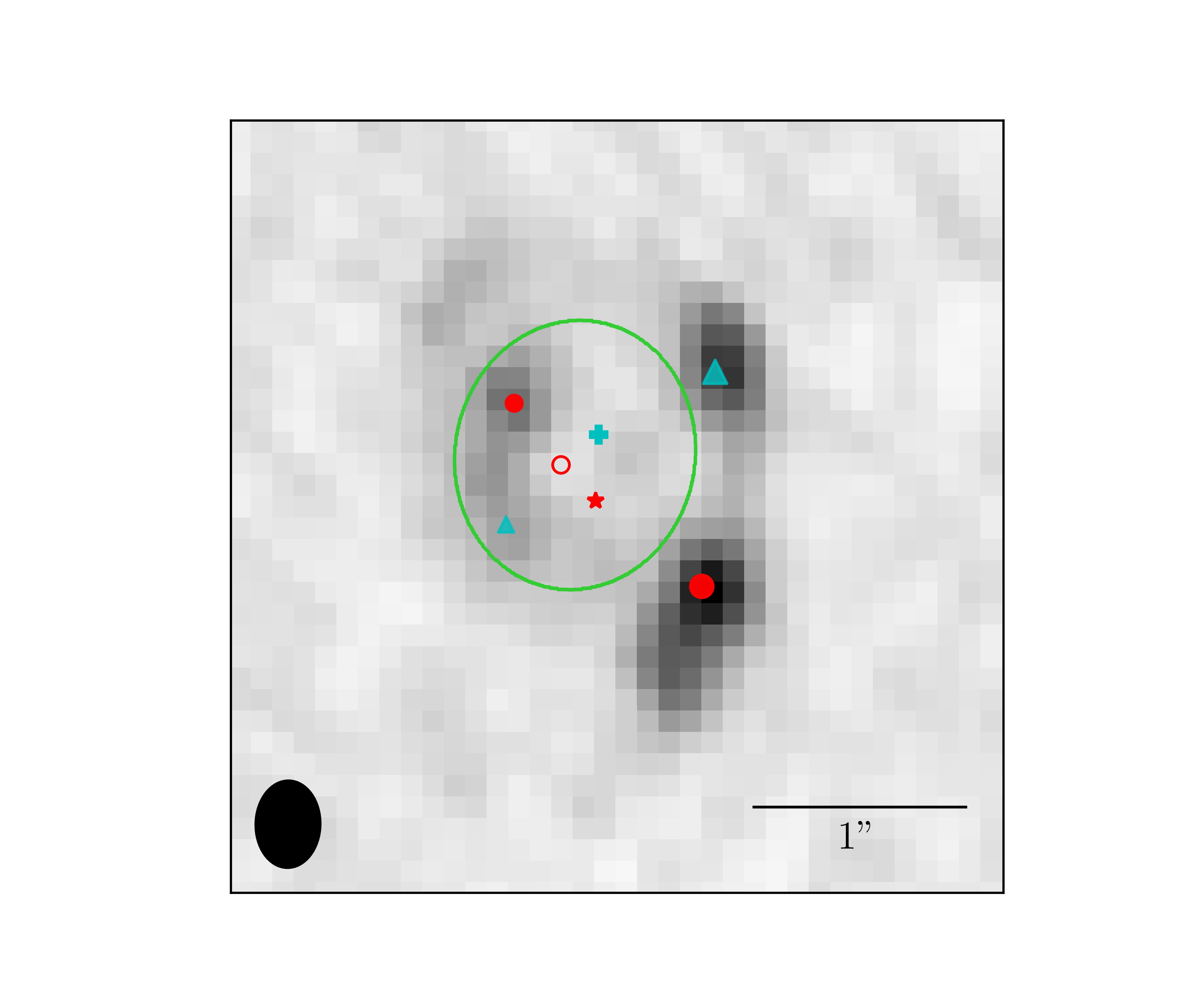}}
    \caption[J2329$-$1258 Processing Details]{Left: VLA X-band observation of J2329$-$1258, with \textit{Gaia} positions of quasar images A and B overlaid. 
    Center: The same system in combined S-band and C-band, from VLA project 19A-176 (P.I. Mao). 
    Right: Stacked $2-12\,GHz$ image of J2329$-$1258 with the optical-based lens model of \citet{shajib21}. The model's critical curve is shown in green, and pairs of features which correspond according to the lens model are shown, with the quasar images still in red circles and the extended emission in cyan triangles. Source positions for the quasar and extended emission are shown as a star and plus, respectively.}
    \label{fig:2329lensmodel}
\end{figure*}

\subsubsection{J2329$-$1258} \label{2329}

J2329-0734 was discovered by \citet{2329disc}, who used a WISE $W1 - W2$ color cut to select potential blended quasar pairs and crossmatched with the ATLAS survey.
Candidates were checked for consistency in putative image colors and visually inspected before spectroscopic follow-up, which confirmed this object as a lensed quasar at $z=1.31$.
Our X-band observations detected the brighter image, as well as extended emission coming from just above that image.
We detected a noise bump at the position of the other quasar image, but our image fitting procedure favored extended rather than point-source emission at this location.
We do not attempt to set a limit on the flux ratio in this system due to this extended emission.
To further investigate the nature of this source, we turned to archival data from VLA project 19A-176 (P.I. Mao), who obtained {A-configuration} observations of the object in the S and C bands.
{These} data also shows pointlike features at the quasar image location and even more diffuse emission than the X-band data.
A 2-term Multi-Frequency Synthesis \citep{mfs} image created from visibility-space stacking both our data and the archival data is shown in the bottom {right} panel of Figure \ref{fig:lensfig}.
Due to our only matching one quasar image, our statistical chance of random coincidence from our observations is much lower, at only $8\times 10^{-5}$.

\begin{figure}
    \centering
    \includegraphics[trim={70pt, 36pt, 56pt, 38pt}, clip, width=\textwidth]{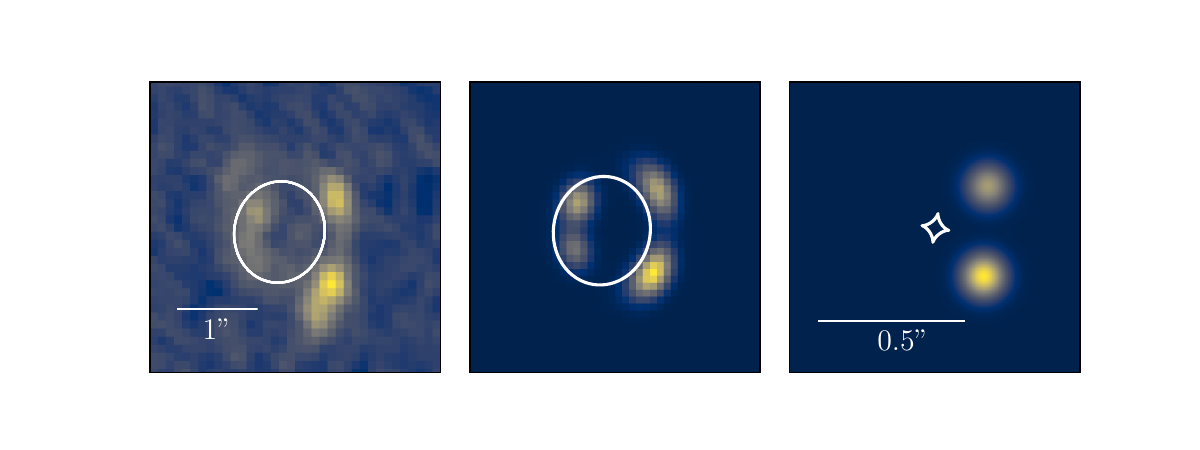}
    \caption[J2329$-$1258 source schematic]{A simplified possible source plane flux distribution for the source J$2329-1258$. Top Left: VLA image of the source as shown in Figures \ref{fig:lensfig} and \ref{fig:2329lensmodel}. The lensing critical curve from \citep{shajib21} is shown in white. Top Right: Simplified lensed flux distribution created with \texttt{Lenstronomy} and convolved to the beam of our observations. Bottom: Unconvolved source plane flux distribution corresponding to the simplified lens flux distribution. The caustic is shown in white.}
    \label{fig:schematic}
\end{figure}

To further investigate the nature of the extended emission present in this lens system, we utilized a lens model created by \citet{shajib21}. This model, constructed using $K$-band ($2.2\,\mu m$) Adaptive Optics observations on the Keck Telescope's NIRC2 instrument, only incorporates near-infrared data and thus is an independent test for our radio observations. 
Figure \ref{fig:2329lensmodel} shows our X-band data, the archival S and C band data, and their combination, as well as the critical curve of \citet{shajib21}'s lens model.
We propagate the locations of image $A$ and the northeast extended component through the lens model and plot their predicted positions.
Image $A$'s counterpart is located at image $B$, as expected, and the northeast component's counterimage is predicted to appear at the location of the southwest component.  
A schematic representation of how J2329$-$1258 could appear in the source plane is shown in Figure \ref{fig:schematic}. We note that this is not a fitted source plane reconstruction and is intended only as an illustration of one possible source-plane flux distribution.
This extended lensed emission may be useful for a gravitational imaging analysis similar to that of \citet{2019MNRAS.483.2125S} with VLBI follow-up. 
However, given the faintness of this source, such an analysis may not be possible without the enhanced sensitivity of the next generation of radio telescopes \citep[priv. comm.]{mckeancomm}.

\subsection{Non-Lensing Results}

\begin{figure*}
    \centering
    \subfigure{\includegraphics[trim = {36mm, 12mm, 33mm, 12mm}, clip, width=0.24\textwidth]{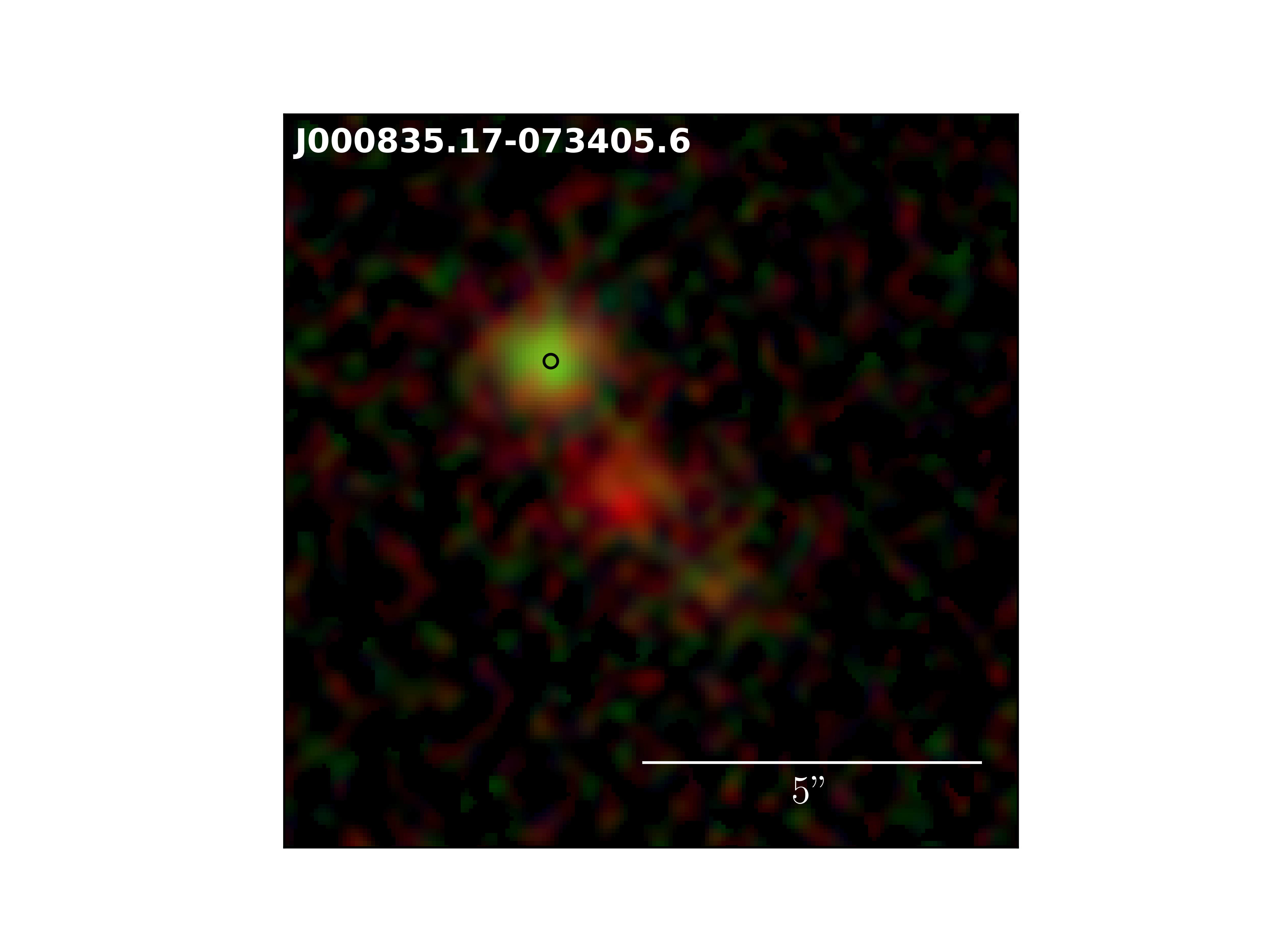}}
   % \hspace{4mm}
    \subfigure{\includegraphics[trim = {36mm, 12mm, 33mm, 12mm}, clip, width=0.24\textwidth]{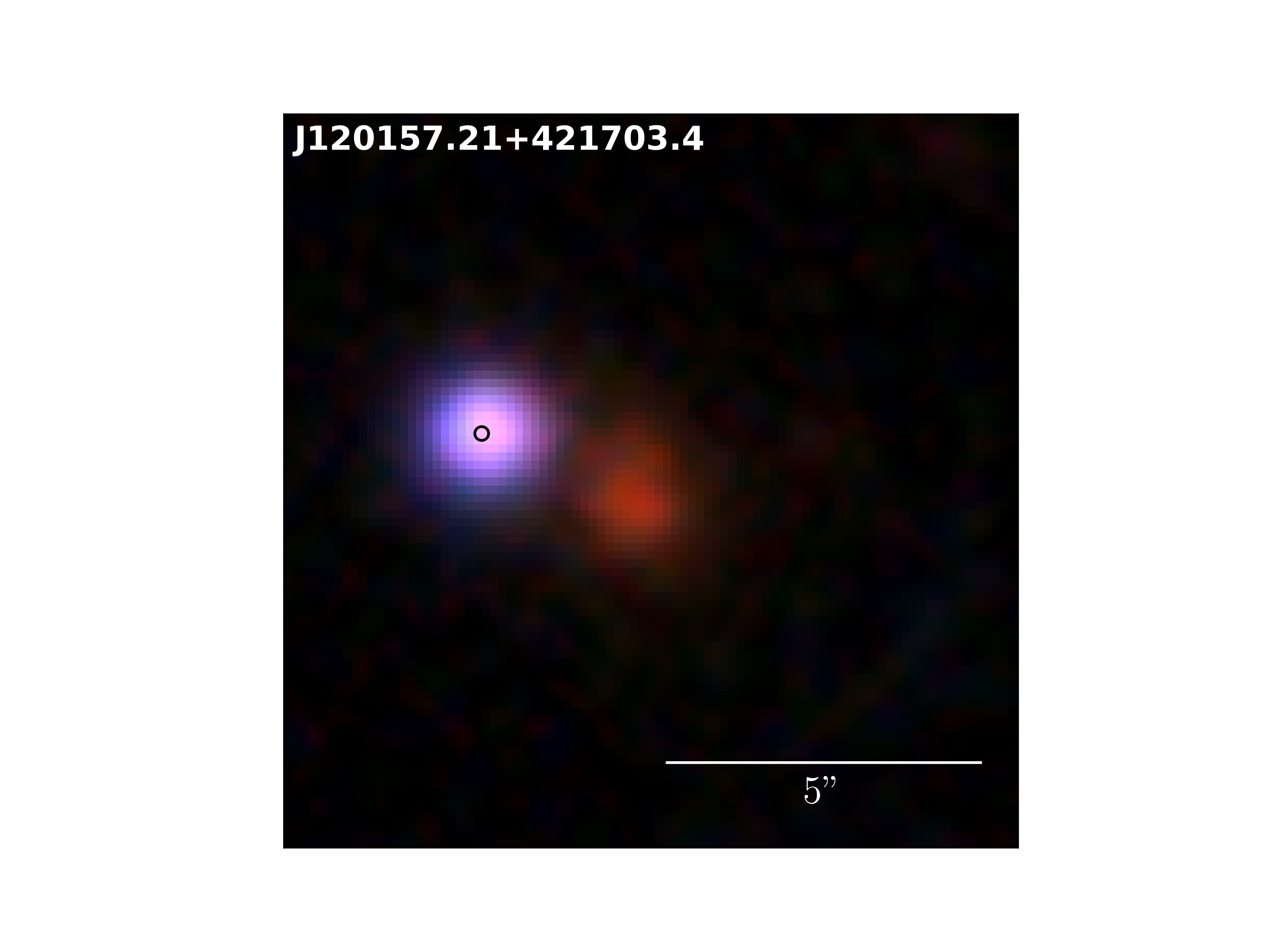}}
    \subfigure{\includegraphics[trim = {36mm, 12mm, 33mm, 12mm}, clip, width=0.24\textwidth]{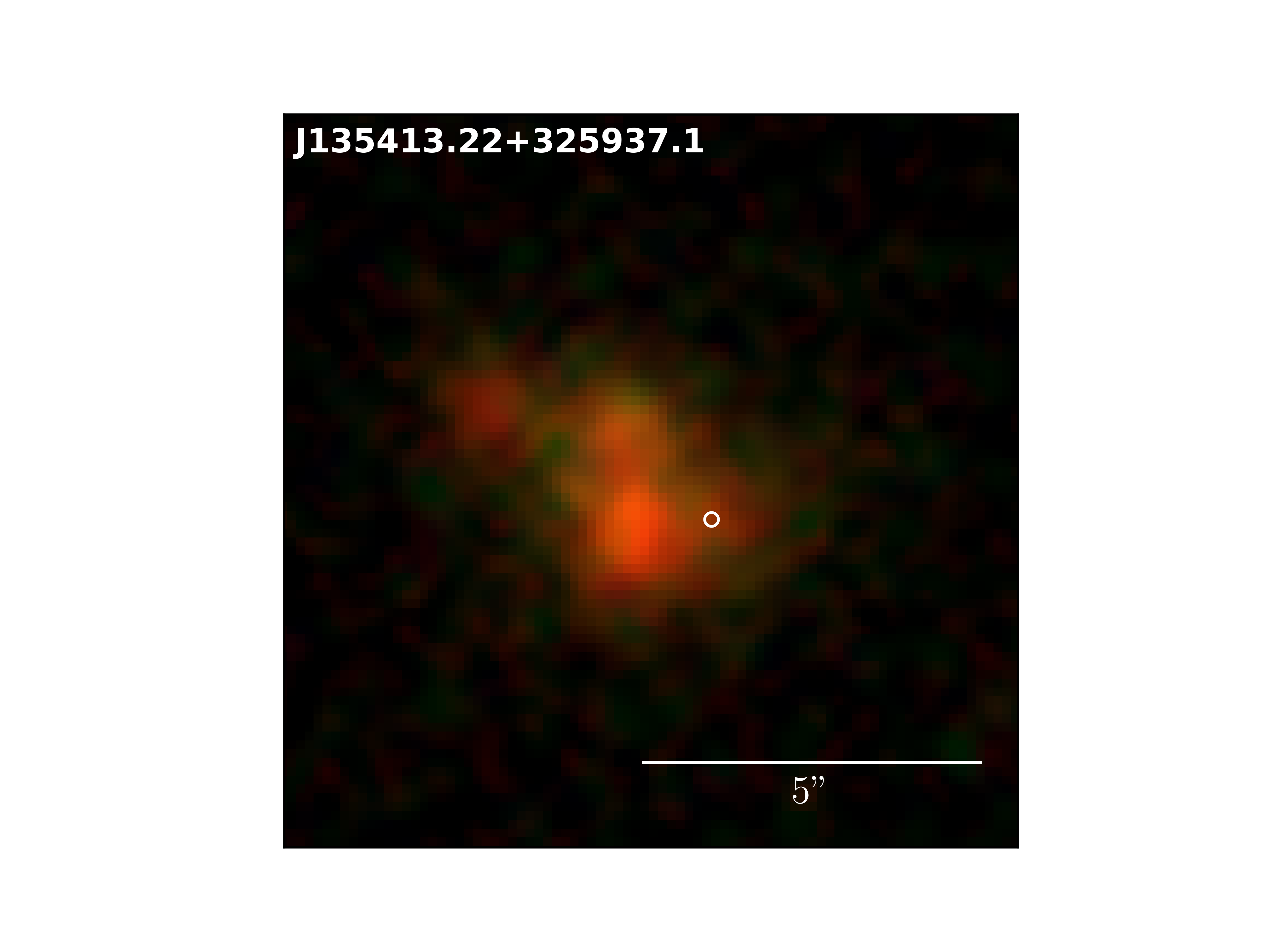}}
   % \hspace{4mm}
    \subfigure{\includegraphics[trim = {36mm, 12mm, 33mm, 12mm}, clip, width=0.24\textwidth]{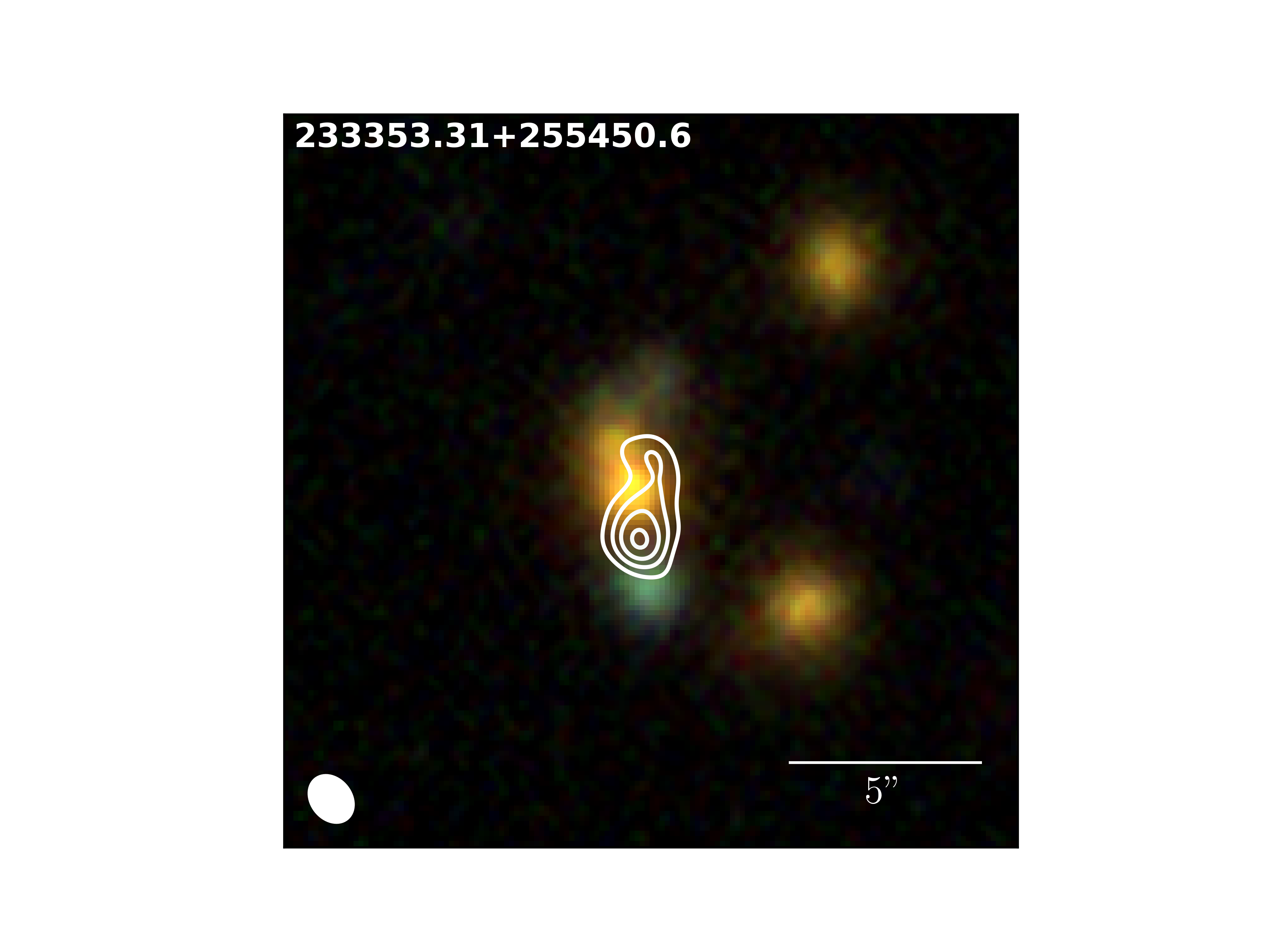}}
    \caption[Non-lensing Results]{Non-Lensing Targets. Top row, left to right: J0008$-$0734 (Section \ref{0008}),  J120157+421703 (Section \ref{120157}).
    Bottom row, left to right: J135413+325937 (Section \ref{135413}), J233353+255450 (Section \ref{2333}). 
    The location of the point-source radio detection is shown by a circle for all panels except for J233353+255450 where we detected extended emission. For J2333353+255450 we show radio contours {corresponding} to 3-,5-,7-, and 9-$\sigma$ flux densities in the uv-tapered radio image.}
    \label{fig:nonlens}
\end{figure*}

In this subsection we discuss the observations of targets that aren't consistent with the radio source being lensed, with those objects shown in Figure \ref{fig:nonlens}.

\subsubsection{J0008$-$0734} \label{0008}

This source was identified as a potential radio lens using the JB07 method, and was singled out for observation due to the relatively bright VLASS detection, green color ($g-r=0.79$, $r-z=0.36$) of the potential source, and possible counterimage in DECaLS. 
However, our X-band follow-up revealed only a $2.37\,$mJy point source coincident with the optical quasar, $2.39''$ away from the lens, and no counterimage.
Our VLA observations of this target have an rms noise of $25\,\mu\text{Jy}\,\text{beam}^{-1}$, and at the $5\sigma$ level we should be sensitive to point sources brighter than $125\,\mu$Jy.
That we detect no radio counterimage suggest that if there were such a counterimage, the flux ratio of the lensed radio source would be a seemingly unrealistic $>20$.
Moreover, the optical flux ratio of the sources immediately north-east and south-west of the the LRG is $\approx8$, so should this be a lensed source then there would be a substantial discrepancy between the optical and radio flux ratios.
While it is not impossible that this source is a lensed quasar, our observations don't support such a conclusion, and we posit that these are likely two unrelated sources.

\subsubsection{J120157+421703} \label{120157}

This source was identified as a possible radio lens using the JB07 method.
The DECaLS image of this source shows a possible very faint arc to the lower right of the LRG. 
The VLA X-band data showed a $7.0$ mJy point source coincident with the optical point source from DECaLS, $2".55'$ from the lens, but no counterimage.
This presents two possibilities when taking the possible arc into account: either the arc is simply an image artifact or other phenomenon and there is no lensing present at all, or the quasar is at or near the lens redshift and is therefore not strongly lensed.

\subsubsection{J135413+326937} \label{135413}

This source was identified as a potential radio lens using the JB07 method.
The X-band observations show a $2.8$ mJy point source offset $1".18$ from the LRG and coincident with the VLASS detection, but no counterimage. 

\subsubsection{J233353+255450} \label{2333}

This source was identified as a possible radio lens using the JB07 method.
Our initial X-band data reduction showed hints of extended emission near the lens location, and so we re-imaged the data with a $1''$ $uv$-plane taper to increase sensitivity at the cost of resolution.
We found an extended $1.7\,$mJy source located between the supposed lens and source, which we interpret as a radio lobe from the LRG rather than a lensed radio source, a hypothesis that is consistent with the steep spectrum ($\alpha=-0.8$) we measure from the VLASS and X-band flux densities.

\subsection{Other Results}

\begin{figure}
    \centering
    \subfigure{\includegraphics[trim = {36mm, 12mm, 32mm, 12mm}, clip, width=0.4\textwidth]{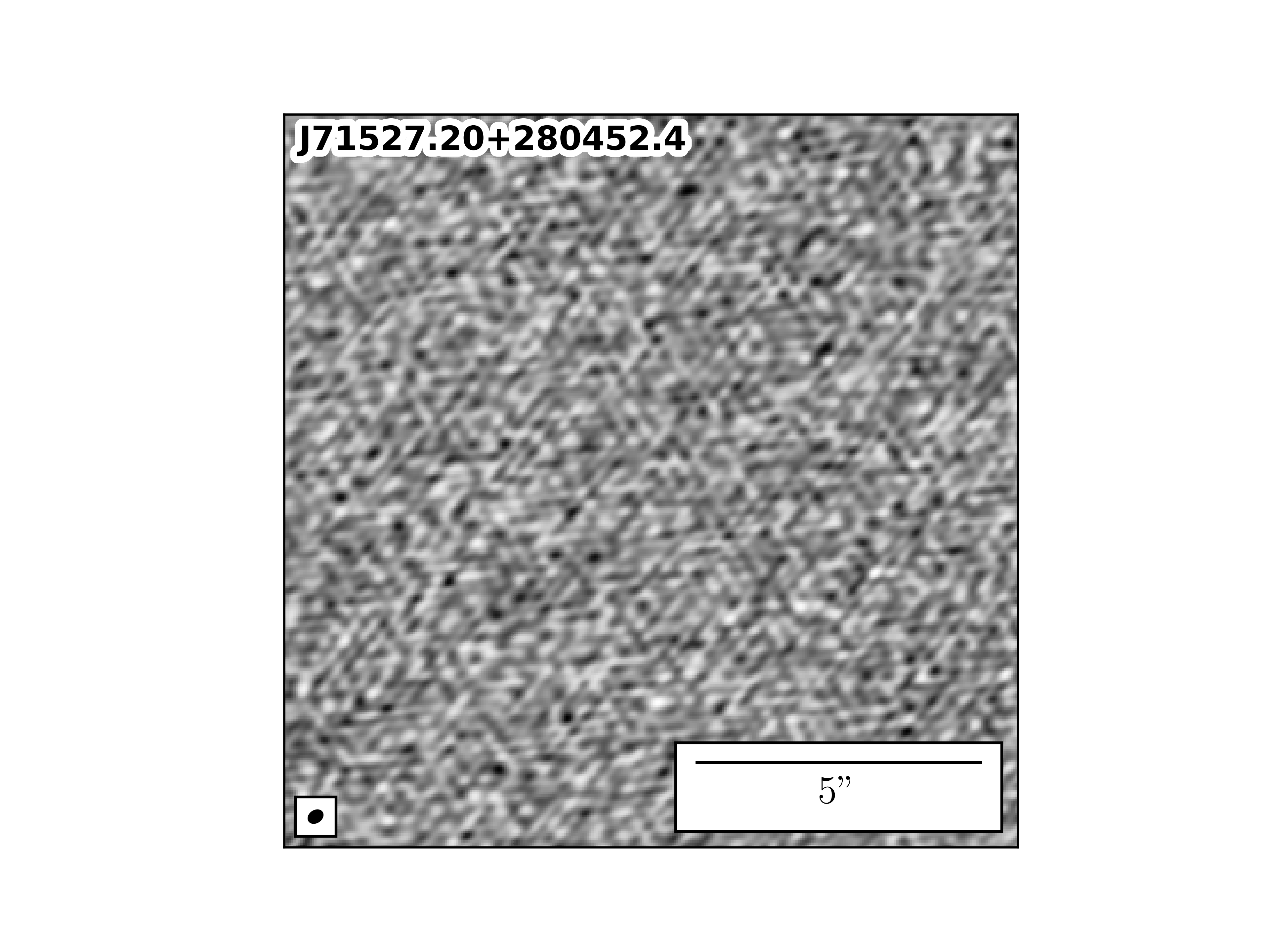}}
    \subfigure{\includegraphics[trim = {36mm, 12mm, 32mm, 12mm}, clip, width=0.4\textwidth]{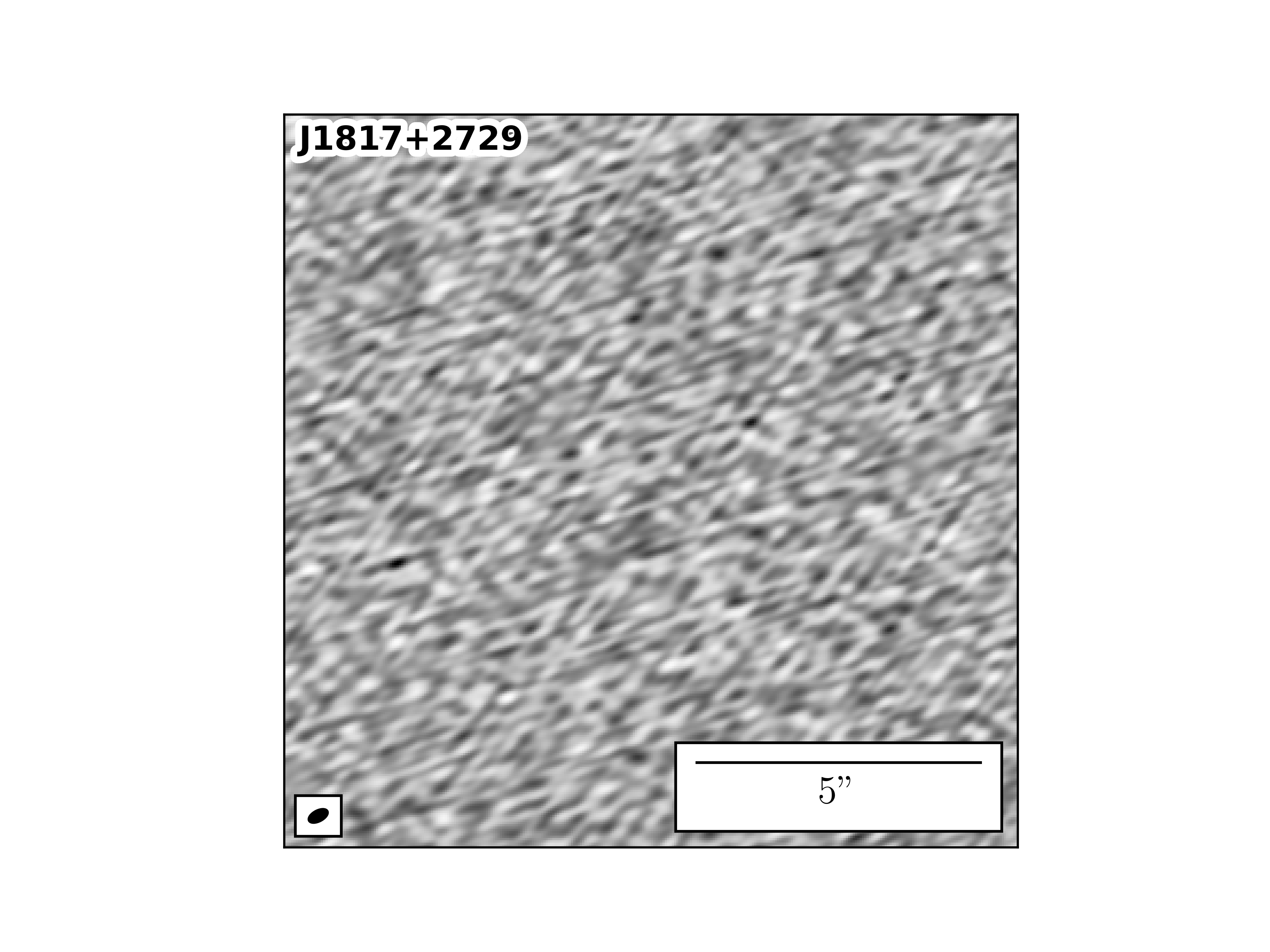}}
    \caption[Non-Detections]{The two non-detections from our observations. Top: J171527+280452 (Section \ref{1715}), and Bottom: J1817+2729 (Section \ref{1817}).
    }
    \label{fig:nondetection}
\end{figure}

\subsubsection{J171527+280452} \label{1715}

This source was identified as a possible radio lens using the JB07 method. 
However, we detected no significant emission in our X-band observations, and the radio map of this source is shown in the bottom panel of Figure \ref{fig:nondetection}.
Based on the $2.6$mJy VLASS epoch 1 flux of the source and a $3\sigma$ nondetection threshold, we estimate a spectral index between 3 and 10 GHz for this source of -2.2, much higher than its VLASS-NVSS spectral index of -0.31.
The source shows no significant variability between VLASS epochs 1 and 2, leading us to suspect the target is either a peaked-spectrum compact source which is undetected at 10GHz, or an extended source which we do not detect due to resolution or sensitivity.
In either case, we cannot rule out the possibility of lensing.

\subsubsection{DES J0412$-$2646} \label{0412}

\begin{figure}
    \centering
    \includegraphics[trim = {36mm, 12mm, 33mm, 12mm}, clip, width=0.45\columnwidth]{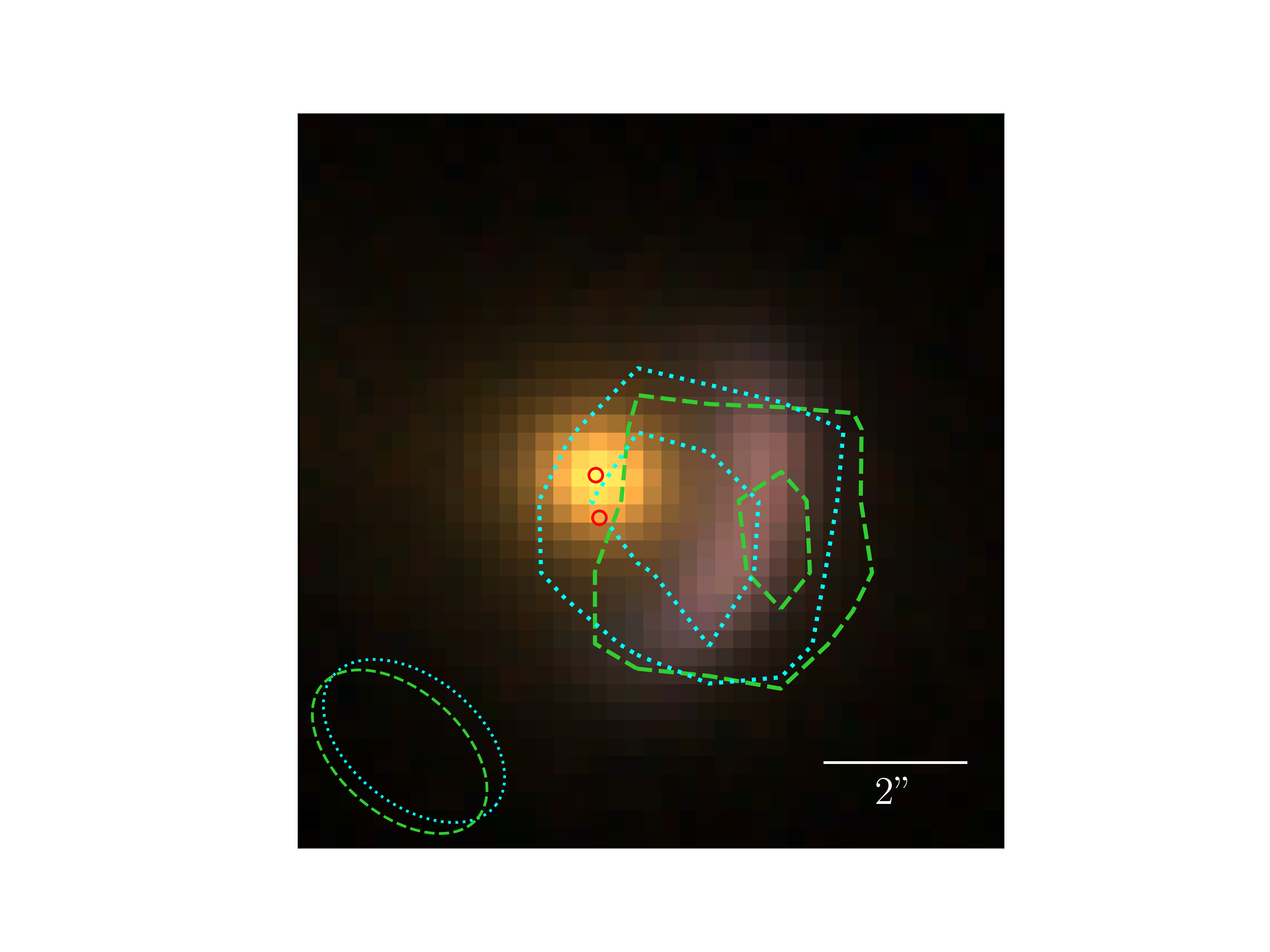}
    \caption[Target J0412$-$2646]{DECaLS $grz$ image of target J0412$-$2646. 4- and 6-$\sigma$ contours are overlaid for VLASS epochs 1 (dashed green lines) and 2 (dotted cyan). The locations of the two radio components from by our observations (see Table \ref{tab:photometry}) are shown in red.}
    \label{fig:0412}
\end{figure}

This source was identified as a lensed galaxy by \citet{Jacobs2019} using a Neural Network-based search of DES. While VLASS images from both epochs seem to be centered away from the lens, our follow-up data shows a $160\,\mu$Jy point source at the location of the lens galaxy and a $260\,\mu$Jy diffuse component to the south of that, possibly indicative of a core+jet or core+lobe morphology.
Figure \ref{fig:0412} shows a DECaLS image of this source with the locations of our VLA detections and contours of two VLASS epochs.
While one epoch has the peak of emission located on top of the arc, the other places it between the arc and the lens galaxy.
It is possible diffuse emission from the source galaxy is responsible for shifting the VLASS detection over, and that this emission is too low surface brightness for or resolved out of our observations at 10 GHz.
However, further observations would be needed to address this hypothesis.

%%%%%%%%%%%%%%%%%%%%%%%%%%%%%%%%
%%%%%%%%%%%%%%%%%%%%%%%%%%%%%%%%
\section{The Known Population of Lensed Radio Sources} \label{sec:discussion}
\subsection{Variability and Spectral Indices of Lensed Radio Sources}

Until recently, only a handful of lensed radio sources were known, with most of these being identified through dedicated searches such as CLASS and JVAS.
The advent of deep and high resolution wide-area sky surveys such as VLASS is now resulting in more detections and correct associations of radio emission from lensed systems, especially lensed quasars.
Additionally, the latest generation of optical surveys with high astrometric precision, such as Gaia, are allowing for the identification of hundreds of new lensed quasars \citep[e.g.,][]{Jacobs2019}.
The result is such that there are now $\approx 80$ lensed radio sources known, more than double the number known less than a decade ago \citep{McKean2015}.
We list all the published gravitational lenses with emission detected at frequencies lower than 100 GHz ($\lambda > 3\,$mm) in Table \ref{tab:allknown}.
This cutoff was chosen to correspond roughly with both the point where dust begins to dominate the SED of a normal galaxy rather than synchrotron emission \citep{condon92} and the highest observable frequencies of the ngVLA \citep{Carilli2015}.
In this Section of the paper we use these $80$ objects to broadly characterise the observational properties of the lensed radio source population.

\begin{figure}
    \centering
    \subfigure{\includegraphics[trim = 0pt 0pt 11pt 12pt, clip, width=0.5\textwidth]{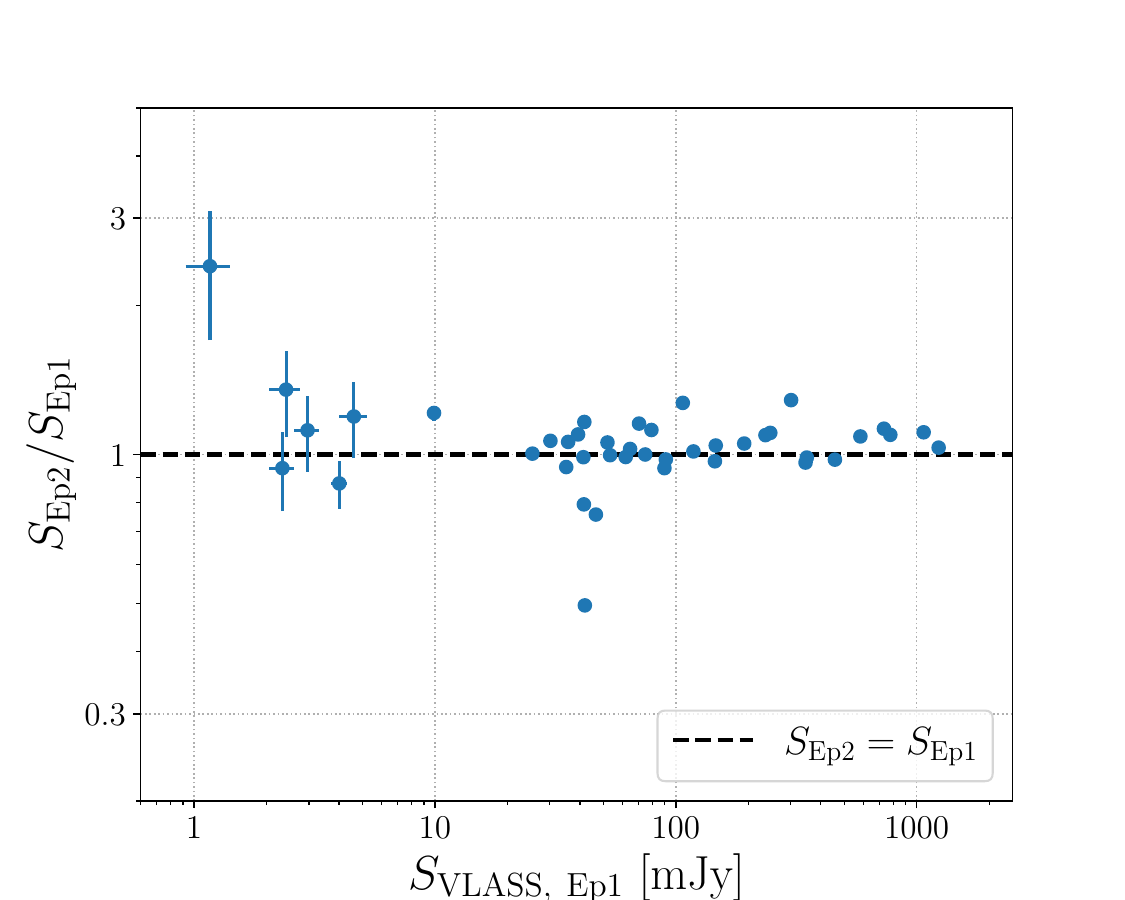}}
    \subfigure{\includegraphics[trim=1mm 0 11pt 12.5pt, clip, width=0.46\textwidth]{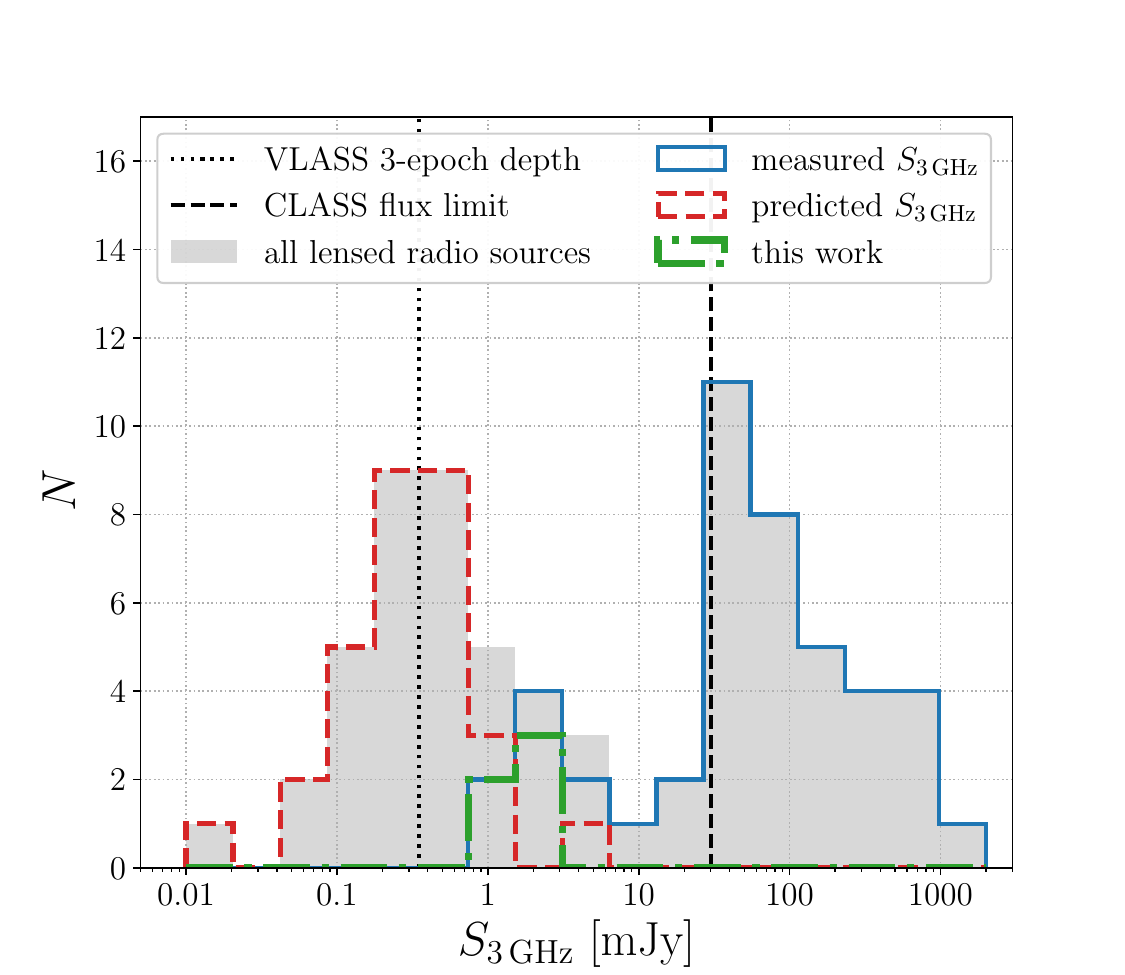}}
    \caption[VLASS Variability and Fluxes of Lensed Radio Sources]{Left: The variability of lensed radio sources in VLASS ($S_{\text{Epoch 1}}/S_{\text{Epoch 2}}$) as a function of brightness in Epoch 1, with the black dashed line denoting zero variability between the two epochs. The slight but systematic trend for brighter fluxes in epoch 2 is likely driven by the limited quality of the VLASS ``Quick Look'' images. Right: Distributions of integrated $3\,$GHz flux densities for lensed radio sources (grey solid histogram). Flux measurements are taken from VLASS where possible (solid blue line) and estimated using spectral index information otherwise (red dashed line). The green dot-dashed line shows the five previously unreported lenses we identified with VLASS in this work. The dashed black vertical line shows the flux limit of the CLASS survey, highlighting the additional sources that can be identified by combining optical and radio information rather than just relying on a dedicated flux-limited radio search. The black dotted line shows the $350\,\mu$Jy point source depth that VLASS will reach after three epochs.}
    \label{fig:fluxdist}
\end{figure}

Some previous dedicated searches for lensed radio sources have specifically looked for flat-spectrum radio sources \citep[e.g.,][]{JB07, Myers2003}.
In principle such a strategy should reduce contamination from the lobes of radio galaxies that can appear offset from their host galaxies, often LRGs, and thus potentially mimic a lensed object in catalog space.
With a reasonably large sample of lensed radio sources now in hand we can potentially explore the spectral index distribution of the population.
Doing so has several benefits, the spectral index can i) provide insights into the type of source being lensed (e.g., quasar, lobe-dominated radio galaxy etc.); ii) potentially guide future search strategies for lensed radio sources; and iii) be used to show the flux distribution of lensed radio sources at a single observer-frame frequency, as opposed to comparing flux densities from different observations at e.g., $1.4\,$GHz and $10\,$GHz.

%\begin{figure}
%    \centering
%    \includegraphics[width=\columnwidth]{variability_ratio_logy.pdf}
%    \caption[Variability of lensed radio sources]{The variability of lensed radio sources in VLASS ($S_{\text{Epoch 1}}/S_{\text{Epoch 2}}$) as a function of brightness in Epoch 1, with the black dashed line denoting zero variability between the two epochs.
%    The slight but systematic trend for brighter fluxes in epoch 2 is likely driven by the limited quality of the VLASS `Quick Look' images.
%    % \KB{Larger axis labels.}
%    }
%    \label{fig:vlass-variability}
%\end{figure}

Ideally the spectral index for radio sources should be calculated using flux measurements from different frequencies obtained at the same time to avoid the potential for source variability biasing the measurement. 
For their $8$ lensed quasars observed with the Australia Telescope Compact Array, \citet{dobie23} provide contemporaneous measurements at $5.5\,$GHz and $9\,$GHz which we use to calculate the spectral index for these sources.
For the remaining sources we do not have contemporaneous multi-band flux density measurements, and are thus dependent on measurements that might be subject to variability.
Using the catalogs from the first two epochs of VLASS \citep[][B. Sebastian et al. in prep.]{Gordon2021} we characterise the variability of the $42$ lensed sources detected in the first epoch of VLASS over timescales of $\sim2$ years in the left panel of Figure \ref{fig:fluxdist}.
With only a few exceptions, most lensed radio show little variability between Epochs 1 and 2 of VLASS, with a median and standard deviation for $S_{\text{Ep 2}}/S_{\text{Ep 1}}$ of $1.0$ and $0.2$ respectively.
Knowing that most lensed radio sources aren't strongly variable strengthens the argument for using flux density measurements taken at different times to estimate the spectral index of these sources.
For $22$ lensed radio sources we have flux density measurements from both VLASS ($3\,$GHz) and FIRST ($1.4\,$GHz).
{Of these, $9$ were selected to be flat spectrum as part of CLASS.
Because these $9$ sources were selected on their spectral index it is inappropriate to include them in a characterization of the spectral index distribution of the parent population of lensed radio sources, and we remove these from consideration here. 
For the $21$ objects with spectral index information and where spectral index wasn't used in their selection as lensed sources, we find the median spectral index to be $\alpha_{\text{median}} = -0.73$, similar to the typical spectral index for the general radio source population \citep[e.g.,][]{Condon1998, Gordon2021}.
If instead we include the CLASS sources in the spectral index distribution of lensed radio sources, we note that the median spectral index is only slightly flatter at $\alpha = -0.69$, suggesting that these bright flat spectrum sources don't dominate the lensed radio source population.}

%%%%%%%%%%%%%%%%%%%%%%%%%%%%%%%%
\subsection{Future Searches}

Notably, $100\,\%$ of our VLASS detected targets that are lensed optical quasars have radio emission from the lensed source, suggesting that lensed quasars conincident with legacy detections in radio surveys present an efficient approach to identifying candidate lensed radio sources.
Moreover, those lensed radio sources detected in flux-limited surveys are likely the most scientifically useful targets due to their typically higher brightness than many sources detected through blind, deep radio observations of lensed optical quasars.
With a suite of deep wide-area optical and near-IR imaging surveys from ground and space, such as the Vera C. Rubin Observatory \citep{Ivezic2019}, the Nancy Grace Roman Space Telescope \citep{Spergel2015}, and the Euclid telescope \citep{Laureijs2011}, coming online over the next few years, thousands of lensed quasars will be discovered \citep[e.g.][]{Yue2022}.
It is interesting to consider how many of these sources will have complementary radio observations.
% Some of the new lensed quasars will have already been detected by current radio surveys, but their lensing status remains unknown due to the multi-arcsecond PSF of such projects. 
Some of the new lensed quasars may have already been detected at radio wavelengths, but due to the multi-arcsecond PSF of current wide-area radio surveys their status as lensed radio sources remains unknown.
Moreover, for a multi-epoch survey such as VLASS, the ability to combine the individual observations from each epoch enables deeper imaging than one epoch of observations alone, increasing their power as a legacy reference catalog to identify radio emission from newly discovered lensed quasars. 
After the end of the planned survey, combined three-epoch VLASS images are expected to have a point source depth of $S_{3\,\text{GHz}}\approx 350\,\mu$Jy \citep{VLASS}, substantially deeper than the $\sim 1\,$mJy depth of the Quick Look images from a single epoch currently available\footnote{A recently proposed fourth VLASS epoch would push the point source sensitivity of combined images down to $300\,\mu$Jy \citep{Nyland2023}.}.

To make predictions for the number of lensed radio sources that might be detected in VLASS we first determine the $3\,$GHz flux density distribution of the known lensed radio sources.
For bright lensed sources within the VLASS footprint we take the $3\,$GHz flux density measurement from the VLASS Epoch 1 Quick Look catalog \citep{Gordon2021}.
For those sources too faint to be detected by VLASS or lying outside the survey footprint, we estimate their $3\,$GHz flux density by extrapolating from available measurements at other frequencies using their measured spectral index where available.
For those sources without a spectral index we assume $\alpha=-0.7$ in line with the typical spectral index for the lensed radio source population.
Where published flux densities for individual lensed images are used, these are summed to provide a total flux for the lensed system, a better approximation of what will observed by a single VLASS beam.
We note here that we do not estimate the $3\,$GHz flux density for PSS 2322$+$1944, as the observed $45\,$GHz emission is attributed to CO($J=2\rightarrow1$) line emission rather than being continuum emission \citep{2008ApJ...686..851R}, and thus extrapolating to $3\,$GHz based on an assumed spectral index is inappropriate in this instance.

%\begin{figure}
%    \centering
%    \includegraphics[trim=1mm 1mm 0 0 clip, width=\columnwidth]{s3ghz_dist.pdf}
%    \caption[VLASS fluxes of lensed radio sources]{Distributions of integrated $3\,$GHz flux densities for lensed radio sources (grey solid histogram).
%    Flux measurements are taken from VLASS where possible (solid blue line) and estimated using spectral index information otherwise (red dashed line).
%    The green dot-dashed line shows the five previously unreported lenses we identified with VLASS in this work.
%    The dashed black vertical line shows the flux limit of the CLASS survey, highlighting the additional sources that can be identified by combining optical and radio information rather than just relying on a dedicated flux-limited radio search.
%    The black dotted line shows the $350\,\mu$Jy point source depth that VLASS will reach after three epochs.
%    }
%    \label{fig:fluxdist}
%\end{figure}

The $S_{3\,\text{GHz}}$ distribution for lensed radio sources is shown in the right panel of Figure \ref{fig:fluxdist}, with predicted and measured flux densities shown by the red dashed and blue solid lines respectively.
An important feature of that figure is the apparent bimodality of the radio flux distribution of lensed sources.
This can be explained by the two broad selection approaches used over the years.
The brighter peak (centered around $100\,$mJy) is mostly the result of the targeted searches for lensed radio sources conducted by CLASS, JVAS, and MG-VLA \citep{MGVLA}.
Indeed, the flux limited nature of these searches is evident in the flux distribution as the sudden drop in sources below \text{$S_{3\,\text{GHz}}\approx30\,$mJy}.
The fainter peak (centered around $300\,\mu$Jy) is the result of radio observations of newly identified lensed quasars in optical imaging. 
Notably, about half of these objects should be detectable in future multi-epoch combined VLASS images, providing a potential pathway to more efficient target selection for future in depth radio observations.
Large numbers of lensed radio sources will be detected in forthcoming optical surveys.
For instance \citet{Yue2022} predict $2,400$ lensed quasars will be identified in the Legacy Survey of Space and Time \citep[LSST,][]{Ivezic2019}, $\sim1,000$ of which will be at depths detectable by current optical surveys.
Approximately $14,000\,\text{deg}^{2}$ ($70\,$\%) of the $20,000\,\text{deg}^{2}$ footprint of LSST will be covered by VLASS, it follows that hundreds of the lensed sources may be detectable in the final-depth VLASS images.

In this work we have focused on using VLASS to identify lensed radio sources, and indeed the high resolution and time domain aspects of the survey provide unique advantages over previous radio surveys for this challenge.
The next generation of radio telescopes however will be even more well suited for identifying lensed radio sources.
The high angular resolution and survey speeds of the Square Kilometer Array (SKA) and the ngVLA will enable the ready identification of the multiple images of radio sources separated on sub arcsecond scales. 
This can provide two key advantages over current approaches.
First, not being dependent on optical observations to identify the lensing configuration has the potential to identify systems where the lensed galaxy has an intrinsically high radio-to-optical luminosity such that it is only detected in radio.
Second, in systems where only the lensed background object is radio loud, the low level of radio contamination may allow for more tightly constrained lens models than would be possible from optical observations where light from the foreground lens galaxy may become problematic.

%%%%%%%%%%%%%%%%%%%%%%%%%%%%%%%

\section{Conclusions} \label{sec:summary}

We report first results from a pilot study seeking to efficiently identify strongly lensed radio sources by combining wide-area optical and radio survey data.
We find that a high fraction of optically selected lensed quasars with radio counterparts in VLASS at mJy-level flux densities are in fact high-confidence lensed radio sources.
The results here suggest that large samples of radio strong lenses could be efficiently identified via targeted follow-up of radio counterparts to lenses found in near-future optical and NIR imaging surveys with the Vera C. Rubin Observatory, Euclid, and Nancy Grace Roman Space Telescope.
Importantly, the radio lens systems from VLASS are bright enough to allow detailed characterization.
Our findings are reinforced by complementary recent results from \citet{dobie23} and \citet{Jackson2024}.

We observed 11 radio lens candidates based on two selection methods.
The method based on \citet{JB07} aiming to discover entirely new lens systems yielded no new radio lenses.
%, was less successful and .
However, given the rarity of gravitational lensing in general, this result was not unexpected, and we note that JB07 themselves found no candidates among a larger follow-up sample.
A successful catalog-based method would require a more sophisticated approach than the one we utilized, and such an approach will become much more necessary in the future thanks to upcoming large and deep surveys in both the radio and optical.

The second method, which utilized existing catalogs of lensed quasars and galaxy-scale arcs, was much more successful.
Five out of the five existing lensed quasars we observed had radio emission from the quasars, rather than the lens, and in only one case did the lens galaxy also emit in the radio.
Furthermore, our single lensed galaxy target is still a possible radio lens given the mismatch between VLASS and VLA positions, although its emission seems to be much fainter than suggested by the VLASS epoch 1 data.
These results suggest that survey-resolution radio emission from lensed quasar systems is more likely to come from the quasar rather than the lens, and presents a possible method to identify more lensed radio sources in the future.

Our candidate selection for this method utilized a list of lensed quasars published in 2019, containing 220 systems. Since then, the publication of hundreds of lensed quasar \citep[e.g.][]{lemon2023, he23} and galaxy-galaxy lens \citep[e.g.][]{dawes23, zaborowski23} candidates has greatly expanded the number of possible targets, suggesting that a new search incorporating the same methodology is likely to discover many more systems.

%\vspace{1em}
%\authorcomment1{
During the final preparations of this manuscript, \citet{Jackson2024} reported independent observations for a sample of radio lens candidates, including 3 of the 4 previously unreported radio lens systems presented in Section \ref{sec:p1results}, as well as an additional 30 sources not considered here.
These two works underscore the opportunities for expanding the catalog of known lensed radio sources through target selection based on lenses identified at other wavelengths.
Similar to \citet{dobie23}, we both find that radio emission from systems involving optically-selected lensed quasars is typically dominated by emission from the lensed quasar rather than the main deflector galaxy.
Our target selection differs from \citet{dobie23} and \citet{Jackson2024} in that we required a spatially coincident VLASS source, and thus all of the new radio lenses discussed here have integrated flux density brighter than $\sim1$ mJy at 3 GHz (Figure~\ref{fig:fluxdist}). 

\chapter{An Expanded Search for Lensed Radio Sources}
\label{sec:chapter4}
\paragraph{The content of this chapter}
will appear, in slightly different form, in an upcoming paper, to be submitted to AAS publishing in Summer 2026. 

In the pilot study conducted in \citet[][the previous Chapter]{martinez2025}, we found that matching known optical lensed quasars with VLASS catalog data was an effective way to locate new lensed radio quasars.
In light of the promise showed by that study, we use improved methods to observe 25 more optical lenses and candidates with the VLA, confirming 10 of them as new radio lenses. 

\section{Target Selection} \label{sec:target}

For this work, we expanded both sides of this matching process from the previous study.
For the lensed quasar catalog, we used an early version of the Strong Lensing Database\footnote{\url{https://sled.amnh.org}} \citep[SLED,][]{SLED}.
This list included not only the $\sim300$ spectroscopically confirmed lensed quasars but also over 1000 candidate lenses from recent searches such as \citet{he23, dawes23}. 
These lenses are largely not spectroscopically confirmed, and so may in fact be observational quasar pairs or other confounders such as star+galaxy pairs. 
As a result, we expect a lower purity from these unconfirmed samples than from the confirmed lenses.

On the radio side, we chose to increase our sensitivity by stacking the existing epochs of VLASS in the image plane.
Due to the third epoch being incomplete at the time of the search, only two epochs were available for some targets.
The stacking procedure was as follows:
First, cutouts around the targets were acquired using the CIRADA database.
Then, the images from each epoch were deconvolved using the Python \texttt{spectral-cube} package, and reconvolved to the smallest common beam between them using the \texttt{radio-beam} package.
The images were then stacked using the \texttt{reproject} library, and forced photometry was run on the stack using \texttt{photutils}.
This stacking method is not without its downsides, but it was serviceable to detect radio emission from a target lens down to $\sim0.7$mJy/beam. 
For more details on the image-plane stacking procedure and its advantages and disadvantages, we refer the reader to our upcoming wide-area stacking paper, \citet[][\textit{in prep}]{vlassstack}.

Our lensed quasar catalog contained 1590 lenses and candidates within the VLASS footprint.
Of these, 60 were already known radio lenses, and 30 more had unpublished high-resolution radio observations in the NRAO archive. 
We ran the stacked forced-photometry procedure on the remaining 1500 and visually inspected all candidates with $\geq 3\sigma$ radio detections.
At this stage, candidates were eliminated if the radio emission seemed spurious or from an obvious radio galaxy (for example, a double or triple source morphology with no optical counterparts), or if the source had been confirmed as a non-lens in the literature.
We were left with 38 candidate radio lenses.

\section{VLA Observations} \label{sec:c4obs}

Due to our observing program's priority, only 25 of our targets were observed by the VLA in Semester 24B (the remaining candidates have been approved as filler observations for Semester 26A).
Each source was observed in C-band ($4-8$GHz) in full-polarization\footnote{While the cross-hand polarizations were observed, we did not calculate polarization terms and thus report no polarization information for these targets.} spectral line mode, for a total bandwidth of $\sim4$GHz.
Additionally, observations used real-time blanking and the mixed 3/8 bit quantizer in order to minimize the effects of RFI.
Observation blocks consisted of standard setup and requantizer gain scans, followed by observations of one of the NRAO's standard Flux Density Calibrators.
After this scan (which was also used for bandpass calibration), observations alternated between a complex gain calibrator and the target assuming an atmospheric stability/calibration cycle time of 8 minutes.
Table \ref{table:obs} shows the gain calibrators used for each target, as well as the total time on-source, which was calculated in the following way.
Beginning with the forced photometry from stacked VLASS, we propagated the S-band flux density to C-band assuming a spectral index\footnote{ we adopt the convention relating flux density, $S$, and frequency, $\nu$, by $S\propto \nu^{\alpha}$} of $\alpha=-0.8$, corresponding to that of the general radio source population \citep[e.g.,][]{Condon1998, Gordon2021}.
Under the assumption that this flux was entirely due to the two lensed images, the putative C-band flux was then split according to the optical flux ratio of the two components, taken from existing literature.
As quasars can vary in both optical and radio, to be safe, we further divided the fainter image's flux by 2,  and in cases where a flux ratio was not reported, a ratio of $10:1$ was assumed.
Finally, we aimed for a signal-to-noise ratio (SNR) of 5 for the fainter image and thus set our noise floor accordingly.
This flux was then used for in the NRAO's exposure time calculator\footnote{\url{https://obs.vla.nrao.edu/ect/}} with an assumed bandwidth of 3.5GHz to account for RFI.
Short required exposure times from the calculator were padded to 1 minute in order to account for any technical difficulties.

\subsection{Calibration and Imaging}\label{imaging}
VLA data was processed in two ways, both using the NRAO's Science-Ready Data Products (SRDP) program.
The SRDP uses a VLA pipeline to automatically apply flux, bandpass, and complex gain calibrations to the raw data, producing a calibrated Measurement Set (MS) of the science targets that is ready for imaging.
When possible, the pipeline images and self-calibrates the MS to increase dynamic range.
This imaging was carried out for 13 of our 25 targets, and all of these were self-calibrated.
The SRDP images for these observations were $2^{14}$ pixels to a side, and go out nearly to the first primary beam null.

Our other 12 observed targets were not imaged in the SRDP pipeline but were still calibrated.
These were imaged using the resources of the Center for High Throughput Computing (CHTC) at the University of Wisconsin $-$ Madison.
We used the \texttt{MPICASA} package to enable parallel deconvolution and calibration in order to speed up imaging times.
\texttt{MPICASA}, while powerful, is lacking implementation of some capabilities, and thus faceted imaging and A-projection were unavailable.
Typical CHTC imaging runs were conducted on 8 CPU cores, using Briggs (\texttt{robust} = 0.5) weighting, W-projection with 128 planes, and 2-Taylor-term Multi-Frequency Synthesis.
In most cases, pixel size was set to $0''.05$, which was at least 1/3 the minor axis of every image's PSF, and image size was at least 2' across.

Given the $\sim 13'$ primary beam of VLA C-band observations, bright sources typically appeared as contaminants in our images, and these were deconvolved separately as outlier fields. 
In addition, 3 sources had phase calibrators which were resolved by VLA C-band, and these calibrators were imaged and extensively self-calibrated before their amplitude and phase solutions were transferred to the target field.
These sources are noted in Table \ref{table:obs}, as are the sources where target-field self-calibration was successful.

 \subsection{Analysis}\label{analysis}

A defining feature of radio images is the presence of correlated noise between pixels, as opposed to  the typical Poisson or Gaussian noise present in CCD images.
Therefore, it is typically desirable to conduct analysis in the $uv$ plane, where measurements are independent.
\newpage

{\footnotesize
 \onehalfspacing
\begin{ThreePartTable}
\begin{TableNotes}[flushleft]
    \setlength{\labelsep}{0pt}
    \item[a] \label{obs-a}{This target was self-calibrated using the CHTC. All SRDP-processed targets were self-calibrated by NRAO.}
    \item[b] \label{obs-b}{This calibrator was flagged as slightly resolved by the NRAO and was self-calibrated on CHTC before phase transferring to the target field.}
    \item[c] \label{obs-c}{The NRAO's known flux model of 3C 286 was used for phase calibration rather than treating it as a point source.}
    \item[] Note. --- The ``Known Lens'' column indicates whether the target is a spectroscopically confirmed optical lens, according to prior literature. Times are rounded to the nearest 0.25 minutes.
  \end{TableNotes}
\begin{longtable}[c]{ccclccl}
\caption[Details of Radio Lens Candidate Observations]{Details of radio lens candidate observations} \label{table:obs} \\
\hline \hline
{Target} & {Known Lens} & {$S_{3 \text{GHz}}$} & {Calibrator} & {Obs. Date} & {Time on Source} & {Processing}\\
 { } & { } & {[mJy]} & { } & { } & {[min]} & { } \\
 \hline
\endfirsthead
\caption[]{\textit{(continued)}} \\
\hline \hline
{Target} & {Known Lens} & {$S_{3 \text{GHz}}$} & {Calibrator} & {Obs. Date} & {Time on Source} & {Processing}\\
 { } & { } & {[mJy]} & { } & { } & {[min]} & { } \\
 \hline
\endhead
\hline
\endfoot
\hline
\insertTableNotes
\endlastfoot
J0050$-$1740 & Y & 0.41 & J0050$-$0929 & 2025 Jan 27 & 26.75 & SRDP \\
J0055$-$1212 & N & 12.27 & J0050$-$0929 & 2025 Jan 27 & 1.25 & SRDP \\
J0122$+$7838 & N & 2.41 & J0217$+$7349 & 2025 Jan 19 & 1.25 & CHTC \\
J0138$+$4841 & N & 1.65 & J0136$+$4751 & 2025 Jan 19 & 1.25 & CHTC\tnotex{obs-a} \\
J0156$-$2751 & Y & 0.46 & J0145$-$2733 & 2025 Jan 27 & 4.0 & SRDP \\
J0242$-$1002 & N & 0.46 & J0241$-$0815 & 2024 Nov 18 & 2.0 & CHTC\tnotex{obs-a} \\
J0336$-$3244 & N & 0.59 & J0402$-$3147 & 2025 Jan 21 & 71.5 & CHTC \\
J0347$-$2154 & Y & 0.18 & J0340$-$2119 & 2025 Jan 12 & 44.0 & CHTC \\
J0416$+$7428 & Y & 0.43 & J0410$+$7656 & 2025 Jan 19 & 9.75 & CHTC\tnotex{obs-a} \\
J0501$-$0733 & N & 0.46 & J0501$-$0159\tnotex{obs-b} & 2024 Nov 18 & 6.0 & CHTC \\
J0821$+$0735 & N & 0.85 & J0831$+$0429 & 2024 Dec 13 & 4.5 & SRDP \\
J0833$-$0721 & N & 0.46 & J0902$-$1415 & 2024 Dec 13 & 9.5 & SRDP \\
J0909$-$0749 & N & 2.57 & J0902$-$1415 & 2024 Dec 13 & 1.5 & SRDP \\
J0916$-$2848 & N & 0.23 & J0956$+$2618 & 2025 Jan 8 & 49.0 & SRDP \\
J0920$+$2241 & N & 4.58 & J0854$+$2006 & 2024 Dec 19 & 1.75 & SRDP \\
J0921$+$3020 & Y & 3.46 & J0915$+$2933 & 2024 Dec 19 & 1.5 & SRDP \\
J0926$+$3059 & N & 14.42 & J0915$+$2933 & 2024 Dec 19 & 1.25 & SRDP \\
J0940$+$2131 & N & 0.52 & J0956$+$2515 & 2024 Dec 19 & 4.0 & SRDP \\
J1111$+$3804 & Y & 4.47 & J1104$+$3812 & 2024 Dec 19 & 1.5 & SRDP \\
J1326$+$3020 & Y & 0.25 & 3C 286\tnotex{obs-c} & 2025 Jan 24 & 29.25 & SRDP \\
J2015$+$0707 & Y & 0.61 & J2025$+$0316 & 2024 Nov 4 & 14.0 & CHTC\tnotex{obs-a} \\
J2205$+$1019 & Y & 0.38 & J2147$+$0929 & 2024 Nov 4 & 6.0 & CHTC \\
J2232$+$1315 & N & 0.68 & J2232$+$1143\tnotex{obs-b} & 2024 Oct 30 & 65.5 & CHTC \\
J2308$+$3201 & Y & 0.58 & J2301$+$3726\tnotex{obs-b} & 2025 Jan 19 & 3.0 & CHTC \\
J2324$-$1225 & N & 0.65 & J2331$-$1556 & 2024 Nov 4 & 2.5 & CHTC
\end{longtable}
\end{ThreePartTable}}

\noindent Unfortunately, these measurements are computationally expensive, especially in long observations.
As our aim is primarily to determine the lensing nature of systems, and accurate fluxes and astrometry are secondary concerns, we elect to follow \citet{dobie23} and \citet{martinez2025} and use an image plane analysis.

We employed the \texttt{PyBDSF} program, a Python-based interactive source finder built for radio observations, to measure our VLA images.
\texttt{PyBDSF} identifies islands of significant flux above background using a sliding box method to account for spatial variations.
These islands are fit with Gaussian blobs, which are hierarchically grouped into components and sources.
The program's performance is highly sensitive to choice of sliding box size, and in some cases (such as extreme sidelobes around bright sources) this was changed from the automatically calculated default.
A few of our targets have relatively low-SNR detections, and \texttt{PyBDSF} did not detect some of them.
These cases are noted in the following sections, and we measured these components manually using the Python package \texttt{photutils} \citep{photutils}.

\subsubsection{Cross-matching Statistics}
Table \ref{table:comps} shows flux and position measurements for each of our observed sources.
In cases where a component corresponds to a known optical source, we also report a frequentist statistic quantifying chance of a random alignment.
We use the same procedure as in \citet[][Chapter \ref{sec:chapter3}]{martinez2025}, but this time report the negative logarithm of the probability, such that a higher number in the table means a greater chance of true correspondence.
Additionally, while in our previous work the radio sources were all matched to \textit{Gaia} sources, we couldn't achieve this for every object in these observations, and fell back to DESI Legacy Surveys and then PAN-STARRS in those cases\footnote{as these surveys have reduced astrometric accuracy compared to \textit{Gaia}, the probability of random chance is generally greater for those objects.}.
The matching survey used is given in the final column of Table \ref{table:comps}.

\subsubsection{Astrometric Accuracy}
In general, the absolute astrometry of a VLA observation is roughly 1/10 the synthesized beam size for a normally calibrated image.
Self calibration of a source may introduce an offset to absolute astrometry, though it should preserve relative measurements (i.e. distance between sources).
Wideband and wide-field effects can also introduce astrometric errors at the edge of large fields when proper measures such as A-Projection are not taken.
As our fields are typically small, and our lensed quasar targets coincide with the telescope pointing, we expect projection effects to be negligible for target measurements.
Finally, CASA's \texttt{tclean} routine is known to contain an unexplained $\sim50$ milliarcsecond systematic astrometry error\footnote{see \url{https://casadocs.readthedocs.io/en/v6.7.3/notebooks/introduction.html}}, an effect comparable with the expected standard positional error previously mentioned.

To assuage astrometric concerns, we matched all our measured \texttt{PyBDSF} sources with the \textit{Gaia} DR3 catalogue \citep{Gaiadr3}, finding 30 which were measured in both regular and self-calibrated images.
We found no significant difference between the two catalogs' angular separations from those \textit{Gaia} sources when the match was closer than 50 milliarcseconds and conclude that the self-calibration effect is subdominant to the systematic errors.
With these results in mind, in Table \ref{table:comps}, we report only the source fitting error from \texttt{pybdsf}, and ignore projection, self-calibration, and systematic \texttt{tclean} error.
Additionally, positions given are those from the self-calibrated image when available due to better flux measurements, therefore we must caution the reader against using these positions as ``ground truth'' for very high precision astrometric exercises.

\section{Results} \label{sec:results}

\begin{figure} 
     \centering
     \subfigure{\includegraphics[trim={\mosleftclip, \mosbotclip, \mosrightclip, \mostopclip}, clip, width=\mosaicwidth]{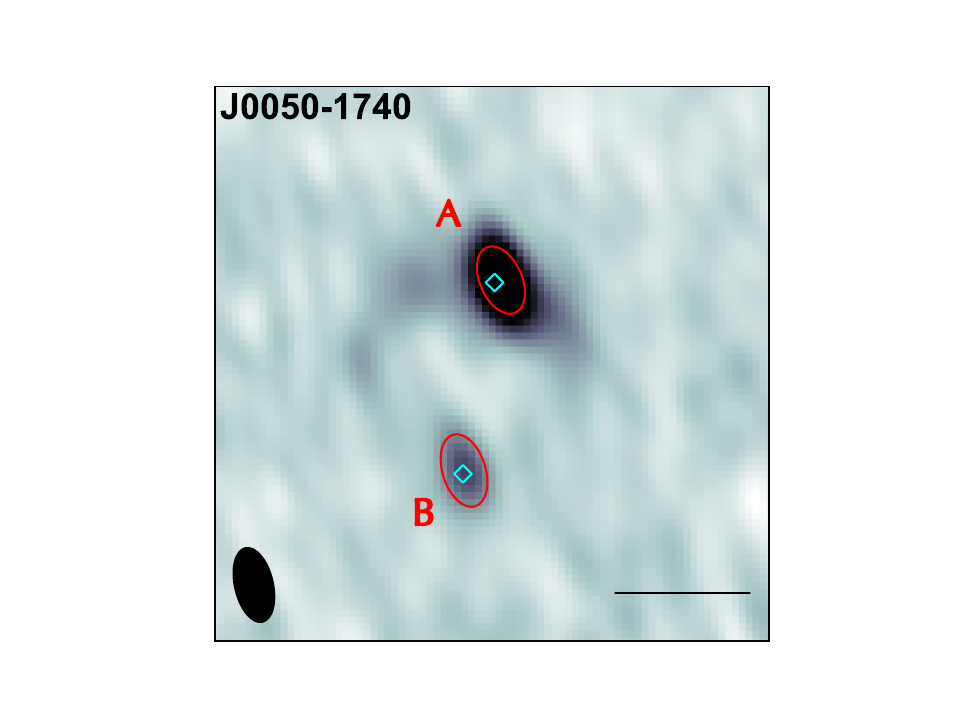}}
     \subfigure{\includegraphics[trim={\mosleftclip, \mosbotclip, \mosrightclip, \mostopclip}, clip, width=\mosaicwidth]{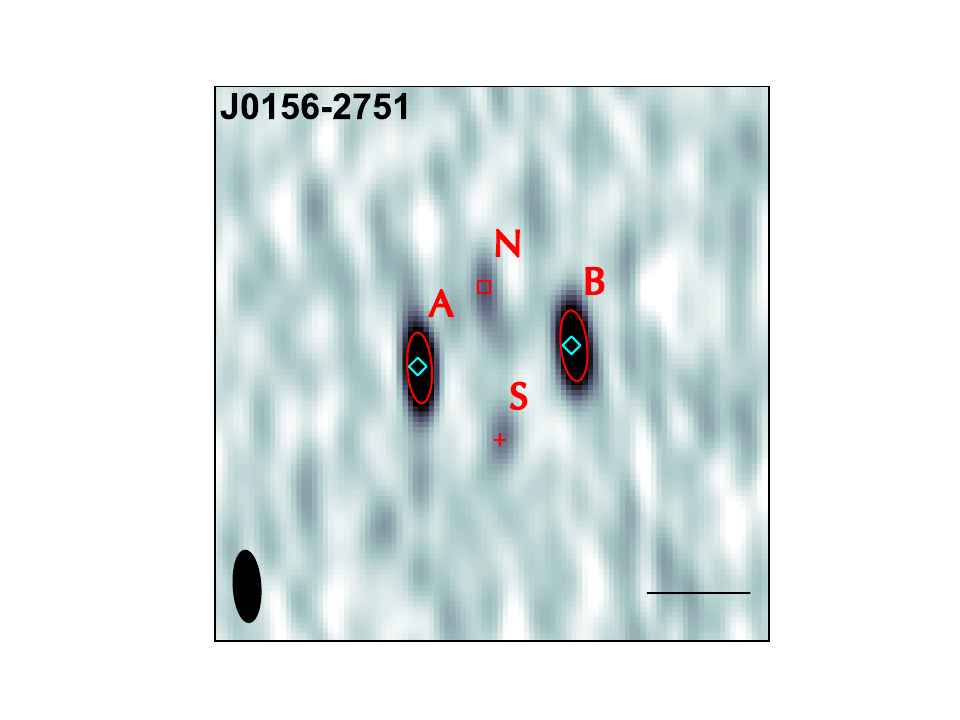}}
     \subfigure{\includegraphics[trim={\mosleftclip, \mosbotclip, \mosrightclip, \mostopclip}, clip, width=\mosaicwidth]{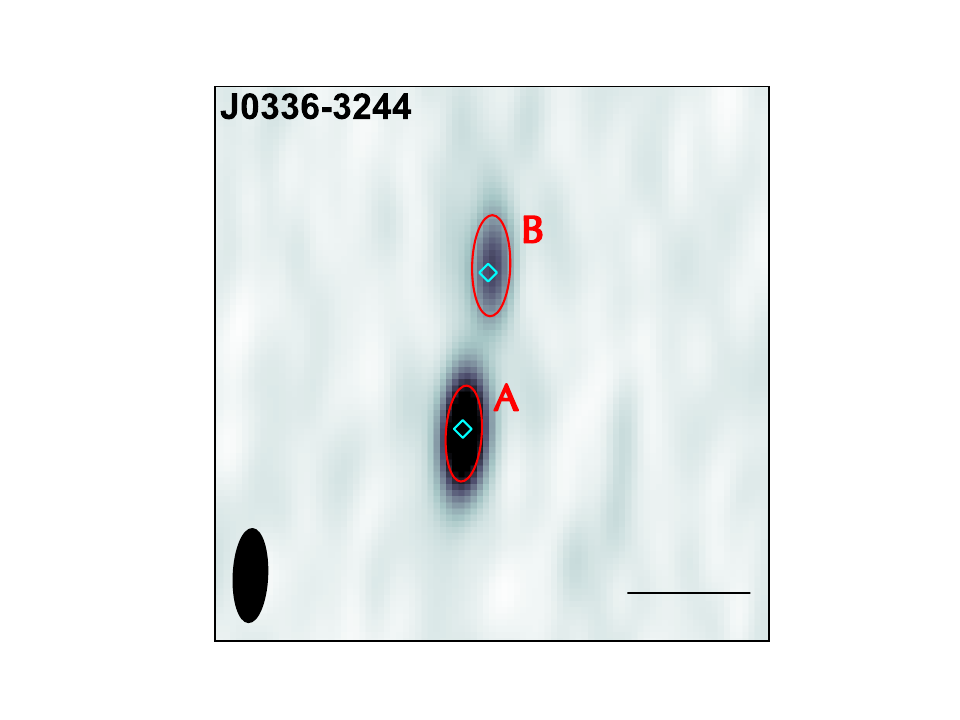}}
     \subfigure{\includegraphics[trim={\mosleftclip, \mosbotclip, \mosrightclip, \mostopclip}, clip, width=\mosaicwidth]{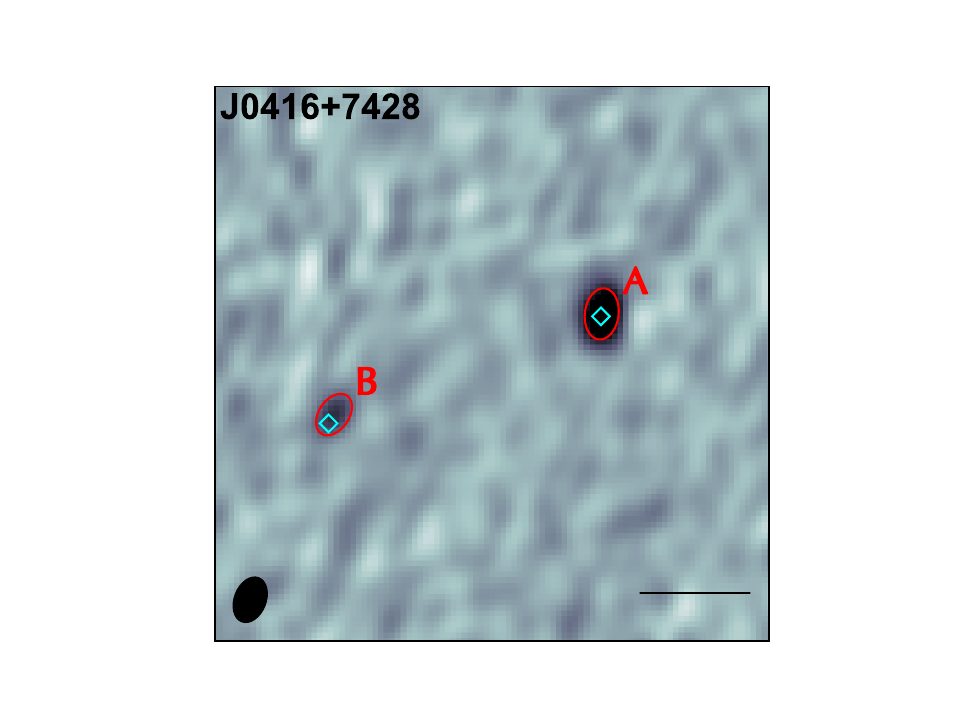}}
     \subfigure{\includegraphics[trim={\mosleftclip, \mosbotclip, \mosrightclip, \mostopclip}, clip, width=\mosaicwidth]{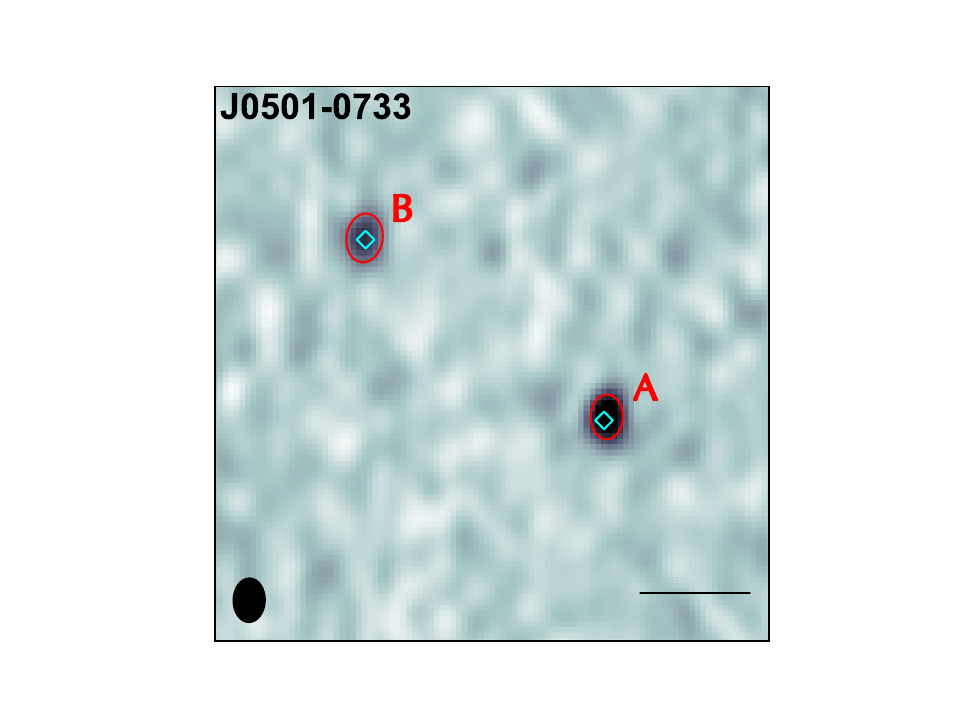}}
     \subfigure{\includegraphics[trim={\mosleftclip, \mosbotclip, \mosrightclip, \mostopclip}, clip, width=\mosaicwidth]{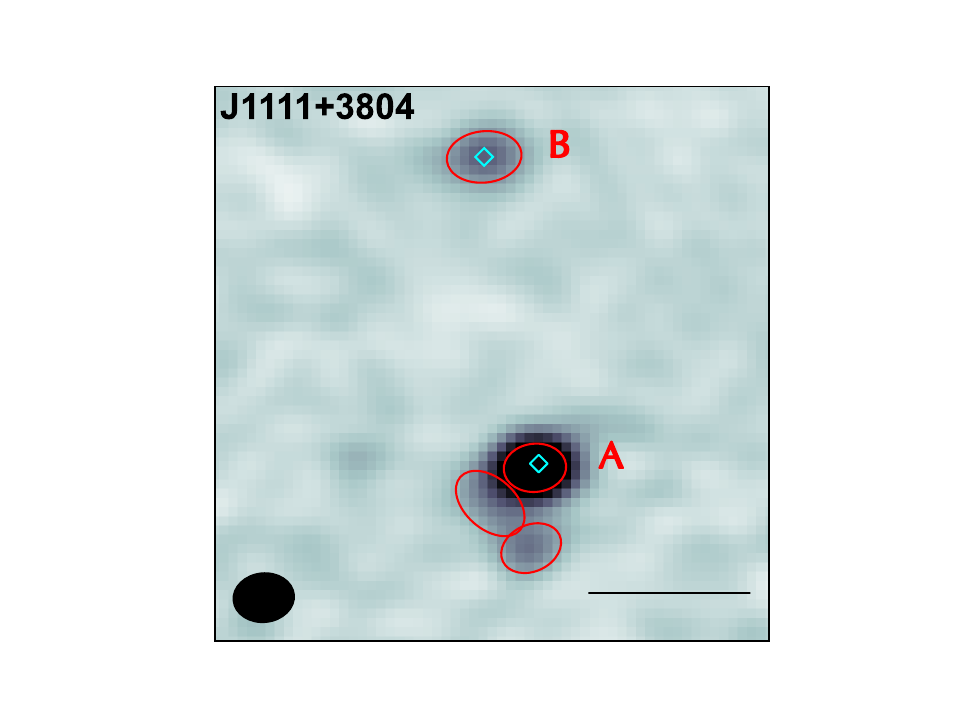}}
     \subfigure{\includegraphics[trim={\mosleftclip, \mosbotclip, \mosrightclip, \mostopclip}, clip, width=\mosaicwidth]{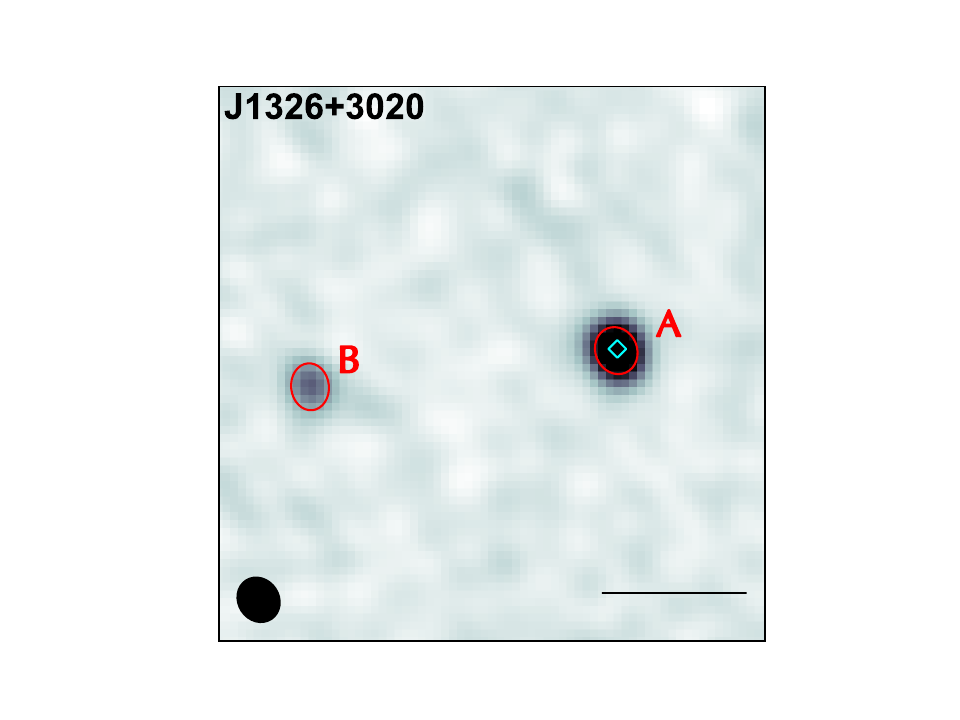}}
     \subfigure{\includegraphics[trim={\mosleftclip, \mosbotclip, \mosrightclip, \mostopclip}, clip, width=\mosaicwidth]{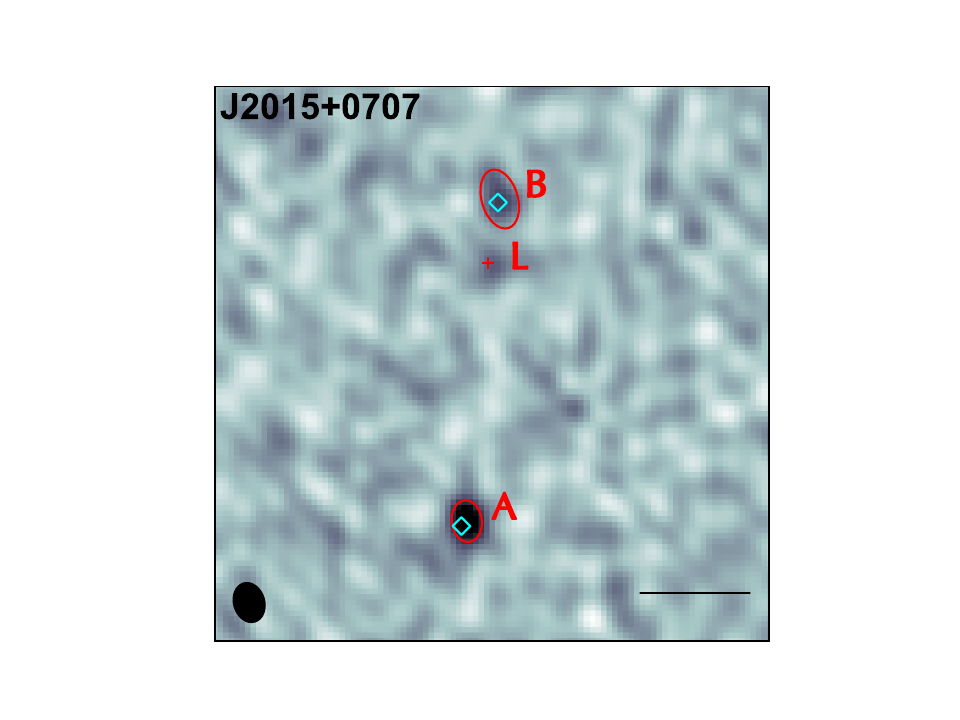}}
     \subfigure{\includegraphics[trim={\mosleftclip, \mosbotclip, \mosrightclip, \mostopclip}, clip, width=\mosaicwidth]{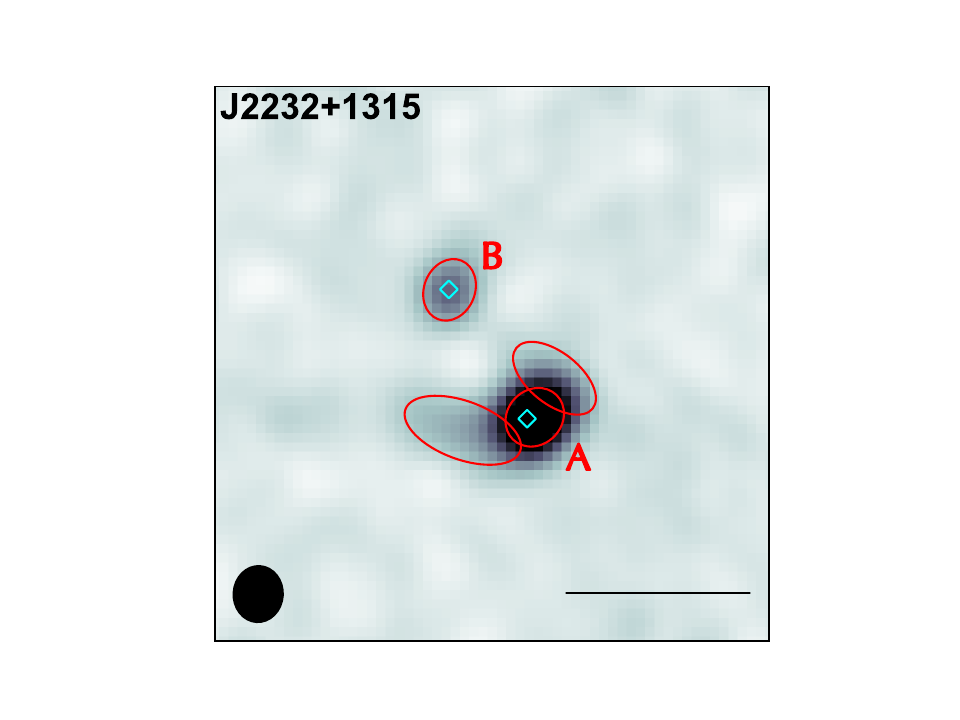}}
     \subfigure{\includegraphics[trim={\mosleftclip, \mosbotclip, \mosrightclip, \mostopclip}, clip, width=\mosaicwidth]{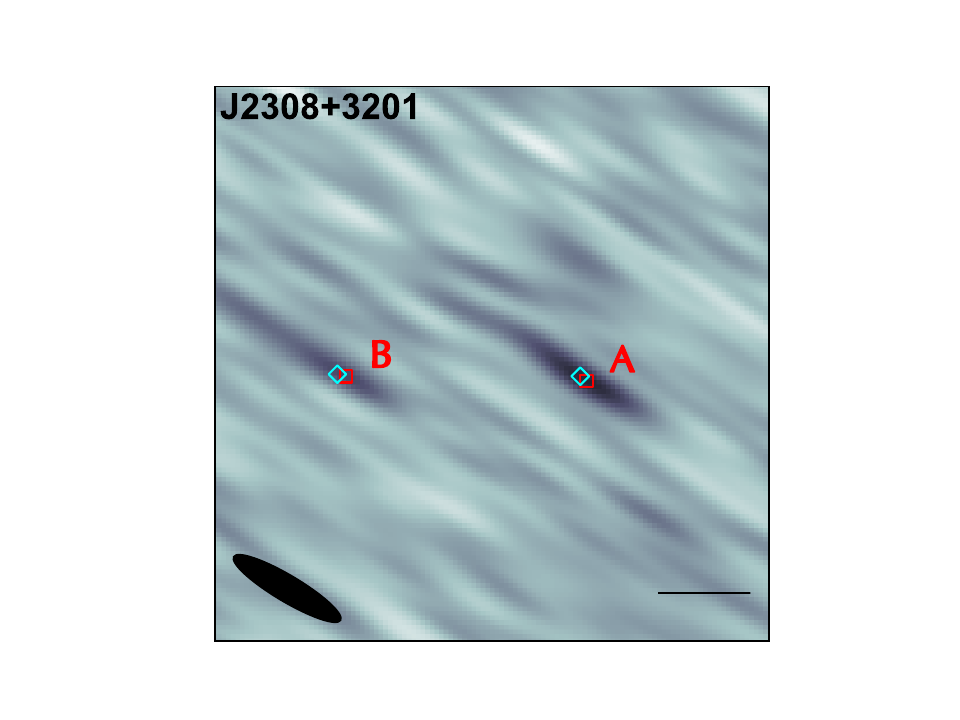}}
     
     \caption[Radio Lenses]{New radio lenses found as a result of this search. Labeled points correspond to entries in Table \ref{table:comps}. Ellipses show the size and orientation of \texttt{pyBDSF} components, while blue diamonds are at the coordinates of the cross-matched optical sources. A red square (corresponding to the ``a'' note in the Table) represents a significant flux island that was not fit by \texttt{pyBDSF}, and a red plus (Table note ``b'') is a source that with a forced photometry aperture placed by hand. Image scaling is linear, but was chosen arbitrarily to bring out features discussed in the individual source paragraphs. The beam size is shown in the bottom left of each image, and a 1 arcsecond scalebar is shown at bottom right.}
     \label{fig:newlenses}
     \end{figure}

\subsection{New Radio Lenses}
These targets had significant 6GHz emission from both quasar images, and we classify them as radio lenses.
In most cases, the lens system has already been spectroscopically confirmed, and we will argue for a lensing hypothesis when this is not the case.

\subsubsection{J0050$-$1740} \label{J0050}
This lensed quasar, also known as HE 0047$-$1756, was found by \citet{wisotzki04} in a brightness-based lensed quasar survey.
A $1''.44$ separation double with $z_{lens}=0.407$ and $z_{source}=1.678$, J0050$-$1740 is a target for time-delay cosmography, with $t_{AB}=-10.4\pm 3.5$ days \citep{cosmograil2020}.
As such, the lens system is well-studied in optical and near-infrared, and is known to exhibit significant microlensing \citep{2017A&A...597A..49G, 2014ApJ...797...61R}.
\citet{wisotzki04} found hints of an Einstein Ring in their NIR discovery image, which was confirmed by \citet{ding17} and \citet{shajib21} with high-resolution follow-up using HST WFC3-UVIS and Keck NIRC2, respectively.
Our VLA image of the lens similarly shows hints of this ring, which could indicate star formation from the quasar host galaxy \citep{condon92}. 
However, given the limited resolution of the observations, AGN activity, and lack of longe-wavelength IR measurements, we cannot rule out another cause of the feature, such as a compact radio jet.
We also report a radio flux ratio of $2.93\pm0.43$, noting that the $A$ component may be overestimated due to contributions from the extended emission.

\subsubsection{J0156$-$2751} \label{J0156}
\citet{lemon2023} confirmed this system as a doubly lensed quasar at $z=2.97$, and it was also identified as a candidate by \citet{he23}.
We identify both these images, and \texttt{pyBDSF} additionally picked out a $5.4\sigma$ noise bump between the images, labeled $N$ in Figure \ref{fig:newlenses}.
Closer inspection also revealed a second $4.3\sigma$ bump which was not identified (reported as $S$) which could represent a counterimage of $N$, assuming both are true detections.
Additionally, the radio flux ratio of the confirmed images $(A/B = 1.01\pm0.1)$ is consistent with that of the extra components $(N/S = 1.25\pm0.4)$.
As these two components have no optical counterpart, we are left with three options: $(1)$ the additional detections are spurious and represent a chance alignment of noise; $(2)$ we are detecting lensed emission from a second source (perhaps a jet hotspot of the quasar core); or $(3)$ we have detected a third and fourth image of the quasar itself.

\subsubsection{J0336$-$3244}\label{J0336}

\citet{dawes23} reports this target as a B-grade lens candidate. Our VLA observation detect two radio point souces at a separation of $1''.38$, suggesting this is indeed a radio lens.
While the $A/B=2.33\pm0.2$ radio flux ratio we measure is very different from the \textit{Gaia} flux ratio of $\sim10$, we ascribe this to optical variability.
However, as this source is not spectroscopically confirmed as a lensed quasar in the optical, we cannot rule out the possibility of a chance alignment or binary quasar scenario in this case.

\subsubsection{J0416$+$7428}\label{J0416}
\citet{lemon2023} spectroscopically confirmed this target as a $2''.64$ separation double at $z=0.9$.
The lens galaxy appears in PAN-STARRS imaging to be part of a galaxy group, and we detect several radio sources near this lens system. We also detect both quasar images, with radio flux ratio $3.10\pm0.8$.
\citet{lemon2023} also note the brightness $(i_{mag} = 16.22)$ and low redshift $(z=0.098)$ of the lens galaxy, the second-lowest redshift for any known gravitational lens.

\subsubsection{J0501$-$0733} \label{J0501}
\citet{he23} rated this lens between grade A and C, and its Legacy Surveys image shows hints of excess $g$-band emission around the quasar images.
The calibrator used in the observation of this target was flagged as possibly extended and so was self-calibrated before transferring phases to the target visibilities.
Additionally, a bright radio source was caught at the very edge of the primary beam which we were unable to separately clean, leading to increased noise in our final radio image.
Our VLA image detects two radio components with separation $2''.7$ coincident with the \textit{Gaia} positions of the quasars.
We also measure a flux ratio of $1.50\pm0.3$, consistent with the \textit{Gaia} flux ratio of $1.62$. 
Therefore, we consider this system a probable lens, although we recommend future spectroscopic follow-up to conclusively confirm this hypothesis.

\subsubsection{J1111$+$3804} \label{J1111}
This $z=3.020$ double was confirmed by \citet{chan22}, who note a redder continuum in the fainter image due to lens galaxy flux contamination.
The system has image separation $2''.0$ and appears extended in VLASS. 
We observe two components at 6GHz corresponding to the two quasar images, with a flux ratio of $3.35\pm0.5$ -- more extreme than the \textit{Gaia} flux ratio of $1.7$, a discrepancy which could be caused by variability in either observation band.
We also detect additional faint components around image A, which were initially grouped by \texttt{pyBDSF} into a compound source.
In Table \ref{table:comps}, we have only reported information about the Gaussian corresponding to the brightest component, in order to make astrometry and flux measurements more accurate.
We note that deeper radio observations will be necessary to gain a better understanding of this extended emission and search for its counterimage.

\begin{figure*}
	\includegraphics[trim={8.5mm, 16.5mm, 19.5mm, 21mm}, clip, width=\textwidth]{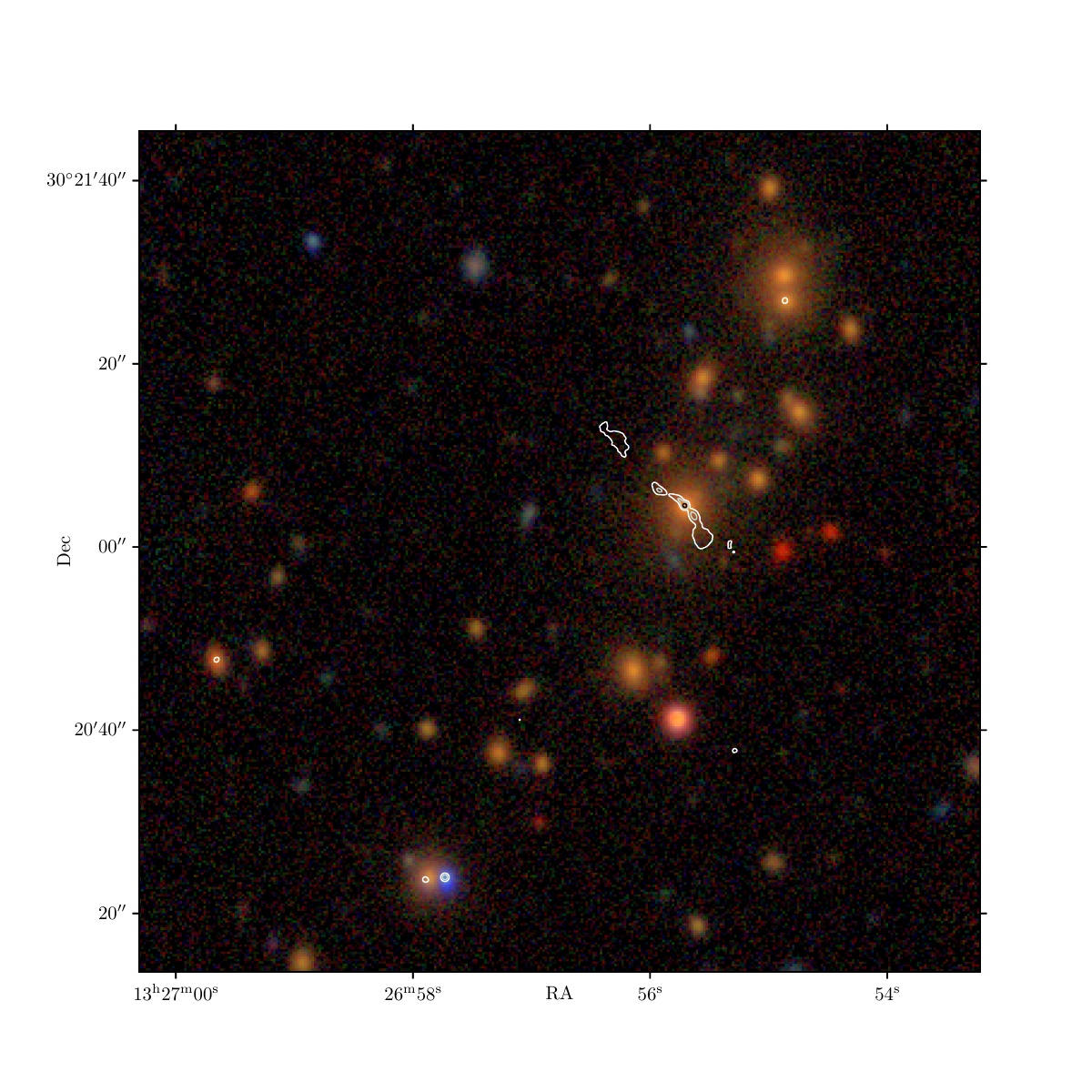}
	\caption[DECam Image of J1326$+$3020]{DECam $grz$ image of lens system J1326$+$3020 (lower left) and galaxy cluster RCS J132655$+$3021.1, with 6GHz VLA contours. Contour levels are 32, 141, and 631 $\mu$Jy.}
	\label{fig:1326}
\end{figure*}

\subsubsection{J1326$+$3020} \label{J1326}
\citet{lemon2023} spectroscopically confirmed this $2''.11$ separation double as a $z=1.852$ quasar lensed by a $z=0.339$ red galaxy.
The fainter image is only visible in the Legacy Surveys residual image due to its blending with the lens galaxy, and our C-band observations give a radio flux ratio of $4.12\pm0.5$.
This system is also notable for its association with $z=0.362$ galaxy cluster RCS J132655$+$3021.1 \citep{RCS}, the central galaxy of which is a known NVSS and FIRST source, and appears in the latter as a double source \citep{2011ApJ...734..103G}.
A Dark Energy Camera (DECam) optical $grz$ image of the cluster's central region with radio contours is shown in Figure \ref{fig:1326}.
The radio data in the figure is from the non-self-calibrated SRDP image due to artifacting in the self-calibrated image near the brightest source, and has been slightly smoothed to a $0''.5$ circular beam in the image plane to bring out the BCG's radio lobes.
In addition to detecting resolving the BCG into a FR$-$I morphology, we also find fainter radio emission from two other cluster members, as well as two sources with no optical counterpart in the same field.
We note that any future lens modeling of this system will need to incorporate an external contribution from the cluster. 

\subsubsection{J2015$+$0707} \label{J2015}
This $z=2.59$ double has separation $2''.93$ and was confirmed by \citet{lemon2023}.
We detect both images in C-band, with a flux ratio of $1.32\pm0.5$.
We also note a $3.6\sigma$ noise bump between the two images which could be emission from the lens galaxy, we label this $L$ in Figure \ref{fig:newlenses}.

\subsubsection{J2232$+$1315} \label{J2232}
\citet{he23} gave this candidate a B/C, and it appears as a faint but significant source in the stacked VLASS observations.
Our VLA data of the field was tagged for an extended complex gain calibrator, which we self-calibrated and transferred to the science target.
Not only do we detect both quasars at 3GHz, we also pick up significant extended emission surrounding the brighter object, suggesting the presence of an additional gravitationally lensed component.
As with J1111$+$3804 (\ref{J1111}), we report only the flux and position of the brightest Gaussian of the compound \texttt{pyBDSF} source, and recover a flux ratio of $3.68\pm0.3$.
The corresponding \textit{Gaia} flux ratio of 11.7 is much more extreme, a discrepancy we ascribe to variability in optical or radio.
This lens system also appears in a relatively crowded radio field, and we pick up multiple other AGN in our C-band data.
While this target is not a spectroscopically confirmed lensed quasar, we believe the C-band observations strongly support the lensing hypothesis and recommend spectroscopic follow-up as well as high-resolution optical follow-up to ascertain the origin of the extended radio emission.

\subsubsection{J2308$+$3201} \label{J2308}
This $z=2.30$, $2''.63$ separation double was confirmed by \citet{lemon2023}. 
As our complex gain calibrator was flagged as extended, we self-calibrated it and transferred phases to the target field.
We marginally detect both quasar images, but in part due to the highly elongated beam, \texttt{PyBDSF} was unable to model the two images. 
We therefore report the peak flux of the \texttt{PyBDSF} in \ref{table:comps}, which corresponds to the total flux assuming a point source.
From this we get a radio flux ratio of $1.42\pm0.34$.

{\footnotesize
 \onehalfspacing
\begin{ThreePartTable}
\begin{TableNotes}[flushleft]
    \setlength{\labelsep}{0pt}
    \item[a] \label{fluxes-a}This source was not fit to a model with \texttt{pybdsf}, but it was detected as a significant flux island. The position and flux measurements reported correspond to the peak pixel value in the island.
    \item[b] \label{fluxes-b}This source has no optical survey counterpart detected and thus we report no matching statistics.
    \item[c] \label{fluxes-c}This source was forced based on no optical survey, and the initial aperture was placed by eye. The coordinates given are that of the centroid of that aperture as reported by \texttt{photutils}. The flux and flux errors are found with the same procedure as the other forced sources.
    \item[] Note. --- Lensed quasar images are indicated by capital letters, while components for sources we judged as inconclusive or non-lenses are labeled otherwise. Components labeled ``FA$_i$'' (Forced Aperture) are forced photometry results, and the coordinates given are those of an optical source from the survey in the final column. As such we do not report position errors or matching statistics for these sources. Fluxes for these are given by the maximum pixel value in a one-beam aperture at the given coordinates. Flux error reported is the RMS of a nearby, empty field in the image.
  \end{TableNotes}
\begin{longtable}{llrcrcccc}
\caption[Radio Components for the 25 Lens Candidates Observed.]{Radio Components for the 25 lens candidates observed.}  \label{table:comps} \\
\hline \hline
{Target} & {Comp.} & \multicolumn{1}{c}{RA} & {$\sigma$RA} & \multicolumn{1}{c}{Dec} & {$\sigma$Dec} & {Flux Density} & {$-$log$(P)$} & {Survey}\\
{ } & { } & \multicolumn{1}{c}{[deg]} & {[mas]} & \multicolumn{1}{c}{[deg]} & {[mas]} & {[$\mu$Jy]} & { } & { }\\
\hline 
\endfirsthead
\caption[]{\textit{(continued)}} \\
\hline \hline
{Target} & {Comp.} & \multicolumn{1}{c}{RA} & {$\sigma$RA} & \multicolumn{1}{c}{Dec} & {$\sigma$Dec} & {Flux Density} & {$-$log$(P)$} & {Survey}\\
{ } & { } & \multicolumn{1}{c}{[deg]} & {[mas]} & \multicolumn{1}{c}{[deg]} & {[mas]} & {[$\mu$Jy]} & { } & { }\\
\hline
\endhead
\hline
\endfoot
\hline
\insertTableNotes
\endlastfoot
J0050$-$1740 & A & 12.61591751 & 3.7 & $-$17.66920928 & 6.6 & $208\pm9$ & 4.58 & Gaia \\
 & B & 12.61599660 & 11 & $-$17.66959867 & 23 & $71\pm10$ & 6.99 & Gaia \\
J0055$-$1212 & AGN (3) & 13.95363133 & 0.27 & $-$12.20260349 & 0.73 & $8872\pm47$ & 3.75 & LS DR10 \\
 & 1 & 13.95431861 & 11 & $-$12.20245379 & 34 & $526\pm75$ & $-$ & $-$ \\
 & 2 & 13.95414927 & 3.5 & $-$12.20253504 & 7.2 & $1248\pm58$ & $-$ & $-$ \\
 & 4 & 13.95250219 & 14 & $-$12.20306645 & 46 & $268\pm59$ & $-$ & $-$ \\
 & FA$_1$ & 13.95341484 & $-$ & $-$12.20299857 & $-$ & $21 \pm  27$& $-$ & Gaia \\
J0122$+$7838 & AGN (1) & 20.73333857 & 6.4 & 78.64853915 & 4.1 & $1212\pm54$ & 5.24 & Gaia \\
 & FA$_1$& 20.73290667 & $-$ & 78.64910643 & $-$ & $36\pm27$ & $-$ & Gaia \\
J0138$+$4841 & West & 24.59136123 & 13 & 48.69631020 & 6 & $571\pm54$ & 4.92 & Gaia \\
 & East & 24.59165029 & 56 & 48.69633268 & 40 & $260\pm78$ & 4.43 & Gaia \\
J0156$-$2751 & A & 29.10417233 & 2.2 & $-$27.85618522 & 17 & $302\pm22$ & 5.31 & Gaia \\
 & B & 29.10370400 & 3.1 & $-$27.85612565 & 18 & $300\pm23$ & 5.22 & Gaia \\
 & North\tnotex{fluxes-a} \tnotex{fluxes-b} & 29.10397659 & $-$ & $-$27.85596833 & $-$ & $76\pm14$ & $-$ & $-$ \\
 & South\tnotex{fluxes-c} & 29.10392755 & $-$ & $-$27.85637795 & $-$ & $61\pm14$ & $-$ & $-$ \\
J0242$-$1002 & AGN (1) & 40.68885340 & 8.6 & $-$10.04898595 & 11 & $219\pm20$ & 5.32 & Gaia \\
 & FA$_1$& 40.68839322 & $-$ & $-$10.04945710 & $-$ &$ 34\pm11$ & $-$ & Gaia \\
J0336$-$3244 & A & 54.08165875 & 1.2 & $-$32.74111627 & 7.4 & $242\pm7$ & 4.61 & Gaia \\
 & B & 54.08158526 & 3.2 & $-$32.74073776 & 20 & $104\pm7$ & 4.61 & Gaia \\
J0347$-$2154 & FA$_1$ & 56.76889649 & $-$ & $-$21.90924050 & $-$ & $12\pm4$ & $-$ & Gaia \\
 & FA$_2$ & 56.76900444 & $-$ & $-$21.90960374 & $-$ & $17\pm4$ & $-$ & Gaia \\
J0416$+$7428 & A & 64.19670016 & 6.6 & 74.48268560 & 14 & $207\pm21$ & 5.48 & Gaia \\
 & B & 64.19920738 & 20 & 74.48243340 & 25 & $67\pm16$ & 4.27 & Gaia \\
J0501$-$0733 & A & 75.37257311 & 5.6 & $-$7.55197541 & 11 & $217\pm20$ & 5.60 & Gaia \\
 & B & 75.37318534 & 13 & $-$7.55152713 & 22 & $145\pm24$ & 5.01 & Gaia \\
J0821$+$0735 & AGN (1) & 125.43068941 & 1.6 & 7.59608189 & 1.9 & $921\pm21$ & 5.68 & Gaia \\
 & FA$_1$ & 125.43019027 & $-$ & 7.59587138 & $-$ & $34\pm12$ & $-$ & LS DR10 \\
J0833$-$0721 & West & 128.47320848 & 15 & $-$7.35186937 & 25 & $50\pm12$ & 5.31 & Gaia \\
 & FA$_1$ & 128.47286509 & $-$ & $-$7.35185338 & $-$ & $31\pm8$ & $-$ & Gaia \\
J0909$-$0749 & AGN (1) & 137.49468854 & 0.79 & $-$7.81787727 & 0.42 & $6665\pm80$ & 5.54 & Gaia \\
 & FA$_1$ & 137.49446645 & $-$ & $-$7.81785177 & $-$ & $30\pm24$ & $-$ & Gaia \\
J0916$-$2848 & Diffuse & 139.16461400 & 92 & $-$28.81425932 & 95 & $208\pm34$ & 3.01 & PS1 DR2 \\
J0920$+$2241 & AGN (1) & 140.21141923 & 5.6 & 22.68715904 & 3.8 & $2530\pm55$ & 5.55 & Gaia \\
 & FA$_1$ & 140.21178993 & $-$ & 22.68715532 & $-$ & $70\pm26$ & $-$ & Gaia \\
J0921$+$3020 & Lens (1) & 140.26873068 & 5.9 & 30.34228263 & 2.8 & $1427\pm39$ & 2.84 & Gaia \\
 & 2 & 140.26818334 & 5.8 & 30.34176079 & 2.5 & $1213\pm36$ & 4.70 & Gaia \\
J0926$+$3059 & AGN (1) & 141.64395401 & 0.26 & 30.99609066 & 0.62 & $12472\pm73$ & 5.52 & Gaia \\
 & FA$_1$ & 141.64364099 & $-$ & 30.99554202 & $-$ & $61\pm26$ & $-$ & LS DR10 \\
J0940$+$2131 & FA$_1$ & 145.09796763 & $-$ & 21.52820946 & $-$ & $60\pm14$ & $-$ & Gaia \\
 & FA$_2$ & 145.09783871 & $-$ & 21.52804392 & $-$ & $26\pm14$ & $-$ & Gaia \\
J1111$+$3804 & A & 167.77719650 & 3.5 & 38.07341726 & 2.1 & $1123\pm37$ & 4.88 & Gaia \\
 & B & 167.77730703 & 19 & 38.07395003 & 9.7 & $335\pm46$ & 2.85 & LS DR9 \\
J1326$+$3020 & A & 201.74054543 & 2.0 & 30.33998242 & 2.4 & $301\pm9$ & 5.69 & Gaia \\
 & B\tnotex{fluxes-b} & 201.74122710 & 5.9 & 30.33991294 & 9.2 & $73\pm8$ & $-$ & $-$ \\
J2015$+$0707 & A & 303.80375478 & 8.8 & 7.11664124 & 16 & $138\pm21$ & 4.28 & Gaia \\
 & B & 303.80367161 & 29 & 7.11744819 & 57 & $104\pm33$ & 4.52 & Gaia \\
 & Lens\tnotex{fluxes-c} & 303.80369490 & $-$ & 7.11728190 & $-$ & $44\pm12$ & $-$ & $-$ \\
J2205$+$1019 & AGN (1) & 331.41607183 & 14 & 10.33061562 & 13 & $149\pm28$ & 4.75 & Gaia \\
 & FA$_1$ & 331.41643669 & $-$ & 10.33074800 & $-$ & $27\pm17$ & $-$ & Gaia \\
J2232$+$1315 & A & 338.16448793 & 1.8 & 13.25502746 & 1.8 & $298\pm6$ & 5.07 & Gaia \\
 & B & 338.16461948 & 5.2 & 13.25521901 & 6.9 & $81\pm6$ & 5.15 & Gaia \\
J2308$+$3201 & A\tnotex{fluxes-a} & 347.07736426 & $-$ & 32.02927515 & $-$ & $101\pm14$ & 4.14 & Gaia \\
 & B\tnotex{fluxes-a} & 347.07821616 & $-$ & 32.02928903 & $-$ & $71\pm14$ & 4.00 & Gaia \\
J2324$-$1225 & AGN (1) & 351.20595331 & 4.8 & $-$12.43186968 & 7.7 & $1113\pm50$ & 6.27 & Gaia \\
 & FA$_1$ & 351.20555944 & $-$ & $-$12.43226718 & $-$ & $34\pm27$ & $-$ & LS DR10 \\

\end{longtable}
\end{ThreePartTable}
}

\subsection{Isolated Radio Quasars}
The majority of these sources were dominated by a bright point source corresponding to a \textit{Gaia} object, with no appreciable flux at the other putative image position.
As mentioned above, we used \texttt{photutils} to force photometry for those positions, and we report the peak flux within an aperture the size of the synthesized beam, centered on the optical position.
We interpret the resulting flux as real or not depending its SNR, and in many cases the flux ratio implied by the forced measurement is far too extreme to be real.

\begin{figure*} 
     \centering
     \subfigure{\includegraphics[trim={\mosleftclip, \mosbotclip, \mosrightclip, \mostopclip}, clip, width=\mosaicwidth]{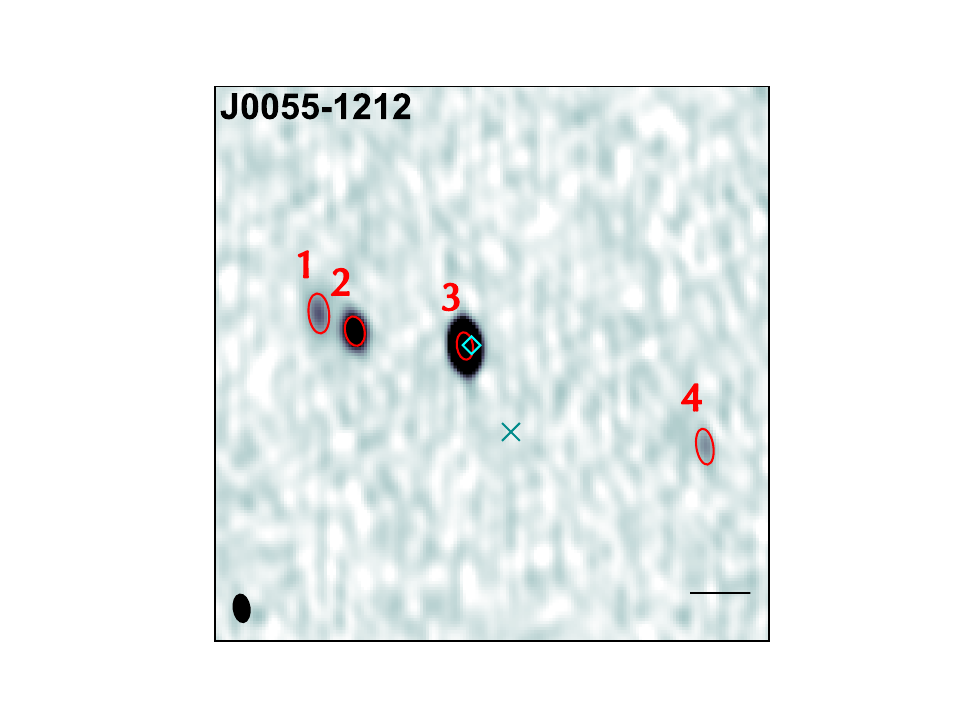}}
     \subfigure{\includegraphics[trim={\mosleftclip, \mosbotclip, \mosrightclip, \mostopclip}, clip, width=\mosaicwidth]{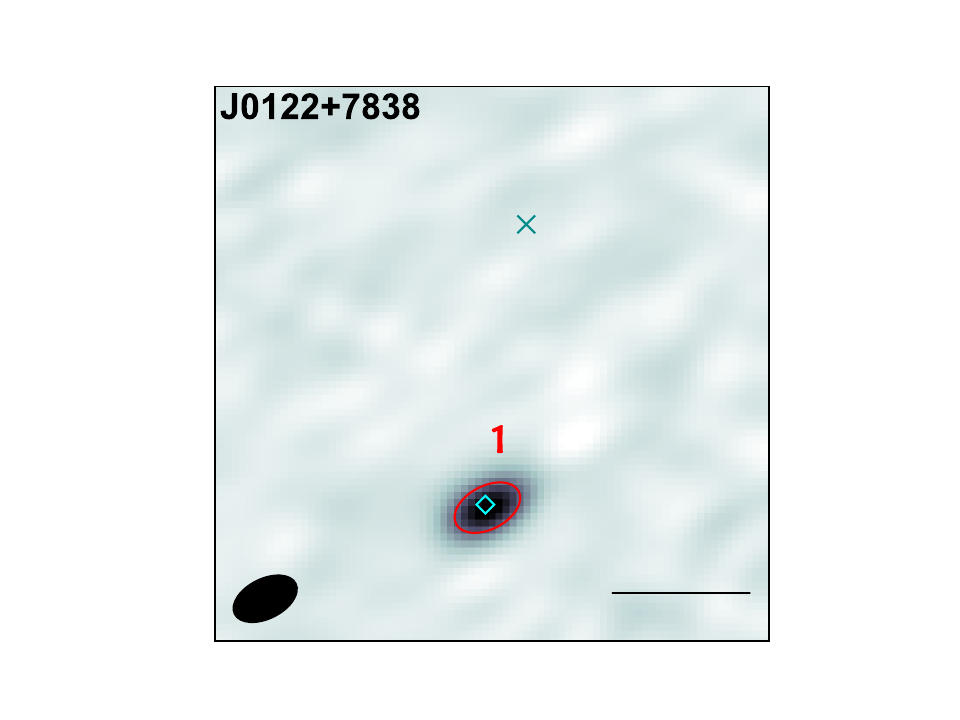}}
     \subfigure{\includegraphics[trim={\mosleftclip, \mosbotclip, \mosrightclip, \mostopclip}, clip, width=\mosaicwidth]{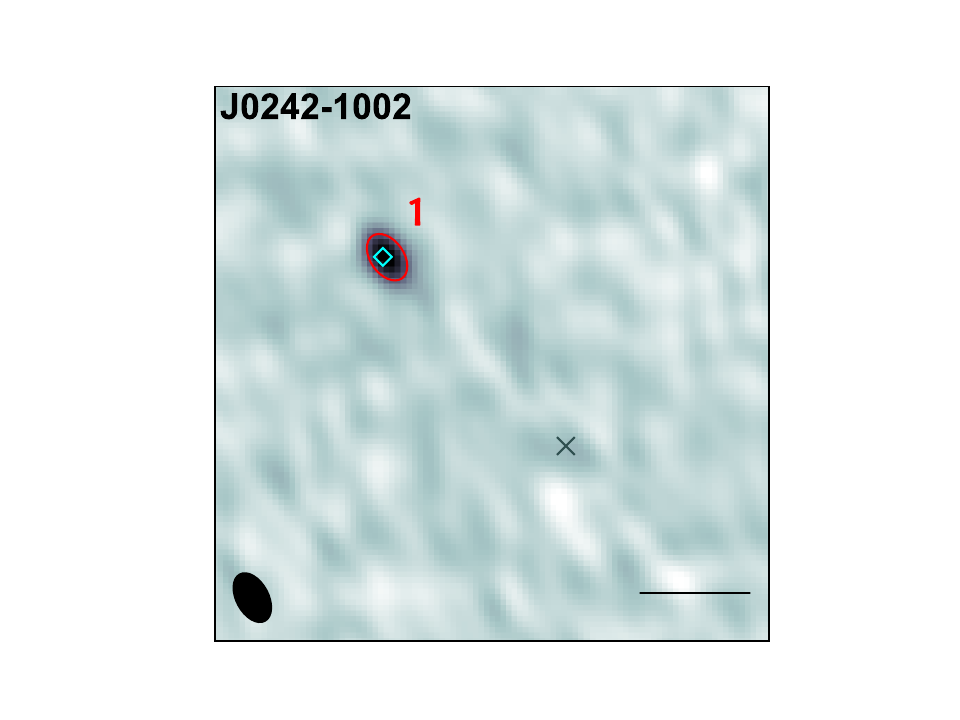}}
     \subfigure{\includegraphics[trim={\mosleftclip, \mosbotclip, \mosrightclip, \mostopclip}, clip, width=\mosaicwidth]{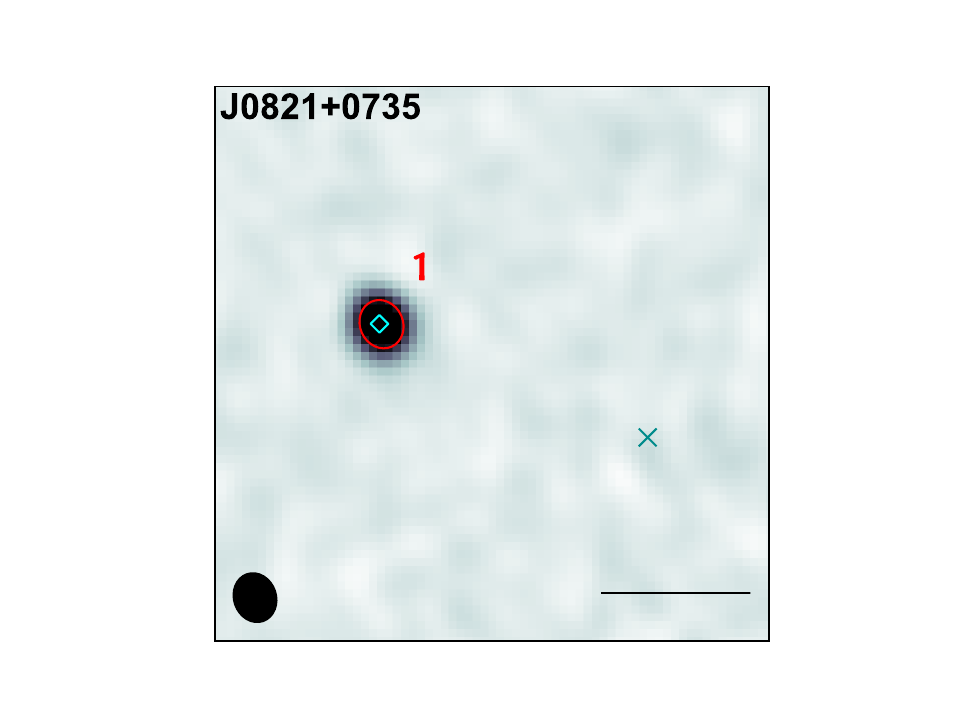}}
     \subfigure{\includegraphics[trim={\mosleftclip, \mosbotclip, \mosrightclip, \mostopclip}, clip, width=\mosaicwidth]{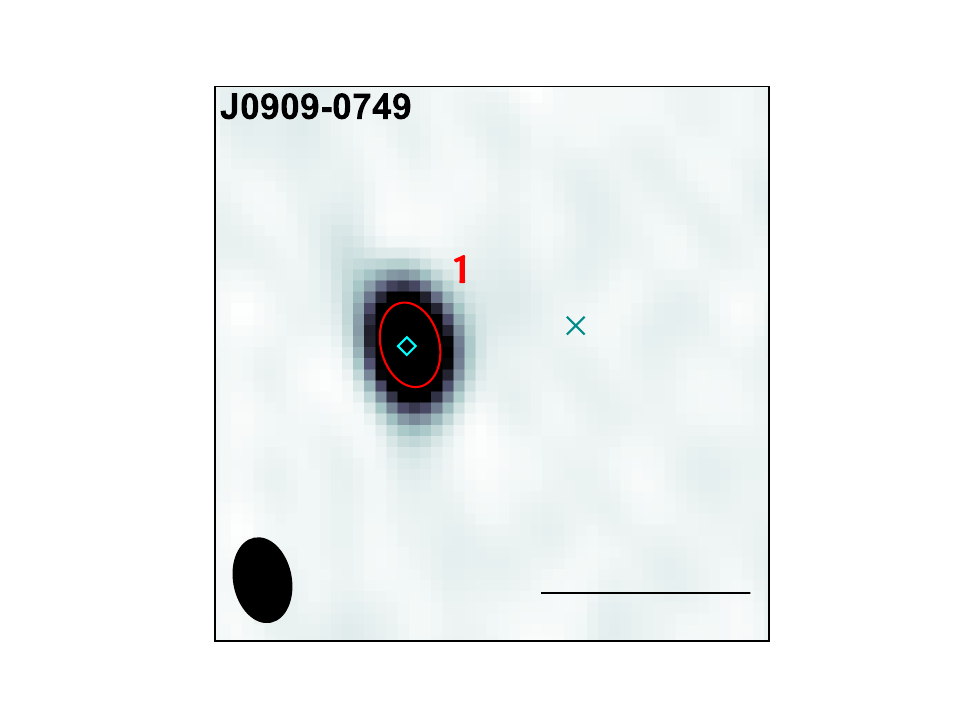}}
     \subfigure{\includegraphics[trim={\mosleftclip, \mosbotclip, \mosrightclip, \mostopclip}, clip, width=\mosaicwidth]{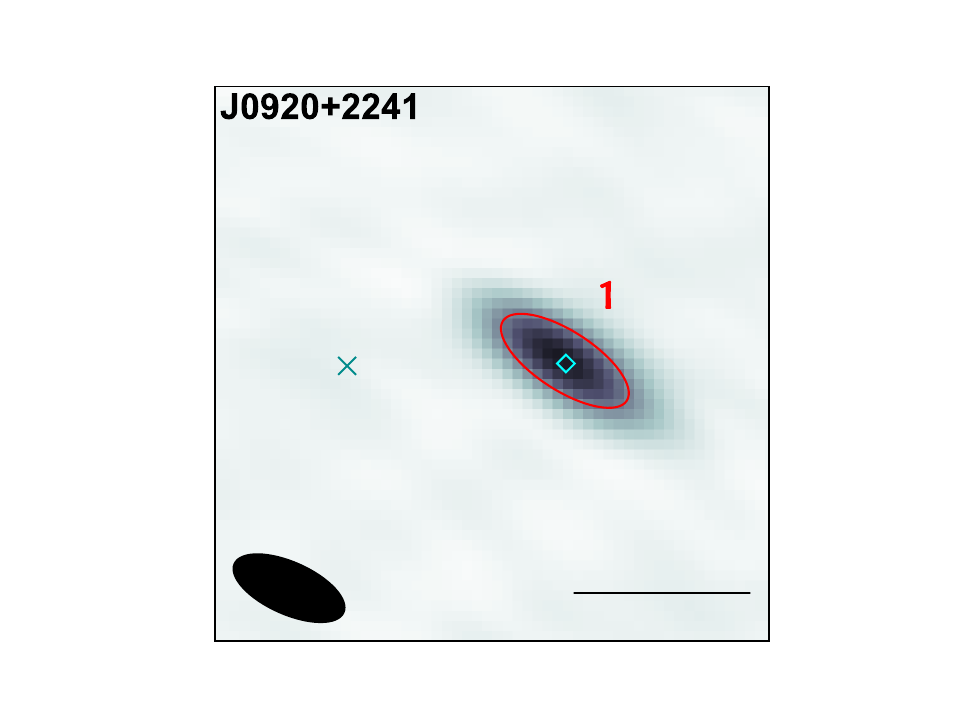}}
     \subfigure{\includegraphics[trim={\mosleftclip, \mosbotclip, \mosrightclip, \mostopclip}, clip, width=\mosaicwidth]{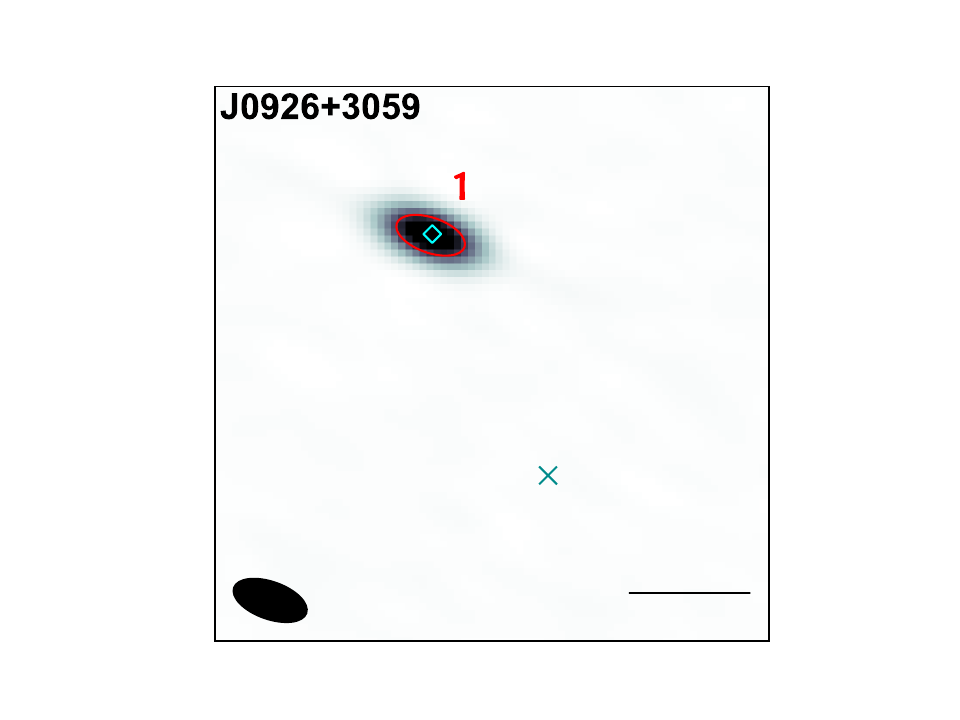}}
     \subfigure{\includegraphics[trim={\mosleftclip, \mosbotclip, \mosrightclip, \mostopclip}, clip, width=\mosaicwidth]{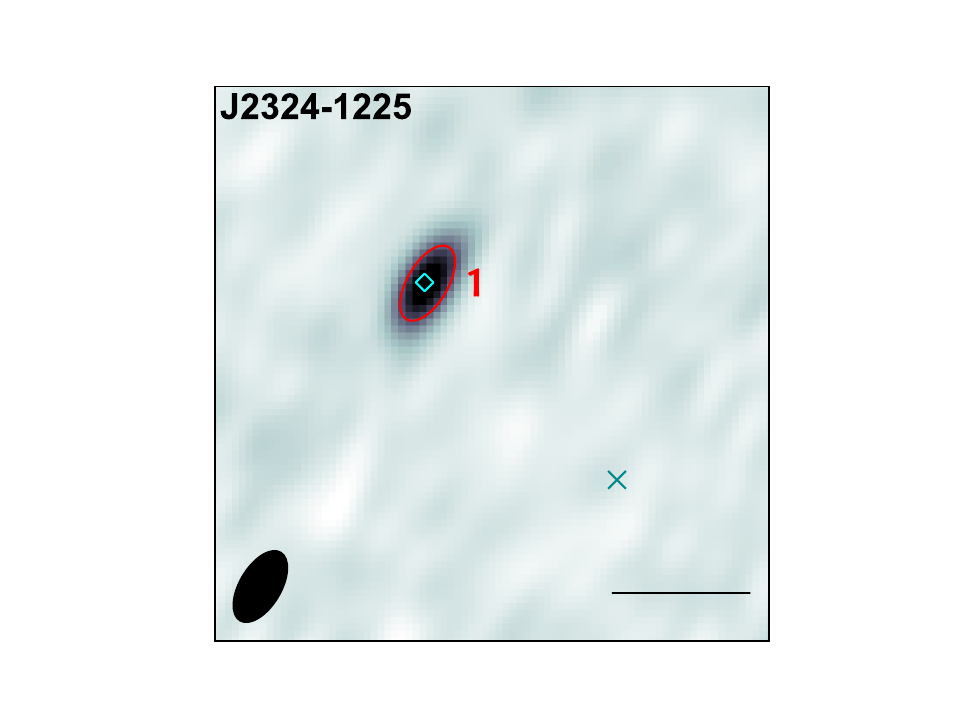}}

     \caption[Single Quasars]{Isolated Single Quasars observed by the VLA. The mark-up scheme is the same as in Figure \ref{fig:newlenses}, with the addition of non-detections for forced apertures (denoted ``FA'' in Table \ref{table:comps}). The blue ``X'' marks the position of the forced aperture, corresponding to an optical survey object.}
     \label{fig:iso}
\end{figure*}

\subsubsection{J0055$-$1212} \label{J0055}
This target, an unconfirmed lensed system, was given a grade between ``B'' and ``A'' by \citet{he23} based on a visual inspection following a catalog search in Legacy Surveys DR9.
Despite having similar optical colors, our VLA data contained strong radio emission from the position of one object and no emission from the other.
We also find additional radio components on both sides of that source that do not correspond to any object in the Legacy Surveys.
These could be from an over-resolved radio jet coming from the main source, though the complete lack of any flux between the hotspots makes this unconvincing.
On the other hand, the alignment of the additional sources suggests some association rather than a chance overdensity of unrelated extragalactic radio sources.
While the system could still be a gravitational lens with radio-quiet source and radio-loud lens, we find no hint of a lens in Legacy Survey Imaging, and the radio-optical alignment suggests this is an optical quasar double or star-quasar association with one radio source, rather than a true lensed quasar.

\subsubsection{J0122$+$7838} \label{J0122}

\citet{rusu19} identified this target as a quasar pair using a catalog-based search incorporating WISE and PAN-STARRS data.
In that work, the lens was given a score of C based on visual inspection and was not followed up spectroscopically.
Our radio follow-up shows emission corresponding with the southern quasar of the pair, and no emission associated with the northern one. 
Forced photometry of the northern component 
We therefore classify this object a non-lens quasar pair.
\subsubsection{J0242$-$1002}\label{J0242}

\citet{dawes23} reports this target as a ``C''-grade lens candidate from a catalog-based search designed to find quasar pairs in DESI Legacy Surveys \citep{Dey2019} data.
Additionally, the brighter of the two quasars in the system has a DESI DR1 redshift of 1.6650 \citep{DESIdr1}, although a warning flag in that data suggests the redshift fit may not be completely reliable.
In our observations, nearby source NVSS J024303$-$100020 was caught at the edge of the primary beam below 7GHz and used to self-calibrate those spectral windows.
We detected only the brighter of the two quasars in the radio data as a point source, and forced photometry at the other quasar location gives a $3.1\sigma$ noise bump, which we interpret as a non-detection.

\subsubsection{J0821$+$1735} \label{J0821}
This target was rated grade ``A'' by \citet{he23}, and the two candidate images have vey similar optical and NIR colors.
However, we only detect one object at C-band, a point source with flux 0.9 mJy, and the forced aperture at the other quasar location gives a $2.8\sigma$ noise bump, a probable non-detection.
We designate the target an optical double quasar and not a lens.

\subsubsection{J0909$-$0749} \label{J0909}
Similar to J0833$-$0721 (Section \ref{J0833}), \citet{lemon2023} list this target as ``unclassified'' due to lack of lensing galaxy, as well as significantly different optical continua.
We detect one quasar image as a slightly extended 6.6 mJy source, and forced photometry at the other quasar's location gives a $1.25\sigma$ peak flux of 31$\mu$Jy, which we interpret as a non-detection.
Therefore, we conclude that this system is not lensed, but is instead an optical quasar pair, and possibly a candidate binary quasar given the similar redshifts.

\subsubsection{J0920$+$2241}\label{J0920}
\citet{dawes23} gave this candidate a ``C'' grade, and an SDSS DR16 image of the brighter quasar gives a redshift of $z=1.36$. 
However, we only detect 6GHz emission coincident with the optically fainter quasar, a marginally resolved 2.5 mJy source.
The brighter quasar registered as a $2.7\sigma$ bump when photometrically forced, and we thus designate this system an optical quasar pair.

\subsubsection{J0926$+$3059} \label{J0926}
\citet{chan22} identified this candidate lens using a catalog-based color search, and its brighter component was measured by SDSS to have $z=2.257$.
In that analysis, the system was labeled ``inconclusive'', but our VLA data can confidently reject this system as two distinct quasars.
We detect a 12 mJy source coincident with the brighter optical source, and a $2.3\sigma$ peak in the \texttt{photutils} aperture. 

\subsubsection{J2324$-$1225} \label{J2324}
\citet{rusu19} lists this source as a ``C'' grade candidate after a catalogue-based search, and note that only one of the two candidate images is detected by \textit{Gaia}.
We also only detect that source in our C-band data, as a 1mJy, marginally resolved source.
Using the Legacy Surveys DR10 position, we find a $1.3\sigma$ peak at the other candidate image position.
Therefore, we conclude this system is likely not a lens.

\subsection{Inconclusive Results and Non-Detected Sources}
\label{sec:inconclusive}
These targets, with the exception of J0940$+$2131, all have spectroscopic data that at very least suggests the lensing hypothesis, but were not confidently detected as radio lenses due to low SNR or odd morphology.
We cannot rule them out as lensed quasars, but also cannot confirm them as radio lenses.

\begin{figure*} \label{fig:inconclusive}
     \centering
     \subfigure{\includegraphics[trim={\mosleftclip, \mosbotclip, \mosrightclip, \mostopclip}, clip, width=\mosaicwidth]{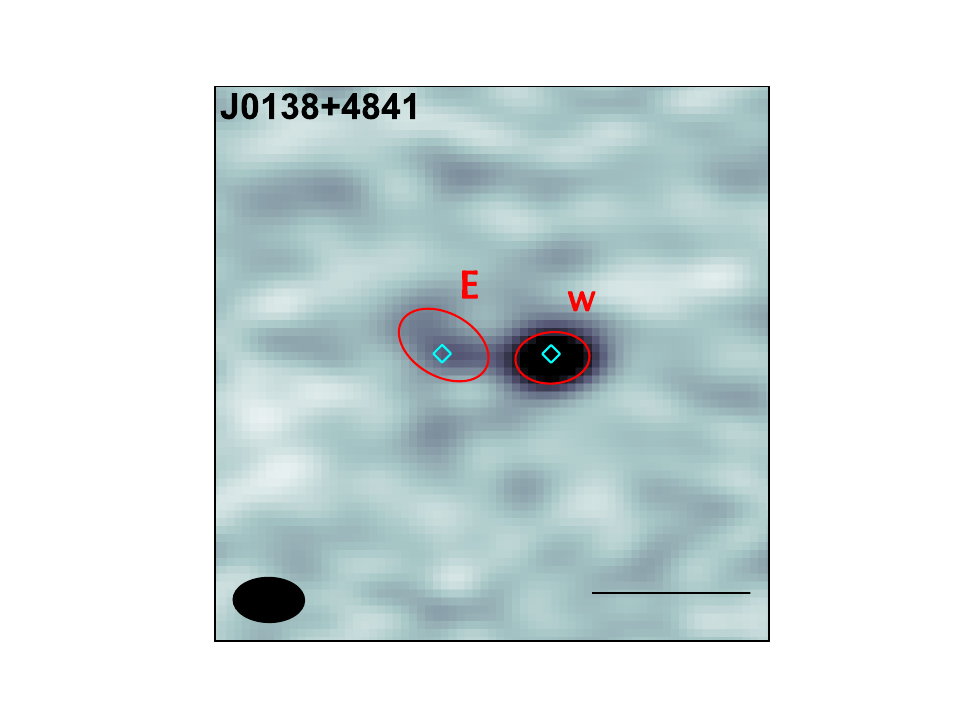}}
     \subfigure{\includegraphics[trim={\mosleftclip, \mosbotclip, \mosrightclip, \mostopclip}, clip, width=\mosaicwidth]{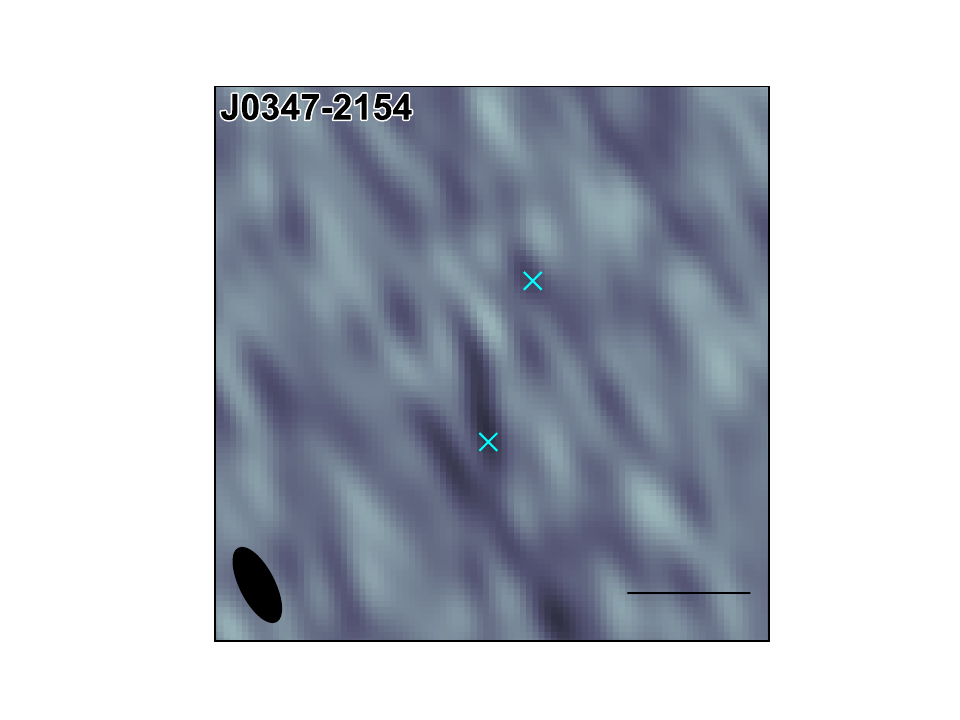}}
     \subfigure{\includegraphics[trim={\mosleftclip, \mosbotclip, \mosrightclip, \mostopclip}, clip, width=\mosaicwidth]{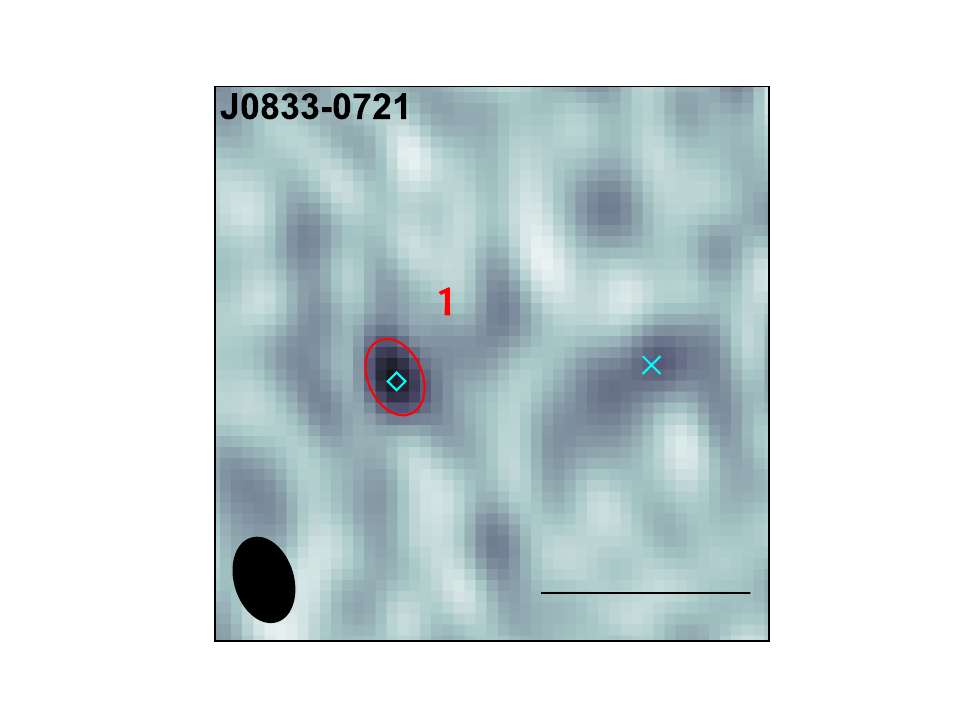}}
     \subfigure{\includegraphics[trim={\mosleftclip, \mosbotclip, \mosrightclip, \mostopclip}, clip, width=\mosaicwidth]{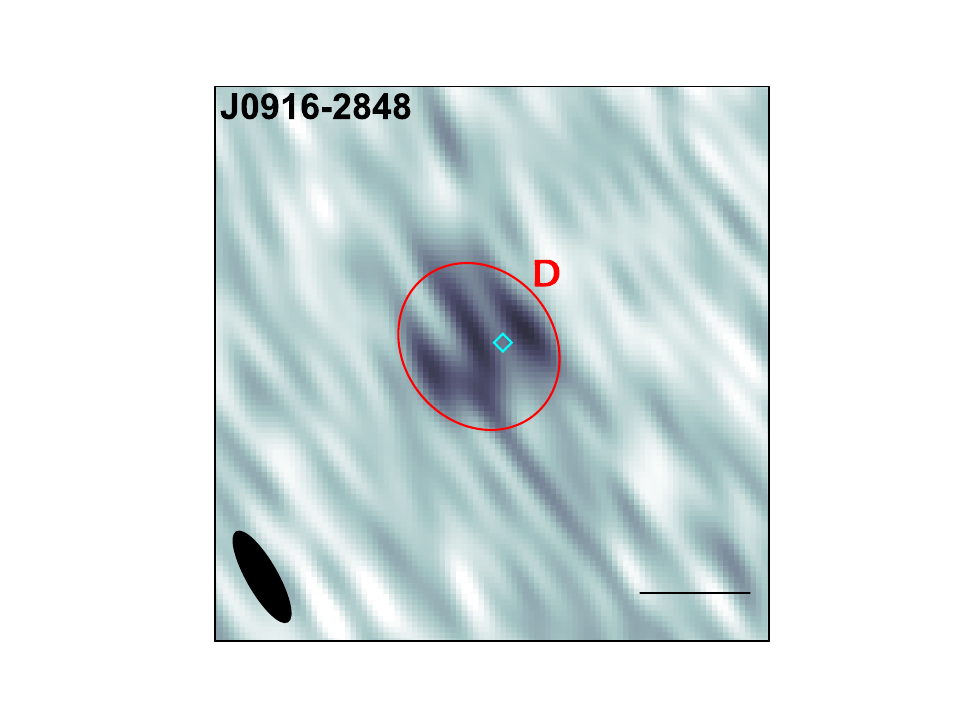}}
     \subfigure{\includegraphics[trim={\mosleftclip, \mosbotclip, \mosrightclip, \mostopclip}, clip, width=\mosaicwidth]{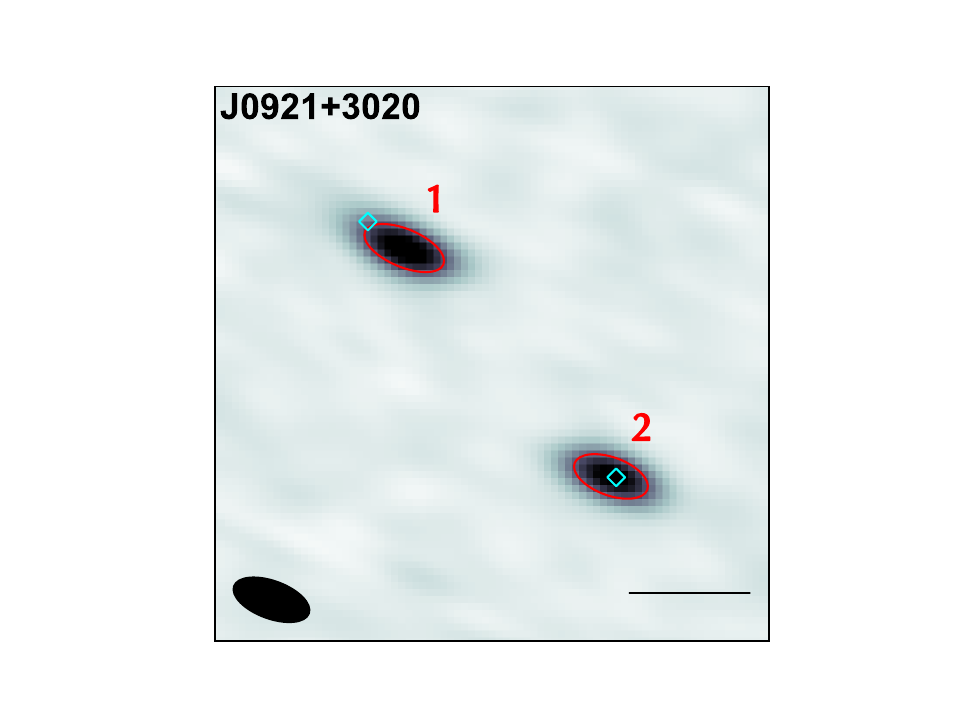}}
     \subfigure{\includegraphics[trim={\mosleftclip, \mosbotclip, \mosrightclip, \mostopclip}, clip, width=\mosaicwidth]{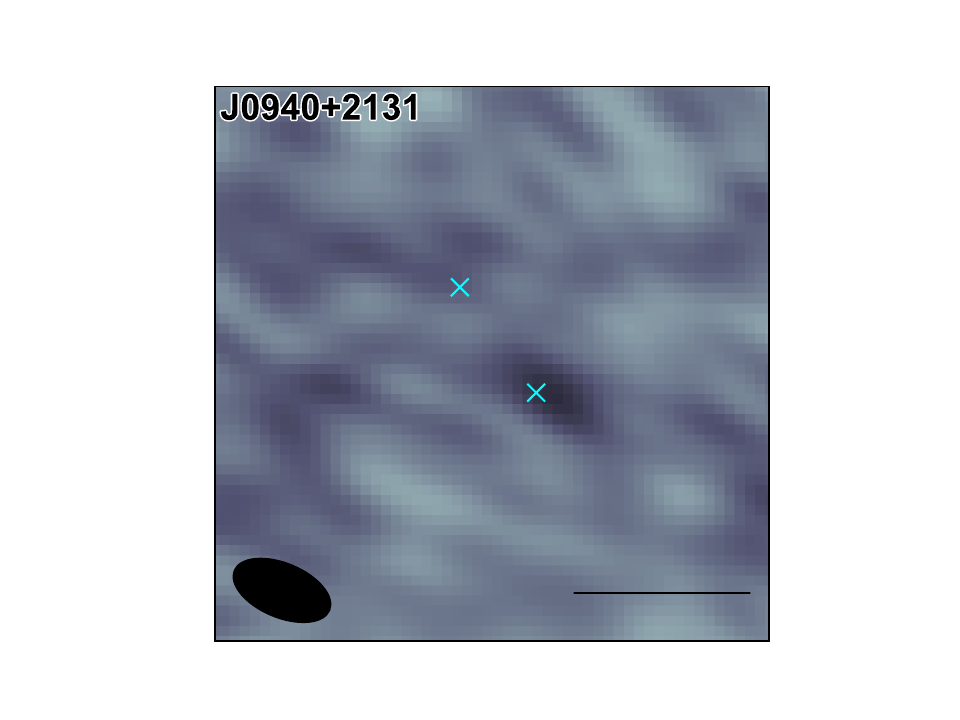}}
     \subfigure{\includegraphics[trim={\mosleftclip, \mosbotclip, \mosrightclip, \mostopclip}, clip, width=\mosaicwidth]{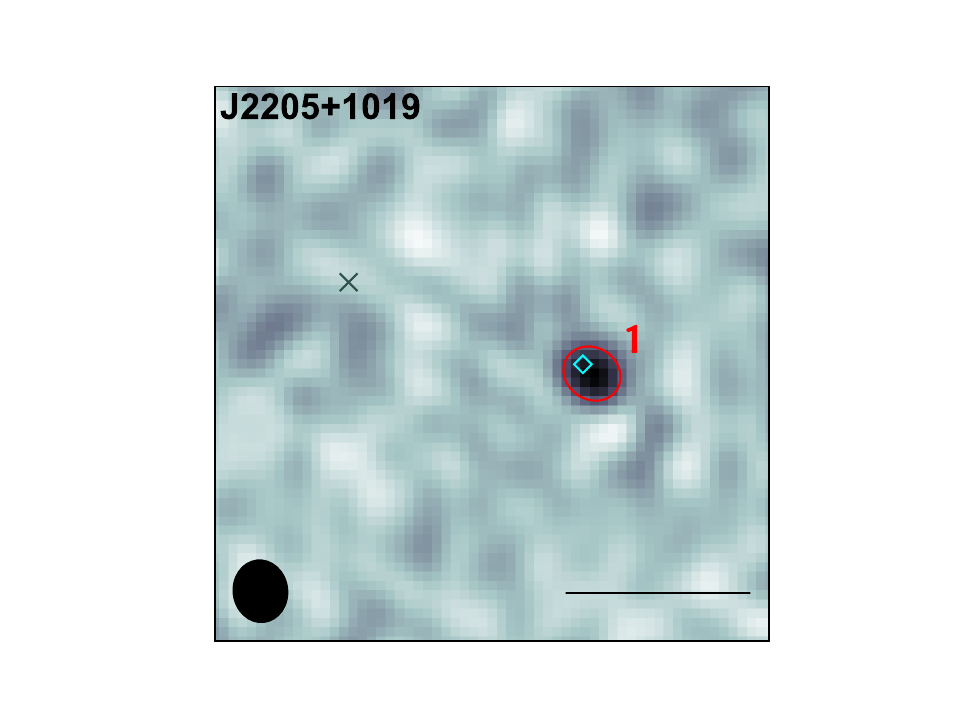}}
     \caption[Inconclusive/Non-Detected Sources]{Inconclusive sources and non-detections. The mark-up scheme is the same as in Figures \ref{fig:newlenses} and \ref{fig:iso}.}
\end{figure*}

\subsubsection{J0138$+$4841} \label{J0138}

This $0''.69$ separation pair was spectroscopically followed up by \citet{lemon2023}, who also noted its presence in VLASS epoch 1. 
While the spectra in that analysis were similar and matched the pair's \textit{Gaia} flux ratio, the lack of high-resolution imaging data for the source prevented a conclusive identification as a lens.
In our VLA follow-up data, the nearby source NVSS J013841$+$483608 was present at the edge of the antenna primary beam below $5.5$Ghz and interfered with imaging, but we were ultimately able to deconvolve it separately and use it for self-calibration of those frequencies.
We found two components, coincident with the \textit{Gaia} positions of the two quasars.
However, given the marginal detection of the fainter component, as well as the continued lack of higher-resolution optical data identifying a lensing galaxy, we label the source ``unclassified.''

\subsubsection{J0347$-$2154}\label{J0347}

Target J0347$-$2154 was spectroscopically confirmed by \citet{lemon2023} as a doubly lensed quasar.
However, we find no corresponding detection at 6GHz, after a 44 minute observation with an image RMS of 4$\mu$Jy.
Our VLASS stack cutout gave a flux of 176$\mu$Jy at 3GHz, but upon re-inspecting the image, the detection seems to be spurious.
We also fail to detect the lens system at lower frequency with NVSS and RACS, which seems to confirm that the quasars being lensed are radio-quiet.
We admit that this source probably should not have been observed in the first place.

\subsubsection{J0833$-$0721} \label{J0833}
\citet{lemon2023} found this target to be a ``Unclassified Quasar Pair'', as both images had very similar spectra in follow-up, but no hint of a lensing galaxy was observed. 
We detect the brighter quasar at a low level of significance, and forced photometry of the fainter quasar gives a $3.8\sigma$ noise bump and realistic implied flux ratio of $A/B=1.6$.
Therefore, while we cannot rule out a lensing origin to the system, we also cannot rule out a binary quasar system or other nearly-identical scenario, and thus leave this system unclassified.

\subsubsection{J0916$-$2848} \label{J0916}
This is lens is also ``unclassified'' by \citet{lemon2023}, who observe different optical continua but similar line profiles between the two quasars, and find no trace of a lens galaxy in survey data.
Analysis of our 49 minute VLA observation on this source recovers one \texttt{BDSF} component which is nearly circular and $1''.5$ in diameter, with an integrated flux of $\sim 200\mu$Jy. 
This component seems to be partially over-resolved by our A-config observations, and given the highly elongated beam used in this observation, we cannot say much more about this detection.
We leave this object ``unclassified''.

\subsubsection{J0921$+$3020} \label{J0921}
This $z=3.33$ double was identified as a $2''.93$ separation lensed quasar by \citet{lemon2023}, and is resolved into two components in VLASS. 
However, while one of those components is coincident with the brighter quasar image, the other radio and optical components do not match nearly as well, with the fainter optical image lying on the other side of the $z=0.428$ lens galaxy from the radio source. 
Our VLA observations find one unresolved source coincident with the bright quasar image, but the other is significantly offset from the \textit{Gaia} position and appears on top of the lens galaxy.
The Legacy Surveys image for this source suffers from blending, and there is no confident optical detection of the lens galaxy.
However, we were able to approximate its position with the peak $z$-band pixel (assuming the quasar counterimage is concentrated in the $g$ band), which is a closer but still unconfident match to the radio position.
One possible explanation for the anomalous radio emission is a radio-emitting lens galaxy, which could blend with the fainter quasar image's flux and produce a single unresolved image offset from either source.
This hypothesis is supported by \texttt{pyBDSF} measuring the component in question with a $\sim0.2''$ deconvolved major axis, but this is in no way conclusive.
While this object is confidently an optical lens, we leave the question of its nature as a radio lens unresolved.

\subsubsection{J0940$+$2131} \label{J0940}
This target was marked grade A by \citet{dawes23}.
While our stacked VLASS image analysis gave a promising combined image flux of 500 $\mu$Jy, \texttt{pyBDSF} detected neither optical object.
Using forced photometry, we find a tentative $4.3\sigma$ detection of one of the optical images, and a much less believable $1.8\sigma$ peak for the other.
Due to the weakness of this detection we cannot rule out lensing for this system, leaving it ``inconclusive.''

\subsubsection{J2205$+$1019}\label{J2205}
This system was listed as a ``probable'' lens by \citet{lemon2023}, who give a quasar redshift of 1.78 and an unusually low lens redshift of 0.108. 
We detect a component at C-band which seems to be coincident with the fainter quasar, but a forced aperture on the brighter one gives only a $1.6\sigma$ peak.
However, the optical data for this system is not good enough to discern the fainter image from the lens galaxy, and so it is possible we are actually detecting radio emission from the lens galaxy, and the lensed quasars are radio quiet. 
We therefore label this system an ``inconclusive'' lens, but note it is probably not a radio lens.

\section{Discussion} \label{sec:disc}

\subsection{Radio Observations as Lens Confirmation}
\begin{figure} %UNITS FOR TIMES!!!
    \centering%                  left bottom right top
    \includegraphics[trim={10pt, 10pt, 10pt, 10pt}, clip, width=0.45\textwidth]{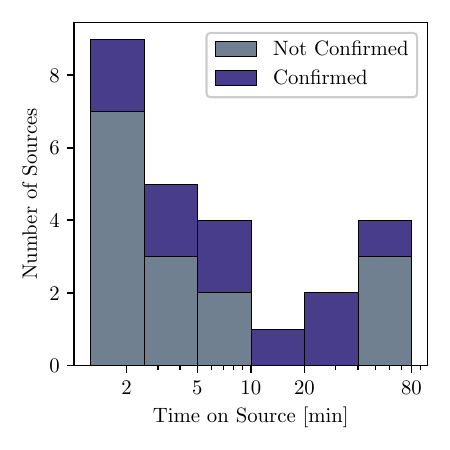}
    \caption[Histogram of Time on Source for Each Target]{Histogram of on-source times for confirmed lenses and non-confirmed candidates. Bins are equally spaced logarithmically, illustrating that the majority of our observations required less than five minutes on source.}
    \label{fig:lenstimes}
    \end{figure}

Our method of radio follow-up of optical lenses has turned out to be an effective tool in disambiguating the nature of lens candidates.
Of the 15 observed candidates without spectroscopic confirmation, 3 were found to be lenses and 4 to be inconclusive, with the remaining 8 confident single-quasar detections.
Figure \ref{fig:lenstimes} shows a histogram of time on source for each of the targets. 
Our VLA observations of these 15 non-confirmed sources took a total of 223 minutes on source, but this is heavily skewed by the three longest observations -- 12 of these sources were observed in 37 minutes.
By comparison, the time necessary for spectroscopic confirmation of a lensed quasar by an optical telescope is typically greater than 10 minutes per target \citep[e.g.,][]{lemon2023}.  
It is however important to note that radio observations are much better as a source of evidence \textit{against} lensing than as they are \textit{for} lensing.
In the former case, a candidate optical lens is revealed to be a bright radio quasar with no counterpart, while in the latter the radio detection of both candidate optical lensed images could be the result of a visual binary of radio quasars or dual AGN rather than a true lens.
Additionally, radio continuum observations provide no way to determine the redshift of the lensed images, meaning that a spectroscopic follow-up would still be necessary if knowing the lensing nature of a system is crucial.
Finally, the fraction of lens candidates with coincident VLASS detection is still about 1 in 100, and so such a disambiguating method is unrealistic for an overwhelming majority of lens candidates. 
These factors mean that radio follow-up of optical lensing candidates will never replace spectroscopic confirmation, but in cases with many lens candidates, it could still provide a cheap measurement of accuracy of the lens search method.

\chapter{Conclusion}
\label{sec:conclusion}
\section{Viability of Astrometry-based Dark matter Constraints}

In Chapter \ref{sec:chapter2}, we used simulations to probe the spread of magnification ratios in gravitational lenses with added dark matter substructure.
Our results pointed to ratios of transfer matrix determinants being similar in constraining power to ratios of image fluxes, the standard analysis method.
A more comprehensive analysis, e.g. testing additional image configurations and a more realistic radio source morphology, would be necessary to confirm this result, but assuming this is successful, we can essentially copy over population-based predictions from flux ratios to this method.
For example, \citet{2024MNRAS.530.2960N} predict a rejection of $M_{hm} \geq 10^7 M_\odot$ given 3\% measurements of flux ratios for 31 lenses, and we could expect a similar result with the same number of radio lenses, provided the measurement precision is there.

Unfortunately it isn't presently clear if that population of radio lenses exists, and while there are more than 30 quadruply lensed radio quasars (see Table \ref{tab:allknown}), VLBI observations of these would be necessary to both ascertain the presence of features usable in the determinant ratio method and to attempt the analysis.
Many of the lenses from the JVAS and CLASS surveys have existing VLBI data from the course of those projects, and these lenses represent the most promising candidates.

However, the real strength of this method is as an independent check on an optical/NIR flux ratio.
As a probe of a different wavelength (and thus size scale), a lens with measurements of both the flux ratio and determinant ratio should be more constraining in general than one with just the flux ratios, and the latter method will provide an effective check to the former. 
 A candidate lens for such an analysis would be J2105$+$6345, which we observed in Chapter \ref{sec:chapter3} and which has also been observed by the JWST for a warm dark matter constraint by \citet{gilman25}.

\section{Lenses Available for Direct Gravitational Imaging}

The sensitivity of the ngVLA is highly dependent on the visibility weighting scheme or ``taper,'' especially when using the largest subarrays.
At 1 milliarcsecond resolution, band 2 of ngVLA (3.5-12 GHz) will achieve 3.5$\mu$Jy/beam with a 1-hour resolution, but this can be improved by over a factor of ten by moving to higher frequency or lower resolution, putting robust detections of few-$\mu$Jy sources well within reach.
This may open up fainter sources to either direct gravitational imaging or more statistical methods, like that presented in Chapter \ref{sec:chapter2}.

In particular, the number of sources available for direct gravitational imaging is subject to strict brightness and geometry requirements.
The two targets for which the procedure has been successful were both discovered in the 1900s and have peak fluxes in excess of 1mJy/beam, even at VLBA resolution.
Both sources also sport impressive arcs which are detected at high signal-to-noise.
To get an idea of the amount of extended emission necessary we examined the global VLBI image of JVAS B1938$+$666 from \citet{b1938vlbi}.
We found that the primary arc in that system (component ``C'' in \citet{1997MNRAS.289..450K}) was spanned by about 50 resolution elements, each with a signal-to-noise of at least 100, and with a total flux at 1.7 GHz of 400 mJy.
The global VLBI observations of MG J0751$+$2716 had a similar SNR across the extended emission and is a similar integrated brightness despite a lower peak flux.

On the other hand, the new lenses presented in this document are much fainter -- for example, J2232$+$1315 (\ref{J2232}) has a brightest image with total (including extended emission) flux of 440 $\mu$Jy\footnote{at 5 GHz as opposed to B1938+666's 1.7 -- however the ngVLA observations to follow up this source would also need to take place at that frequency due to the telescope's sensitivity curve.}, a factor of nearly 1000 less than B1938$+$666.
Supposing a similar situation to the latter lens, where that flux is split evenly across 50 resolution elements (already unlikely given the core- or hotspot-dominated nature of many radio AGN), a sensitivity of 0.088 $\mu$Jy/beam would be required to achieve the same 100 SNR in each element.
At 5 milliarcsecond resolution, this may be achievable in a 3 hour observation using the entire 244-antenna ngVLA\footnote{Calculated using the NRAO's ngECT version 1.0, \url{https://ngect.nrao.edu}.}.
In a more realistic flux distribution scenario where emission is dominated by the core or a few bright points, the addition of other radio observatories to ngVLA would be necessary, as well as a longer observation block. 

Regardless, a sub-mJy lens like J2232+1315 likely represents the lower brightness limit of lenses that could even be theoretically used for a direct gravitational imaging study -- at least for the current and immediate next generation of radio telescopes.
The upshot to this is that whatever imaging candidates are out there, they ought to be bright enough to be detected as unresolved sources in VLASS.
This condition is necessary, not sufficient -- the majority of the lenses presented in previous Chapters would not be suitable for direct gravitational imaging.
Thus, new radio sources like those in Chapters \ref{sec:chapter3} and \ref{sec:chapter4} are essential to filling out the rest of this candidate population.

\section{Selecting More New Radio Lenses}

The method piloted in Chapter \ref{sec:chapter3} and utilized to great effect in Chapter \ref{sec:chapter4} is a reasonably effective way to find radio lenses, and extending it to future surveys in both radio and optical promises to locate many more.
On the optical side, deeper imaging from the Rubin LSST and Euclid surveys promise tens to hundreds of thousands of new lenses, with the latter survey already reporting hundreds \citep{2025A&A...696A.214P, 2025arXiv250315324E}.
For the sub-population of these that are lensed quasars, a crossmatch to stacked VLASS could produce dozens of new radio lenses.
We reiterate that these sources are already present and catalogued in VLASS -- the optical data is simply not good enough to identify them as lenses.
Assuming 200 new quad quasar lenses suitable for a flux ratio analysis in the first years of these upcoming surveys \citep{OM10}, and additionally supposing one tenth of those are radio-emitting, the method outlined in Chapter \ref{sec:chapter2} could be used in at least 20 lenses, provided VLBI follow-up finds suitable radio geometry.
That analysis would provide both a check of the magnification ratio probed by the optical/NIR measurement and an additional constraint based on a different size scale.

Generalizing to radio lenses of any type rather than those useful for Dark Matter constraints, the next decade will see many new sources from entirely radio-based observations thanks to the upcoming SKA and iLOTTS surveys.
The former's Mid survey will be able to match to LSST lenses, and its $0''.04-0''.7$ resolution should make dedicated follow-up to confirm lensing unnecessary, with the lensed images being directly visible in the survey data itself \citep{SKA}.
The latter, which will operate at lower frequency but still with high resolution, will introduce an entirely new population of lensed star-forming galaxies with low-frequency synchrotron emission \citep{iLOFAR, iLOTTS}.
These objects will also be directly visible in the survey data itself, but we expect their surface brightnesses to be too low and spectra too steep to be suitable for gravitational imaging.

\appendix
\chapter{Derivations and Calculations}
\paragraph{In this Appendix } 
we review the derivation of the lensing observables from General Relativity (GR), beginning with the Einstein Field Equations.
The treatment here will broadly follow that of \citet{galaxiesbook} and \citet{carrolbook}, with additional elements from other sources noted in the text, and will focus on the calculation of geodesics in linearized gravity. 
The aim here is not to be comprehensive by any means, but to provide some relativistic background to a reader with passing familiarity, and as such we will gloss over the majority of the tensor calculus details (see the above references for much more thorough treatments).
We will use Einstein summation notation unless otherwise noted, and the $(-+++)$ sign convention.
While factors of $c$ are included in the main text, here we set $c=1$.

\section{An Extremely Brief Introduction to General Relativity}

The theory of GR is often summed up with the elegant Einstein Field Equations $G_{\mu\nu} = 8\pi G T_{\mu\nu}$, a statement which is opaque at best --
in reality, we need many ingredients in order to start working through a problem such as the bending of light in a weak gravitational field.
In general, when finding a geodesic in a given spacetime, we first need the \textbf{metric} for that spacetime $g_{\mu\nu}$.
The metric has myriad uses in GR, but in this context its most important role is defining the notion of ``shortest distance'' between two points.
For example, Minkowski space, which describes a perfectly flat universe with no matter at all, has the following metric $\eta_{\mu\nu}$\footnote{$\eta$ as a metric signifier is reserved specifically for Minkowski space.}:
\begin{equation}
\renewcommand{\arraystretch}{0.6}
\eta_{\mu\nu} = \begin{pmatrix}
-1 & \!0 & 0 & 0 \\
0 & \!1 & 0 & 0 \\
0 & \!0 & 1 & 0 \\
0 & \!0 & 0 & 1 \\
\end{pmatrix}.
\end{equation}
For a given metric $g_{\mu\nu}$ there is also an inverse metric $g^{\mu\nu}$ with $g_{\mu\lambda}g^{\lambda\nu} = \delta^{\nu}_{\mu}$, with $\delta$ the Kronecker delta, or identity matrix.
If the metric is defined as a matrix as above, its inverse is the matrix inverse.
The metric and inverse metric can be used to raise and lower indices of a given vector or tensor through the summing of repeated indices, i.e. $g^{\mu\nu}h_{\mu\rho}=h^\nu_\rho$ for arbitrary $h$.
We can also obtain the familiar notion of \textbf{trace} with the metric, notated like so: $\text{tr}(h) = g^{\mu\nu}h_{\mu\nu} = h^\mu_\mu$
Another more space-efficient way to define the metric is with the \textbf{line element} $ds$:
\begin{equation}
\label{eqn:lineelement}
ds^2 = g_{\mu\nu}\text{d}x^{\mu}\text{d}x^{\nu}.
\end{equation}
In Minkowski space, then, we have $ds^2 = -\text{d}t^2 + \text{d}x^2 + \text{d}y^2 + \text{d}z^2$, which looks like typical Euclidean 3-dimensional space if you ignore the time component.
As in Euclidean space, the line element can be used to calculate distances, but once we move beyond flat spacetime the procedure becomes much more complicated than a simple $\Delta s = \int ds$.

Recall that in GR, what we experience as the force of gravity is the curvature of spacetime.
Just as an elementary kinematics problem accounts for the acceleration due to gravity, all our calculations of paths must take curvature into account.
To generalize both the motion of objects in ``freefall'' and the ``straight line'' paths light takes through space, we introduce the \textbf{geodesic equation}:
\begin{equation}
\label{eqn:geodesic}
\frac{d^2x^{\mu}}{d\lambda^2} + \Gamma^{\mu}_{\rho\epsilon}\frac{dx^{\rho}}{d\lambda}\frac{dx^{\epsilon}}{d\lambda} = 0.
\end{equation}
The geodesic equation describes paths $x(\lambda)$ through spacetime, with a generic $\lambda$ parameterizing the curve\footnote{For a null geodesic, we cannot use the proper time to parameterize as it becomes singular at the speed of light.}.
The $\Gamma$ in the formula is the \textbf{Christoffel symbol}, which is a correction used to ensure derivatives of vectors transform properly in curved space.
It is related to the partial derivatives of the metric's elements (denoted like $\partial_\mu$) in this way:
\begin{equation} \label{eqn:Chris}
\Gamma^{\lambda}_{\mu\nu} = \frac{1}{2}g^{\lambda \epsilon}\left(\partial_\mu g_{\nu\epsilon} + \partial_\nu g_{\epsilon\mu}-\partial_\epsilon g_{\mu\nu}\right).
\end{equation}

In the situation where we already have a well-defined metric, the geodesic equation is all we need to study motion in a given spacetime.
However, if we have a more generic case (as we will see in the next section) we must derive the metric's specific form.
In that case, we must introduce a few more quantities, the first being the \textbf{Riemann Tensor}:
\begin{equation} \label{eqn:riemann}
R^\lambda_{\epsilon\mu\nu} = \partial_\mu\Gamma^\lambda_{\nu\epsilon}-\partial_\nu\Gamma^\lambda_{\mu\epsilon}+\Gamma^\lambda_{\mu\delta}\Gamma^\delta_{\nu\epsilon}-\Gamma^\lambda_{\nu\delta}\Gamma^\delta_{\mu\epsilon}.
\end{equation}
The Riemann tensor can be thought of as the commutator of partial derivatives\footnote{actually ``covariant derivatives'', a generalization of the partial derivative for curved spaces.}, and it vanishes in flat space.
As the Riemann tensor involves only Christoffel symbols and their derivatives, it is completely determined by the metric and its derivatives.
We ``contract'' the Riemann tensor (by summing over indices Einstein-wise) to create two more quantities, the \textbf{Ricci Tensor} $R_{\mu\nu}$ and \textbf{Ricci Scalar} $R$:
\begin{equation}\label{eqn:ricci}
R_{\mu\nu} = R^\lambda_{\mu\lambda\nu}; \quad R = g^{\mu\nu}R_{\mu\nu}.
\end{equation}
The final piece to discuss is the \textbf{stress-energy tensor} $T^{\mu\nu}$, which is a generalization of a single particle's 4-momentum to a macroscopic system.
This tensor is determined based on the properties of the matter in a given situation.
In cosmology, it is a good approximation to treat matter (galaxies, dust, dark matter, etc) as a perfect fluid with energy density (in the rest frame) $\rho$ and isotropic pressure $p$.
In the ``comoving'' frame we will work in for cosmology, the tensor will depend on the metric, but it takes on a diagonal form when one index is raised:
\begin{equation} \label{eqn:stressenergy}
\renewcommand{\arraystretch}{0.6}
T^{\mu}_{\nu} = T_{\lambda\nu}g^{\lambda\mu}\begin{pmatrix}
-\rho & 0 & 0 & 0 \\
0 & p & 0 & 0 \\
0 & 0 & p & 0 \\
0 & 0 & 0 & p \\
\end{pmatrix}.
\end{equation}
For cosmological purposes, we will define the \textbf{equation of state} $w=p/\rho$.
With all that out of the way, we are finally able to fully introduce the full \textbf{Einstein Field Equations} (EFE's):
\begin{equation} \label{eqn:EFE}
R_{\mu\nu}-\frac{1}{2}Rg_{\mu\nu} + \Lambda g_{\mu\nu}= 8\pi G\,T_{\mu\nu}.
\end{equation}
The first two terms on the left hand side of the equation are sometimes combined to make the ``Einstein Tensor'' $G_{\mu\nu}$, and the $\Lambda$ is of course the infamous \textbf{cosmological constant}.

\section{Flat FLRW Spacetime} \label{sec:flatflrw}

Before we examine light's deflection around massive bodies, we first need to consider the shape of space as we know it today.
Gravitational lensing takes place in an expanding universe that can be described by the following Friedman-Lema\^{i}tre-Robertson-Walker (FLRW) metric:
\begin{equation} \label{eqn:flrw}
ds^2 = -\text{d}t^2 + a(t)^2\left(\frac{\text{d}r^2}{1-k r^2}+r^2d\Omega^2\right),
\end{equation}
where $a(t)$ is a dimensionless ``scale factor'' and $k$ is the curvature ($+1$, $-1$, and $0$ values of $k$ mean a closed, open, and flat universe, respectively).
The FLRW universe is spatially homogeneous and isotropic -- note the d$\Omega$ in the line element -- and thus the scale factor mainly dictates cosmological distances and leaves angles unchanged.
While it isn't totally necessary when considering lensing, we will consider the EFE's for a flat $(k=0)$ universe in order to introduce some concepts related to the $\Lambda$CDM cosmology.
It turns out we can incorporate non-zero curvature as an energy component later, so the flat case will be adequate for now\footnote{see \citet{carrolbook} and \citet{galaxiesbook} for an approach using a general $k$, as well as the derivation of the metric from the assumption of homogeneity.}.

The Flat FLRW metric is given by:
\begin{equation} \label{eqn:flrwflat}
ds^2 = -\text{d}t^2 + a^2(t)(\text{d}t^2 + \text{d}x^2 + \text{d}y^2 + \text{d}z^2),
\end{equation}
a diagonal matrix with $(-1, a^2, a^2, a^2)$ on the diagonal.
We can calculate the Christoffel symbols for this metric, most of which go to zero thanks to its diagonal nature and lack of spatial derivatives.
Taking Latin indices to be spatial only ($1-3$), the two non-zero symbols are:
\begin{equation} \label{eqn:flatchris}
\Gamma^0_{ij} = \delta_{ij}a\dot{a}; \quad \Gamma^i_{0j} = \Gamma^i_{j0} = \delta_{ij}\frac{\dot{a}}{a}.
\end{equation}
The Riemann Tensor is much more tedious to calculate, but it turns out that it too only has a few nonzero components.
Up to antisymmetry $(R^\lambda_{\epsilon\mu\nu} = -R^\lambda_{\epsilon\nu\mu})$ they are:
\begin{equation}
R^0_{i0i} = a\ddot{a}; \quad R^i_{00i} = \frac{\ddot{a}}{a}; \quad R^i_{jij} = (1-\delta_{ij})\dot{a}^2.
\end{equation}
From here, contracting to the Ricci Tensor is relatively simple:
\begin{equation}
R_{00} = -3\frac{\ddot{a}}{a}; \quad R_{0i}=R_{i0}=0; \quad R_{ij}=\delta_{ij}(a\ddot{a} + 2\dot{a}^2),
\end{equation}
as is construction of the Ricci Scalar:
\begin{equation}
R = g^{\mu\nu}R_{\mu\nu} = \frac{6}{a^2}(a\ddot{a} + \dot{a}^2).
\end{equation}

The last ingredient needed for the EFE's is the cosmological constant $\Lambda$, for which we will follow the example of \citet{dodelson} and move into the stress-energy tensor.
This allows for an easier treatment of dynamical dark energy, where the equation of state parameter $w$ varies with time.
In fact, rather than consider one 	``cosmological fluid'' with a single density and pressure, we can treat each component of the universe separately, such that our stress-energy tensor takes the form:
\begin{equation}
T^\mu_\nu = T_M + T_R + T_\Lambda + (\ldots),
\end{equation}
omitting the identical $^\mu_\nu$ indices for the components for simplicity's sake.
Here, we essentially treat the total mass-energy content of the universe as a set of coexisting fluids with different properties, although in practice at most times one component dominates over the others, and $T^\mu_\nu$ will retain its diagonal form as in \ref{eqn:stressenergy}.
The Cosmological Constant contribution takes the form $T_\Lambda = -\frac{\Lambda}{8\pi G}\delta^\mu_\nu$, which also means we have set its equation of state to be $w_\Lambda=-1$.

We can now solve the Einstein Field Equations (\ref{eqn:EFE}), beginning with the $00$ term:
\begin{equation}\label{eqn:1fried}
\left(\frac{\dot{a}}{a}\right)^2 = \frac{8\pi G}{3}\rho,
\end{equation}
a relation also called the \textbf{First Friedman Equation}.
As both sides of our EFE's are diagonal, we are left with the $ii$ terms:
\begin{equation}
2a\ddot{a} - \dot{a}^2 = -8\pi G p,
\end{equation}
which we can substitute into \ref{eqn:1fried} to produce the \textbf{Second Friedman Equation}:
\begin{equation} \label{eqn:2fried}
\frac{\ddot{a}}{a} = -\frac{4\pi G}{3}(\rho+3p).
\end{equation}
These equations govern the universe's largest scales, and are used in greater detail in \ref{sec:LCDM}.

\subsection{Density Parameters and Spatial Curvature}

We can rewrite \ref{eqn:1fried} in order to introduce some convenient parameters.
Dividing the constants, we have:
\begin{equation}
\frac{3H^2}{8\pi G} = \rho = \rho_{cr},
\end{equation}
where $\dot{a}^2/a^2 = H$, the Hubble Parameter.
This introduces $\rho_{cr}$, the \textbf{critical density}, which is simply the density of a Euclidean universe.
This definition may seem tautological, but it is motivated by the possibility of a non-flat universe.
While the flat universe is supported by late-time measurements such as  \citep{DESI24}, the most recent Planck CMB analysis \citep{planck18} which supports an open ($k=-1$) universe, and the possibility of curvature shouldn't simply be ignored.
Rather than re-do calculations for each possible value of $k$, we instead can re-introduce the spatial curvature as a component of the stress-energy tensor as we did for the Cosmological Constant.

The Friedman equations for general $k$ are largely unchanged from the flat case, apart from the addition of a new term in the first equation:
\begin{equation}
\left(\frac{\dot{a}}{a}\right)^2 = H^2 = \frac{8\pi G}{3}\rho - \frac{k}{a^2}.
\end{equation}
Incorporating our definition of $\rho_{cr}$, we have:
\begin{equation} \label{eqn:omegaK}
\frac{k}{H^2a^2} = \frac{\rho}{\rho_{cr}} - 1 = \Omega - 1.
\end{equation}
$\Omega$, the \textbf{density parameter}, is 1 when the universe is Euclidean.
Cosmologists typically break up $\Omega$ into components just as we broke up $T^\mu_\nu$ before, so we can define quantities like $\Omega_m$ and $\Omega_R$, with $\Omega_i =\rho_i/\rho_{cr}$.
For convenience, we define $\Omega_k$ to be the left side of \ref{eqn:omegaK}, so our total density parameter is the sum of all individual densities and can be a number other than 1.
Convention also typically sets $\Omega_i = \Omega_i(t_0)$, including $\Omega_k = k/H_0^2$. 

The component $\Omega_i$ evolve different according to their equations of state $w = p/\rho$.
As we are considering cosmic ``fluids'', they must follow the \textbf{continuity equation}:
\begin{equation}\label{eqn:continuity}
\dot{\rho} = -3\frac{a}{\dot{a}}(\rho + p).
\end{equation}
Using our definition of equation of state, we can write the general solution to \ref{eqn:continuity} as:
\begin{equation}\label{eqn:stateevo}
\rho \propto a^{-3(1+w)}.
\end{equation}
As previously mentioned, we set $w_\Lambda = -1$, but it could evolve with time.
Nonrelativistic matter (both baryonic and dark) can be considered to be pressureless, and so $w_m = 0$, while radiation has $w_r = 1/3$ (this is due to the tracelessness of the electromagnetic stress-energy tensor).
Neutrinos, which have very small but nonzero mass, act like radiation in the early universe but like matter today, and so their (small) contribution to $\rho$ is also time-varying.

Incorporating \ref{eqn:stateevo} for the different components into \ref{eqn:omegaK}, the first Friedman Equation in the present universe becomes:
\begin{equation}
H^2 = H_0^2(\Omega_ra^{-4} + \Omega_ma^{-3} + \Omega_ka^{-2}+\Omega_{\Lambda}).
\end{equation}

\section{General Relativity in the Weak-Field Limit} \label{sec:appdxlens}

We are interested in the deflections of light around massive bodies in space due to gravity, but nearly all gravitational lensing events only involve relatively weak fields (the exception being things like, for example, the Event Horizon Telescope's images of black hole shadows).
In this case, it is sufficient to add a linear perturbation to the metric that describes the space with no lens in it.
For us, this is FLRW space, but we are able to use Minkowski space as our base metric to perturb without issue, for two reasons.
Firstly, we are concerned with light bending on distance scales no larger than a galaxy cluster, which is much smaller than the cosmological scale of the FLRW universe.
While the light ultimately travels a cosmological distance, the time spent being deflected is small compared to the total travel time from source to observer.
This is essentially the thin lens approximation, which will come up again later.
Secondly, the isotropy of FLRW spacetime implies no bending of light based on spacetime's largest scales, especially in our Euclidean or near-Euclidean universe.
The scale factor will of course change the energy of the light via cosmological redshift, but the lensing observables will be the same between either background metric, assuming we use the correct distances between source and lens.
For a complete treatment of weak gravitational fields in the FLRW metric, see \citet{bertschinger}.

We now introduce the metric of \textbf{linearized gravity}, aka Minkowski spacetime with linear perturbations:
\begin{equation}
g_{\mu\nu} = \eta_{\mu\nu} + h_{\mu\nu}; \quad |h_{\mu\nu}| \ll 1.
\end{equation}
As we are looking at linear perturbations only, we will be discarding any $h^2$ or second derivative terms, a fact we first use when finding the inverse metric $g^{\mu\nu}$.
Recalling that the metric is effectively a $4\times 4$ matrix $g = \eta + h$, we can use the fact that the Taylor expansion for $(1+x)^{-1}$ holds for matrices to get:
\begin{equation}
(\eta + h)^{-1} = (\eta + \eta\eta^{-1}h)^{-1} = \eta^{-1}(\delta+\eta^{-1}h)^{-1} \approx \eta^{-1}(\delta - \eta^{-1}h) = \eta^{-1}-\eta^{-1}\eta^{-1}h,
\end{equation}
recalling the Kronecker delta $\delta$ is equivalent to the identity matrix in this case.
Translating back into the element notation of tensors, we define $h^{\mu\nu}$ as follows:
\begin{equation}
g^{\mu\nu} = \eta^{\mu\nu}\! - \eta^{\mu\lambda}\eta^{\nu\rho}h_{\lambda\rho} = \eta^{\mu\nu}\! - h^{\mu\nu}.
\end{equation}
The multiplications by the inverse Minkowski matrix in the definition of the inverse perturbation are solely for raising the indices, and in the standard basis (where $\eta_{\mu\nu}$ takes its familiar diagonal form) the entries of $h_{\mu\nu}$ and $h^{\mu\nu}$ are equal.
Note also that we are able to raise and lower the indices of $h$ using $\eta$ thanks to our perturbative definition, this property will become useful when setting up the EFE's for this metric.

We next need the Christoffel symbols for the perturbed metric.
Recalling that Minkowski space is homogeneous and isotropic in all directions, including time, and so all its derivatives are zero.
Therefore, when using Equation \ref{eqn:Chris}, we only need to look at derivatives of the perturbation.
Furthermore, we can split the inverse metric and discard any $h\partial h$-type terms as they are second order in $h$ and beyond our expansion.
\begin{equation}
\label{eqn:pertChris}
\Gamma^{\lambda}_{\mu\nu} = \frac{1}{2}\eta^{\lambda\epsilon}\left(\partial_\mu h_{\nu\epsilon} + \partial_\nu h_{\epsilon\mu} - \partial_\epsilon h_{\mu\nu}\right) + \mathcal{O}(h^2).
\end{equation}
From here, it may initially seem that we have everything we need to solve the geodesic equation.
However, we have only a generic form of a metric, which could be valid for any linear order perturbation of flat space.
It would be much better to relate our perturbation $h$ to something like the classical gravitational potential $\Phi$, and so we must continue on to obtain the full Einstein Field Equations.

Of the four terms defining the Riemann tensor (Equation \ref{eqn:riemann}), we can immediately throw out the last two, as they are clearly $\mathcal{O}(h^2)$.
As before, the derivatives of $\eta$ all cancel, and so the tensor we get is composed of the second derivatives of $h$:
\begin{equation}
R^\lambda_{\epsilon\mu\nu} = \frac{1}{2}\eta^{\lambda\rho}(\partial_\mu\partial\nu h_{\epsilon\rho} + \partial_\mu\partial_\epsilon h_{\rho\nu} - \partial_\mu\partial_\rho h_{\nu\epsilon}) - \frac{1}{2}\eta^{\lambda\rho}(\partial_\nu\partial_\mu h_{\epsilon\rho} + \partial_\nu\partial_\epsilon h_{\rho\mu} - \partial_\nu\partial_\rho h_{\mu\epsilon}).
\end{equation}
We can combine these first terms and use the equality of mixed partial derivatives $\partial_\mu\partial_\nu = \partial_\nu\partial_\mu$ for the final form of $R^\lambda_{\epsilon\mu\nu}$,
\begin{equation}
R^\lambda_{\epsilon\mu\nu} = \frac{1}{2}\eta^{\lambda\rho}(\partial_\mu\partial_\epsilon h_{\rho\nu} - \partial_\mu\partial_\rho h_{\nu\epsilon} - \partial_\nu\partial_\epsilon h_{\rho\mu} + \partial_\nu\partial_\rho h_{\mu\epsilon}).
\end{equation}
We proceed by constructing the Ricci tensor and scalar from \ref{eqn:ricci}.
To keep indices consistent with the previous equation, we will call the Ricci Tensor $R_{\epsilon\nu}$ rather than $R_{\mu\nu}$, but this relabeling doesn't change the calculation.
Here, we can use $\eta$ to raise indices of $h$ when applicable, and use our above definition of trace:
\begin{equation}
R_{\epsilon\nu} = R^\lambda_{\epsilon\lambda\nu} = \frac{1}{2}(\partial_\lambda\partial_\epsilon h^{\lambda}_{\nu} - \partial_\lambda\partial_\rho\eta^{\lambda\rho}h_{\nu\epsilon} - \partial_\nu\partial_\epsilon \text{tr}(h) + \partial_\nu\partial_\rho h^\rho_\epsilon).
\end{equation}
Contracting the remaining indices gives the scalar Ricci curvature, where once again we ignore $h\partial\partial h$ terms:
\begin{equation}
R = g^{\epsilon\nu}R_{\epsilon\nu} = \frac{1}{2}(\partial_\lambda\partial_\epsilon h^{\epsilon\lambda} - \partial_\lambda\partial_\rho \eta^{\lambda\rho}\text{tr}(h) - \partial_\nu\partial_\epsilon\eta^{\epsilon\nu}\text{tr}(h) + \partial_\nu\partial_\rho h^{\rho\nu}) + \mathcal{O}(h^2).
\end{equation}
We clean this up by relabeling all our summed dummy indices for a final expression of $R$:
\begin{equation}
R = \partial_\mu\partial_\nu h^{\mu\nu} - \partial_\mu\partial_\nu\eta^{\mu\nu}\text{tr}(h).
\end{equation}
We can now at long last write the EFE's (Equation \ref{eqn:EFE}) for a general linear perturbation to Minkowski Space, by throwing out our final $\mathcal{O}(h^2)$ terms, including the $\Lambda h_{\mu\nu}$ term (we assume the effect of the cosmological constant to be small for the perturbation and thus only include $\Lambda \eta_{\mu\nu}$):
\begin{equation}
\label{eqn:linearEFE}
\frac{1}{2}(\partial_\epsilon\partial_\mu h^\epsilon_\nu - \partial_\epsilon\partial_\rho\eta^{\epsilon\rho}h_{\mu\nu} - \partial_\mu\partial_\nu\text{tr}(h) + \partial_\nu\partial_\epsilon h^\epsilon_\mu - \partial_\epsilon\partial_\rho h^{\epsilon\rho}\eta_{\mu\nu} + \partial_\epsilon\partial_\rho\eta^{\epsilon\rho}\text{tr}(h)\eta_{\mu\nu}) + \Lambda\eta_{\mu\nu} = 8\pi GT_{\mu\nu}.
\end{equation}
We will use the diagonal stress-energy tensor defined in \ref{eqn:stressenergy}, where we are in the rest frame ``cosmological fluid'' is not moving.
This approximation is more than sufficient for our linear approximation, and eventually, we will simplify it further by removing the pressure terms (this corresponds to not considering relativistic matter like neutrinos).

\subsection{Gauge Invariance}

Recall that the symmetry of the EFE's gives us ten equations to solve, rather than the sixteen implied by the $4\times4$ nature of the tensors.
However, four of the ten parameters in these equations are simply our coordinates ($x^i$ and the proper time $\tau$), which we are allowed to transform without changing coordinates.
This is a form of \textbf{gauge invariance} analogous to the one from Maxwell's Equations, but it turns out to be more tricky due to the non-commutativity of our tensors.
In our linearized metric, we inherit this gauge invariance, which we will exploit to make solving the EFE's easier.
As with much of the above, this treatment is heavily abridged and the reader is referred to \citet{carrolbook} and \cite{bertschinger} for more details.

To make the field equations easier to deal with, we will split $h$ into ``Tensor'', ``Vector'', and ``Scalar'' parts, which correspond to $h_{ij}=h_{ji}$, $h_{0i}=h_{i0}$, and $h_{00}$ respectively, and recalling the Einstein convention that latin indices correspond only to spatial dimensions.
This is an example of a group theoretic ``irreducible representation,'' and the three parts will only transform into themselves under rotations, similarly to how a quantum mechanical 2-electron system decomposes into singlet and triplet states.
Using some highly suggestive functional forms, we define the decomposed components as follows:
\begin{equation}
h_{00}=-2\Phi; \quad h_{0i} = w_i; \quad h_{ij} = 2s_{ij} - 2\Psi\delta_{ij}.
\end{equation}
Note that we have separated the tensor part into trace-free and trace components $s_{ij}$ and $\Psi$, respectively:
\begin{equation}
s_{ij} = \frac{1}{2}(h_{ij} - \frac{1}{3}\delta^{ab}h_{ab}\delta_{ij}); \quad \Psi = -\frac{1}{6}\delta^{ij}h_{ij}.
\end{equation}
We could now plug these definitions back into the various symbols and definitions from the previous section, but in the interest of space we will instead immediately define our gauge and solve the EFE's from there (see \citet{carrolbook} for those detailed quantities).
We will use a gauge which is analogous to electromagnetism's Coulomb Gauge, referred to as the \textbf{transverse gauge} in \citet{carrolbook}, and the \textbf{Poisson gauge} in \citet{bertschinger} and \citet{galaxiesbook}.
It is defined as follows:
\begin{equation}
\partial_i s^{ij} = 0; \quad \partial_i w^i = \partial_i h^{0i} = 0,
\end{equation}
recalling that repeated indices are still summed in Einstein notation, and so this gauge takes the divergences of $w$ and $s$ to be zero.
Incorporating these conditions into \ref{eqn:linearEFE} will give us solvable equations which we can use to find our weak-field Newtonian metric (be very careful of the minus sign in the definition of $\Psi$ when taking traces of $h$).
We will also reintroduce the familiar Laplace operator $\nabla^2 = \partial_i\partial_i\delta^{ii}$ to get our new EFE's:
\begin{equation}
\begin{gathered}[t]
2\nabla^2\Psi - \Lambda = 8\pi G T_{00}; \quad \partial_0\partial_i\Psi - \frac{1}{2}\nabla^2w_i = 8\pi G T_{0i}; \\
(\delta_{ij}\nabla^2 - \partial_i\partial_j)(\Phi - \Psi) - \partial_0(\partial_i w_j + \partial_j w_i) + 2\partial_0^2\Psi\delta_{ij} - \partial_\epsilon\partial_\rho\eta^{\epsilon\rho}s_{ij} + \Lambda\delta_{ij} = 8\pi G T_{ij}.
\end{gathered}
\end{equation}
We will now simplify these equations in three ways. First, recall that in the Newtonian limit, matter is moving non-relativistically, and so the pressure from the ``cosmological fluid'' will be much less than the fluid's energy density, which allows us to set the spatial diagonal of $T$ to zero, leaving us with one finite term $T_{00} = \rho$ (this is the same as setting matter's equation of state $w_m=0$).
Taking this a step further, we can restrict ourselves to only discussing static sources of gravity (a very reasonable assumption for gravitational lensing), allowing for the removal of all time derivatives of $\Phi$, $\Psi$, $w_i$, and $s_{ij}$.
Finally, we drop the cosmological constant terms, as we are concerned with relatively small structures (i.e. galaxies and galaxy clusters) that have densities much greater than the energy of $\Lambda$.
Putting this all together, our equations are much simpler:
\begin{equation}
\begin{gathered}[t]
\nabla^2\Psi = 4\pi G\rho; \quad \nabla^2 w_i = 0; \quad(\delta_{ij}\nabla^2 - \partial_i\partial_j)(\Phi - \Psi)  = \nabla^2 s_{ij}.
\end{gathered}
\end{equation}
The first of these equations is clearly the Poisson equation for gravity, meaning that we have properly recovered the Newtonian limit.
The second, along with our divergence-free gauge for $w_i$, sets the vector components of the perturbation to zero.
Finally, we can take the trace of the third equation (again being very careful about signs in the $\Psi$ term), to get the following:
\begin{equation}
\nabla^2(\Phi - \Psi) = 0.
\end{equation}
If we assume our perturbations are well-behaved and decay to zero at infinity (like the gravitational field of a galaxy), this sets $\Phi = \Psi$, and we can rewrite our scalar equation with the familiar variable; $\nabla^2\Phi = 4\pi G\rho$.
This also implies $\nabla^2s_{ij}=0$, allowing us to zero out the $s_{ij}$ terms given the gauge as we did the $w_i$ ones.
Our final task is to recover our metric after making all the simplifications from earlier.
Since all off-diagonal terms have vanished, our metric $g_{\mu\nu} = \eta_{\mu\nu} + h_{\mu\nu}$ has become:
\begin{equation}
\label{eqn:Nmetric}
ds^2 = -(1 + 2\Phi)\text{d}t^2 + (1 - 2\Phi)(\text{d}x^2 + \text{d}y^2 + \text{d}z^2),
\end{equation}
with the perturbation taking the simple form $h_{\mu\nu} = -2\Phi\delta_{\mu\nu}$.
When considering nonrelativistic matter, our newfound metric could be used with the geodesic equation \ref{eqn:geodesic} to recover the law of universal gravitation (for this derivation, see \citet{galaxiesbook}).
Instead, however, we will turn our attention to relativistic particles, specifically light, and study the effect a weak gravitational field has on those trajectories.

\section{Null Geodesics}

\begin{figure}
	\centering
	\includegraphics[trim={0pt, 0pt, 0pt, 0pt}, clip, width=\textwidth]{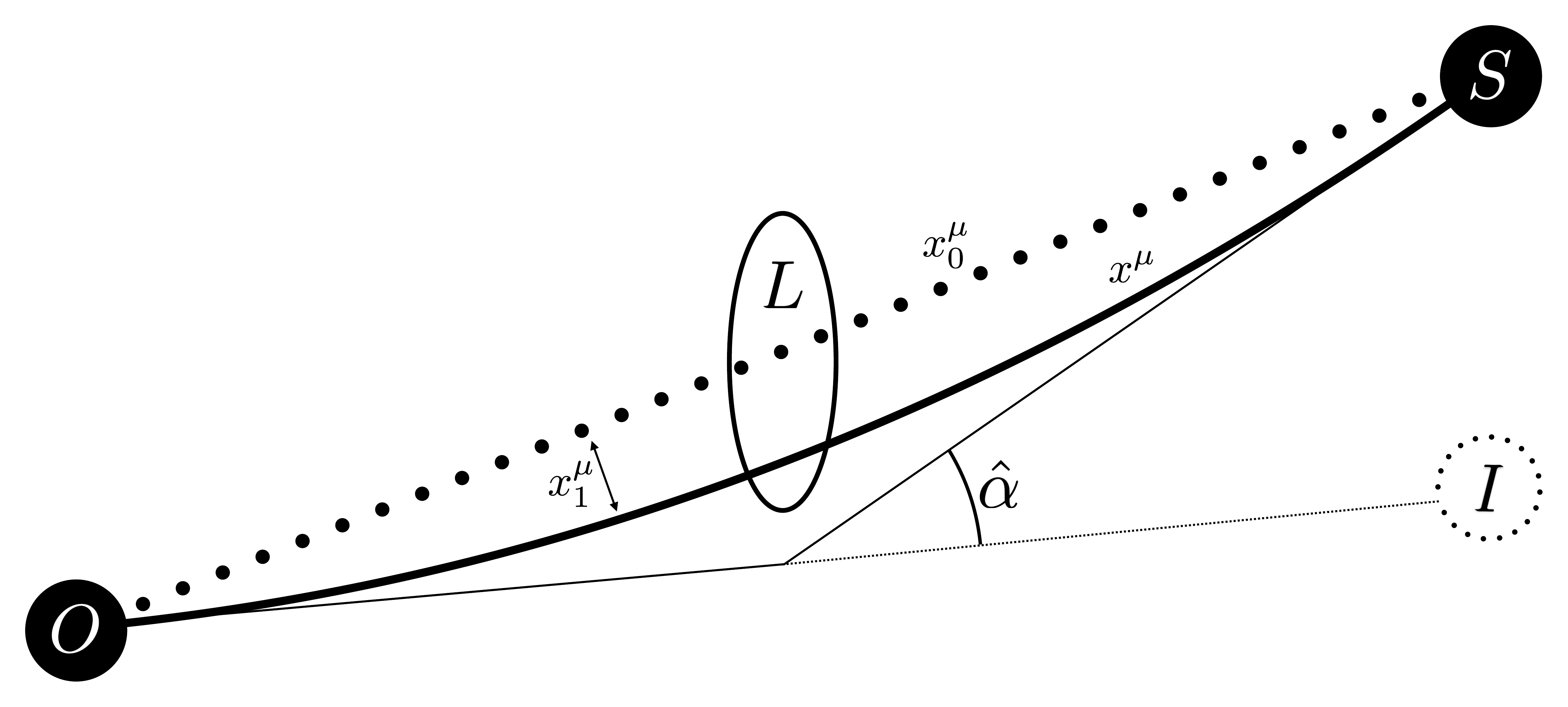}
	\caption[Gravitational Light Deflection]{Gravitational light deflection. Light travels from the source ($S$) through the lens ($L$) to the observer ($O$), appearing to the latter at the image location ($I$). The unperturbed path $x_0^\mu(\lambda)$ is shown with a dotted line while the final, perturbed path $x^\mu(\lambda)$ is a solid line. The perturbation $x_1^\mu(\lambda)$ is shown at one point but changes throughout.}
	 \label{fig:lens_gr}
\end{figure}

The primary difficulty with working with light rays in General Relativity arises from the inability to parameterize geodesics with the proper time, as is standard with other sorts of motion.
Therefore, we introduce a new parameter (often described as the ``affiine'' parameter) $\lambda$ such that $x^\mu$ becomes $x^\mu(\lambda)$.
Since the geodesic equation needs Christoffel symbols in our weak-field metric, we calculate those first from \ref{eqn:pertChris}. As before, we will consider static gravitational fields (treating galaxies and clusters as static from the point of view of light passing near them) and drop all $\frac{\text{d}\Phi}{\text{d}t}$ terms\footnote{There are lot of them, so as a mnemonic it may be useful to remember that the Christoffel symbols in this situation are zero if there are an odd number of zeros in the indices.}.
\begin{equation}
\label{eqn:weakChris}
\begin{gathered}
\Gamma^0_{00} = \Gamma^0_{ij} = \Gamma^i_{0j} = 0 \\
\Gamma^0_{i0} = \Gamma^i_{00} = \partial_i \Phi \\
\Gamma^i_{jk} = (\delta_{jk}\partial_i - \delta_{ik}\partial_j - \delta_{ij}\partial_k)\Phi.\
\end{gathered}
\end{equation}
From here we will once again benefit from a perturbative approach.
As light travels on null geodesics, our metric \ref{eqn:Nmetric} gives us a formula for the line element \ref{eqn:lineelement}, which is identically zero:
\begin{equation}
\label{eqn:nullcond}
0 = g_{\mu\nu}\frac{\text{d}x^\mu}{\text{d}\lambda}\frac{\text{d}x^\nu}{\text{d}\lambda}.
\end{equation}
This happens to be the relativistic version of the ``eikonal equation'' from geometric optics \citep{landaufields}.
Recall that for a (locally) plane wave, the \textbf{wave vector} $k$ determines the direction of the ``light ray'', and has the form $k^\mu = \frac{\text{d}x^\mu}{\text{d}\lambda}$.
We will treat the wave vector perturbatively by splitting our light path $x^\mu = x_0^\mu(\lambda) + x_1^\mu(\lambda)$, with $|x_0 | \gg |x_1 |$
and set $k^{\mu} = \frac{{d}x^\mu_0}{{d}\lambda}$ and $l^{\mu} = \frac{{d}x^\mu_1}{d\lambda}$.\footnote{From now on, $k$ will refer to this zeroth-order component and not the full wave vector.}
We will see that the zeroth-order term $k$ corresponds to the light's path in the non-perturbed metric (corresponding to no mass present), with $l$ encoding the lensing deflection, as illustrated in Figure \ref{fig:lens_gr}.
As before, we will use Minkowski space for our background, but this approach also works in a general FLRW spacetime \citep{1993ApJ...415..459P, 1996ApJ...458...46P}.

We will consider the zeroth and first order terms separately in our two equations, \ref{eqn:geodesic} and \ref{eqn:nullcond}.
At zeroth order, the geodesic equation can be rewritten as $\frac{dk^\mu}{d\lambda} = 0$, as the unperturbed metric $\eta_{\mu\nu}$ has all of its Christoffel symbols zero.
Recalling that $\lambda$ is our stand-in for time, this tells us that light in Minkowski space has no acceleration, and thus travels in a ``straight line''.
\ref{eqn:nullcond} at zeroth order simply tells us $\eta_{\mu\nu}k^\mu k^\nu = 0$, meaning that the $(k^0)^2 = \delta_{ij}k^i k^j \equiv k^2$ (the wave vector cannot be zero).
Here we have defined the scalar unperturbed wave vector $k$ for convenience's sake.
The first order equations are less trivial, and we will again drop any higher-order terms (i.e. $l^2$, $\Gamma l$, $hl$, and $h^2$):
Then, the null geodesic condition gives us:
\begin{equation}
\label{eqn:nullpert} %my dearest nullpert...
0 = (\eta_{\mu\nu} + h_{\mu\nu})(k^\mu + l^\mu)(k^\nu + l^\nu) = 2\eta_{\mu\nu}k^\mu l^\nu + h_{\mu\nu}k^\mu k^\nu + \mathcal{O}(l^2\text{, etc.}),
\end{equation}
and, incorporating \ref{eqn:weakChris}, the geodesic equation produces:
\begin{equation}
\frac{dk^\mu}{d\lambda} + \frac{dl^\mu}{d\lambda}  = -\Gamma^\mu_{\rho\epsilon}(k^\rho + l^\rho)(k^\epsilon + l^\epsilon) = -\Gamma^\mu_{\rho\epsilon}k^\rho k^\epsilon + \mathcal{O}(l^2\text{, etc.}).
\end{equation}
At this point we treat the space and time components of the trajectory separately and enter the nonzero Christoffel symbols (recall that $\frac{dk^\mu}{d\lambda} = 0$ so the ``acceleration'' side of the geodesic equation only includes the perturbation).
The zero component is now:
\begin{equation}
\frac{dl^0}{d\lambda} =-2\partial_i \Phi k^0 k^i = -2k(\vec{\nabla}\Phi \cdot \vec{k}),
\end{equation}
with the gradient $\vec{\nabla}$ defined normally, and $\vec{k}$ denoting the spatial parts of $k^\mu$
(We will slowly move away from Einstein notation from here as the non-tensor formulation is more standard in lensing literature).
Integrating this equation (and setting $l^0 = 0$ at 0) gives us $l^0$:
\begin{equation}
l^0 = -2k\int (\vec{k} \cdot \vec{\nabla}\Phi) d\lambda = -2k\int \vec{\nabla}\Phi \cdot d\vec{x} = -2k\Phi,
\end{equation}
recalling that $\vec{k} = \frac{d\vec{x}}{d\lambda}$.
We can rewrite our first-order null geodesic condition \ref{eqn:nullpert} and then substitute in $l^0$ in a more vector-friendly form like so:
\begin{equation}
0 = -kl^0 + \vec{k}\cdot\vec{l} - 2k^2\Phi = 2k^2\Phi +  \vec{k}\cdot\vec{l} - 2k^2\Phi =  \vec{k}\cdot\vec{l},
\end{equation}
which tells us that to first order, the perturbation to the background path is orthogonal to that background path.

The spatial components of $l$ require very careful bookkeeping with indices, but eventually produce:
\begin{equation}
\frac{dl^i}{d\lambda} = -k^2(\partial_i 2\Phi) + 2k_i \partial_j \Phi k^j,
\end{equation}
which can be expressed in vector form as
\begin{equation}
\frac{d\vec{l}}{d\lambda} = -2k^2 \vec{\nabla}\Phi + 2\vec{k}(\vec{\nabla}\Phi \cdot \vec{k}) = -2k^2 \vec{\nabla}_\perp\Phi.
\end{equation}
The ``transverse gradient'' $\vec{\nabla}_\perp\Phi = \vec{\nabla}\Phi - \vec{k}(\vec{\nabla}\Phi \cdot \vec{k}) / k^2$ is the gradient of $\Phi$ minus its projection along the background path $\vec{k}$. 
We can now at long last introduce the \textbf{deflection angle} $\hat{\alpha}$, defined as the magnitude of the spatial perturbation to the path $\Delta \vec{l}$ divided by the scalar unperturbed wavevector, and shown in Figure \ref{fig:lens_gr}.
\begin{equation}
\hat{\alpha} =  -\frac{\Delta\vec{l}}{k} = \frac{1}{k}\int\frac{d\vec{l}}{d\lambda}d\lambda = -2k\int\vec{\nabla}_\perp \Phi d\lambda.
\end{equation}
Here, the negative sign is due to the fact that we look ``backwards'' from redshift 0 to the source position.
The final step is to rescale the affine parameter $\lambda$ to $s = k\lambda$, which allows our integral to be over physical spatial distance.
\begin{equation}
\label{eqn:deflection}
\hat{\alpha} = 2\int \vec{\nabla}_\perp \Phi ds.
\end{equation}
Of note is the factor of 2 in \ref{eqn:deflection}.
It arises from the presence of $\Phi$ in both the spatial and time parts of the perturbed metric,\footnote{Recalling $\Psi$ from the previous section, $2\Phi$ generalizes to $\Phi+\Psi$ in situations where the two are not equal, as is the case in some extensions of GR. For a detailed treatment that keeps the two separate until this step, see \citet{galaxiesbook}.} and is a purely relativistic effect.
In fact, the lensing of stars around the sun was a successful test of GR in the early 1900s, as the pure Newtonian prediction (found by taking a limit as mass goes to zero) lacks that factor of 2.

\section{The Thin Lens and Fermat Potential} \label{sec:thinlens}

\begin{figure} 
	\centering
	\includegraphics[trim={0pt, 0pt, 0pt, 0pt}, clip, width=\textwidth]{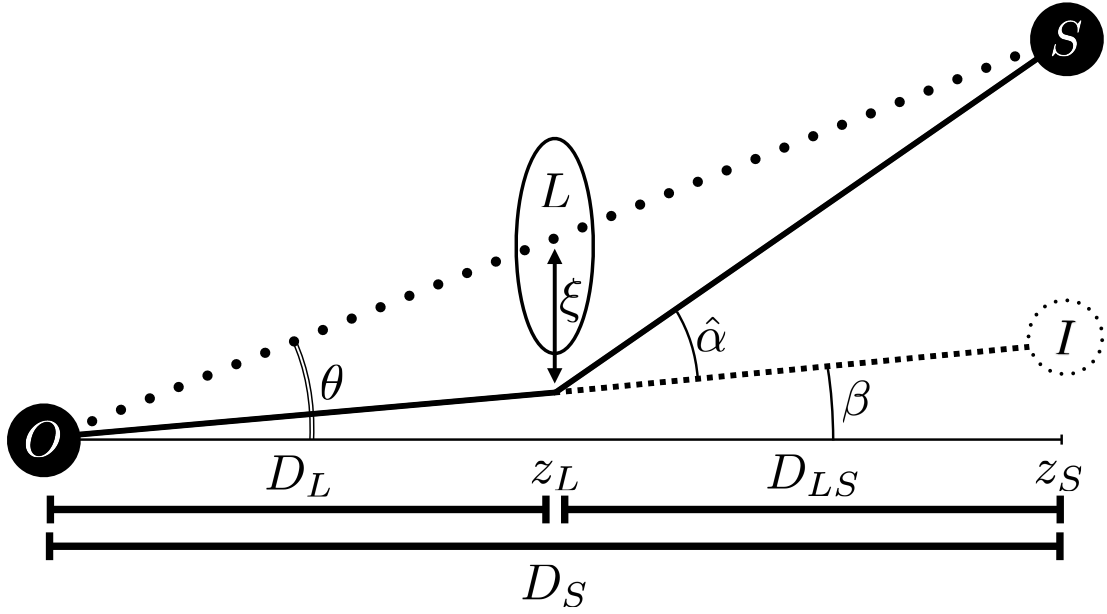}
	\caption[The Thin Lens Approximation (Again)]{Diagram of a thin lens. Light travels from the source ($S$) and is deflected at the lens ($L$) redshift, appearing to the observer ($O$) at image location ($I$). The x-axis in the diagram is arbitrary, and the angles are two-dimensional vectors in practice. Compare to \ref{fig:lens_gr}, where the deflection smoothy varies along the geodesic. (Reproduced from \ref{sec:lensing})}
	\label{fig:lens_schema}
\end{figure}

The canonical approach to lensing observables relies on assuming the majority of the lensing happens at one spot along the path light takes from source to observer.
This turns out to be a very good assumption, and while ``thick'' lenses have been studied \citep{kovner87, 1988MNRAS.233..265P}, they are more applicable to examining large-scale structure effects and weak lensing \citep{2005MNRAS.358...39Y, 1991A&A...248..349B}.
Imagining a non-lensed geodesic as a straight line from source to observer, the thin lens simply breaks that line into two such that the light now is at a distance $\xi$ when it passes by the lens.
The situation is illustrated in Figure \ref{fig:lens_schema}, and in terms of observables we can only measure $\beta$ and (via spectroscopy) the distances.
There must be an accompanying \textbf{geometric time delay}, as the perturbed path is longer.
We can measure this directly if we know the distances involved, and this delay is given by:
\begin{equation} \label{eqn:geom_delay}
\Delta t_{geom} = \frac{D_L D_S}{2D_{LS}}|\vec{\theta} - \vec{\beta} |^2.
\end{equation}
 For a good derivation of this, see Chapter 15.2 of \citet{galaxiesbook}.

There is also a \textbf{gravitational time delay} (also called Shapiro delay) arising from the perceived slowing of time by the gravitational field from the perspective of an observer far from it.
We want to find the difference in time between a geodesic with no field and one with a field, which (recalling our definition of $x^\mu(\lambda)$) is written simply as:
\begin{equation}
\Delta t = \int l^0 d\lambda  = -2k\int \Phi d\lambda,
\end{equation}
as the unperturbed path will simply be the integral of $k$. 
As in the deflection (\ref{eqn:deflection}), we change variables to $s$ to get the final form of this time delay: 
\begin{equation}
\label{shapiro}
\Delta t_{grav} = -2\int \Phi ds.
\end{equation}

In classical optics, Fermat's principle states that light takes the path that minimizes the arrival time.
This nearly holds true in General Relativity as well, with one major modification: light takes path(s) that \textit{extremize} the arrival time.
In other words, if we had some function $\tau(\vec{\beta})$ that defined the possible light travel times for different $\vec{\beta}$ on the sky, we can expect images to form where $\nabla\tau = 0$.
The $\tau$ in question is, of course, related to the total time delay (the unperturbed arrival time will be constant length and can be ignored if we just care about image formation, and it is also not observable) -- but we need to get it into a more useful form.
By first rescaling and then parameterizing its integral, we define the \textbf{lensing potential} from the geometric time delay:
\begin{equation}\label{eqn:lensing_pot}
\psi(\vec{\theta}) = \frac{D_{LS}}{D_L D_S}\Delta t_{grav} = \frac{2D_{LS}}{D_L D_S}\int\Phi(D_L\vec{\theta}, z) \text{d}z.
\end{equation}
$\phi$ is a dimensionless scalar function, and alternately motivated by our formula for the gravitational deflection (\ref{eqn:deflection}) -- the gradient and integral can be swapped such that the deflection angle is given by:
\begin{equation}
\hat{\alpha} = \frac{D_S D_L}{D_{LS}}\nabla\psi(\vec{\theta}).
\end{equation}

The distances involved in our time delays need some special consideration before we combine them.
For nearby objects (such as galactic microlensing events), we can use the typical Euclidean distance, but at cosmological scales we need to take into account the changes in scale factor between source, lens, and observer.
As the lensing observables involve angles, the \textbf{angular diameter distance} is the correct choice, and is found in a flat universe by the following formula:
\begin{equation}
D_A(z_1, z_2) = \frac{1+z_1}{1+z_2}\int^{z_2}_{z_1}\frac{dz'}{H(z')}.
\end{equation}
The above formula is simply the comoving distance (\ref{eqn:comoving}) generalized to start at a redshift other than 0, with an additional prefactor of $a_2/a_1$\footnote{the angular diameter has a more complex form in a curved universe, but we are fine to assume a flat one here.}.
Note that this means $D_L + D_{LS} \neq D_S$!
Finally, we must account for cosmological time dilation, as the time delay formulae above are in the frame of the lens (at redshift $z_L$) -- to shift to the frame of the observer an additional prefactor of $(1+z_L)$ is necessary.

We can now define the aforementioned $\tau$, the \textbf{Fermat potential}:
\begin{equation} \label{eqn:fermat_pot}
\tau(\vec{\beta}, \vec{\theta}) = \frac{1}{2}|\vec{\theta} - \vec{\beta}|^2 - \psi(\vec{\theta}) = \frac{1}{(1+z_L)}\frac{D_{LS}}{D_L D_S}\Delta t.
\end{equation}
%maybe this can go up in the text body?
The Fermat potential describes the (scaled) arrival-time surface, and examining it can provide useful information about image positions and other properties.
Firstly, if we imagine empty space $(\psi = 0)$, the potential is quadratic, with one minimum and no other extrema, and the image will simply form at the source position.
The addition of mass via the lensing can ``lift'' portions of the potential from this quadratic, creating saddle points and maxima and leading to the formation of multiple images.
In short, images form when $\nabla \tau = 0$.

\subsection{Critical Density and Convergence}
We will briefly return to the lensing potential $\psi$ to introduce some convenient notation.
\ref{eqn:lensing_pot} depends on the gravitational potential, but we can use Poisson's equation to work in terms of density instead.
Consider the Laplacian of the lensing potential, $\nabla^2\psi$, which depends only on the projected 2D mass density:
\begin{equation}
\nabla^2\psi = \frac{D_{LS}D_L}{D_S}2\int\text{d}z\nabla^2\Phi(D_L\vec{\theta},z) =  \frac{D_{LS}D_L}{D_S}2\int\text{d}z4\pi G \rho(D_L\vec{\theta},z) =  \frac{D_{LS}D_L}{D_S}8\pi G \Sigma(D_L\vec{\theta}).
\end{equation}
The above calculation handwaves a few steps (see \citep{galaxiesbook} for more rigour) and relies on the thin lens approximation, but in essence we switch the order of differentiation and integration, invoke Gauss's law for gravity, and integrate over $z$.
$D_L$ moves to the top of the prefactor thanks to a change to angular variables in the Laplacian.
In doing this, we've found that $\psi$ itself follows Poisson's (2D) equation, commonly written as:
\begin{equation} \label{eqn:lenspoisson}
\nabla^2\psi = \frac{\partial^2\psi}{\partial\theta_1\partial\theta_1} + \frac{\partial^2\psi}{\partial\theta_2\partial\theta_2} = 2\kappa(\vec{\theta}).
\end{equation}
We define the \textbf{lensing convergence} $\kappa = \Sigma/\Sigma_{cr}$ as the ratio of the projected mass density to the ``lensing critical density,'' \footnote{in lensing literature, simply the ``critical density'', the specification here is to disambiguate from the cosmological critical density $\rho_{cr}$.} which can be understood as the density necessary for strong lensing given a source-lens configuration,
\begin{equation}
\Sigma_{cr} = \frac{1}{4\pi G}\frac{D_S}{D_{LS}D_L}.
\end{equation}
The solution to \ref{eqn:lenspoisson} is simple to obtain by using the Green's function for the 2D Laplacian:
\begin{equation}
\psi(\vec{\theta}) = \frac{1}{\pi}\int \kappa (\vec{\theta}') \ln (\vec{\theta} - \vec{\theta}') \text{d}^2\vec{\theta}',
\end{equation}
the familiar form of the lensing potential as it appears in \ref{sec:lensing}.
\chapter{Transfer Matrix Elements with Image Rotation} \label{sec:transformation}

\paragraph{This appendix presents a partial}
derivation of the transfer matrix elements used in Chapter \ref{sec:chapter2}, with the addition of a differential rotation component.
As before, the non-reference images $i$ all have typical thin-lens transfer matrices.
However, we add rotation to the reference image (image ``1'').
\begin{equation}
\mathcal{A}_1 = \begin{pmatrix}
1 - \kappa - \gamma_1 & -\gamma_2 + \omega \\
-\gamma_2 - \omega & 1 - \kappa + \gamma_1\\
\end{pmatrix}; \quad
\mathcal{A}_{i\neq 1} =  \begin{pmatrix}
1 - \kappa - \gamma_1 & -\gamma_2 \\ 
-\gamma_2 & 1 - \kappa + \gamma_1\\
\end{pmatrix}.
\end{equation}
As before, we factor $1-(\kappa)$ from both matrices and introduce reduced shear $g_i = \gamma_i/(1-\kappa)$.
In $\mathcal{A}_1$, we define the ``reduced rotation'' $w = \omega/(1-\kappa)$.
Just as $g_1$ and $g_2$ are the observables in the typical point-matching scheme, $w$ rather than $\omega$ is here due to the MST.
The rewritten amplification matrices take the form:
\begin{equation}
\mathcal{A}_1 = (1-\kappa)\begin{pmatrix}
1-g_1 & -g_2 + w \\
-g_2 - w & 1 + g_1\\
\end{pmatrix}; \quad
\mathcal{A}_{i\neq 1} = (1 - \kappa)\begin{pmatrix}
1 - g_1 & -g_2 \\ 
-g_2 & 1 + g_1\\
\end{pmatrix}.
\end{equation}
From here, we broadly follow \citet{wagnertessore} by constructing transfer matrices $\mathcal{T}_i$ = $\mathcal{A}_i^{-1}\mathcal{A}_1$.
Their results can be re-obtained by setting $w=0$.

Due to the inversion of $\mathcal{A}_i$, all elements of $\mathcal{T}_i$ have the prefactor $(1-\kappa_1)\mu_i = f_i / (1 - g_{i,1}^2 - g_{i_2}^2)$.
They are: 
\begin{equation}
\begin{aligned}
\mathcal{T}_{11} &= (1-\kappa_1)\mu_i\left[(1+g_{i,1})(1-g_{1,1} - g_{i,2}(g_{1,2}+w)\right]  \\
\mathcal{T}_{12} &= (1-\kappa_1)\mu_i\left[g_{i,2}(1+g_{1,1})+(1+g_{i,1})(-g_{1,2}+w)\right] \\
\mathcal{T}_{21} &= (1-\kappa_1)\mu_i\left[(1 - g_{1,1})g_{i,2} - (1-g_{i,1})(g_{1,2}+w)\right] \\
 \mathcal{T}_{22} &= (1-\kappa_1)\mu_i\left[(1-g_{i,1})(1+g_{1,1}) + g_{i,2}(-g_{1,2}+w)\right], \\
\end{aligned}
\end{equation}
where the first subscript of each $g$ gives the image (1 or $i$) and the second the index of the shear component (1 or 2).
To solve these equations, \citet{wagnertessore} introduce four quantities $\{a_i, b_i, c_i, d_i\}$ for each non-reference image, composed of elements of $\mathcal{T}$.
In the $w\neq0$ case, those quantities are:
\begin{equation}\label{eqn:rotationeqs}
\begin{aligned}
a_i &= \mathcal{T}_{11} - \mathcal{T}_{22} = 2(1-\kappa_1)\mu_i\left(g_{i,1} - g_{1,1} - g_{i,2}w\right) \\
b_i &= \mathcal{T}_{21} + \mathcal{T}_{12} = 2(1-\kappa_1)\mu_i\left(g_{i,2} - g_{1,2} + g_{i,1}w\right) \\
c_i &= \mathcal{T}_{21} - \mathcal{T}_{12} = 2(1-\kappa_1)\mu_i\left(g_{i,1}g_{1,2} - g_{i,2}g_{1,1} - w\right) \\
d_i &= \mathcal{T}_{11} + \mathcal{T}_{22} = 2(1-\kappa_1)\mu_i\left(1 - g_{i,1}g_{1,1} - g_{i,2}g_{1,2}\right). \\
\end{aligned}
\end{equation}
The final two quantities above, $c_i$ and $d_i$, are the curl and divergence of the image map, and they interact with $w$ in predictable ways -- the quantity vanishes for the latter, while appearing on its own in the former.
\citet{wagnertessore} present solutions for both the convergence ratios and all reduced shears in the $w=0$ case, where three images (i.e., images $1,i,j$) are necessary to solve the 8 equations exactly.
In our case, we need four images to additionally constrain $w$, and the resulting system of 12 nonlinear equations has resisted solution with \texttt{Mathematica}.
A numerical solution may be possible, but to solve the system analytically\footnote{assuming it has a solution in the first place}, a different parameterization is probably necessary.
Some evidence to this fact is that the relation $g_{1,2}a_i - g_{1,1}b_i = c_i$, which is true always when $w=0$, implies $g_{1,1}g_{i,1} = g_{1,2}g_{i,2}$ when $w\neq0$.
This further implies that the ratio $g_1/g_2$ is identical for all four images, which is certainly not true for a generic lens.

Finally, the magnifications $\mu = 1/|\mathcal{A}|$ in this framework require slightly special care.
While the non-reference images have the same form as normal, $\mu_i = ((1-\kappa_i)(1-g_{i,1}^2 - g_{i2}^2))^{-1}$, the reference image with rotation has magnification:
\begin{equation}
\mu_1 = \frac{1}{|\mathcal{A}|} = \frac{1}{(1-\kappa_1)(1 - g_{1,1}^2 - g_{1,2}^2 + w^2)}.
\end{equation}
However, the rotational component enters into the magnification only at second order, and so can probably be avoided as $w$ should be small.
The perturbative approach may also be a promising avenue for solving \ref{eqn:rotationeqs}.

\chapter{The Known Radio Gravitational Lenses}

\paragraph{This table appeared in} \citet{martinez2025}, Chapter \ref{sec:chapter3} of this document, and listed the known lensed radio sources up to shortly before that article's submission.
We have reproduced it here with the following changes: 
First, to reflect the ``zig-zag'' findings of \citet{2025A&A...694A.300D}, the entry for PS J1721+8842 is listed as 6 images rather than 4+2.
Second, radio lenses from \citet{Jackson2024} and Chapter \ref{sec:chapter4} have also been added to bring the list closer to completeness.
We have also noted the lenses from that paper which were also studied by \citet{dobie23} and \citet{Jackson2024}.
In the former paper, the targets where only one lensed image was marginally detected have been omitted, but those with one confident lensed quasar detection and one non-detection (due to a high flux ratio) are included.
While we measured stacked 3-epoch VLASS fluxes for all Chapter \ref{sec:chapter4} lens candidates in Table \ref{table:obs}, we only include the results of a single-epoch crossmatch with the Epoch 1 catalogue of \citet{Gordon2021} for consistency with the rest of the table.
We also only include the confirmed radio lenses from that Chapter, though an argument could be made to include several of the ambiguous cases from \ref{sec:inconclusive}.

%\newgeometry{width=150mm,top=35mm,bottom=25mm,bindingoffset=6mm, headsep=10pt}
{\scriptsize
 \onehalfspacing
\begin{landscape}
\begin{ThreePartTable}
\begin{TableNotes}[flushleft]
    \setlength{\labelsep}{0pt}
    \item[a]\label{allknown-a} {Source is outside of VLASS footprint ($\delta < 40^{\circ}$) or otherwise masked}
    \item[b]\label{allknown-b} {Position of lens unreliable/unknown, position of brightest source image given instead}
    \item[c]\label{allknown-c}{Listed as a strong candidate for lensing but not spectroscopically confirmed}
    \item[d]\label{allknown-d}{Lens system bright in radio but at too low resolution to confirm emission from source}
    \item[e]\label{allknown-e}{Only one radio component detected, but confidently associated with a single confirmed lensed image}
    \item[] {Note. --- Objects are ordered by lens discovery year. (1): The name given to object in its discovery paper.
(2): Discovery method for the lens system, using the following key: JCP - Bright, flat-spectrum source search, as seen in the JVAS \citep{1999MNRAS.307..225K}, CLASS \citep{2003MNRAS.341...13B}, and PMN \citep{2000AJ....120.2868W} lens surveys; RADIO - Other radio-based lens search or serendipitous discovery; XMATCH - Joint optical+radio search; OPTICAL - Lens system discovered by an optical search and confirmed as a radio source later; GAIA - Lens discovered specifically utilizing \textit{Gaia} data and confirmed as a radio source later. RA, Dec: Coordinates are J2000 and correspond to the lens deflector in each system, unless otherwise noted. Many close quasar lenses have faint or blended lenses with poor astrometry -- see table note \ref{allknown-b}. 
(3): Total flux from the nearest component to the lens coordinates within $5''$, using the \citet{Gordon2021} VLASS quick-look catalog. Non-detections are marked $<1.0$mJy corresponding to that catalog's limiting flux.
(4): Number of images of the radio AGN visible in the system. Sources where the radio emission is from a lensed radio lobe rather than an AGN core are marked ``Lobe'', and those where the emission is from a lensed high redshift, ultra-luminous star-forming galaxy are marked ``SFG''. When multiple of these sources are present they are all noted.
(5): Maximum image separation for lensed AGN cores. For SFG and Lobe sources the Einstein radius is given. 
(6): When multiple references are given, the first corresponds to the discovery of radio emission and the (original) discovery of lensing at another wavelength is listed after the semicolon. \citet{martinez2025}, the paper this table originally appeared in, has been abbreviated \citetalias{martinez2025} and entries labeled ``This work'' were found in Chapter \ref{sec:chapter4}.
(7): VLASS Epoch 1 source associated with this lens system, if one matched within $5''$.}
  \end{TableNotes}
\begin{longtable}{lcccccclc}
\caption[List of Published Radio Gravitational Lenses]{List of published radio gravitational lenses.}  \label{tab:allknown} \\
\hline \hline

{Name} & {Method} & {RA} & {Dec} & {$S_{\text{VLASS}}$} & {Images} & {Sep.} & {References} & {VLASS Component} \\ 
{} & {} & \multicolumn{1}{c}{[deg]} &\multicolumn{1}{c}{[deg]}& {[mJy]} & {} & {[arcsec]}& {} & {} \\
\hline \endfirsthead
\caption[]{\textit{(continued)}} \\
\hline \hline
{Name} & {Method} & {RA} & {Dec} & {$S_{\text{VLASS}}$} & {Images} & {Sep.} & {References} & {VLASS Component} \\ 
{} & {} & \multicolumn{1}{c}{[deg]} &\multicolumn{1}{c}{[deg]}& {[mJy]} & {} & {[arcsec]}& {} & {} \\
\hline
\endhead
\hline
\endfoot
\hline
\insertTableNotes
\endlastfoot
QSO B0957+5608 & RADIO & 150.3369 & 55.8974 & 317.483& 2 & 6.17 & \citet{walsh79} & J100120.93+555355.8 \\
PG B1115+080 & OPTICAL & 169.57062 & 7.7663 & $<1.0$ & 4 & 2.43 & \citet{2021MNRAS.508.4625H}; \citet{1980Natur.285..641W} &  \\
MG B2016+112 & RADIO & 304.8253 & 11.4537 & 93.026 & 3 & 2.56 & \citet{1984Sci...223...46L} & J201918.00+112712.2 \\
B2237+0305 & OPTICAL & 340.125975 & 3.358508 & $<1.0$ & 4 & 1.78 & \citet{1996AJ....112..897F}; \citet{1985AJ.....90..691H} &  \\
MG B1131+0456 & RADIO & 172.9854 & 4.9302 & 261.992 & 2 & 2.2 & \citet{1988Natur.333..537H} & J113156.44+045549.5 \\
PKS B1830$-$211 & RADIO & 278.4164 & $-$21.0609 &\tnotex{allknown-a} & 3 & 0.99 & \citet{1988MNRAS.231..229P} &  \\
B1413 + 117 & OPTICAL & 213.9426 & 11.4953 & 3.487 & 4 & 1.35 & \citet{2023MNRAS.524.3671Z}; \citet{1988Natur.334..325M} & J141546.22+112943.7 \\
MG B1654+1346 & RADIO & 253.6741 & 13.7726 & 205.031 & Lobe & 2.0 & \citet{1989AJ.....97.1283L} & J165441.79+134621.4 \\
MG B0414+0534 & RADIO & 63.6571 & 5.5786 & 930.234 & 4 & 2.4 & \citet{1992AJ....104..968H} & J041437.74+053443.0 \\
JVAS B1422+231 & JCP & 216.1587 & 22.9335 & 676.326 & 4 & 1.3 & \citet{1992MNRAS.259P...1P} & J142438.11+225600.7 \\
JVAS B0218+35.7 & JCP & 35.2729\tnotex{allknown-b} & 35.9372 & 1073.526 & 2 & 0.335 & \citet{1993MNRAS.261..435P} & J022105.46+355613.8 \\
MG B1549+3047 & RADIO & 237.3014 & 30.7879 & 508.064 & Lobe & 2.0 & \citet{1993AJ....105..847L} & J154912.55+304714.9 \\
CLASS B1600+434 & JCP & 240.4187\tnotex{allknown-b} & 43.2798 & 40.532 & 2 & 1.4 & \citet{1995MNRAS.274L..25J} & J160140.50+431647.2 \\
CLASS B1608+656 & JCP & 242.3082 & 65.5413 & 35.675 & 4 & 2.27 & \citet{1995ApJ...447L...5M} & J160914.03+653228.1 \\
FSC 10214+4724 & OPTICAL & 156.1437 & 47.1531 & $<1.0$ & 4/SFG & 1.0 & \citet{2013MNRAS.434.3322D}; \citet{1995ApJ...449L..29G} &  \\
MG B0751+2716 & RADIO & 117.923 & 27.2755 & 304.438 & 4 & 0.8 & \citet{1997AJ....114...48L} & J075141.53+271631.8 \\
JVAS B1938+666 & JCP & 294.6055\tnotex{allknown-b} & 66.8148 & 398.267 & 4 & 1.02 & \citet{1997MNRAS.289..450K} & J193825.26+664852.8 \\
RX J0911+0551 & OPTICAL & 137.86479\tnotex{allknown-b} & 5.848 & $<1.0$ & 4 & 3.25 & \citet{2015MNRAS.454..287J}; \citet{1997AaA...317L..13B} &  \\
CLASS B0712+472 & JCP & 109.0152 & 47.1474 & 26.245 & 4 & 1.46 & \citet{1998MNRAS.296..483J} & J071603.59+470850.1 \\
FBQ B0951+2635 & XMATCH & 147.84412 & 26.58725 & $<1.0$ & 2 & 1.1 & \citet{1998AJ....115.1371S} &  \\
CLASS B1933+503 & JCP & 293.6293 & 50.4232 & 79.109 & 4 & 1.52 & \citet{1998MNRAS.301..310S} & J193430.92+502523.3 \\
APM B08279+5255 & OPTICAL & 127.9235\tnotex{allknown-b} & 52.75486 & $<1.0$ & 3 & 0.38 & \citet{1999AJ....118.1922I}; \citet{1998ApJ...505..529I} &  \\
JVAS B1030+074 & JCP & 158.3918\tnotex{allknown-b} & 7.1906 & 300.708 & 2 & 1.65 & \citet{1998MNRAS.300..649X} & J103334.02+071126.3 \\
CLASS B1127+385 & JCP & 172.5007 & 38.2005 & 35.965 & 2 & 0.7 & \citet{1999MNRAS.303..727K} & J113000.14+381203.1 \\
CLASS B1152+199 & JCP & 178.8264 & 19.6615 & 53.896 & 2 & 1.56 & \citet{1999AJ....117.2565M} & J115518.32+193942.0 \\
CLASS B1359+154 & JCP & 210.3981\tnotex{allknown-b} & 15.2237 & 45.277 & 6 & 1.71 & \citet{1999AJ....117.2565M} & J140135.54+151324.8 \\
CLASS B1555+375 & JCP & 239.2998\tnotex{allknown-b} & 37.36 & 34.178 & 4 & 0.42 & \citet{1999AJ....118..654M} & J155711.95+372136.0 \\
CLASS B2045+265 & JCP & 311.8349 & 26.7339 & 36.293 & 4 & 1.9 & \citet{1999AJ....117..658F} & J204720.27+264402.4 \\
JVAS B2114+022 & JCP & 319.2116 & 2.4297 & 127.486 & 2 & 2.56 & \citet{1999MNRAS.307..225K} & J211650.76+022546.7 \\
HS B2209+1914 & OPTICAL & 332.87625\tnotex{allknown-b} & 19.4869 & 2.024 & 2 & 1.04 & \citetalias{martinez2025}/\citet{Jackson2024}; \citet{hamburg} & J221130.31+192913.3 \\
HE B0230$-$2130 & OPTICAL & 38.13792\tnotex{allknown-b} & $-$21.29056 & $<1.0$ & 4 & 2.05 & \citet{Jackson2024}; \citet{W99} & \\
CLASS B0128+437 & JCP & 22.8059\tnotex{allknown-b} & 43.9703 & 61.221 & 4 & 0.55 & \citet{2000MNRAS.319L...7P} & J013113.45$+$435812.9 \\
PMN J1838$-$3427 & JCP & 279.6187 & $-34.4618$ & 214.667 & 2 & 0.99 & \citet{2000AJ....120.2868W} & J183828.50$-$342741.2 \\
CLASS B0739+366 & JCP & 115.7132\tnotex{allknown-b} & 36.5788 & 30.526 & 2 & 0.53 & \citet{2001AJ....121..619M} & J074251.20+363443.6 \\
FIRST J0816+5003 & XMATCH & 124.1618 & 50.0688 & 64.939 & Lobe & 2.0 & \citet{2001ApJ...547...60L} & J081638.73+500407.2 \\
FIRST J0823+3906\tnotex{allknown-c} & XMATCH & 125.8496 & 39.11 & 56.205 & Lobe & 5.0 & \citet{2001ApJ...547...60L} & J082323.65+390638.4 \\
FIRST J1622+3531\tnotex{allknown-c} & XMATCH & 245.6239 & 35.5257 & 102.976 & Lobe & 3.0 & \citet{2001ApJ...547...60L} & J162229.77+353134.3 \\
PMN J2004$-$1349 & JCP & 301.0294 & $-$13.8252 & 22.097 & 2 & 1.13 & \citet{2001AJ....121.1223W} & J200407.05$-$134931.0 \\
CLASS B2319+051 & JCP & 350.4201 & 5.4602 & 68.925 & 2 & 1.36 & \citet{2001AJ....122..591R} & J232140.81+052737.3 \\
PMN J0134$-$0931 & JCP & 23.6486\tnotex{allknown-b} & $-$9.5175 & 636.323 & 5 & 0.68 & \citet{2002ApJ...564..143W} & J013435.67$-$093102.7 \\
CLASS B0445+123 & JCP & 72.0916\tnotex{allknown-b} & 12.4654 & 31.086 & 2 & 1.35 & \citet{2003MNRAS.338..957A} & J044822.00+122755.5 \\
FIRST J1004+1229 & XMATCH & 151.1037 & 12.4894 & 8.624 & 2 & 1.54 & \citet{2002AJ....123.2925L} & J100424.87+122922.5 \\
PMN J1632$-$0033 & JCP & 248.2403 & $-$0.5559 & 167.349 & 3 & 1.47 & \citet{2002AJ....123...10W} & J163257.68$-$003320.9 \\
HE B0435$-$1223 & OPTICAL & 69.56198 & $-$12.28739 & $<1.0$ & 4 & 2.54 & \citet{2015MNRAS.454..287J}; \citet{2002AaA...395...17W} &  \\
HS B0810+2554 & OPTICAL & 123.38054 & 25.75092 & $<1.0$ & 4 & 0.91 & \citet{2015MNRAS.454..287J}; \citet{2002AaA...382L..26R} &  \\
CLASS B0631+519 & JCP & 98.8013\tnotex{allknown-b} & 51.9505 & 46.425 & 2 & 1.16 & \citet{2003MNRAS.341...13B} & J063512.35+515701.2 \\
CLASS B0850+054 & JCP & 133.2232 & 5.2543 & 78.061 & 2 & 0.68 & \citet{2003MNRAS.338.1084B} & J085253.57+051515.8 \\
CLASS B2108+213 & JCP & 317.7256 & 21.5162 & 36.473 & 2 & 4.57 & \citet{2003MNRAS.341...13B}; \citet{McKean2005} & J211054.07+213058.8 \\
RXS J1131$-$1231 & OPTICAL & 172.96461 & $-$12.53289 & 4.008 & 4/SFG & 3.23 & \citet{Wucknitz:2009xu}; \citet{2003AaA...406L..43S} & J113151.53$-$123158.0 \\
SDSS J1004+4112 & OPTICAL & 151.14546 & 41.21189 & $<1.0$ & 4 & 14.62 & \citet{2011ApJ...739L..28J}; \citet{2003Natur.426..810I} &  \\
SDSS J0924+0219 & OPTICAL & 141.2325771 & 2.3234747 & $<1.0$ & 4 & 1.81 & \citet{2015MNRAS.454..287J}; \citet{2003AJ....126..666I} &  \\
HE B0047$-$1756 & OPTICAL & 12.6158\tnotex{allknown-b} & $-$17.6693 & $<1.0$ & 2 & 1.44 & This work; \citet{wisotzki04} & \\
FOV J0743+1553\tnotex{allknown-c} & XMATCH & 115.9744 & 15.8903 & 47.366 & Lobe & 1.8 & \citet{2005AJ....130.1977H} & J074353.85+155324.8 \\
SDSS J1259+1241\tnotex{allknown-d} & OPTICAL & 194.9811138\tnotex{allknown-b} & 12.69751076 & $<1.0$ & 2 & 3.5 & \citet{dobie23}; \citet{2006AJ....131....1H} &  \\
SDSS J1353+1138 & OPTICAL & 208.27637 & 11.63439 & $<1.0$ & 2 & 1.41 & \citet{Jackson2024}; \citet{2006AJ....131.1934I} & \\
CLASS J0316+4328 & JCP & 49.2122\tnotex{allknown-b} & 43.472 & 126.464 & 2 & 0.5 & \citet{2007MNRAS.381L..55B} & J031650.88+432819.2 \\
PSS J2322+1944 & OPTICAL & 350.5298 & 19.7397 & $<1.0$ & SFG & 1.5 & \citet{2008ApJ...686..851R} &  \\
SDSS J1226-0006 & OPTICAL & 186.53358 & $-$0.10064 & $<1.0$ & 2 & 1.21 & \citet{Jackson2024}; \citet{2008AJ....135..496I} & \\
ULAS J2343-0050\tnotex{allknown-e} & OPTICAL & 355.79983 & 0.84272 & $<1.0$ & 2 & 1.4 & \citet{Jackson2024}; \citet{2008MNRAS.387..741J} & \\
ULAS J0820+0810\tnotex{allknown-e} & OPTICAL & 245.06687 & 8.20494 & $<0.1$ & 2 & 2.3 & \citet{Jackson2024}; \citet{2009MNRAS.398.1423J} & \\
SDSS J1339+1310 & OPTICAL & 204.77974 & 13.17768 & $<1.0$ & 2 & 1.69 & \citet{Jackson2024}; \citet{2009AJ....137.4118I} & \\
SDSS 1258+1657 & OPTICAL & 194.58019\tnotex{allknown-b} & 16.95491 & $<1.0$ & 2 & 1.28 & \citet{Jackson2024}; \citet{2009AJ....137.4118I} & \\
SDSS J1054+2733 & OPTICAL & 163.67025 & 27.55178 & $<1.0$ & 2 & 1.27 & \citet{Jackson2024}; \citet{2010AJ....139.1614K} & \\
SDSS J1349+1227 & OPTICAL & 207.37492 & 12.45236 & $<1.0$ & 2 & 3.0 & \citet{Jackson2024}; \citet{2010AJ....139.1614K} & \\
SDSS J0946+1835\tnotex{allknown-e} & OPTICAL & 146.52041\tnotex{allknown-b} & 18.59495 & $<1.0$ & 2 & 3.06 & \citet{Jackson2024}; \citet{2010AJ....140..370M} & \\
ULAS J1405+0959 & OPTICAL & 228.910408 & 15.193039 & $<1.0$ & 2 & 1.95 & \citet{Jackson2024}; \citet{2012MNRAS.419.2014J} & \\
SDSS 1320+1644 & OPTICAL & 200.248 & 16.73408 & $<1.0$ & 2 & 8.6 & \citet{Jackson2024}; \citet{2013ApJ...765..139R} & \\
SDSS J1515+1511 & OPTICAL & 228.910408 & 15.193039 & $<1.0$ & 2 & 1.99 & \citet{Jackson2024}; \citet{2014AJ....147..153I} & \\
SDSS J1128+2402 & OPTICAL & 172.077117 & 24.038236 & $<1.0$ & 2 & 0.844 & \citet{Jackson2024}; \citet{2014AJ....147..153I} & \\
SDSS J0818+0601 & OPTICAL & 124.62692 & 6.02722 & $<1.0$ & 2 & 1.15 & \citet{Jackson2024}; \citet{2016MNRAS.456.1595M} & \\
SDSS J1442+4055 & OPTICAL & 220.7279 & 40.926494 & $<1.0$ & 2 & 2.1 & \citet{Jackson2024}; \citet{2016MNRAS.456.1948S} & \\
WISE J2329$-$1258 & OPTICAL & 352.491 & $-$12.98306 & 1.013 & 2 & 1.26 & \citetalias{martinez2025}/\citet{Jackson2024}; \citet{2329disc} & J232957.86$-$125859.1 \\
PS J1721+8842 & GAIA & 260.43437 & 88.70599 & 1.848 & 6 & 4.03 & \citet{2021MNRAS.508L..64M}; \citet{Lemon2018} & J172146.08+884221.9 \\
PS J0140+4107 & GAIA & 25.2042\tnotex{allknown-b} & 41.1333 & $<1.0$ & 2 & 1.44 & \citet{Jackson2024}; \citet{Lemon2018} & \\
J0146$-$1133 & GAIA & 26.63691 & $-$11.56113 & $<1.0$ & 2 & 1.69 & \citet{Jackson2024}; \citet{Lemon2018} & \\
PS J1831+5447 & GAIA & 277.8636 & 54.79965 & 12.434 & 2 & 2.32 & \citet{Jackson2024}; \citet{Lemon2018} & J183127.08+544759.9 \\
J0941+0518 & GAIA & 145.34378 & 5.30664 & $<1.0$ & 2 & 5.4 & \citet{Jackson2024}; \citet{Lemon2018} & \\
PS J2124+1632 & GAIA & 321.07029 & 16.53841 & $<1.0$ & 2 & 3.02 & \citet{Jackson2024}; \citet{Lemon2018} & \\
PS J2305+3714\tnotex{allknown-e} & GAIA & 346.48273 & 37.23932 & $<1.0$ & 2 & 2.2 & \citet{Jackson2024}; \citet{Lemon2018} & \\
PS J2332-1852 & GAIA & 353.08034 & $-$18.86853 & $<1.0$ & 2 & 1.97 & \citet{Jackson2024}; \citet{Lemon2018} & \\
GRAL J1131$-$4419\tnotex{allknown-d} & GAIA & 172.750041\tnotex{allknown-b} & $-$44.3330556 &\tnotex{allknown-a} & 4 & 1.7 & \citet{dobie23}; \citet{2018AaA...616L..11K} &  \\
WGD J2038$-$4008\tnotex{allknown-d} & GAIA & 309.511278\tnotex{allknown-b} & $-$40.137107 &\tnotex{allknown-a} & 4 & 2.87 & \citet{dobie23}; \citet{2018MNRAS.479.4345A} &  \\
J0235$-$2433 & GAIA & 38.86431 & $-$24.55356 & $<1.0$ & 2 & 2.05 & \citet{Jackson2024}; \citet{2018MNRAS.479.4345A} & \\
DES J0245$-$0556 & GAIA & 41.356506\tnotex{allknown-b} & $-$5.950145 & $<1.0$ & 2 & 1.9 & \citet{Jackson2024}; \citet{2018MNRAS.479.4345A} & \\
% MJV 1255+1158\tnotex{allknown-c} & RADIO & 193.874\tnotex{allknown-b} & 11.9816 & 36.024 & 2 & 0.46 & \citet{2019MNRAS.483.2125S} & J125529.76+115854.2 \\
% MJV J1330+3141\tnotex{allknown-c} & RADIO & 202.5398\tnotex{allknown-b} & 31.6846 & 38.521 & 2 & 0.54 & \citet{2019MNRAS.483.2125S} & J133009.54+314104.5 \\
J0013+5119 & GAIA & 3.348077\tnotex{allknown-b} & 51.3183 & 2.097 & 2 & 1.89 & \citetalias{martinez2025}/\citet{Jackson2024}; \citet{Lemon19} & J001323.53+511905.9 \\
J1537$-$3010\tnotex{allknown-d} & GAIA & 234.355598 & $-$30.171335 & $<1.0$ & 4 & 3.3 & \citet{dobie23}; \citet{Lemon19} &  \\
GRAL J1817+2729 & GAIA & 274.378545\tnotex{allknown-b} & 27.494468 & 2.575 & 4 & 1.8 & \citetalias{martinez2025}/\citet{dobie23}; \citet{Lemon19} & J181730.82+272940.2 \\
J0228+3953 & GAIA & 37.046244 & 39.88536 & $<1.0$ & 2 & 1.57 & \citet{Jackson2024}; \citet{Lemon19} & \\
J1949+7732 & GAIA & 297.40117 & 77.54416 & $<1.0$ & 2 & 1.59 & \citet{Jackson2024}; \citet{Lemon19} & \\
J2145+6345 & GAIA & 326.2713 & 63.7614461 & 1.182 & 4 & 2.07 & \citetalias{martinez2025}; \citet{Lemon19} & J214505.20+634541.1 \\
GRAL J0248+1913 & GAIA & 42.2031\tnotex{allknown-b} & 19.22528 & $<1.0$ & 4 & 1.76 & \citet{dobie23}; \citet{2019AaA...622A.165D} &  \\
GRAL J0659+1629 & GAIA & 104.766823\tnotex{allknown-b} & 16.485772 & $<1.0$ & 4 & 5.2 & \citet{dobie23}; \citet{2019AaA...622A.165D} &  \\
GRAL J0530$-$3730\tnotex{allknown-d} & GAIA & 82.6541\tnotex{allknown-b} & 82.6541 & $<1.0$ & 3 & 1.04 & \citet{dobie23}; \citet{2019AaA...622A.165D} &  \\
GRAL J2014$-$3024 & GAIA & 303.7258333\tnotex{allknown-b} & $-$30.4144444 & $<1.0$ & 4 & 2.5 & \citet{dobie23}; \citet{2019AaA...622A.165D} &  \\
GRAL J0246$-$1845\tnotex{allknown-d} & GAIA & 41.5508333\tnotex{allknown-b} & $-$18.7514167 & $<1.0$ & 2 & 1.0 & \citet{dobie23}; \citet{2019arXiv191208977K} &  \\
GRAL J0346+2154 & GAIA & 56.5458 & 21.9124 & $<1.0$ & 2 & 0.99 & \citet{dobie23}; \citet{2019arXiv191208977K} &  \\
GRAL J0818+0601 & GAIA & 124.6269582\tnotex{allknown-b} & 6.027244393 & $<1.0$ & 2 & 1.15 & \citet{dobie23}; \citet{2019arXiv191208977K} &  \\
GRAL J1556$-$1352\tnotex{allknown-d} & GAIA & 239.23375\tnotex{allknown-b} & $-$13.8694722 & $<1.0$ & 2 & 0.96 & \citet{dobie23}; \citet{2019arXiv191208977K} &  \\
GRAL J2343+0435 & GAIA & 355.8775\tnotex{allknown-b} & 4.5994444 & $<1.0$ & 2 & 1.23 & \citet{dobie23}; \citet{2019arXiv191208977K} &  \\
DES J0229+0320\tnotex{allknown-d} & GAIA & 37.49255525\tnotex{allknown-b} & 37.49255525 & $<1.0$ & 2 & 2.14 & \citet{dobie23}; \citet{2020MNRAS.494.3491L} &  \\
GRAL J0607$-$2152\tnotex{allknown-d} & GAIA & 91.795\tnotex{allknown-b} & $-$21.8713889 & $<1.0$ & 4 & 1.7 & \citet{dobie23}; \citet{2021ApJ...921...42S} &  \\
GRAL J0608+4229 & GAIA & 92.1725\tnotex{allknown-b} & 42.4936111 & $<1.0$ & 4 & 1.3 & \citet{dobie23}; \citet{2021ApJ...921...42S} &  \\
GRAL J0818$-$2613\tnotex{allknown-d} & GAIA & 124.6179167\tnotex{allknown-b} & $-$26.2236111 & $<1.0$ & 4 & 6.2 & \citet{dobie23}; \citet{2021ApJ...921...42S} &  \\
GRAL J1651$-$0417 & GAIA & 252.7720833\tnotex{allknown-b} & $-$4.2902778 & $<1.0$ & 4 & 10.1 & \citet{dobie23}; \citet{2021ApJ...921...42S} &  \\
GRAL J2103$-$0850 & GAIA & 315.8708333\tnotex{allknown-b} & $-$8.8469444 & $<1.0$ & 4 & 1.0 & \citet{dobie23}; \citet{2021ApJ...921...42S} &  \\
SDSS J0823+2418 & GAIA & 125.9211496 & 24.3015122 & $<1.0$ & 2 & 0.64 & \citet{2023ApJ...956..117G}; \citet{2021ApJ...921...42S} & \\
UNIONSJ1111+3804 & OPTICAL & 167.7773\tnotex{allknown-b} & 38.0736 & 4.473 & 2 & 1.96 & This work; \citet{chan22} & J111106.53+380424.6 \\
J0156$-$2751 & GAIA & 29.1039\tnotex{allknown-b} & -27.8562 & $<1.0$ & 2 & 1.5 & This work; \citet{lemon2023} & \\
DECALS J0336$-$3244\tnotex{allknown-c} & OPTICAL & 54.08159\tnotex{allknown-b} & $-$32.74075 & $<1.0$ & 2 & 0.69 & This work; \citet{2023ApJS..269...61D} & \\
J0416+7428 & GAIA & 64.1972\tnotex{allknown-b} & 74.4827 & $<1.0$ & 2 & 2.64 & This work; \citet{lemon2023} & \\
J1326+3020 & GAIA & 201.7411 & 30.3400 & $<1.0$ & 2 & 2.11 & This work; \citet{lemon2023} & \\
J2015+0707 & GAIA & 303.80369 & 7.11728 & $<1.0$ & 2 & 2.93 & This work; \citet{lemon2023} & \\
J230818+320145 & GAIA & 347.0777\tnotex{allknown-b} & 32.0294 & $<1.0$ & 2 & 2.63 & This work; \citet{lemon2023} & \\
J050129$-$073307\tnotex{allknown-c} & OPTICAL & 75.37257429\tnotex{allknown-b} & $-$7.551979509 & $<1.0$ & 2 & 2.70 & This work; \citet{he23} & \\
J223239+131518\tnotex{allknown-c} & OPTICAL & 338.1644925\tnotex{allknown-b} & 13.25502128 & $<1.0$ & 2 & 1.02 & This work; \citet{he23} & \\
\end{longtable}
\end{ThreePartTable}
\end{landscape}
}
%\restoregeometry

\bibliographystyle{aasjournal_old}

%\setcitestyle{numbers}
\bibliography{main.bib}{}

@ARTICLE{rusu19,
       author = {{Rusu}, Cristian E. and {Berghea}, Ciprian T. and {Fassnacht}, Christopher D. and {More}, Anupreeta and {Seman}, Erica and {Nelson}, George J. and {Chen}, Geoff C. -F.},
        title = "{A search for gravitationally lensed quasars and quasar pairs in Pan-STARRS1: spectroscopy and sources of shear in the diamond 2M1134-2103}",
      journal = {MNRAS},
         year = 2019,
        month = jul,
       volume = {486},
       number = {4},
        pages = {4987-5007},
          doi = {10.1093/mnras/stz1142},
archivePrefix = {arXiv},
       eprint = {1803.07175},
 primaryClass = {astro-ph.GA},
       adsurl = {https://ui.adsabs.harvard.edu/abs/2019MNRAS.486.4987R}
}

@ARTICLE{2023ApJS..269...61D,
       author = {{Dawes}, C. and {Storfer}, C. and {Huang}, X. and {Aldering}, G. and {Cikota}, Aleksandar and {Dey}, Arjun and {Schlegel}, D.~J.},
        title = "{Finding Multiply Lensed and Binary Quasars in the DESI Legacy Imaging Surveys}",
      journal = {\apjs},
         year = 2023,
        month = dec,
       volume = {269},
       number = {2},
          eid = {61},
        pages = {61},
          doi = {10.3847/1538-4365/ad015a},
archivePrefix = {arXiv},
       eprint = {2208.06356},
 primaryClass = {astro-ph.CO},
       adsurl = {https://ui.adsabs.harvard.edu/abs/2023ApJS..269...61D}
}

@ARTICLE{wisotzki04,
       author = {{Wisotzki}, L. and {Schechter}, P.~L. and {Chen}, H. -W. and {Richstone}, D. and {Jahnke}, K. and {S{\'a}nchez}, S.~F. and {Reimers}, D.},
        title = "{HE 0047-1756: A new gravitationally lensed double QSO}",
      journal = {\aap},
         year = 2004,
        month = may,
       volume = {419},
        pages = {L31-L34},
          doi = {10.1051/0004-6361:20040131},
archivePrefix = {arXiv},
       eprint = {astro-ph/0403475},
 primaryClass = {astro-ph},
       adsurl = {https://ui.adsabs.harvard.edu/abs/2004A&A...419L..31W}
}

@ARTICLE{2017A&A...597A..49G,
       author = {{Giannini}, E. and {Schmidt}, R.~W. and {Wambsganss}, J. and {Alsubai}, K. and {Andersen}, J.~M. and {Anguita}, T. and {Bozza}, V. and {Bramich}, D.~M. and {Browne}, P. and {Calchi Novati}, S. and {Damerdji}, Y. and {Diehl}, C. and {Dodds}, P. and {Dominik}, M. and {Elyiv}, A. and {Fang}, X. and {Figuera Jaimes}, R. and {Finet}, F. and {Gerner}, T. and {Gu}, S. and {Hardis}, S. and {Harps{\o}e}, K. and {Hinse}, T.~C. and {Hornstrup}, A. and {Hundertmark}, M. and {Jessen-Hansen}, J. and {J{\o}rgensen}, U.~G. and {Juncher}, D. and {Kains}, N. and {Kerins}, E. and {Korhonen}, H. and {Liebig}, C. and {Lund}, M.~N. and {Lundkvist}, M.~S. and {Maier}, G. and {Mancini}, L. and {Masi}, G. and {Mathiasen}, M. and {Penny}, M. and {Proft}, S. and {Rabus}, M. and {Rahvar}, S. and {Ricci}, D. and {Scarpetta}, G. and {Sahu}, K. and {Sch{\"a}fer}, S. and {Sch{\"o}nebeck}, F. and {Skottfelt}, J. and {Snodgrass}, C. and {Southworth}, J. and {Surdej}, J. and {Tregloan-Reed}, J. and {Vilela}, C. and {Wertz}, O. and {Zimmer}, F.},
        title = "{MiNDSTEp differential photometry of the gravitationally lensed quasars WFI 2033-4723 and HE 0047-1756: microlensing and a new time delay}",
      journal = {\aap},
         year = 2017,
        month = jan,
       volume = {597},
          eid = {A49},
        pages = {A49},
          doi = {10.1051/0004-6361/201527422},
archivePrefix = {arXiv},
       eprint = {1610.03732},
 primaryClass = {astro-ph.GA},
       adsurl = {https://ui.adsabs.harvard.edu/abs/2017A&A...597A..49G}
}

@ARTICLE{2014ApJ...797...61R,
       author = {{Rojas}, K. and {Motta}, V. and {Mediavilla}, E. and {Falco}, E. and {Jim{\'e}nez-Vicente}, J. and {Mu{\~n}oz}, J.~A.},
        title = "{Strong Chromatic Microlensing in HE0047-1756 and SDSS1155+6346}",
      journal = {\apj},
         year = 2014,
        month = dec,
       volume = {797},
       number = {1},
          eid = {61},
        pages = {61},
          doi = {10.1088/0004-637X/797/1/61},
archivePrefix = {arXiv},
       eprint = {1409.8246},
 primaryClass = {astro-ph.GA},
       adsurl = {https://ui.adsabs.harvard.edu/abs/2014ApJ...797...61R}
}

@ARTICLE{cosmograil2020,
       author = {{Millon}, M. and {Courbin}, F. and {Bonvin}, V. and {Paic}, E. and {Meylan}, G. and {Tewes}, M. and {Sluse}, D. and {Magain}, P. and {Chan}, J.~H.~H. and {Galan}, A. and {Joseph}, R. and {Lemon}, C. and {Tihhonova}, O. and {Anderson}, R.~I. and {Marmier}, M. and {Chazelas}, B. and {Lendl}, M. and {Triaud}, A.~H.~M.~J. and {Wyttenbach}, A.},
        title = "{COSMOGRAIL. XIX. Time delays in 18 strongly lensed quasars from 15 years of optical monitoring}",
      journal = {\aap},
         year = 2020,
        month = aug,
       volume = {640},
          eid = {A105},
        pages = {A105},
          doi = {10.1051/0004-6361/202037740},
archivePrefix = {arXiv},
       eprint = {2002.05736},
 primaryClass = {astro-ph.CO},
       adsurl = {https://ui.adsabs.harvard.edu/abs/2020A&A...640A.105M}
}

@ARTICLE{shajib21,
       author = {{Shajib}, Anowar J. and {Molina}, Eden and {Agnello}, Adriano and {Williams}, Peter R. and {Birrer}, Simon and {Treu}, Tommaso and {Fassnacht}, Christopher D. and {Morishita}, Takahiro and {Abramson}, Louis and {Schechter}, Paul L. and {Wisotzki}, Lutz},
        title = "{High-resolution imaging follow-up of doubly imaged quasars}",
      journal = {\mnras},
         year = 2021,
        month = may,
       volume = {503},
       number = {2},
        pages = {1557-1567},
          doi = {10.1093/mnras/stab532},
archivePrefix = {arXiv},
       eprint = {2011.01971},
 primaryClass = {astro-ph.GA},
       adsurl = {https://ui.adsabs.harvard.edu/abs/2021MNRAS.503.1557S}
}

@ARTICLE{ding17,
       author = {{Ding}, Xuheng and {Liao}, Kai and {Treu}, Tommaso and {Suyu}, Sherry H. and {Chen}, Geoff C. -F. and {Auger}, Matthew W. and {Marshall}, Philip J. and {Agnello}, Adriano and {Courbin}, Frederic and {Nierenberg}, Anna M. and {Rusu}, Cristian E. and {Sluse}, Dominique and {Sonnenfeld}, Alessandro and {Wong}, Kenneth C.},
        title = "{H0LiCOW. VI. Testing the fidelity of lensed quasar host galaxy reconstruction}",
      journal = {\mnras},
         year = 2017,
        month = mar,
       volume = {465},
       number = {4},
        pages = {4634-4649},
          doi = {10.1093/mnras/stw3078},
archivePrefix = {arXiv},
       eprint = {1610.08504},
 primaryClass = {astro-ph.GA},
       adsurl = {https://ui.adsabs.harvard.edu/abs/2017MNRAS.465.4634D}
}

@ARTICLE{condon92,
       author = {{Condon}, J.~J.},
        title = "{Radio emission from normal galaxies.}",
      journal = {\araa},
         year = 1992,
        month = jan,
       volume = {30},
        pages = {575-611},
          doi = {10.1146/annurev.aa.30.090192.003043},
       adsurl = {https://ui.adsabs.harvard.edu/abs/1992ARA&A..30..575C}
}

@ARTICLE{he23,
       author = {{He}, Zizhao and {Li}, Nan and {Cao}, Xiaoyue and {Li}, Rui and {Zou}, Hu and {Dye}, Simon},
        title = "{Discovering strongly lensed quasar candidates with catalogue-based methods from DESI Legacy Surveys}",
      journal = {\aap},
         year = 2023,
        month = apr,
       volume = {672},
          eid = {A123},
        pages = {A123},
          doi = {10.1051/0004-6361/202245484},
archivePrefix = {arXiv},
       eprint = {2301.11080},
 primaryClass = {astro-ph.CO},
       adsurl = {https://ui.adsabs.harvard.edu/abs/2023A&A...672A.123H}
}

@ARTICLE{chan22,
       author = {{Chan}, J.~H.~H. and {Lemon}, C. and {Courbin}, F. and {Gavazzi}, R. and {Cl{\'e}ment}, B. and {Millon}, M. and {Paic}, E. and {Rojas}, K. and {Savary}, E. and {Vernardos}, G. and {Cuillandre}, J. -C. and {Fabbro}, S. and {Gwyn}, S. and {Hudson}, M.~J. and {Kilbinger}, M. and {McConnachie}, A.},
        title = "{Discovery of strongly lensed quasars in the Ultraviolet Near Infrared Optical Northern Survey (UNIONS)}",
      journal = {\aap},
         year = 2022,
        month = mar,
       volume = {659},
          eid = {A140},
        pages = {A140},
          doi = {10.1051/0004-6361/202142389},
archivePrefix = {arXiv},
       eprint = {2110.09535},
 primaryClass = {astro-ph.GA},
       adsurl = {https://ui.adsabs.harvard.edu/abs/2022A&A...659A.140C}
}

@ARTICLE{RCS,
       author = {{Gladders}, Michael D. and {Yee}, H.~K.~C.},
        title = "{The Red-Sequence Cluster Survey. I. The Survey and Cluster Catalogs for Patches RCS 0926+37 and RCS 1327+29}",
      journal = {\apjs},
         year = 2005,
        month = mar,
       volume = {157},
       number = {1},
        pages = {1-29},
          doi = {10.1086/427327},
archivePrefix = {arXiv},
       eprint = {astro-ph/0411075},
 primaryClass = {astro-ph},
       adsurl = {https://ui.adsabs.harvard.edu/abs/2005ApJS..157....1G}
}

@ARTICLE{1993ApJ...415..459P,
       author = {{Pyne}, Ted and {Birkinshaw}, Mark},
        title = "{Null Geodesics in Perturbed Spacetimes}",
      journal = {\apj},
         year = 1993,
        month = oct,
       volume = {415},
        pages = {459},
          doi = {10.1086/173178},
archivePrefix = {arXiv},
       eprint = {astro-ph/9303020},
 primaryClass = {astro-ph},
       adsurl = {https://ui.adsabs.harvard.edu/abs/1993ApJ...415..459P}
}

@ARTICLE{1996ApJ...458...46P,
       author = {{Pyne}, Ted and {Birkinshaw}, Mark},
        title = "{Beyond the Thin Lens Approximation}",
      journal = {\apj},
         year = 1996,
        month = feb,
       volume = {458},
        pages = {46},
          doi = {10.1086/176791},
archivePrefix = {arXiv},
       eprint = {astro-ph/9504060},
 primaryClass = {astro-ph},
       adsurl = {https://ui.adsabs.harvard.edu/abs/1996ApJ...458...46P}
}

@book{galaxiesbook,
    title={Dynamics and Astrophysics of Galaxies},
    author={{Bovy}, J.},
    year={2026},
    publisher={Princeton University Press},
    address= {Princeton, NJ}
}

@MISC{bertschinger,
       author = {{Bertschinger}, Edmund},
        title = "{Cosmological dynamics}",
 howpublished = {Technical Report, Massachusetts Inst. of Tech. Cambridge, MA United States Dept. of Physics.},
         year = 1995,
        month = jan,
        pages = {22249},
          doi = {10.48550/arXiv.astro-ph/9503125},
archivePrefix = {arXiv},
       eprint = {astro-ph/9503125},
 primaryClass = {astro-ph},
       adsurl = {https://ui.adsabs.harvard.edu/abs/1995STIN...9622249B}
}

@BOOK{carrolbook,
       author = {{Carroll}, Sean M.},
        title = "{Spacetime and geometry. An introduction to general relativity}",
         year = 2004,
         publisher = {Addison Wesley},
       adsurl = {https://ui.adsabs.harvard.edu/abs/2004sgig.book.....C}
}

@BOOK{landaufields,
       author = {{Landau}, Lev Davidovich and {Lifshitz}, E.~M.},
        title = "{The classical theory of fields}",
         year = 1975,
         publisher = {Pergamon Press},
       adsurl = {https://ui.adsabs.harvard.edu/abs/1975ctf..book.....L}
}

@BOOK{radiohist,
       author = {{Malphrus}, Benjamin K.},
        title = "{The history of radio astronomy and the National Radio Astronomy Observatory : evolution toward big science}",
         year = 1996,
       adsurl = {https://ui.adsabs.harvard.edu/abs/1996hra..book.....M},
       publisher={Krieger Publishing Company}
}

@INPROCEEDINGS{trimblelens,
       author = {{Trimble}, Virginia},
        title = "{The First Lenses}",
    booktitle = {Gravitational Lensing: Recent Progress and Future Goals},
         year = 2001,
       editor = {{Brainerd}, Tereasa G. and {Kochanek}, Christopher S.},
       series = {Astronomical Society of the Pacific Conference Series},
       volume = {237},
        month = jan,
        pages = {1},
       adsurl = {https://ui.adsabs.harvard.edu/abs/2001ASPC..237....1T}
}

@INPROCEEDINGS{gabaud,
       author = {{Valls-Gabaud}, David},
        title = "{The conceptual origins of gravitational lensing}",
    booktitle = {Albert Einstein Century International Conference},
         year = 2006,
       editor = {{Alimi}, Jean-Michel and {F{\"u}zfa}, Andr{\'e}},
       series = {American Institute of Physics Conference Series},
       volume = {861},
        month = nov,
    publisher = {AIP},
        pages = {1163-1163},
          doi = {10.1063/1.2399715},
archivePrefix = {arXiv},
       eprint = {1206.1165},
 primaryClass = {physics.hist-ph},
       adsurl = {https://ui.adsabs.harvard.edu/abs/2006AIPC..861.1163V}
}

@ARTICLE{eddington,
       author = {{Dyson}, F.~W. and {Eddington}, A.~S. and {Davidson}, C.},
        title = "{A Determination of the Deflection of Light by the Sun's Gravitational Field, from Observations Made at the Total Eclipse of May 29, 1919}",
      journal = {Philosophical Transactions of the Royal Society of London Series A},
         year = 1920,
        month = jan,
       volume = {220},
        pages = {291-333},
          doi = {10.1098/rsta.1920.0009},
       adsurl = {https://ui.adsabs.harvard.edu/abs/1920RSPTA.220..291D}
}

@ARTICLE{einsteinlens,
       author = {{Einstein}, Albert},
        title = "{Lens-Like Action of a Star by the Deviation of Light in the Gravitational Field}",
      journal = {Science},
         year = 1936,
        month = dec,
       volume = {84},
       number = {2188},
        pages = {506-507},
          doi = {10.1126/science.84.2188.506},
       adsurl = {https://ui.adsabs.harvard.edu/abs/1936Sci....84..506E}
}

@ARTICLE{zwickylens,
       author = {{Zwicky}, F.},
        title = "{Nebulae as Gravitational Lenses}",
      journal = {Physical Review},
         year = 1937,
        month = feb,
       volume = {51},
       number = {4},
        pages = {290-290},
          doi = {10.1103/PhysRev.51.290},
       adsurl = {https://ui.adsabs.harvard.edu/abs/1937PhRv...51..290Z}
}

@BOOK{dodelson,
       author = {{Dodelson}, Scott and {Schmidt}, Fabian},
        title = "{Modern Cosmology}",
         year = 2020,
          doi = {10.1016/C2017-0-01943-2},
          publisher = {Academic Press},
       adsurl = {https://ui.adsabs.harvard.edu/abs/2020moco.book.....D}
}

@INPROCEEDINGS{bhex,
       author = {{Johnson}, Michael D. and {Akiyama}, Kazunori and {Baturin}, Rebecca and {Bilyeu}, Bryan and {Blackburn}, Lindy and {Boroson}, Don and {C{\'a}rdenas-Avenda{\~n}o}, Alejandro and {Chael}, Andrew and {Chan}, Chi-kwan and {Chang}, Dominic and {Cheimets}, Peter and {Chou}, Cathy and {Doeleman}, Sheperd S. and {Farah}, Joseph and {Galison}, Peter and {Gamble}, Ronald and {Gammie}, Charles F. and {Gelles}, Zachary and {G{\'o}mez}, Jos{\'e} L. and {Gralla}, Samuel E. and {Grimes}, Paul and {Gurvits}, Leonid I. and {Hadar}, Shahar and {Haworth}, Kari and {Hada}, Kazuhiro and {Hecht}, Michael H. and {Honma}, Mareki and {Houston}, Janice and {Hudson}, Ben and {Issaoun}, Sara and {Jia}, He and {Jorstad}, Svetlana and {Kauffman}, Jens and {Kovalev}, Yuri Y. and {Kurczynski}, Peter and {Lafon}, Robert E. and {Lupsasca}, Alexandru and {Lehmensiek}, Robert and {Ma}, Chung-Pei and {Marrone}, Daniel P. and {Marscher}, Alan P. and {Melnick}, Gary and {Narayan}, Ramesh and {Niinuma}, Kotaro and {Noble}, Scott C. and {Palmer}, Eric J. and {Palumbo}, Daniel C.~M. and {Paritsky}, Lenny and {Peretz}, Eliad and {Pesce}, Dominic and {Plavin}, Alexander and {Quataert}, Eliot and {Rana}, Hannah and {Ricarte}, Angelo and {Roelofs}, Freek and {Shtyrkova}, Katia and {Sinclair}, Laura C. and {Small}, Jeffrey and {Kumara}, Sridharan Tirupati and {Srinivasan}, Ranjani and {Strominger}, Andrew and {Tiede}, Paul and {Tong}, Edward and {Wang}, Jade and {Weintroub}, Jonathan and {Wielgus}, Maciek and {Wong}, George},
        title = "{The Black Hole Explorer: motivation and vision}",
    booktitle = {Space Telescopes and Instrumentation 2024: Optical, Infrared, and Millimeter Wave},
         year = 2024,
       editor = {{Coyle}, Laura E. and {Matsuura}, Shuji and {Perrin}, Marshall D.},
       series = {Society of Photo-Optical Instrumentation Engineers (SPIE) Conference Series},
       volume = {13092},
        month = aug,
          eid = {130922D},
        pages = {130922D},
          doi = {10.1117/12.3019835},
archivePrefix = {arXiv},
       eprint = {2406.12917},
 primaryClass = {astro-ph.IM},
       adsurl = {https://ui.adsabs.harvard.edu/abs/2024SPIE13092E..2DJ}
}

@ARTICLE{radioastron,
       author = {{Valtonen}, Mauri J. and {Dey}, Lankeswar and {Zola}, Staszek and {Gupta}, Alok C. and {Kishore}, Shubham and {Gopakumar}, Achamveedu and {Wiita}, Paul J. and {Gu}, Minfeng and {Nilsson}, Kari and {Zhang}, Zhongli and {Hudec}, Rene and {Matsumoto}, Katsura and {Drozdz}, Marek and {Ogloza}, Waldemar and {Berdyugin}, Andrei V. and {Reichart}, Daniel E. and {Mugrauer}, Markus and {Pursimo}, Tapio and {Ciprini}, Stefano and {Nakaoka}, Tatsuya and {Uemura}, Makoto and {Imazawa}, Ryo and {Zejmo}, Michal and {Kouprianov}, Vladimir V. and {Davidson}, Jr., James W. and {Sadun}, Alberto and {{\v{S}}trobl}, Jan and {Jel{\'\i}nek}, Martin and {Susobhanan}, Abhimanyu},
        title = "{Identifying the Secondary Jet in the RadioAstron Image of OJ 287}",
      journal = {\apj},
         year = 2025,
        month = oct,
       volume = {992},
       number = {1},
          eid = {110},
        pages = {110},
          doi = {10.3847/1538-4357/ae057e},
archivePrefix = {arXiv},
       eprint = {2510.06744},
 primaryClass = {astro-ph.HE},
       adsurl = {https://ui.adsabs.harvard.edu/abs/2025ApJ...992..110V}
}

@ARTICLE{2024MNRAS.529.2428D,
       author = {{Deane}, Roger P. and {Radcliffe}, Jack F. and {Njeri}, Ann and {Akoto-Danso}, Alexander and {Bernardi}, Gianni and {Smirnov}, Oleg M. and {Beswick}, Rob and {Garrett}, Michael A. and {Jarvis}, Matt J. and {Whittam}, Imogen H. and {Bourke}, Stephen and {Paragi}, Zsolt},
        title = "{The VLBA CANDELS GOODS-North Survey - I. survey design, processing, data products, and source counts}",
      journal = {\mnras},
         year = 2024,
        month = apr,
       volume = {529},
       number = {3},
        pages = {2428-2442},
          doi = {10.1093/mnras/stae253},
archivePrefix = {arXiv},
       eprint = {2401.12298},
 primaryClass = {astro-ph.GA},
       adsurl = {https://ui.adsabs.harvard.edu/abs/2024MNRAS.529.2428D}
}

@ARTICLE{DESI24,
       author = {{Adame}, A.~G. and {Aguilar}, J. and {Ahlen}, S. and {Alam}, S. and {Alexander}, D.~M. and {Alvarez}, M. and {Alves}, O. and {Anand}, A. and {Andrade}, U. and {Armengaud}, E. and {Avila}, S. and {Aviles}, A. and {Awan}, H. and {Bahr-Kalus}, B. and {Bailey}, S. and {Baltay}, C. and {Bault}, A. and {Behera}, J. and {BenZvi}, S. and {Bera}, A. and {Beutler}, F. and {Bianchi}, D. and {Blake}, C. and {Blum}, R. and {Brieden}, S. and {Brodzeller}, A. and {Brooks}, D. and {Buckley-Geer}, E. and {Burtin}, E. and {Calderon}, R. and {Canning}, R. and {Carnero Rosell}, A. and {Cereskaite}, R. and {Cervantes-Cota}, J.~L. and {Chabanier}, S. and {Chaussidon}, E. and {Chaves-Montero}, J. and {Chen}, S. and {Chen}, X. and {Claybaugh}, T. and {Cole}, S. and {Cuceu}, A. and {Davis}, T.~M. and {Dawson}, K. and {de la Macorra}, A. and {de Mattia}, A. and {Deiosso}, N. and {Dey}, A. and {Dey}, B. and {Ding}, Z. and {Doel}, P. and {Edelstein}, J. and {Eftekharzadeh}, S. and {Eisenstein}, D.~J. and {Elliott}, A. and {Fagrelius}, P. and {Fanning}, K. and {Ferraro}, S. and {Ereza}, J. and {Findlay}, N. and {Flaugher}, B. and {Font-Ribera}, A. and {Forero-S{\'a}nchez}, D. and {Forero-Romero}, J.~E. and {Frenk}, C.~S. and {Garcia-Quintero}, C. and {Gazta{\~n}aga}, E. and {Gil-Mar{\'\i}n}, H. and {Gontcho a Gontcho}, S. and {Gonzalez-Morales}, A.~X. and {Gonzalez-Perez}, V. and {Gordon}, C. and {Green}, D. and {Gruen}, D. and {Gsponer}, R. and {Gutierrez}, G. and {Guy}, J. and {Hadzhiyska}, B. and {Hahn}, C. and {Hanif}, M.~M.~S. and {Herrera-Alcantar}, H.~K. and {Honscheid}, K. and {Howlett}, C. and {Huterer}, D. and {Ir{\v{s}}i{\v{c}}}, V. and {Ishak}, M. and {Juneau}, S. and {Kara{\c{c}}ayl{\i}}, N.~G. and {Kehoe}, R. and {Kent}, S. and {Kirkby}, D. and {Kremin}, A. and {Krolewski}, A. and {Lai}, Y. and {Lan}, T.-W. and {Landriau}, M. and {Lang}, D. and {Lasker}, J. and {Le Goff}, J.~M. and {Le Guillou}, L. and {Leauthaud}, A. and {Levi}, M.~E. and {Li}, T.~S. and {Linder}, E. and {Lodha}, K. and {Magneville}, C. and {Manera}, M. and {Margala}, D. and {Martini}, P. and {Maus}, M. and {McDonald}, P. and {Medina-Varela}, L. and {Meisner}, A. and {Mena-Fern{\'a}ndez}, J. and {Miquel}, R. and {Moon}, J. and {Moore}, S. and {Moustakas}, J. and {Mueller}, E. and {Mu{\~n}oz-Guti{\'e}rrez}, A. and {Myers}, A.~D. and {Nadathur}, S. and {Napolitano}, L. and {Neveux}, R. and {Newman}, J.~A. and {Nguyen}, N.~M. and {Nie}, J. and {Niz}, G. and {Noriega}, H.~E. and {Padmanabhan}, N. and {Paillas}, E. and {Palanque-Delabrouille}, N. and {Pan}, J. and {Penmetsa}, S. and {Percival}, W.~J. and {Pieri}, M.~M. and {Pinon}, M. and {Poppett}, C. and {Porredon}, A. and {Prada}, F. and {P{\'e}rez-Fern{\'a}ndez}, A. and {P{\'e}rez-R{\`a}fols}, I. and {Rabinowitz}, D. and {Raichoor}, A. and {Ram{\'\i}rez-P{\'e}rez}, C. and {Ramirez-Solano}, S. and {Rashkovetskyi}, M. and {Ravoux}, C. and {Rezaie}, M. and {Rich}, J. and {Rocher}, A. and {Rockosi}, C. and {Roe}, N.~A. and {Rosado-Marin}, A. and {Ross}, A.~J. and {Rossi}, G. and {Ruggeri}, R. and {Ruhlmann-Kleider}, V. and {Samushia}, L. and {Sanchez}, E. and {Saulder}, C. and {Schlafly}, E.~F. and {Schlegel}, D. and {Schubnell}, M. and {Seo}, H. and {Shafieloo}, A. and {Sharples}, R. and {Silber}, J. and {Slosar}, A. and {Smith}, A. and {Sprayberry}, D. and {Tan}, T. and {Tarl{\'e}}, G. and {Taylor}, P. and {Trusov}, S. and {Ure{\~n}a-L{\'o}pez}, L.~A. and {Vaisakh}, R. and {Valcin}, D. and {Valdes}, F. and {Vargas-Maga{\~n}a}, M. and {Verde}, L. and {Walther}, M. and {Wang}, B. and {Wang}, M.~S. and {Weaver}, B.~A. and {Weaverdyck}, N. and {Wechsler}, R.~H. and {Weinberg}, D.~H. and {White}, M. and {Yu}, J. and {Yu}, Y. and {Yuan}, S. and {Y{\`e}che}, C. and {Zaborowski}, E.~A. and {Zarrouk}, P. and {Zhang}, H. and {Zhao}, C. and {Zhao}, R. and {Zhou}, R. and {Zhuang}, T.},
        title = "{DESI 2024 VI: cosmological constraints from the measurements of baryon acoustic oscillations}",
      journal = {\jcap},
         year = 2025,
        month = feb,
       volume = {2025},
       number = {2},
          eid = {021},
        pages = {021},
          doi = {10.1088/1475-7516/2025/02/021},
archivePrefix = {arXiv},
       eprint = {2404.03002},
 primaryClass = {astro-ph.CO},
       adsurl = {https://ui.adsabs.harvard.edu/abs/2025JCAP...02..021A}
}

@ARTICLE{planck18,
       author = {{Planck Collaboration} and {Aghanim}, N. and {Akrami}, Y. and {Ashdown}, M. and {Aumont}, J. and {Baccigalupi}, C. and {Ballardini}, M. and {Banday}, A.~J. and {Barreiro}, R.~B. and {Bartolo}, N. and {Basak}, S. and {Battye}, R. and {Benabed}, K. and {Bernard}, J.-P. and {Bersanelli}, M. and {Bielewicz}, P. and {Bock}, J.~J. and {Bond}, J.~R. and {Borrill}, J. and {Bouchet}, F.~R. and {Boulanger}, F. and {Bucher}, M. and {Burigana}, C. and {Butler}, R.~C. and {Calabrese}, E. and {Cardoso}, J.-F. and {Carron}, J. and {Challinor}, A. and {Chiang}, H.~C. and {Chluba}, J. and {Colombo}, L.~P.~L. and {Combet}, C. and {Contreras}, D. and {Crill}, B.~P. and {Cuttaia}, F. and {de Bernardis}, P. and {de Zotti}, G. and {Delabrouille}, J. and {Delouis}, J.-M. and {Di Valentino}, E. and {Diego}, J.~M. and {Dor{\'e}}, O. and {Douspis}, M. and {Ducout}, A. and {Dupac}, X. and {Dusini}, S. and {Efstathiou}, G. and {Elsner}, F. and {En{\ss}lin}, T.~A. and {Eriksen}, H.~K. and {Fantaye}, Y. and {Farhang}, M. and {Fergusson}, J. and {Fernandez-Cobos}, R. and {Finelli}, F. and {Forastieri}, F. and {Frailis}, M. and {Fraisse}, A.~A. and {Franceschi}, E. and {Frolov}, A. and {Galeotta}, S. and {Galli}, S. and {Ganga}, K. and {G{\'e}nova-Santos}, R.~T. and {Gerbino}, M. and {Ghosh}, T. and {Gonz{\'a}lez-Nuevo}, J. and {G{\'o}rski}, K.~M. and {Gratton}, S. and {Gruppuso}, A. and {Gudmundsson}, J.~E. and {Hamann}, J. and {Handley}, W. and {Hansen}, F.~K. and {Herranz}, D. and {Hildebrandt}, S.~R. and {Hivon}, E. and {Huang}, Z. and {Jaffe}, A.~H. and {Jones}, W.~C. and {Karakci}, A. and {Keih{\"a}nen}, E. and {Keskitalo}, R. and {Kiiveri}, K. and {Kim}, J. and {Kisner}, T.~S. and {Knox}, L. and {Krachmalnicoff}, N. and {Kunz}, M. and {Kurki-Suonio}, H. and {Lagache}, G. and {Lamarre}, J.-M. and {Lasenby}, A. and {Lattanzi}, M. and {Lawrence}, C.~R. and {Le Jeune}, M. and {Lemos}, P. and {Lesgourgues}, J. and {Levrier}, F. and {Lewis}, A. and {Liguori}, M. and {Lilje}, P.~B. and {Lilley}, M. and {Lindholm}, V. and {L{\'o}pez-Caniego}, M. and {Lubin}, P.~M. and {Ma}, Y.-Z. and {Mac{\'\i}as-P{\'e}rez}, J.~F. and {Maggio}, G. and {Maino}, D. and {Mandolesi}, N. and {Mangilli}, A. and {Marcos-Caballero}, A. and {Maris}, M. and {Martin}, P.~G. and {Martinelli}, M. and {Mart{\'\i}nez-Gonz{\'a}lez}, E. and {Matarrese}, S. and {Mauri}, N. and {McEwen}, J.~D. and {Meinhold}, P.~R. and {Melchiorri}, A. and {Mennella}, A. and {Migliaccio}, M. and {Millea}, M. and {Mitra}, S. and {Miville-Desch{\^e}nes}, M.-A. and {Molinari}, D. and {Montier}, L. and {Morgante}, G. and {Moss}, A. and {Natoli}, P. and {N{\o}rgaard-Nielsen}, H.~U. and {Pagano}, L. and {Paoletti}, D. and {Partridge}, B. and {Patanchon}, G. and {Peiris}, H.~V. and {Perrotta}, F. and {Pettorino}, V. and {Piacentini}, F. and {Polastri}, L. and {Polenta}, G. and {Puget}, J.-L. and {Rachen}, J.~P. and {Reinecke}, M. and {Remazeilles}, M. and {Renzi}, A. and {Rocha}, G. and {Rosset}, C. and {Roudier}, G. and {Rubi{\~n}o-Mart{\'\i}n}, J.~A. and {Ruiz-Granados}, B. and {Salvati}, L. and {Sandri}, M. and {Savelainen}, M. and {Scott}, D. and {Shellard}, E.~P.~S. and {Sirignano}, C. and {Sirri}, G. and {Spencer}, L.~D. and {Sunyaev}, R. and {Suur-Uski}, A.-S. and {Tauber}, J.~A. and {Tavagnacco}, D. and {Tenti}, M. and {Toffolatti}, L. and {Tomasi}, M. and {Trombetti}, T. and {Valenziano}, L. and {Valiviita}, J. and {Van Tent}, B. and {Vibert}, L. and {Vielva}, P. and {Villa}, F. and {Vittorio}, N. and {Wandelt}, B.~D. and {Wehus}, I.~K. and {White}, M. and {White}, S.~D.~M. and {Zacchei}, A. and {Zonca}, A.},
        title = "{Planck 2018 results. VI. Cosmological parameters}",
      journal = {\aap},
         year = 2020,
        month = sep,
       volume = {641},
          eid = {A6},
        pages = {A6},
          doi = {10.1051/0004-6361/201833910},
archivePrefix = {arXiv},
       eprint = {1807.06209},
 primaryClass = {astro-ph.CO},
       adsurl = {https://ui.adsabs.harvard.edu/abs/2020A&A...641A...6P}
}

@INPROCEEDINGS{hubblelaw,
       author = {{Way}, M.~J.},
        title = "{Dismantling Hubble's Legacy?}",
    booktitle = {Origins of the Expanding Universe: 1912-1932},
         year = 2013,
       editor = {{Way}, M.~J. and {Hunter}, D.},
       series = {Astronomical Society of the Pacific Conference Series},
       volume = {471},
        month = apr,
        pages = {97},
          doi = {10.48550/arXiv.1301.7294},
archivePrefix = {arXiv},
       eprint = {1301.7294},
 primaryClass = {physics.hist-ph},
       adsurl = {https://ui.adsabs.harvard.edu/abs/2013ASPC..471...97W}
}

@ARTICLE{guth,
       author = {{Guth}, Alan H.},
        title = "{Inflationary universe: A possible solution to the horizon and flatness problems}",
      journal = {\prd},
         year = 1981,
        month = jan,
       volume = {23},
       number = {2},
        pages = {347-356},
          doi = {10.1103/PhysRevD.23.347},
       adsurl = {https://ui.adsabs.harvard.edu/abs/1981PhRvD..23..347G}
}

@ARTICLE{2025NewAR.10101733A,
       author = {{Alexander}, D.~M. and {Hickox}, R.~C. and {Aird}, J. and {Combes}, F. and {Costa}, T. and {Habouzit}, M. and {Harrison}, C.~M. and {Leng}, R.~I. and {Morabito}, L.~K. and {Uckelman}, S.~L. and {Vickers}, P.},
        title = "{What drives the growth of black holes: A decade of progress}",
      journal = {\nar},
         year = 2025,
        month = dec,
       volume = {101},
          eid = {101733},
        pages = {101733},
          doi = {10.1016/j.newar.2025.101733},
archivePrefix = {arXiv},
       eprint = {2506.19166},
 primaryClass = {astro-ph.GA},
       adsurl = {https://ui.adsabs.harvard.edu/abs/2025NewAR.10101733A}
}

@ARTICLE{urry,
       author = {{Urry}, C. Megan and {Padovani}, Paolo},
        title = "{Unified Schemes for Radio-Loud Active Galactic Nuclei}",
      journal = {\pasp},
         year = 1995,
        month = sep,
       volume = {107},
        pages = {803},
          doi = {10.1086/133630},
archivePrefix = {arXiv},
       eprint = {astro-ph/9506063},
 primaryClass = {astro-ph},
       adsurl = {https://ui.adsabs.harvard.edu/abs/1995PASP..107..803U}
}

@ARTICLE{1988MNRAS.233..265P,
       author = {{Padmanabhan}, T. and {Subramanian}, Kandaswamy},
        title = "{The focusing equation, caustics and the condition for multiple imaging by thick gravitational lenses}",
      journal = {\mnras},
         year = 1988,
        month = jul,
       volume = {233},
        pages = {265-284},
          doi = {10.1093/mnras/233.2.265},
       adsurl = {https://ui.adsabs.harvard.edu/abs/1988MNRAS.233..265P}
}

@ARTICLE{kovner87,
       author = {{Kovner}, Israel},
        title = "{The Thick Gravitational Lens: A Lens Composed of Many Elements at Different Distances}",
      journal = {\apj},
         year = 1987,
        month = may,
       volume = {316},
        pages = {52},
          doi = {10.1086/165179},
       adsurl = {https://ui.adsabs.harvard.edu/abs/1987ApJ...316...52K}
}

@ARTICLE{2005MNRAS.358...39Y,
       author = {{Yoshida}, Hiroshi and {Nakamura}, Kouji and {Omote}, Minoru},
        title = "{The continuous limit of the multiple lens effect and the optical scalar equation}",
      journal = {\mnras},
         year = 2005,
        month = mar,
       volume = {358},
       number = {1},
        pages = {39-48},
          doi = {10.1111/j.1365-2966.2005.08698.x},
archivePrefix = {arXiv},
       eprint = {astro-ph/0407040},
 primaryClass = {astro-ph},
       adsurl = {https://ui.adsabs.harvard.edu/abs/2005MNRAS.358...39Y}
}

@ARTICLE{1991A&A...248..349B,
       author = {{Bartelmann}, M. and {Schneider}, P.},
        title = "{Gravitational lensing by large-scale structures}",
      journal = {\aap},
         year = 1991,
        month = aug,
       volume = {248},
       number = {2},
        pages = {349-353},
       adsurl = {https://ui.adsabs.harvard.edu/abs/1991A&A...248..349B}
}

@ARTICLE{blandfordnarayan,
       author = {{Blandford}, Roger and {Narayan}, Ramesh},
        title = "{Fermat's Principle, Caustics, and the Classification of Gravitational Lens Images}",
      journal = {\apj},
         year = 1986,
        month = nov,
       volume = {310},
        pages = {568},
          doi = {10.1086/164709},
       adsurl = {https://ui.adsabs.harvard.edu/abs/1986ApJ...310..568B}
}

@ARTICLE{hooperDM,
       author = {{Bertone}, Gianfranco and {Hooper}, Dan},
        title = "{History of dark matter}",
      journal = {Reviews of Modern Physics},
         year = 2018,
        month = oct,
       volume = {90},
       number = {4},
          eid = {045002},
        pages = {045002},
          doi = {10.1103/RevModPhys.90.045002},
archivePrefix = {arXiv},
       eprint = {1605.04909},
 primaryClass = {astro-ph.CO},
       adsurl = {https://ui.adsabs.harvard.edu/abs/2018RvMP...90d5002B}
}

@ARTICLE{swart,
       author = {{de Swart}, J.~G. and {Bertone}, G. and {van Dongen}, J.},
        title = "{How dark matter came to matter}",
      journal = {Nature Astronomy},
         year = 2017,
        month = mar,
       volume = {1},
          eid = {0059},
        pages = {0059},
          doi = {10.1038/s41550-017-0059},
archivePrefix = {arXiv},
       eprint = {1703.00013},
 primaryClass = {astro-ph.CO},
       adsurl = {https://ui.adsabs.harvard.edu/abs/2017NatAs...1E..59D}
}

@ARTICLE{zwickycoma,
       author = {{Zwicky}, F.},
        title = "{Die Rotverschiebung von extragalaktischen Nebeln}",
      journal = {Helvetica Physica Acta},
         year = 1933,
        month = jan,
       volume = {6},
        pages = {110-127},
       adsurl = {https://ui.adsabs.harvard.edu/abs/1933AcHPh...6..110Z}
}

@ARTICLE{smithcoma,
       author = {{Smith}, Sinclair},
        title = "{The Mass of the Virgo Cluster}",
      journal = {\apj},
         year = 1936,
        month = jan,
       volume = {83},
        pages = {23},
          doi = {10.1086/143697},
       adsurl = {https://ui.adsabs.harvard.edu/abs/1936ApJ....83...23S}
}

@ARTICLE{rubinford,
       author = {{Rubin}, Vera C. and {Ford}, Jr., W. Kent},
        title = "{Rotation of the Andromeda Nebula from a Spectroscopic Survey of Emission Regions}",
      journal = {\apj},
         year = 1970,
        month = feb,
       volume = {159},
        pages = {379},
          doi = {10.1086/150317},
       adsurl = {https://ui.adsabs.harvard.edu/abs/1970ApJ...159..379R}
}

@ARTICLE{21cmDM,
       author = {{Rogstad}, D.~H. and {Shostak}, G.~S. and {Rots}, A.~H.},
        title = "{Aperture synthesis study of neutral hydrogen in the galaxies NGC 6946 and IC 342.}",
      journal = {\aap},
         year = 1973,
        month = jan,
       volume = {22},
        pages = {111-119},
       adsurl = {https://ui.adsabs.harvard.edu/abs/1973A&A....22..111R}
}

@MISC{baumann,
       author = {{Baumann}, Daniel},
        title = "{Cosmology: Part III Mathematical Tripos}",
        year = 2018,
        howpublished = {Lecture Notes, Cambridge University, Cambridge, U.K.},
}

@ARTICLE{1984ApJ...287..538G,
       author = {{Gorenstein}, M.~V. and {Shapiro}, I.~I. and {Rogers}, A.~E.~E. and {Cohen}, N.~L. and {Corey}, B.~E. and {Porcas}, R.~W. and {Falco}, E.~E. and {Bonometti}, R.~J. and {Preston}, R.~A. and {Rius}, A. and {Whitney}, A.~R.},
        title = "{The milli-arcsecond images of Q0957+56.1.}",
      journal = {\apj},
         year = 1984,
        month = dec,
       volume = {287},
        pages = {538-548},
          doi = {10.1086/162712},
       adsurl = {https://ui.adsabs.harvard.edu/abs/1984ApJ...287..538G}
}

@INCOLLECTION{trimbleDM,
       author = {{Trimble}, Virginia},
        title = "{History of Dark Matter in Galaxies}",
    booktitle = {Planets, Stars and Stellar Systems. Volume 5: Galactic Structure and Stellar Populations},
         year = 2013,
         publisher={Springer Science+Business Media},
       editor = {{Oswalt}, Terry D. and {Gilmore}, Gerard},
       volume = {5},
        pages = {1091},
          doi = {10.1007/978-94-007-5612-0_21},
       adsurl = {https://ui.adsabs.harvard.edu/abs/2013pss5.book.1091T}
}

@ARTICLE{RIME,
       author = {{Smirnov}, O.~M.},
        title = "{Revisiting the radio interferometer measurement equation. I. A full-sky Jones formalism}",
      journal = {\aap},
         year = 2011,
        month = mar,
       volume = {527},
          eid = {A106},
        pages = {A106},
          doi = {10.1051/0004-6361/201016082},
archivePrefix = {arXiv},
       eprint = {1101.1764},
 primaryClass = {astro-ph.IM},
       adsurl = {https://ui.adsabs.harvard.edu/abs/2011A&A...527A.106S}
}

@BOOK{thompsonbook,
       author = {{Thompson}, A. Richard and {Moran}, James M. and {Swenson}, George W.},
        title = "{Interferometry and synthesis in radio astronomy}",
         year = 1986,
         publisher = {Wiley-Interscience},
       adsurl = {https://ui.adsabs.harvard.edu/abs/1986isra.book.....T}
}

@INPROCEEDINGS{selfcalcornwell,
       author = {{Cornwell}, Tim and {Fomalont}, Edward B.},
        title = "{Self-Calibration}",
    booktitle = {Synthesis Imaging in Radio Astronomy},
         year = 1989,
       editor = {{Perley}, Richard A. and {Schwab}, Frederic R. and {Bridle}, Alan H.},
       series = {Astronomical Society of the Pacific Conference Series},
       volume = {6},
        month = jan,
        pages = {185},
       adsurl = {https://ui.adsabs.harvard.edu/abs/1989ASPC....6..185C}
}

@ARTICLE{facet,
       author = {{Cornwell}, T.~J. and {Perley}, R.~A.},
        title = "{Radio-interferometric imaging of very large fields. The problem of non-coplanar arrays.}",
      journal = {\aap},
         year = 1992,
        month = jul,
       volume = {261},
        pages = {353-364},
       adsurl = {https://ui.adsabs.harvard.edu/abs/1992A&A...261..353C}
}

@ARTICLE{wproj,
       author = {{Cornwell}, T.~J. and {Golap}, K. and {Bhatnagar}, S.},
        title = "{The Noncoplanar Baselines Effect in Radio Interferometry: The W-Projection Algorithm}",
      journal = {IEEE Journal of Selected Topics in Signal Processing},
         year = 2008,
        month = nov,
       volume = {2},
       number = {5},
        pages = {647-657},
          doi = {10.1109/JSTSP.2008.2005290},
archivePrefix = {arXiv},
       eprint = {0807.4161},
 primaryClass = {astro-ph},
       adsurl = {https://ui.adsabs.harvard.edu/abs/2008ISTSP...2..647C}
}

@ARTICLE{JB07,
       author = {{Jackson}, N. and {Browne}, I.~W.~A.},
        title = "{Improving efficiency in radio surveys for gravitational lenses}",
      journal = {\mnras},
         year = 2007,
        month = jan,
       volume = {374},
       number = {1},
        pages = {168-175},
          doi = {10.1111/j.1365-2966.2006.11126.x},
archivePrefix = {arXiv},
       eprint = {astro-ph/0609818},
 primaryClass = {astro-ph},
       adsurl = {https://ui.adsabs.harvard.edu/abs/2007MNRAS.374..168J}
}

@ARTICLE{Lemon19,
       author = {{Lemon}, Cameron A. and {Auger}, Matthew W. and {McMahon}, Richard G.},
        title = "{Gravitationally lensed quasars in Gaia - III. 22 new lensed quasars from Gaia data release 2}",
      journal = {\mnras},
         year = 2019,
        month = mar,
       volume = {483},
       number = {3},
        pages = {4242-4258},
          doi = {10.1093/mnras/sty3366},
archivePrefix = {arXiv},
       eprint = {1810.04480},
 primaryClass = {astro-ph.GA},
       adsurl = {https://ui.adsabs.harvard.edu/abs/2019MNRAS.483.4242L}
}

@ARTICLE{hamburg,
       author = {{Hagen}, H. -J. and {Engels}, D. and {Reimers}, D.},
        title = "{The Hamburg Quasar Survey. III. Further new bright quasars}",
      journal = {\aaps},
         year = 1999,
        month = feb,
       volume = {134},
        pages = {483-487},
          doi = {10.1051/aas:1999442},
       adsurl = {https://ui.adsabs.harvard.edu/abs/1999A&AS..134..483H}
}

@ARTICLE{Gordon2021,
       author = {{Gordon}, Yjan A. and {Boyce}, Michelle M. and {O'Dea}, Christopher P. and {Rudnick}, Lawrence and {Andernach}, Heinz and {Vantyghem}, Adrian N. and {Baum}, Stefi A. and {Bui}, Jean-Paul and {Dionyssiou}, Mathew and {Safi-Harb}, Samar and {Sander}, Isabel},
        title = "{A Quick Look at the 3 GHz Radio Sky. I. Source Statistics from the Very Large Array Sky Survey}",
      journal = {\apjs},
         year = 2021,
        month = aug,
       volume = {255},
       number = {2},
          eid = {30},
        pages = {30},
          doi = {10.3847/1538-4365/ac05c0},
archivePrefix = {arXiv},
       eprint = {2102.11753},
 primaryClass = {astro-ph.GA},
       adsurl = {https://ui.adsabs.harvard.edu/abs/2021ApJS..255...30G}
}

@ARTICLE{Dey2019,
       author = {{Dey}, Arjun and {Schlegel}, David J. and {Lang}, Dustin and {Blum}, Robert and {Burleigh}, Kaylan and {Fan}, Xiaohui and {Findlay}, Joseph R. and {Finkbeiner}, Doug and {Herrera}, David and {Juneau}, St{\'e}phanie and {Landriau}, Martin and {Levi}, Michael and {McGreer}, Ian and {Meisner}, Aaron and {Myers}, Adam D. and {Moustakas}, John and {Nugent}, Peter and {Patej}, Anna and {Schlafly}, Edward F. and {Walker}, Alistair R. and {Valdes}, Francisco and {Weaver}, Benjamin A. and {Y{\`e}che}, Christophe and {Zou}, Hu and {Zhou}, Xu and {Abareshi}, Behzad and {Abbott}, T.~M.~C. and {Abolfathi}, Bela and {Aguilera}, C. and {Alam}, Shadab and {Allen}, Lori and {Alvarez}, A. and {Annis}, James and {Ansarinejad}, Behzad and {Aubert}, Marie and {Beechert}, Jacqueline and {Bell}, Eric F. and {BenZvi}, Segev Y. and {Beutler}, Florian and {Bielby}, Richard M. and {Bolton}, Adam S. and {Brice{\~n}o}, C{\'e}sar and {Buckley-Geer}, Elizabeth J. and {Butler}, Karen and {Calamida}, Annalisa and {Carlberg}, Raymond G. and {Carter}, Paul and {Casas}, Ricard and {Castander}, Francisco J. and {Choi}, Yumi and {Comparat}, Johan and {Cukanovaite}, Elena and {Delubac}, Timoth{\'e}e and {DeVries}, Kaitlin and {Dey}, Sharmila and {Dhungana}, Govinda and {Dickinson}, Mark and {Ding}, Zhejie and {Donaldson}, John B. and {Duan}, Yutong and {Duckworth}, Christopher J. and {Eftekharzadeh}, Sarah and {Eisenstein}, Daniel J. and {Etourneau}, Thomas and {Fagrelius}, Parker A. and {Farihi}, Jay and {Fitzpatrick}, Mike and {Font-Ribera}, Andreu and {Fulmer}, Leah and {G{\"a}nsicke}, Boris T. and {Gaztanaga}, Enrique and {George}, Koshy and {Gerdes}, David W. and {Gontcho}, Satya Gontcho A. and {Gorgoni}, Claudio and {Green}, Gregory and {Guy}, Julien and {Harmer}, Diane and {Hernandez}, M. and {Honscheid}, Klaus and {Huang}, Lijuan Wendy and {James}, David J. and {Jannuzi}, Buell T. and {Jiang}, Linhua and {Joyce}, Richard and {Karcher}, Armin and {Karkar}, Sonia and {Kehoe}, Robert and {Kneib}, Jean-Paul and {Kueter-Young}, Andrea and {Lan}, Ting-Wen and {Lauer}, Tod R. and {Le Guillou}, Laurent and {Le Van Suu}, Auguste and {Lee}, Jae Hyeon and {Lesser}, Michael and {Perreault Levasseur}, Laurence and {Li}, Ting S. and {Mann}, Justin L. and {Marshall}, Robert and {Mart{\'\i}nez-V{\'a}zquez}, C.~E. and {Martini}, Paul and {du Mas des Bourboux}, H{\'e}lion and {McManus}, Sean and {Meier}, Tobias Gabriel and {M{\'e}nard}, Brice and {Metcalfe}, Nigel and {Mu{\~n}oz-Guti{\'e}rrez}, Andrea and {Najita}, Joan and {Napier}, Kevin and {Narayan}, Gautham and {Newman}, Jeffrey A. and {Nie}, Jundan and {Nord}, Brian and {Norman}, Dara J. and {Olsen}, Knut A.~G. and {Paat}, Anthony and {Palanque-Delabrouille}, Nathalie and {Peng}, Xiyan and {Poppett}, Claire L. and {Poremba}, Megan R. and {Prakash}, Abhishek and {Rabinowitz}, David and {Raichoor}, Anand and {Rezaie}, Mehdi and {Robertson}, A.~N. and {Roe}, Natalie A. and {Ross}, Ashley J. and {Ross}, Nicholas P. and {Rudnick}, Gregory and {Safonova}, Sasha and {Saha}, Abhijit and {S{\'a}nchez}, F. Javier and {Savary}, Elodie and {Schweiker}, Heidi and {Scott}, Adam and {Seo}, Hee-Jong and {Shan}, Huanyuan and {Silva}, David R. and {Slepian}, Zachary and {Soto}, Christian and {Sprayberry}, David and {Staten}, Ryan and {Stillman}, Coley M. and {Stupak}, Robert J. and {Summers}, David L. and {Sien Tie}, Suk and {Tirado}, H. and {Vargas-Maga{\~n}a}, Mariana and {Vivas}, A. Katherina and {Wechsler}, Risa H. and {Williams}, Doug and {Yang}, Jinyi and {Yang}, Qian and {Yapici}, Tolga and {Zaritsky}, Dennis and {Zenteno}, A. and {Zhang}, Kai and {Zhang}, Tianmeng and {Zhou}, Rongpu and {Zhou}, Zhimin},
        title = "{Overview of the DESI Legacy Imaging Surveys}",
      journal = {\aj},
         year = 2019,
        month = may,
       volume = {157},
       number = {5},
          eid = {168},
        pages = {168},
          doi = {10.3847/1538-3881/ab089d},
archivePrefix = {arXiv},
       eprint = {1804.08657},
 primaryClass = {astro-ph.IM},
       adsurl = {https://ui.adsabs.harvard.edu/abs/2019AJ....157..168D}
}

@ARTICLE{Gaia2016,
       author = {{Gaia Collaboration} and {Prusti}, T. and {de Bruijne}, J.~H.~J. and {Brown}, A.~G.~A. and {Vallenari}, A. and {Babusiaux}, C. and {Bailer-Jones}, C.~A.~L. and {Bastian}, U. and {Biermann}, M. and {Evans}, D.~W. and {Eyer}, L. and {Jansen}, F. and {Jordi}, C. and {Klioner}, S.~A. and {Lammers}, U. and {Lindegren}, L. and {Luri}, X. and {Mignard}, F. and {Milligan}, D.~J. and {Panem}, C. and {Poinsignon}, V. and {Pourbaix}, D. and {Randich}, S. and {Sarri}, G. and {Sartoretti}, P. and {Siddiqui}, H.~I. and {Soubiran}, C. and {Valette}, V. and {van Leeuwen}, F. and {Walton}, N.~A. and {Aerts}, C. and {Arenou}, F. and {Cropper}, M. and {Drimmel}, R. and {H{\o}g}, E. and {Katz}, D. and {Lattanzi}, M.~G. and {O'Mullane}, W. and {Grebel}, E.~K. and {Holland}, A.~D. and {Huc}, C. and {Passot}, X. and {Bramante}, L. and {Cacciari}, C. and {Casta{\~n}eda}, J. and {Chaoul}, L. and {Cheek}, N. and {De Angeli}, F. and {Fabricius}, C. and {Guerra}, R. and {Hern{\'a}ndez}, J. and {Jean-Antoine-Piccolo}, A. and {Masana}, E. and {Messineo}, R. and {Mowlavi}, N. and {Nienartowicz}, K. and {Ord{\'o}{\~n}ez-Blanco}, D. and {Panuzzo}, P. and {Portell}, J. and {Richards}, P.~J. and {Riello}, M. and {Seabroke}, G.~M. and {Tanga}, P. and {Th{\'e}venin}, F. and {Torra}, J. and {Els}, S.~G. and {Gracia-Abril}, G. and {Comoretto}, G. and {Garcia-Reinaldos}, M. and {Lock}, T. and {Mercier}, E. and {Altmann}, M. and {Andrae}, R. and {Astraatmadja}, T.~L. and {Bellas-Velidis}, I. and {Benson}, K. and {Berthier}, J. and {Blomme}, R. and {Busso}, G. and {Carry}, B. and {Cellino}, A. and {Clementini}, G. and {Cowell}, S. and {Creevey}, O. and {Cuypers}, J. and {Davidson}, M. and {De Ridder}, J. and {de Torres}, A. and {Delchambre}, L. and {Dell'Oro}, A. and {Ducourant}, C. and {Fr{\'e}mat}, Y. and {Garc{\'\i}a-Torres}, M. and {Gosset}, E. and {Halbwachs}, J. -L. and {Hambly}, N.~C. and {Harrison}, D.~L. and {Hauser}, M. and {Hestroffer}, D. and {Hodgkin}, S.~T. and {Huckle}, H.~E. and {Hutton}, A. and {Jasniewicz}, G. and {Jordan}, S. and {Kontizas}, M. and {Korn}, A.~J. and {Lanzafame}, A.~C. and {Manteiga}, M. and {Moitinho}, A. and {Muinonen}, K. and {Osinde}, J. and {Pancino}, E. and {Pauwels}, T. and {Petit}, J. -M. and {Recio-Blanco}, A. and {Robin}, A.~C. and {Sarro}, L.~M. and {Siopis}, C. and {Smith}, M. and {Smith}, K.~W. and {Sozzetti}, A. and {Thuillot}, W. and {van Reeven}, W. and {Viala}, Y. and {Abbas}, U. and {Abreu Aramburu}, A. and {Accart}, S. and {Aguado}, J.~J. and {Allan}, P.~M. and {Allasia}, W. and {Altavilla}, G. and {{\'A}lvarez}, M.~A. and {Alves}, J. and {Anderson}, R.~I. and {Andrei}, A.~H. and {Anglada Varela}, E. and {Antiche}, E. and {Antoja}, T. and {Ant{\'o}n}, S. and {Arcay}, B. and {Atzei}, A. and {Ayache}, L. and {Bach}, N. and {Baker}, S.~G. and {Balaguer-N{\'u}{\~n}ez}, L. and {Barache}, C. and {Barata}, C. and {Barbier}, A. and {Barblan}, F. and {Baroni}, M. and {Barrado y Navascu{\'e}s}, D. and {Barros}, M. and {Barstow}, M.~A. and {Becciani}, U. and {Bellazzini}, M. and {Bellei}, G. and {Bello Garc{\'\i}a}, A. and {Belokurov}, V. and {Bendjoya}, P. and {Berihuete}, A. and {Bianchi}, L. and {Bienaym{\'e}}, O. and {Billebaud}, F. and {Blagorodnova}, N. and {Blanco-Cuaresma}, S. and {Boch}, T. and {Bombrun}, A. and {Borrachero}, R. and {Bouquillon}, S. and {Bourda}, G. and {Bouy}, H. and {Bragaglia}, A. and {Breddels}, M.~A. and {Brouillet}, N. and {Br{\"u}semeister}, T. and {Bucciarelli}, B. and {Budnik}, F. and {Burgess}, P. and {Burgon}, R. and {Burlacu}, A. and {Busonero}, D. and {Buzzi}, R. and {Caffau}, E. and {Cambras}, J. and {Campbell}, H. and {Cancelliere}, R. and {Cantat-Gaudin}, T. and {Carlucci}, T. and {Carrasco}, J.~M. and {Castellani}, M. and {Charlot}, P. and {Charnas}, J. and {Charvet}, P. and {Chassat}, F. and {Chiavassa}, A. and {Clotet}, M. and {Cocozza}, G. and {Collins}, R.~S. and {Collins}, P. and {Costigan}, G. and {Crifo}, F. and {Cross}, N.~J.~G. and {Crosta}, M. and {Crowley}, C. and {Dafonte}, C. and {Damerdji}, Y. and {Dapergolas}, A. and {David}, P. and {David}, M. and {De Cat}, P. and {de Felice}, F. and {de Laverny}, P. and {De Luise}, F. and {De March}, R. and {de Martino}, D. and {de Souza}, R. and {Debosscher}, J. and {del Pozo}, E. and {Delbo}, M. and {Delgado}, A. and {Delgado}, H.~E. and {di Marco}, F. and {Di Matteo}, P. and {Diakite}, S. and {Distefano}, E. and {Dolding}, C. and {Dos Anjos}, S. and {Drazinos}, P. and {Dur{\'a}n}, J. and {Dzigan}, Y. and {Ecale}, E. and {Edvardsson}, B. and {Enke}, H. and {Erdmann}, M. and {Escolar}, D. and {Espina}, M. and {Evans}, N.~W. and {Eynard Bontemps}, G. and {Fabre}, C. and {Fabrizio}, M. and {Faigler}, S. and {Falc{\~a}o}, A.~J. and {Farr{\`a}s Casas}, M. and {Faye}, F. and {Federici}, L. and {Fedorets}, G. and {Fern{\'a}ndez-Hern{\'a}ndez}, J. and {Fernique}, P. and {Fienga}, A. and {Figueras}, F. and {Filippi}, F. and {Findeisen}, K. and {Fonti}, A. and {Fouesneau}, M. and {Fraile}, E. and {Fraser}, M. and {Fuchs}, J. and {Furnell}, R. and {Gai}, M. and {Galleti}, S. and {Galluccio}, L. and {Garabato}, D. and {Garc{\'\i}a-Sedano}, F. and {Gar{\'e}}, P. and {Garofalo}, A. and {Garralda}, N. and {Gavras}, P. and {Gerssen}, J. and {Geyer}, R. and {Gilmore}, G. and {Girona}, S. and {Giuffrida}, G. and {Gomes}, M. and {Gonz{\'a}lez-Marcos}, A. and {Gonz{\'a}lez-N{\'u}{\~n}ez}, J. and {Gonz{\'a}lez-Vidal}, J.~J. and {Granvik}, M. and {Guerrier}, A. and {Guillout}, P. and {Guiraud}, J. and {G{\'u}rpide}, A. and {Guti{\'e}rrez-S{\'a}nchez}, R. and {Guy}, L.~P. and {Haigron}, R. and {Hatzidimitriou}, D. and {Haywood}, M. and {Heiter}, U. and {Helmi}, A. and {Hobbs}, D. and {Hofmann}, W. and {Holl}, B. and {Holland}, G. and {Hunt}, J.~A.~S. and {Hypki}, A. and {Icardi}, V. and {Irwin}, M. and {Jevardat de Fombelle}, G. and {Jofr{\'e}}, P. and {Jonker}, P.~G. and {Jorissen}, A. and {Julbe}, F. and {Karampelas}, A. and {Kochoska}, A. and {Kohley}, R. and {Kolenberg}, K. and {Kontizas}, E. and {Koposov}, S.~E. and {Kordopatis}, G. and {Koubsky}, P. and {Kowalczyk}, A. and {Krone-Martins}, A. and {Kudryashova}, M. and {Kull}, I. and {Bachchan}, R.~K. and {Lacoste-Seris}, F. and {Lanza}, A.~F. and {Lavigne}, J. -B. and {Le Poncin-Lafitte}, C. and {Lebreton}, Y. and {Lebzelter}, T. and {Leccia}, S. and {Leclerc}, N. and {Lecoeur-Taibi}, I. and {Lemaitre}, V. and {Lenhardt}, H. and {Leroux}, F. and {Liao}, S. and {Licata}, E. and {Lindstr{\o}m}, H.~E.~P. and {Lister}, T.~A. and {Livanou}, E. and {Lobel}, A. and {L{\"o}ffler}, W. and {L{\'o}pez}, M. and {Lopez-Lozano}, A. and {Lorenz}, D. and {Loureiro}, T. and {MacDonald}, I. and {Magalh{\~a}es Fernandes}, T. and {Managau}, S. and {Mann}, R.~G. and {Mantelet}, G. and {Marchal}, O. and {Marchant}, J.~M. and {Marconi}, M. and {Marie}, J. and {Marinoni}, S. and {Marrese}, P.~M. and {Marschalk{\'o}}, G. and {Marshall}, D.~J. and {Mart{\'\i}n-Fleitas}, J.~M. and {Martino}, M. and {Mary}, N. and {Matijevi{\v{c}}}, G. and {Mazeh}, T. and {McMillan}, P.~J. and {Messina}, S. and {Mestre}, A. and {Michalik}, D. and {Millar}, N.~R. and {Miranda}, B.~M.~H. and {Molina}, D. and {Molinaro}, R. and {Molinaro}, M. and {Moln{\'a}r}, L. and {Moniez}, M. and {Montegriffo}, P. and {Monteiro}, D. and {Mor}, R. and {Mora}, A. and {Morbidelli}, R. and {Morel}, T. and {Morgenthaler}, S. and {Morley}, T. and {Morris}, D. and {Mulone}, A.~F. and {Muraveva}, T. and {Musella}, I. and {Narbonne}, J. and {Nelemans}, G. and {Nicastro}, L. and {Noval}, L. and {Ord{\'e}novic}, C. and {Ordieres-Mer{\'e}}, J. and {Osborne}, P. and {Pagani}, C. and {Pagano}, I. and {Pailler}, F. and {Palacin}, H. and {Palaversa}, L. and {Parsons}, P. and {Paulsen}, T. and {Pecoraro}, M. and {Pedrosa}, R. and {Pentik{\"a}inen}, H. and {Pereira}, J. and {Pichon}, B. and {Piersimoni}, A.~M. and {Pineau}, F. -X. and {Plachy}, E. and {Plum}, G. and {Poujoulet}, E. and {Pr{\v{s}}a}, A. and {Pulone}, L. and {Ragaini}, S. and {Rago}, S. and {Rambaux}, N. and {Ramos-Lerate}, M. and {Ranalli}, P. and {Rauw}, G. and {Read}, A. and {Regibo}, S. and {Renk}, F. and {Reyl{\'e}}, C. and {Ribeiro}, R.~A. and {Rimoldini}, L. and {Ripepi}, V. and {Riva}, A. and {Rixon}, G. and {Roelens}, M. and {Romero-G{\'o}mez}, M. and {Rowell}, N. and {Royer}, F. and {Rudolph}, A. and {Ruiz-Dern}, L. and {Sadowski}, G. and {Sagrist{\`a} Sell{\'e}s}, T. and {Sahlmann}, J. and {Salgado}, J. and {Salguero}, E. and {Sarasso}, M. and {Savietto}, H. and {Schnorhk}, A. and {Schultheis}, M. and {Sciacca}, E. and {Segol}, M. and {Segovia}, J.~C. and {Segransan}, D. and {Serpell}, E. and {Shih}, I. -C. and {Smareglia}, R. and {Smart}, R.~L. and {Smith}, C. and {Solano}, E. and {Solitro}, F. and {Sordo}, R. and {Soria Nieto}, S. and {Souchay}, J. and {Spagna}, A. and {Spoto}, F. and {Stampa}, U. and {Steele}, I.~A. and {Steidelm{\"u}ller}, H. and {Stephenson}, C.~A. and {Stoev}, H. and {Suess}, F.~F. and {S{\"u}veges}, M. and {Surdej}, J. and {Szabados}, L. and {Szegedi-Elek}, E. and {Tapiador}, D. and {Taris}, F. and {Tauran}, G. and {Taylor}, M.~B. and {Teixeira}, R. and {Terrett}, D. and {Tingley}, B. and {Trager}, S.~C. and {Turon}, C. and {Ulla}, A. and {Utrilla}, E. and {Valentini}, G. and {van Elteren}, A. and {Van Hemelryck}, E. and {van Leeuwen}, M. and {Varadi}, M. and {Vecchiato}, A. and {Veljanoski}, J. and {Via}, T. and {Vicente}, D. and {Vogt}, S. and {Voss}, H. and {Votruba}, V. and {Voutsinas}, S. and {Walmsley}, G. and {Weiler}, M. and {Weingrill}, K. and {Werner}, D. and {Wevers}, T. and {Whitehead}, G. and {Wyrzykowski}, {\L}. and {Yoldas}, A. and {{\v{Z}}erjal}, M. and {Zucker}, S. and {Zurbach}, C. and {Zwitter}, T. and {Alecu}, A. and {Allen}, M. and {Allende Prieto}, C. and {Amorim}, A. and {Anglada-Escud{\'e}}, G. and {Arsenijevic}, V. and {Azaz}, S. and {Balm}, P. and {Beck}, M. and {Bernstein}, H. -H. and {Bigot}, L. and {Bijaoui}, A. and {Blasco}, C. and {Bonfigli}, M. and {Bono}, G. and {Boudreault}, S. and {Bressan}, A. and {Brown}, S. and {Brunet}, P. -M. and {Bunclark}, P. and {Buonanno}, R. and {Butkevich}, A.~G. and {Carret}, C. and {Carrion}, C. and {Chemin}, L. and {Ch{\'e}reau}, F. and {Corcione}, L. and {Darmigny}, E. and {de Boer}, K.~S. and {de Teodoro}, P. and {de Zeeuw}, P.~T. and {Delle Luche}, C. and {Domingues}, C.~D. and {Dubath}, P. and {Fodor}, F. and {Fr{\'e}zouls}, B. and {Fries}, A. and {Fustes}, D. and {Fyfe}, D. and {Gallardo}, E. and {Gallegos}, J. and {Gardiol}, D. and {Gebran}, M. and {Gomboc}, A. and {G{\'o}mez}, A. and {Grux}, E. and {Gueguen}, A. and {Heyrovsky}, A. and {Hoar}, J. and {Iannicola}, G. and {Isasi Parache}, Y. and {Janotto}, A. -M. and {Joliet}, E. and {Jonckheere}, A. and {Keil}, R. and {Kim}, D. -W. and {Klagyivik}, P. and {Klar}, J. and {Knude}, J. and {Kochukhov}, O. and {Kolka}, I. and {Kos}, J. and {Kutka}, A. and {Lainey}, V. and {LeBouquin}, D. and {Liu}, C. and {Loreggia}, D. and {Makarov}, V.~V. and {Marseille}, M.~G. and {Martayan}, C. and {Martinez-Rubi}, O. and {Massart}, B. and {Meynadier}, F. and {Mignot}, S. and {Munari}, U. and {Nguyen}, A. -T. and {Nordlander}, T. and {Ocvirk}, P. and {O'Flaherty}, K.~S. and {Olias Sanz}, A. and {Ortiz}, P. and {Osorio}, J. and {Oszkiewicz}, D. and {Ouzounis}, A. and {Palmer}, M. and {Park}, P. and {Pasquato}, E. and {Peltzer}, C. and {Peralta}, J. and {P{\'e}turaud}, F. and {Pieniluoma}, T. and {Pigozzi}, E. and {Poels}, J. and {Prat}, G. and {Prod'homme}, T. and {Raison}, F. and {Rebordao}, J.~M. and {Risquez}, D. and {Rocca-Volmerange}, B. and {Rosen}, S. and {Ruiz-Fuertes}, M.~I. and {Russo}, F. and {Sembay}, S. and {Serraller Vizcaino}, I. and {Short}, A. and {Siebert}, A. and {Silva}, H. and {Sinachopoulos}, D. and {Slezak}, E. and {Soffel}, M. and {Sosnowska}, D. and {Strai{\v{z}}ys}, V. and {ter Linden}, M. and {Terrell}, D. and {Theil}, S. and {Tiede}, C. and {Troisi}, L. and {Tsalmantza}, P. and {Tur}, D. and {Vaccari}, M. and {Vachier}, F. and {Valles}, P. and {Van Hamme}, W. and {Veltz}, L. and {Virtanen}, J. and {Wallut}, J. -M. and {Wichmann}, R. and {Wilkinson}, M.~I. and {Ziaeepour}, H. and {Zschocke}, S.},
        title = "{The Gaia mission}",
      journal = {\aap},
         year = 2016,
        month = nov,
       volume = {595},
          eid = {A1},
        pages = {A1},
          doi = {10.1051/0004-6361/201629272},
archivePrefix = {arXiv},
       eprint = {1609.04153},
 primaryClass = {astro-ph.IM},
       adsurl = {https://ui.adsabs.harvard.edu/abs/2016A&A...595A...1G}
}

@ARTICLE{DES2016,
       author = {{Dark Energy Survey Collaboration} and {Abbott}, T. and {Abdalla}, F.~B. and {Aleksi{\'c}}, J. and {Allam}, S. and {Amara}, A. and {Bacon}, D. and {Balbinot}, E. and {Banerji}, M. and {Bechtol}, K. and {Benoit-L{\'e}vy}, A. and {Bernstein}, G.~M. and {Bertin}, E. and {Blazek}, J. and {Bonnett}, C. and {Bridle}, S. and {Brooks}, D. and {Brunner}, R.~J. and {Buckley-Geer}, E. and {Burke}, D.~L. and {Caminha}, G.~B. and {Capozzi}, D. and {Carlsen}, J. and {Carnero-Rosell}, A. and {Carollo}, M. and {Carrasco-Kind}, M. and {Carretero}, J. and {Castander}, F.~J. and {Clerkin}, L. and {Collett}, T. and {Conselice}, C. and {Crocce}, M. and {Cunha}, C.~E. and {D'Andrea}, C.~B. and {da Costa}, L.~N. and {Davis}, T.~M. and {Desai}, S. and {Diehl}, H.~T. and {Dietrich}, J.~P. and {Dodelson}, S. and {Doel}, P. and {Drlica-Wagner}, A. and {Estrada}, J. and {Etherington}, J. and {Evrard}, A.~E. and {Fabbri}, J. and {Finley}, D.~A. and {Flaugher}, B. and {Foley}, R.~J. and {Fosalba}, P. and {Frieman}, J. and {Garc{\'\i}a-Bellido}, J. and {Gaztanaga}, E. and {Gerdes}, D.~W. and {Giannantonio}, T. and {Goldstein}, D.~A. and {Gruen}, D. and {Gruendl}, R.~A. and {Guarnieri}, P. and {Gutierrez}, G. and {Hartley}, W. and {Honscheid}, K. and {Jain}, B. and {James}, D.~J. and {Jeltema}, T. and {Jouvel}, S. and {Kessler}, R. and {King}, A. and {Kirk}, D. and {Kron}, R. and {Kuehn}, K. and {Kuropatkin}, N. and {Lahav}, O. and {Li}, T.~S. and {Lima}, M. and {Lin}, H. and {Maia}, M.~A.~G. and {Makler}, M. and {Manera}, M. and {Maraston}, C. and {Marshall}, J.~L. and {Martini}, P. and {McMahon}, R.~G. and {Melchior}, P. and {Merson}, A. and {Miller}, C.~J. and {Miquel}, R. and {Mohr}, J.~J. and {Morice-Atkinson}, X. and {Naidoo}, K. and {Neilsen}, E. and {Nichol}, R.~C. and {Nord}, B. and {Ogando}, R. and {Ostrovski}, F. and {Palmese}, A. and {Papadopoulos}, A. and {Peiris}, H.~V. and {Peoples}, J. and {Percival}, W.~J. and {Plazas}, A.~A. and {Reed}, S.~L. and {Refregier}, A. and {Romer}, A.~K. and {Roodman}, A. and {Ross}, A. and {Rozo}, E. and {Rykoff}, E.~S. and {Sadeh}, I. and {Sako}, M. and {S{\'a}nchez}, C. and {Sanchez}, E. and {Santiago}, B. and {Scarpine}, V. and {Schubnell}, M. and {Sevilla-Noarbe}, I. and {Sheldon}, E. and {Smith}, M. and {Smith}, R.~C. and {Soares-Santos}, M. and {Sobreira}, F. and {Soumagnac}, M. and {Suchyta}, E. and {Sullivan}, M. and {Swanson}, M. and {Tarle}, G. and {Thaler}, J. and {Thomas}, D. and {Thomas}, R.~C. and {Tucker}, D. and {Vieira}, J.~D. and {Vikram}, V. and {Walker}, A.~R. and {Wechsler}, R.~H. and {Weller}, J. and {Wester}, W. and {Whiteway}, L. and {Wilcox}, H. and {Yanny}, B. and {Zhang}, Y. and {Zuntz}, J.},
        title = "{The Dark Energy Survey: more than dark energy - an overview}",
      journal = {\mnras},
         year = 2016,
        month = aug,
       volume = {460},
       number = {2},
        pages = {1270-1299},
          doi = {10.1093/mnras/stw641},
archivePrefix = {arXiv},
       eprint = {1601.00329},
 primaryClass = {astro-ph.CO},
       adsurl = {https://ui.adsabs.harvard.edu/abs/2016MNRAS.460.1270D}
}

@ARTICLE{Bruzewski2021,
       author = {{Bruzewski}, S. and {Schinzel}, F.~K. and {Taylor}, G.~B. and {Petrov}, L.},
        title = "{Radio Counterpart Candidates to Unassociated 4FGL-DR2 Sources}",
      journal = {\apj},
         year = 2021,
        month = jun,
       volume = {914},
       number = {1},
          eid = {42},
        pages = {42},
          doi = {10.3847/1538-4357/abf73b},
archivePrefix = {arXiv},
       eprint = {2102.07397},
 primaryClass = {astro-ph.HE},
       adsurl = {https://ui.adsabs.harvard.edu/abs/2021ApJ...914...42B}
}

@ARTICLE{Lemon2017,
       author = {{Lemon}, Cameron A. and {Auger}, Matthew W. and {McMahon}, Richard G. and {Koposov}, Sergey E.},
        title = "{Gravitationally lensed quasars in Gaia: I. Resolving small-separation lenses}",
      journal = {\mnras},
         year = 2017,
        month = dec,
       volume = {472},
       number = {4},
        pages = {5023-5032},
          doi = {10.1093/mnras/stx2094},
archivePrefix = {arXiv},
       eprint = {1709.08976},
 primaryClass = {astro-ph.GA},
       adsurl = {https://ui.adsabs.harvard.edu/abs/2017MNRAS.472.5023L}
}

@ARTICLE{Lemon2018,
       author = {{Lemon}, Cameron A. and {Auger}, Matthew W. and {McMahon}, Richard G. and {Ostrovski}, Fernanda},
        title = "{Gravitationally lensed quasars in Gaia - II. Discovery of 24 lensed quasars}",
      journal = {\mnras},
         year = 2018,
        month = oct,
       volume = {479},
       number = {4},
        pages = {5060-5074},
          doi = {10.1093/mnras/sty911},
archivePrefix = {arXiv},
       eprint = {1803.07601},
 primaryClass = {astro-ph.GA},
       adsurl = {https://ui.adsabs.harvard.edu/abs/2018MNRAS.479.5060L}
}

@ARTICLE{Jacobs2019,
       author = {{Jacobs}, C. and {Collett}, T. and {Glazebrook}, K. and {Buckley-Geer}, E. and {Diehl}, H.~T. and {Lin}, H. and {McCarthy}, C. and {Qin}, A.~K. and {Odden}, C. and {Caso Escudero}, M. and {Dial}, P. and {Yung}, V.~J. and {Gaitsch}, S. and {Pellico}, A. and {Lindgren}, K.~A. and {Abbott}, T.~M.~C. and {Annis}, J. and {Avila}, S. and {Brooks}, D. and {Burke}, D.~L. and {Carnero Rosell}, A. and {Carrasco Kind}, M. and {Carretero}, J. and {da Costa}, L.~N. and {De Vicente}, J. and {Fosalba}, P. and {Frieman}, J. and {Garc{\'\i}a-Bellido}, J. and {Gaztanaga}, E. and {Goldstein}, D.~A. and {Gruen}, D. and {Gruendl}, R.~A. and {Gschwend}, J. and {Hollowood}, D.~L. and {Honscheid}, K. and {Hoyle}, B. and {James}, D.~J. and {Krause}, E. and {Kuropatkin}, N. and {Lahav}, O. and {Lima}, M. and {Maia}, M.~A.~G. and {Marshall}, J.~L. and {Miquel}, R. and {Plazas}, A.~A. and {Roodman}, A. and {Sanchez}, E. and {Scarpine}, V. and {Serrano}, S. and {Sevilla-Noarbe}, I. and {Smith}, M. and {Sobreira}, F. and {Suchyta}, E. and {Swanson}, M.~E.~C. and {Tarle}, G. and {Vikram}, V. and {Walker}, A.~R. and {Zhang}, Y. and {DES Collaboration}},
        title = "{An Extended Catalog of Galaxy-Galaxy Strong Gravitational Lenses Discovered in DES Using Convolutional Neural Networks}",
      journal = {\apjs},
         year = 2019,
        month = jul,
       volume = {243},
       number = {1},
          eid = {17},
        pages = {17},
          doi = {10.3847/1538-4365/ab26b6},
archivePrefix = {arXiv},
       eprint = {1905.10522},
 primaryClass = {astro-ph.GA},
       adsurl = {https://ui.adsabs.harvard.edu/abs/2019ApJS..243...17J}
}

@ARTICLE{Zhou2020,
       author = {{Zhou}, Rongpu and {Newman}, Jeffrey A. and {Dawson}, Kyle S. and {Eisenstein}, Daniel J. and {Brooks}, David D. and {Dey}, Arjun and {Dey}, Biprateep and {Duan}, Yutong and {Eftekharzadeh}, Sarah and {Gazta{\~n}aga}, Enrique and {Kehoe}, Robert and {Landriau}, Martin and {Levi}, Michael E. and {Licquia}, Timothy C. and {Meisner}, Aaron M. and {Moustakas}, John and {Myers}, Adam D. and {Palanque-Delabrouille}, Nathalie and {Poppett}, Claire and {Prada}, Francisco and {Raichoor}, Anand and {Schlegel}, David J. and {Schubnell}, Michael and {Staten}, Ryan and {Tarl{\'e}}, Gregory and {Y{\`e}che}, Christophe},
        title = "{Preliminary Target Selection for the DESI Luminous Red Galaxy (LRG) Sample}",
      journal = {RNAAS},
         year = 2020,
        month = oct,
       volume = {4},
       number = {10},
          eid = {181},
        pages = {181},
          doi = {10.3847/2515-5172/abc0f4},
archivePrefix = {arXiv},
       eprint = {2010.11282},
 primaryClass = {astro-ph.CO},
       adsurl = {https://ui.adsabs.harvard.edu/abs/2020RNAAS...4..181Z}
}

@ARTICLE{Condon1998,
       author = {{Condon}, J.~J. and {Cotton}, W.~D. and {Greisen}, E.~W. and {Yin}, Q.~F. and {Perley}, R.~A. and {Taylor}, G.~B. and {Broderick}, J.~J.},
        title = "{The NRAO VLA Sky Survey}",
      journal = {\aj},
         year = 1998,
        month = may,
       volume = {115},
       number = {5},
        pages = {1693-1716},
          doi = {10.1086/300337},
       adsurl = {https://ui.adsabs.harvard.edu/abs/1998AJ....115.1693C}
}

@ARTICLE{1817,
       author = {{Delchambre}, L. and {Krone-Martins}, A. and {Wertz}, O. and {Ducourant}, C. and {Galluccio}, L. and {Kl{\"u}ter}, J. and {Mignard}, F. and {Teixeira}, R. and {Djorgovski}, S.~G. and {Stern}, D. and {Graham}, M.~J. and {Surdej}, J. and {Bastian}, U. and {Wambsganss}, J. and {Le Campion}, J. -F. and {Slezak}, E.},
        title = "{Gaia GraL: Gaia DR2 Gravitational Lens Systems. III. A systematic blind search for new lensed systems}",
      journal = {\aap},
         year = 2019,
        month = feb,
       volume = {622},
          eid = {A165},
        pages = {A165},
          doi = {10.1051/0004-6361/201833802},
archivePrefix = {arXiv},
       eprint = {1807.02845},
 primaryClass = {astro-ph.CO},
       adsurl = {https://ui.adsabs.harvard.edu/abs/2019A&A...622A.165D}
}

@ARTICLE{2329disc,
       author = {{Schechter}, Paul L. and {Morgan}, Nicholas D. and {Chehade}, B. and {Metcalfe}, N. and {Shanks}, T. and {McDonald}, Michael},
        title = "{First Lensed Quasar Systems from the VST-ATLAS Survey: One Quad, Two Doubles, and Two Pairs of Lensless Twins}",
      journal = {\aj},
         year = 2017,
        month = may,
       volume = {153},
       number = {5},
          eid = {219},
        pages = {219},
          doi = {10.3847/1538-3881/aa6899},
archivePrefix = {arXiv},
       eprint = {1607.07476},
 primaryClass = {astro-ph.GA},
       adsurl = {https://ui.adsabs.harvard.edu/abs/2017AJ....153..219S}
}

@ARTICLE{Gaiadr3,
       author = {{Gaia Collaboration} and {Vallenari}, A. and {Brown}, A.~G.~A. and {Prusti}, T. and {de Bruijne}, J.~H.~J. and {Arenou}, F. and {Babusiaux}, C. and {Biermann}, M. and {Creevey}, O.~L. and {Ducourant}, C. and {Evans}, D.~W. and {Eyer}, L. and {Guerra}, R. and {Hutton}, A. and {Jordi}, C. and {Klioner}, S.~A. and {Lammers}, U.~L. and {Lindegren}, L. and {Luri}, X. and {Mignard}, F. and {Panem}, C. and {Pourbaix}, D. and {Randich}, S. and {Sartoretti}, P. and {Soubiran}, C. and {Tanga}, P. and {Walton}, N.~A. and {Bailer-Jones}, C.~A.~L. and {Bastian}, U. and {Drimmel}, R. and {Jansen}, F. and {Katz}, D. and {Lattanzi}, M.~G. and {van Leeuwen}, F. and {Bakker}, J. and {Cacciari}, C. and {Casta{\~n}eda}, J. and {De Angeli}, F. and {Fabricius}, C. and {Fouesneau}, M. and {Fr{\'e}mat}, Y. and {Galluccio}, L. and {Guerrier}, A. and {Heiter}, U. and {Masana}, E. and {Messineo}, R. and {Mowlavi}, N. and {Nicolas}, C. and {Nienartowicz}, K. and {Pailler}, F. and {Panuzzo}, P. and {Riclet}, F. and {Roux}, W. and {Seabroke}, G.~M. and {Sordo}, R. and {Th{\'e}venin}, F. and {Gracia-Abril}, G. and {Portell}, J. and {Teyssier}, D. and {Altmann}, M. and {Andrae}, R. and {Audard}, M. and {Bellas-Velidis}, I. and {Benson}, K. and {Berthier}, J. and {Blomme}, R. and {Burgess}, P.~W. and {Busonero}, D. and {Busso}, G. and {C{\'a}novas}, H. and {Carry}, B. and {Cellino}, A. and {Cheek}, N. and {Clementini}, G. and {Damerdji}, Y. and {Davidson}, M. and {de Teodoro}, P. and {Nu{\~n}ez Campos}, M. and {Delchambre}, L. and {Dell'Oro}, A. and {Esquej}, P. and {Fern{\'a}ndez-Hern{\'a}ndez}, J. and {Fraile}, E. and {Garabato}, D. and {Garc{\'\i}a-Lario}, P. and {Gosset}, E. and {Haigron}, R. and {Halbwachs}, J. -L. and {Hambly}, N.~C. and {Harrison}, D.~L. and {Hern{\'a}ndez}, J. and {Hestroffer}, D. and {Hodgkin}, S.~T. and {Holl}, B. and {Jan{\ss}en}, K. and {Jevardat de Fombelle}, G. and {Jordan}, S. and {Krone-Martins}, A. and {Lanzafame}, A.~C. and {L{\"o}ffler}, W. and {Marchal}, O. and {Marrese}, P.~M. and {Moitinho}, A. and {Muinonen}, K. and {Osborne}, P. and {Pancino}, E. and {Pauwels}, T. and {Recio-Blanco}, A. and {Reyl{\'e}}, C. and {Riello}, M. and {Rimoldini}, L. and {Roegiers}, T. and {Rybizki}, J. and {Sarro}, L.~M. and {Siopis}, C. and {Smith}, M. and {Sozzetti}, A. and {Utrilla}, E. and {van Leeuwen}, M. and {Abbas}, U. and {{\'A}brah{\'a}m}, P. and {Abreu Aramburu}, A. and {Aerts}, C. and {Aguado}, J.~J. and {Ajaj}, M. and {Aldea-Montero}, F. and {Altavilla}, G. and {{\'A}lvarez}, M.~A. and {Alves}, J. and {Anders}, F. and {Anderson}, R.~I. and {Anglada Varela}, E. and {Antoja}, T. and {Baines}, D. and {Baker}, S.~G. and {Balaguer-N{\'u}{\~n}ez}, L. and {Balbinot}, E. and {Balog}, Z. and {Barache}, C. and {Barbato}, D. and {Barros}, M. and {Barstow}, M.~A. and {Bartolom{\'e}}, S. and {Bassilana}, J. -L. and {Bauchet}, N. and {Becciani}, U. and {Bellazzini}, M. and {Berihuete}, A. and {Bernet}, M. and {Bertone}, S. and {Bianchi}, L. and {Binnenfeld}, A. and {Blanco-Cuaresma}, S. and {Blazere}, A. and {Boch}, T. and {Bombrun}, A. and {Bossini}, D. and {Bouquillon}, S. and {Bragaglia}, A. and {Bramante}, L. and {Breedt}, E. and {Bressan}, A. and {Brouillet}, N. and {Brugaletta}, E. and {Bucciarelli}, B. and {Burlacu}, A. and {Butkevich}, A.~G. and {Buzzi}, R. and {Caffau}, E. and {Cancelliere}, R. and {Cantat-Gaudin}, T. and {Carballo}, R. and {Carlucci}, T. and {Carnerero}, M.~I. and {Carrasco}, J.~M. and {Casamiquela}, L. and {Castellani}, M. and {Castro-Ginard}, A. and {Chaoul}, L. and {Charlot}, P. and {Chemin}, L. and {Chiaramida}, V. and {Chiavassa}, A. and {Chornay}, N. and {Comoretto}, G. and {Contursi}, G. and {Cooper}, W.~J. and {Cornez}, T. and {Cowell}, S. and {Crifo}, F. and {Cropper}, M. and {Crosta}, M. and {Crowley}, C. and {Dafonte}, C. and {Dapergolas}, A. and {David}, M. and {David}, P. and {de Laverny}, P. and {De Luise}, F. and {De March}, R. and {De Ridder}, J. and {de Souza}, R. and {de Torres}, A. and {del Peloso}, E.~F. and {del Pozo}, E. and {Delbo}, M. and {Delgado}, A. and {Delisle}, J. -B. and {Demouchy}, C. and {Dharmawardena}, T.~E. and {Di Matteo}, P. and {Diakite}, S. and {Diener}, C. and {Distefano}, E. and {Dolding}, C. and {Edvardsson}, B. and {Enke}, H. and {Fabre}, C. and {Fabrizio}, M. and {Faigler}, S. and {Fedorets}, G. and {Fernique}, P. and {Fienga}, A. and {Figueras}, F. and {Fournier}, Y. and {Fouron}, C. and {Fragkoudi}, F. and {Gai}, M. and {Garcia-Gutierrez}, A. and {Garcia-Reinaldos}, M. and {Garc{\'\i}a-Torres}, M. and {Garofalo}, A. and {Gavel}, A. and {Gavras}, P. and {Gerlach}, E. and {Geyer}, R. and {Giacobbe}, P. and {Gilmore}, G. and {Girona}, S. and {Giuffrida}, G. and {Gomel}, R. and {Gomez}, A. and {Gonz{\'a}lez-N{\'u}{\~n}ez}, J. and {Gonz{\'a}lez-Santamar{\'\i}a}, I. and {Gonz{\'a}lez-Vidal}, J.~J. and {Granvik}, M. and {Guillout}, P. and {Guiraud}, J. and {Guti{\'e}rrez-S{\'a}nchez}, R. and {Guy}, L.~P. and {Hatzidimitriou}, D. and {Hauser}, M. and {Haywood}, M. and {Helmer}, A. and {Helmi}, A. and {Sarmiento}, M.~H. and {Hidalgo}, S.~L. and {Hilger}, T. and {H{\l}adczuk}, N. and {Hobbs}, D. and {Holland}, G. and {Huckle}, H.~E. and {Jardine}, K. and {Jasniewicz}, G. and {Jean-Antoine Piccolo}, A. and {Jim{\'e}nez-Arranz}, {\'O}. and {Jorissen}, A. and {Juaristi Campillo}, J. and {Julbe}, F. and {Karbevska}, L. and {Kervella}, P. and {Khanna}, S. and {Kontizas}, M. and {Kordopatis}, G. and {Korn}, A.~J. and {K{\'o}sp{\'a}l}, {\'A}. and {Kostrzewa-Rutkowska}, Z. and {Kruszy{\'n}ska}, K. and {Kun}, M. and {Laizeau}, P. and {Lambert}, S. and {Lanza}, A.~F. and {Lasne}, Y. and {Le Campion}, J. -F. and {Lebreton}, Y. and {Lebzelter}, T. and {Leccia}, S. and {Leclerc}, N. and {Lecoeur-Taibi}, I. and {Liao}, S. and {Licata}, E.~L. and {Lindstr{\o}m}, H.~E.~P. and {Lister}, T.~A. and {Livanou}, E. and {Lobel}, A. and {Lorca}, A. and {Loup}, C. and {Madrero Pardo}, P. and {Magdaleno Romeo}, A. and {Managau}, S. and {Mann}, R.~G. and {Manteiga}, M. and {Marchant}, J.~M. and {Marconi}, M. and {Marcos}, J. and {Marcos Santos}, M.~M.~S. and {Mar{\'\i}n Pina}, D. and {Marinoni}, S. and {Marocco}, F. and {Marshall}, D.~J. and {Martin Polo}, L. and {Mart{\'\i}n-Fleitas}, J.~M. and {Marton}, G. and {Mary}, N. and {Masip}, A. and {Massari}, D. and {Mastrobuono-Battisti}, A. and {Mazeh}, T. and {McMillan}, P.~J. and {Messina}, S. and {Michalik}, D. and {Millar}, N.~R. and {Mints}, A. and {Molina}, D. and {Molinaro}, R. and {Moln{\'a}r}, L. and {Monari}, G. and {Mongui{\'o}}, M. and {Montegriffo}, P. and {Montero}, A. and {Mor}, R. and {Mora}, A. and {Morbidelli}, R. and {Morel}, T. and {Morris}, D. and {Muraveva}, T. and {Murphy}, C.~P. and {Musella}, I. and {Nagy}, Z. and {Noval}, L. and {Oca{\~n}a}, F. and {Ogden}, A. and {Ordenovic}, C. and {Osinde}, J.~O. and {Pagani}, C. and {Pagano}, I. and {Palaversa}, L. and {Palicio}, P.~A. and {Pallas-Quintela}, L. and {Panahi}, A. and {Payne-Wardenaar}, S. and {Pe{\~n}alosa Esteller}, X. and {Penttil{\"a}}, A. and {Pichon}, B. and {Piersimoni}, A.~M. and {Pineau}, F. -X. and {Plachy}, E. and {Plum}, G. and {Poggio}, E. and {Pr{\v{s}}a}, A. and {Pulone}, L. and {Racero}, E. and {Ragaini}, S. and {Rainer}, M. and {Raiteri}, C.~M. and {Rambaux}, N. and {Ramos}, P. and {Ramos-Lerate}, M. and {Re Fiorentin}, P. and {Regibo}, S. and {Richards}, P.~J. and {Rios Diaz}, C. and {Ripepi}, V. and {Riva}, A. and {Rix}, H. -W. and {Rixon}, G. and {Robichon}, N. and {Robin}, A.~C. and {Robin}, C. and {Roelens}, M. and {Rogues}, H.~R.~O. and {Rohrbasser}, L. and {Romero-G{\'o}mez}, M. and {Rowell}, N. and {Royer}, F. and {Ruz Mieres}, D. and {Rybicki}, K.~A. and {Sadowski}, G. and {S{\'a}ez N{\'u}{\~n}ez}, A. and {Sagrist{\`a} Sell{\'e}s}, A. and {Sahlmann}, J. and {Salguero}, E. and {Samaras}, N. and {Sanchez Gimenez}, V. and {Sanna}, N. and {Santove{\~n}a}, R. and {Sarasso}, M. and {Schultheis}, M. and {Sciacca}, E. and {Segol}, M. and {Segovia}, J.~C. and {S{\'e}gransan}, D. and {Semeux}, D. and {Shahaf}, S. and {Siddiqui}, H.~I. and {Siebert}, A. and {Siltala}, L. and {Silvelo}, A. and {Slezak}, E. and {Slezak}, I. and {Smart}, R.~L. and {Snaith}, O.~N. and {Solano}, E. and {Solitro}, F. and {Souami}, D. and {Souchay}, J. and {Spagna}, A. and {Spina}, L. and {Spoto}, F. and {Steele}, I.~A. and {Steidelm{\"u}ller}, H. and {Stephenson}, C.~A. and {S{\"u}veges}, M. and {Surdej}, J. and {Szabados}, L. and {Szegedi-Elek}, E. and {Taris}, F. and {Taylor}, M.~B. and {Teixeira}, R. and {Tolomei}, L. and {Tonello}, N. and {Torra}, F. and {Torra}, J. and {Torralba Elipe}, G. and {Trabucchi}, M. and {Tsounis}, A.~T. and {Turon}, C. and {Ulla}, A. and {Unger}, N. and {Vaillant}, M.~V. and {van Dillen}, E. and {van Reeven}, W. and {Vanel}, O. and {Vecchiato}, A. and {Viala}, Y. and {Vicente}, D. and {Voutsinas}, S. and {Weiler}, M. and {Wevers}, T. and {Wyrzykowski}, {\L}. and {Yoldas}, A. and {Yvard}, P. and {Zhao}, H. and {Zorec}, J. and {Zucker}, S. and {Zwitter}, T.},
        title = "{Gaia Data Release 3. Summary of the content and survey properties}",
      journal = {\aap},
         year = 2023,
        month = jun,
       volume = {674},
          eid = {A1},
        pages = {A1},
          doi = {10.1051/0004-6361/202243940},
archivePrefix = {arXiv},
       eprint = {2208.00211},
 primaryClass = {astro-ph.GA},
       adsurl = {https://ui.adsabs.harvard.edu/abs/2023A&A...674A...1G}
}

@ARTICLE{dobie23,
       author = {{Dobie}, Dougal and {Sluse}, Dominique and {Deller}, Adam and {Murphy}, Tara and {Krone-Martins}, Alberto and {Stern}, Daniel and {Wang}, Ziteng and {Wang}, Yuanming and {B{\oe}hm}, C{\'e}line and {Djorgovski}, S.~G. and {Galluccio}, Laurent and {Delchambre}, Ludovic and {Connor}, Thomas and {den Brok}, Jakob Sebastiaan and {Do Vale Cunha}, Pedro H. and {Ducourant}, Christine and {Graham}, Matthew J. and {Jalan}, Priyanka and {Klioner}, Sergei A. and {Kl{\"u}ter}, Jonas and {Mignard}, Fran{\c{c}}ois and {Negi}, Vibhore and {Petit}, Quentin and {Scarano}, Sergio and {Slezak}, Eric and {Surdej}, Jean and {Teixeira}, Ramachrisna and {Walton}, Dominic J. and {Wambsganss}, Joachim},
        title = "{Gaia GraL: Gaia DR2 gravitational lens systems - VIII. A radio census of lensed systems}",
      journal = {\mnras},
         year = 2024,
        month = mar,
       volume = {528},
       number = {4},
        pages = {5880-5889},
          doi = {10.1093/mnras/stad4002},
archivePrefix = {arXiv},
       eprint = {2311.07836},
 primaryClass = {astro-ph.GA},
       adsurl = {https://ui.adsabs.harvard.edu/abs/2024MNRAS.528.5880D}
}

@ARTICLE{hogbom,
       author = {{H{\"o}gbom}, J.~A.},
        title = "{Aperture Synthesis with a Non-Regular Distribution of Interferometer Baselines}",
      journal = {\aaps},
         year = 1974,
        month = jun,
       volume = {15},
        pages = {417},
       adsurl = {https://ui.adsabs.harvard.edu/abs/1974A&AS...15..417H}
}

@ARTICLE{casa,
       author = {{CASA Team} and {Bean}, Ben and {Bhatnagar}, Sanjay and {Castro}, Sandra and {Donovan Meyer}, Jennifer and {Emonts}, Bjorn and {Garcia}, Enrique and {Garwood}, Robert and {Golap}, Kumar and {Gonzalez Villalba}, Justo and {Harris}, Pamela and {Hayashi}, Yohei and {Hoskins}, Josh and {Hsieh}, Mingyu and {Jagannathan}, Preshanth and {Kawasaki}, Wataru and {Keimpema}, Aard and {Kettenis}, Mark and {Lopez}, Jorge and {Marvil}, Joshua and {Masters}, Joseph and {McNichols}, Andrew and {Mehringer}, David and {Miel}, Renaud and {Moellenbrock}, George and {Montesino}, Federico and {Nakazato}, Takeshi and {Ott}, Juergen and {Petry}, Dirk and {Pokorny}, Martin and {Raba}, Ryan and {Rau}, Urvashi and {Schiebel}, Darrell and {Schweighart}, Neal and {Sekhar}, Srikrishna and {Shimada}, Kazuhiko and {Small}, Des and {Steeb}, Jan-Willem and {Sugimoto}, Kanako and {Suoranta}, Ville and {Tsutsumi}, Takahiro and {van Bemmel}, Ilse M. and {Verkouter}, Marjolein and {Wells}, Akeem and {Xiong}, Wei and {Szomoru}, Arpad and {Griffith}, Morgan and {Glendenning}, Brian and {Kern}, Jeff},
        title = "{CASA, the Common Astronomy Software Applications for Radio Astronomy}",
      journal = {\pasp},
         year = 2022,
        month = nov,
       volume = {134},
       number = {1041},
          eid = {114501},
        pages = {114501},
          doi = {10.1088/1538-3873/ac9642},
archivePrefix = {arXiv},
       eprint = {2210.02276},
 primaryClass = {astro-ph.IM},
       adsurl = {https://ui.adsabs.harvard.edu/abs/2022PASP..134k4501C}
}

@ARTICLE{mfs,
       author = {{Conway}, J.~E. and {Cornwell}, T.~J. and {Wilkinson}, P.~N.},
        title = "{Multi-frequency synthesis : a new technique in radio interferometrie imaging.}",
      journal = {\mnras},
         year = 1990,
        month = oct,
       volume = {246},
        pages = {490},
       adsurl = {https://ui.adsabs.harvard.edu/abs/1990MNRAS.246..490C}
}

@ARTICLE{selfcal,
       author = {{Readhead}, A.~C.~S. and {Wilkinson}, P.~N.},
        title = "{The mapping of compact radio sources from VLBI data.}",
      journal = {\apj},
         year = 1978,
        month = jul,
       volume = {223},
        pages = {25-36},
          doi = {10.1086/156232},
       adsurl = {https://ui.adsabs.harvard.edu/abs/1978ApJ...223...25R}
}

@ARTICLE{VLASS,
       author = {{Lacy}, M. and {Baum}, S.~A. and {Chandler}, C.~J. and {Chatterjee}, S. and {Clarke}, T.~E. and {Deustua}, S. and {English}, J. and {Farnes}, J. and {Gaensler}, B.~M. and {Gugliucci}, N. and {Hallinan}, G. and {Kent}, B.~R. and {Kimball}, A. and {Law}, C.~J. and {Lazio}, T.~J.~W. and {Marvil}, J. and {Mao}, S.~A. and {Medlin}, D. and {Mooley}, K. and {Murphy}, E.~J. and {Myers}, S. and {Osten}, R. and {Richards}, G.~T. and {Rosolowsky}, E. and {Rudnick}, L. and {Schinzel}, F. and {Sivakoff}, G.~R. and {Sjouwerman}, L.~O. and {Taylor}, R. and {White}, R.~L. and {Wrobel}, J. and {Andernach}, H. and {Beasley}, A.~J. and {Berger}, E. and {Bhatnager}, S. and {Birkinshaw}, M. and {Bower}, G.~C. and {Brandt}, W.~N. and {Brown}, S. and {Burke-Spolaor}, S. and {Butler}, B.~J. and {Comerford}, J. and {Demorest}, P.~B. and {Fu}, H. and {Giacintucci}, S. and {Golap}, K. and {G{\"u}th}, T. and {Hales}, C.~A. and {Hiriart}, R. and {Hodge}, J. and {Horesh}, A. and {Ivezi{\'c}}, {\v{Z}}. and {Jarvis}, M.~J. and {Kamble}, A. and {Kassim}, N. and {Liu}, X. and {Loinard}, L. and {Lyons}, D.~K. and {Masters}, J. and {Mezcua}, M. and {Moellenbrock}, G.~A. and {Mroczkowski}, T. and {Nyland}, K. and {O'Dea}, C.~P. and {O'Sullivan}, S.~P. and {Peters}, W.~M. and {Radford}, K. and {Rao}, U. and {Robnett}, J. and {Salcido}, J. and {Shen}, Y. and {Sobotka}, A. and {Witz}, S. and {Vaccari}, M. and {van Weeren}, R.~J. and {Vargas}, A. and {Williams}, P.~K.~G. and {Yoon}, I.},
        title = "{The Karl G. Jansky Very Large Array Sky Survey (VLASS). Science Case and Survey Design}",
      journal = {\pasp},
         year = 2020,
        month = mar,
       volume = {132},
       number = {1009},
          eid = {035001},
        pages = {035001},
          doi = {10.1088/1538-3873/ab63eb},
archivePrefix = {arXiv},
       eprint = {1907.01981},
 primaryClass = {astro-ph.IM},
       adsurl = {https://ui.adsabs.harvard.edu/abs/2020PASP..132c5001L}
}

@ARTICLE{FIRSTfinal,
       author = {{Helfand}, David J. and {White}, Richard L. and {Becker}, Robert H.},
        title = "{The Last of FIRST: The Final Catalog and Source Identifications}",
      journal = {\apj},
         year = 2015,
        month = mar,
       volume = {801},
       number = {1},
          eid = {26},
        pages = {26},
          doi = {10.1088/0004-637X/801/1/26},
archivePrefix = {arXiv},
       eprint = {1501.01555},
 primaryClass = {astro-ph.GA},
       adsurl = {https://ui.adsabs.harvard.edu/abs/2015ApJ...801...26H}
}

@ARTICLE{SDSSDR7,
       author = {{Abazajian}, Kevork N. and {Adelman-McCarthy}, Jennifer K. and {Ag{\"u}eros}, Marcel A. and {Allam}, Sahar S. and {Allende Prieto}, Carlos and {An}, Deokkeun and {Anderson}, Kurt S.~J. and {Anderson}, Scott F. and {Annis}, James and {Bahcall}, Neta A. and {Bailer-Jones}, C.~A.~L. and {Barentine}, J.~C. and {Bassett}, Bruce A. and {Becker}, Andrew C. and {Beers}, Timothy C. and {Bell}, Eric F. and {Belokurov}, Vasily and {Berlind}, Andreas A. and {Berman}, Eileen F. and {Bernardi}, Mariangela and {Bickerton}, Steven J. and {Bizyaev}, Dmitry and {Blakeslee}, John P. and {Blanton}, Michael R. and {Bochanski}, John J. and {Boroski}, William N. and {Brewington}, Howard J. and {Brinchmann}, Jarle and {Brinkmann}, J. and {Brunner}, Robert J. and {Budav{\'a}ri}, Tam{\'a}s and {Carey}, Larry N. and {Carliles}, Samuel and {Carr}, Michael A. and {Castander}, Francisco J. and {Cinabro}, David and {Connolly}, A.~J. and {Csabai}, Istv{\'a}n and {Cunha}, Carlos E. and {Czarapata}, Paul C. and {Davenport}, James R.~A. and {de Haas}, Ernst and {Dilday}, Ben and {Doi}, Mamoru and {Eisenstein}, Daniel J. and {Evans}, Michael L. and {Evans}, N.~W. and {Fan}, Xiaohui and {Friedman}, Scott D. and {Frieman}, Joshua A. and {Fukugita}, Masataka and {G{\"a}nsicke}, Boris T. and {Gates}, Evalyn and {Gillespie}, Bruce and {Gilmore}, G. and {Gonzalez}, Belinda and {Gonzalez}, Carlos F. and {Grebel}, Eva K. and {Gunn}, James E. and {Gy{\"o}ry}, Zsuzsanna and {Hall}, Patrick B. and {Harding}, Paul and {Harris}, Frederick H. and {Harvanek}, Michael and {Hawley}, Suzanne L. and {Hayes}, Jeffrey J.~E. and {Heckman}, Timothy M. and {Hendry}, John S. and {Hennessy}, Gregory S. and {Hindsley}, Robert B. and {Hoblitt}, J. and {Hogan}, Craig J. and {Hogg}, David W. and {Holtzman}, Jon A. and {Hyde}, Joseph B. and {Ichikawa}, Shin-ichi and {Ichikawa}, Takashi and {Im}, Myungshin and {Ivezi{\'c}}, {\v{Z}}eljko and {Jester}, Sebastian and {Jiang}, Linhua and {Johnson}, Jennifer A. and {Jorgensen}, Anders M. and {Juri{\'c}}, Mario and {Kent}, Stephen M. and {Kessler}, R. and {Kleinman}, S.~J. and {Knapp}, G.~R. and {Konishi}, Kohki and {Kron}, Richard G. and {Krzesinski}, Jurek and {Kuropatkin}, Nikolay and {Lampeitl}, Hubert and {Lebedeva}, Svetlana and {Lee}, Myung Gyoon and {Lee}, Young Sun and {French Leger}, R. and {L{\'e}pine}, S{\'e}bastien and {Li}, Nolan and {Lima}, Marcos and {Lin}, Huan and {Long}, Daniel C. and {Loomis}, Craig P. and {Loveday}, Jon and {Lupton}, Robert H. and {Magnier}, Eugene and {Malanushenko}, Olena and {Malanushenko}, Viktor and {Mandelbaum}, Rachel and {Margon}, Bruce and {Marriner}, John P. and {Mart{\'\i}nez-Delgado}, David and {Matsubara}, Takahiko and {McGehee}, Peregrine M. and {McKay}, Timothy A. and {Meiksin}, Avery and {Morrison}, Heather L. and {Mullally}, Fergal and {Munn}, Jeffrey A. and {Murphy}, Tara and {Nash}, Thomas and {Nebot}, Ada and {Neilsen}, Eric H., Jr. and {Newberg}, Heidi Jo and {Newman}, Peter R. and {Nichol}, Robert C. and {Nicinski}, Tom and {Nieto-Santisteban}, Maria and {Nitta}, Atsuko and {Okamura}, Sadanori and {Oravetz}, Daniel J. and {Ostriker}, Jeremiah P. and {Owen}, Russell and {Padmanabhan}, Nikhil and {Pan}, Kaike and {Park}, Changbom and {Pauls}, George and {Peoples}, John, Jr. and {Percival}, Will J. and {Pier}, Jeffrey R. and {Pope}, Adrian C. and {Pourbaix}, Dimitri and {Price}, Paul A. and {Purger}, Norbert and {Quinn}, Thomas and {Raddick}, M. Jordan and {Re Fiorentin}, Paola and {Richards}, Gordon T. and {Richmond}, Michael W. and {Riess}, Adam G. and {Rix}, Hans-Walter and {Rockosi}, Constance M. and {Sako}, Masao and {Schlegel}, David J. and {Schneider}, Donald P. and {Scholz}, Ralf-Dieter and {Schreiber}, Matthias R. and {Schwope}, Axel D. and {Seljak}, Uro{\v{s}} and {Sesar}, Branimir and {Sheldon}, Erin and {Shimasaku}, Kazu and {Sibley}, Valena C. and {Simmons}, A.~E. and {Sivarani}, Thirupathi and {Allyn Smith}, J. and {Smith}, Martin C. and {Smol{\v{c}}i{\'c}}, Vernesa and {Snedden}, Stephanie A. and {Stebbins}, Albert and {Steinmetz}, Matthias and {Stoughton}, Chris and {Strauss}, Michael A. and {SubbaRao}, Mark and {Suto}, Yasushi and {Szalay}, Alexander S. and {Szapudi}, Istv{\'a}n and {Szkody}, Paula and {Tanaka}, Masayuki and {Tegmark}, Max and {Teodoro}, Luis F.~A. and {Thakar}, Aniruddha R. and {Tremonti}, Christy A. and {Tucker}, Douglas L. and {Uomoto}, Alan and {Vanden Berk}, Daniel E. and {Vandenberg}, Jan and {Vidrih}, S. and {Vogeley}, Michael S. and {Voges}, Wolfgang and {Vogt}, Nicole P. and {Wadadekar}, Yogesh and {Watters}, Shannon and {Weinberg}, David H. and {West}, Andrew A. and {White}, Simon D.~M. and {Wilhite}, Brian C. and {Wonders}, Alainna C. and {Yanny}, Brian and {Yocum}, D.~R. and {York}, Donald G. and {Zehavi}, Idit and {Zibetti}, Stefano and {Zucker}, Daniel B.},
        title = "{The Seventh Data Release of the Sloan Digital Sky Survey}",
      journal = {\apjs},
         year = 2009,
        month = jun,
       volume = {182},
       number = {2},
        pages = {543-558},
          doi = {10.1088/0067-0049/182/2/543},
archivePrefix = {arXiv},
       eprint = {0812.0649},
 primaryClass = {astro-ph},
       adsurl = {https://ui.adsabs.harvard.edu/abs/2009ApJS..182..543A}
}

@ARTICLE{FIRST,
       author = {{Becker}, Robert H. and {White}, Richard L. and {Helfand}, David J.},
        title = "{The FIRST Survey: Faint Images of the Radio Sky at Twenty Centimeters}",
      journal = {\apj},
         year = 1995,
        month = sep,
       volume = {450},
        pages = {559},
          doi = {10.1086/176166},
       adsurl = {https://ui.adsabs.harvard.edu/abs/1995ApJ...450..559B}
}

@ARTICLE{SDSS,
       author = {{York}, Donald G. and {Adelman}, J. and {Anderson}, John E., Jr. and {Anderson}, Scott F. and {Annis}, James and {Bahcall}, Neta A. and {Bakken}, J.~A. and {Barkhouser}, Robert and {Bastian}, Steven and {Berman}, Eileen and {Boroski}, William N. and {Bracker}, Steve and {Briegel}, Charlie and {Briggs}, John W. and {Brinkmann}, J. and {Brunner}, Robert and {Burles}, Scott and {Carey}, Larry and {Carr}, Michael A. and {Castander}, Francisco J. and {Chen}, Bing and {Colestock}, Patrick L. and {Connolly}, A.~J. and {Crocker}, J.~H. and {Csabai}, Istv{\'a}n and {Czarapata}, Paul C. and {Davis}, John Eric and {Doi}, Mamoru and {Dombeck}, Tom and {Eisenstein}, Daniel and {Ellman}, Nancy and {Elms}, Brian R. and {Evans}, Michael L. and {Fan}, Xiaohui and {Federwitz}, Glenn R. and {Fiscelli}, Larry and {Friedman}, Scott and {Frieman}, Joshua A. and {Fukugita}, Masataka and {Gillespie}, Bruce and {Gunn}, James E. and {Gurbani}, Vijay K. and {de Haas}, Ernst and {Haldeman}, Merle and {Harris}, Frederick H. and {Hayes}, J. and {Heckman}, Timothy M. and {Hennessy}, G.~S. and {Hindsley}, Robert B. and {Holm}, Scott and {Holmgren}, Donald J. and {Huang}, Chi-hao and {Hull}, Charles and {Husby}, Don and {Ichikawa}, Shin-Ichi and {Ichikawa}, Takashi and {Ivezi{\'c}}, {\v{Z}}eljko and {Kent}, Stephen and {Kim}, Rita S.~J. and {Kinney}, E. and {Klaene}, Mark and {Kleinman}, A.~N. and {Kleinman}, S. and {Knapp}, G.~R. and {Korienek}, John and {Kron}, Richard G. and {Kunszt}, Peter Z. and {Lamb}, D.~Q. and {Lee}, B. and {Leger}, R. French and {Limmongkol}, Siriluk and {Lindenmeyer}, Carl and {Long}, Daniel C. and {Loomis}, Craig and {Loveday}, Jon and {Lucinio}, Rich and {Lupton}, Robert H. and {MacKinnon}, Bryan and {Mannery}, Edward J. and {Mantsch}, P.~M. and {Margon}, Bruce and {McGehee}, Peregrine and {McKay}, Timothy A. and {Meiksin}, Avery and {Merelli}, Aronne and {Monet}, David G. and {Munn}, Jeffrey A. and {Narayanan}, Vijay K. and {Nash}, Thomas and {Neilsen}, Eric and {Neswold}, Rich and {Newberg}, Heidi Jo and {Nichol}, R.~C. and {Nicinski}, Tom and {Nonino}, Mario and {Okada}, Norio and {Okamura}, Sadanori and {Ostriker}, Jeremiah P. and {Owen}, Russell and {Pauls}, A. George and {Peoples}, John and {Peterson}, R.~L. and {Petravick}, Donald and {Pier}, Jeffrey R. and {Pope}, Adrian and {Pordes}, Ruth and {Prosapio}, Angela and {Rechenmacher}, Ron and {Quinn}, Thomas R. and {Richards}, Gordon T. and {Richmond}, Michael W. and {Rivetta}, Claudio H. and {Rockosi}, Constance M. and {Ruthmansdorfer}, Kurt and {Sandford}, Dale and {Schlegel}, David J. and {Schneider}, Donald P. and {Sekiguchi}, Maki and {Sergey}, Gary and {Shimasaku}, Kazuhiro and {Siegmund}, Walter A. and {Smee}, Stephen and {Smith}, J. Allyn and {Snedden}, S. and {Stone}, R. and {Stoughton}, Chris and {Strauss}, Michael A. and {Stubbs}, Christopher and {SubbaRao}, Mark and {Szalay}, Alexander S. and {Szapudi}, Istvan and {Szokoly}, Gyula P. and {Thakar}, Anirudda R. and {Tremonti}, Christy and {Tucker}, Douglas L. and {Uomoto}, Alan and {Vanden Berk}, Dan and {Vogeley}, Michael S. and {Waddell}, Patrick and {Wang}, Shu-i. and {Watanabe}, Masaru and {Weinberg}, David H. and {Yanny}, Brian and {Yasuda}, Naoki and {SDSS Collaboration}},
        title = "{The Sloan Digital Sky Survey: Technical Summary}",
      journal = {\aj},
         year = 2000,
        month = sep,
       volume = {120},
       number = {3},
        pages = {1579-1587},
          doi = {10.1086/301513},
archivePrefix = {arXiv},
       eprint = {astro-ph/0006396},
 primaryClass = {astro-ph},
       adsurl = {https://ui.adsabs.harvard.edu/abs/2000AJ....120.1579Y}
}

@misc{mckeancomm,
    author = "McKean, John",
    date = "2023-12-07",
    year = "2023",
    howpublished = "personal communication"
}

@ARTICLE{lemon2023,
       author = {{Lemon}, C. and {Anguita}, T. and {Auger-Williams}, M.~W. and {Courbin}, F. and {Galan}, A. and {McMahon}, R. and {Neira}, F. and {Oguri}, M. and {Schechter}, P. and {Shajib}, A. and {Treu}, T. and {Agnello}, A. and {Spiniello}, C.},
        title = "{Gravitationally lensed quasars in Gaia - IV. 150 new lenses, quasar pairs, and projected quasars}",
      journal = {\mnras},
         year = 2023,
        month = apr,
       volume = {520},
       number = {3},
        pages = {3305-3328},
          doi = {10.1093/mnras/stac3721},
archivePrefix = {arXiv},
       eprint = {2206.07714},
 primaryClass = {astro-ph.GA},
       adsurl = {https://ui.adsabs.harvard.edu/abs/2023MNRAS.520.3305L}
}

@ARTICLE{dawes23,
       author = {{Dawes}, C. and {Storfer}, C. and {Huang}, X. and {Aldering}, G. and {Cikota}, Aleksandar and {Dey}, Arjun and {Schlegel}, D.~J.},
        title = "{Finding Multiply Lensed and Binary Quasars in the DESI Legacy Imaging Surveys}",
      journal = {\apjs},
         year = 2023,
        month = dec,
       volume = {269},
       number = {2},
          eid = {61},
        pages = {61},
          doi = {10.3847/1538-4365/ad015a},
archivePrefix = {arXiv},
       eprint = {2208.06356},
 primaryClass = {astro-ph.CO},
       adsurl = {https://ui.adsabs.harvard.edu/abs/2023ApJS..269...61D}
}

@ARTICLE{zaborowski23,
       author = {{Zaborowski}, E.~A. and {Drlica-Wagner}, A. and {Ashmead}, F. and {Wu}, J.~F. and {Morgan}, R. and {Bom}, C.~R. and {Shajib}, A.~J. and {Birrer}, S. and {Cerny}, W. and {Buckley-Geer}, E.~J. and {Mutlu-Pakdil}, B. and {Ferguson}, P.~S. and {Glazebrook}, K. and {Lozano}, S.~J. Gonzalez and {Gordon}, Y. and {Martinez}, M. and {Manwadkar}, V. and {O'Donnell}, J. and {Poh}, J. and {Riley}, A. and {Sakowska}, J.~D. and {Santana-Silva}, L. and {Santiago}, B.~X. and {Sluse}, D. and {Tan}, C.~Y. and {Tollerud}, E.~J. and {Verma}, A. and {Carballo-Bello}, J.~A. and {Choi}, Y. and {James}, D.~J. and {Kuropatkin}, N. and {Mart{\'\i}nez-V{\'a}zquez}, C.~E. and {Nidever}, D.~L. and {Castellon}, J.~L. Nilo and {No{\"e}l}, N.~E.~D. and {Olsen}, K.~A.~G. and {Pace}, A.~B. and {Mau}, S. and {Yanny}, B. and {Zenteno}, A. and {Abbott}, T.~M.~C. and {Aguena}, M. and {Alves}, O. and {Andrade-Oliveira}, F. and {Bocquet}, S. and {Brooks}, D. and {Burke}, D.~L. and {Carnero Rosell}, A. and {Carrasco Kind}, M. and {Carretero}, J. and {Castander}, F.~J. and {Conselice}, C.~J. and {Costanzi}, M. and {Pereira}, M.~E.~S. and {de Vicente}, J. and {Desai}, S. and {Dietrich}, J.~P. and {Doel}, P. and {Everett}, S. and {Ferrero}, I. and {Flaugher}, B. and {Friedel}, D. and {Frieman}, J. and {Garc{\'\i}a-Bellido}, J. and {Gruen}, D. and {Gruendl}, R.~A. and {Gutierrez}, G. and {Hinton}, S.~R. and {Hollowood}, D.~L. and {Honscheid}, K. and {Kuehn}, K. and {Lin}, H. and {Marshall}, J.~L. and {Melchior}, P. and {Mena-Fern{\'a}ndez}, J. and {Menanteau}, F. and {Miquel}, R. and {Palmese}, A. and {Paz-Chinch{\'o}n}, F. and {Pieres}, A. and {Malag{\'o}n}, A.~A. Plazas and {Prat}, J. and {Rodriguez-Monroy}, M. and {Romer}, A.~K. and {Sanchez}, E. and {Scarpine}, V. and {Sevilla-Noarbe}, I. and {Smith}, M. and {Suchyta}, E. and {To}, C. and {Weaverdyck}, N. and {Delve Collaboration} and {Des Collaboration}},
        title = "{Identification of Galaxy-Galaxy Strong Lens Candidates in the DECam Local Volume Exploration Survey Using Machine Learning}",
      journal = {\apj},
         year = 2023,
        month = sep,
       volume = {954},
       number = {1},
          eid = {68},
        pages = {68},
          doi = {10.3847/1538-4357/ace4ba},
archivePrefix = {arXiv},
       eprint = {2210.10802},
 primaryClass = {astro-ph.GA},
       adsurl = {https://ui.adsabs.harvard.edu/abs/2023ApJ...954...68Z}
}

@ARTICLE{WISE,
       author = {{Wright}, Edward L. and {Eisenhardt}, Peter R.~M. and {Mainzer}, Amy K. and {Ressler}, Michael E. and {Cutri}, Roc M. and {Jarrett}, Thomas and {Kirkpatrick}, J. Davy and {Padgett}, Deborah and {McMillan}, Robert S. and {Skrutskie}, Michael and {Stanford}, S.~A. and {Cohen}, Martin and {Walker}, Russell G. and {Mather}, John C. and {Leisawitz}, David and {Gautier}, Thomas N., III and {McLean}, Ian and {Benford}, Dominic and {Lonsdale}, Carol J. and {Blain}, Andrew and {Mendez}, Bryan and {Irace}, William R. and {Duval}, Valerie and {Liu}, Fengchuan and {Royer}, Don and {Heinrichsen}, Ingolf and {Howard}, Joan and {Shannon}, Mark and {Kendall}, Martha and {Walsh}, Amy L. and {Larsen}, Mark and {Cardon}, Joel G. and {Schick}, Scott and {Schwalm}, Mark and {Abid}, Mohamed and {Fabinsky}, Beth and {Naes}, Larry and {Tsai}, Chao-Wei},
        title = "{The Wide-field Infrared Survey Explorer (WISE): Mission Description and Initial On-orbit Performance}",
      journal = {\aj},
         year = 2010,
        month = dec,
       volume = {140},
       number = {6},
        pages = {1868-1881},
          doi = {10.1088/0004-6256/140/6/1868},
archivePrefix = {arXiv},
       eprint = {1008.0031},
 primaryClass = {astro-ph.IM},
       adsurl = {https://ui.adsabs.harvard.edu/abs/2010AJ....140.1868W}
}

@ARTICLE{unWISE,
       author = {{Lang}, Dustin},
        title = "{unWISE: Unblurred Coadds of the WISE Imaging}",
      journal = {\aj},
         year = 2014,
        month = may,
       volume = {147},
       number = {5},
          eid = {108},
        pages = {108},
          doi = {10.1088/0004-6256/147/5/108},
archivePrefix = {arXiv},
       eprint = {1405.0308},
 primaryClass = {astro-ph.IM},
       adsurl = {https://ui.adsabs.harvard.edu/abs/2014AJ....147..108L}
}

@ARTICLE{walsh79,
       author = {{Walsh}, D. and {Carswell}, R.~F. and {Weymann}, R.~J.},
        title = "{0957+561 A, B: twin quasistellar objects or gravitational lens?}",
      journal = {\nat},
         year = 1979,
        month = may,
       volume = {279},
        pages = {381-384},
          doi = {10.1038/279381a0},
       adsurl = {https://ui.adsabs.harvard.edu/abs/1979Natur.279..381W}
}

@ARTICLE{1984Sci...223...46L,
       author = {{Lawrence}, C.~R. and {Schneider}, D.~P. and {Schmidt}, M. and {Bennett}, C.~L. and {Hewitt}, J.~N. and {Burke}, B.~F. and {Turner}, E.~L. and {Gunn}, J.~E.},
        title = "{Discovery of a New Gravitational Lens System}",
      journal = {Science},
         year = 1984,
        month = jan,
       volume = {223},
       number = {4631},
        pages = {46-49},
          doi = {10.1126/science.223.4631.46},
       adsurl = {https://ui.adsabs.harvard.edu/abs/1984Sci...223...46L}
}

@ARTICLE{1980Natur.285..641W,
       author = {{Weymann}, Ray J. and {Latham}, David and {Angel}, J. Roger P. and {Green}, Richard F. and {Liebert}, James W. and {Turnshek}, David A. and {Turnshek}, Diane E. and {Tyson}, J. Anthony},
        title = "{The triple QSO PG1115 + 08: another probable gravitational lens}",
      journal = {\nat},
         year = 1980,
        month = jun,
       volume = {285},
       number = {5767},
        pages = {641-643},
          doi = {10.1038/285641a0},
       adsurl = {https://ui.adsabs.harvard.edu/abs/1980Natur.285..641W}
}

@ARTICLE{1996AJ....112..897F,
       author = {{Falco}, E.~E. and {Lehar}, J. and {Perley}, R.~A. and {Wambsganss}, J. and {Gorenstein}, M.~V.},
        title = "{VLA Observations of the Gravitational Lens System Q2237+0305}",
      journal = {\aj},
         year = 1996,
        month = sep,
       volume = {112},
        pages = {897},
          doi = {10.1086/118062},
archivePrefix = {arXiv},
       eprint = {astro-ph/9606048},
 primaryClass = {astro-ph},
       adsurl = {https://ui.adsabs.harvard.edu/abs/1996AJ....112..897F}
}

@ARTICLE{1985AJ.....90..691H,
       author = {{Huchra}, J. and {Gorenstein}, M. and {Kent}, S. and {Shapiro}, I. and {Smith}, G. and {Horine}, E. and {Perley}, R.},
        title = "{2237+0305 : a new and unusual gravitational lens.}",
      journal = {\aj},
         year = 1985,
        month = may,
       volume = {90},
        pages = {691-696},
          doi = {10.1086/113777},
       adsurl = {https://ui.adsabs.harvard.edu/abs/1985AJ.....90..691H}
}

@ARTICLE{1988Natur.333..537H,
       author = {{Hewitt}, J.~N. and {Turner}, E.~L. and {Schneider}, D.~P. and {Burke}, B.~F. and {Langston}, G.~I.},
        title = "{Unusual radio source MG1131+0456: a possible Einstein ring}",
      journal = {\nat},
         year = 1988,
        month = jun,
       volume = {333},
       number = {6173},
        pages = {537-540},
          doi = {10.1038/333537a0},
       adsurl = {https://ui.adsabs.harvard.edu/abs/1988Natur.333..537H}
}

@ARTICLE{1988MNRAS.231..229P,
       author = {{Pramesh Rao}, A. and {Subrahmanyan}, R.},
        title = "{1830-211 - a flat spectrum radio source with double structure.}",
      journal = {\mnras},
         year = 1988,
        month = mar,
       volume = {231},
        pages = {229-236},
          doi = {10.1093/mnras/231.2.229},
       adsurl = {https://ui.adsabs.harvard.edu/abs/1988MNRAS.231..229P}
}

@ARTICLE{2023MNRAS.524.3671Z,
       author = {{Zhang}, Lei and {Zhang}, Zhi-Yu and {Nightingale}, James W. and {Zou}, Ze-Cheng and {Cao}, Xiaoyue and {Tsai}, Chao-Wei and {Yang}, Chentao and {Shi}, Yong and {Wang}, Junzhi and {Xu}, Dandan and {Lin}, Ling-Rui and {Zhou}, Jing and {Li}, Ran},
        title = "{Discovery of a radio jet in the Cloverleaf quasar at z = 2.56}",
      journal = {\mnras},
         year = 2023,
        month = sep,
       volume = {524},
       number = {3},
        pages = {3671-3682},
          doi = {10.1093/mnras/stad2069},
archivePrefix = {arXiv},
       eprint = {2212.07027},
 primaryClass = {astro-ph.GA},
       adsurl = {https://ui.adsabs.harvard.edu/abs/2023MNRAS.524.3671Z}
}

@ARTICLE{1988Natur.334..325M,
       author = {{Magain}, P. and {Surdej}, J. and {Swings}, J. -P. and {Borgeest}, U. and {Kayser}, R.},
        title = "{Discovery of a quadruply lensed quasar: the 'clover leaf H1413 + 117}",
      journal = {\nat},
         year = 1988,
        month = jul,
       volume = {334},
       number = {6180},
        pages = {325-327},
          doi = {10.1038/334325a0},
       adsurl = {https://ui.adsabs.harvard.edu/abs/1988Natur.334..325M}
}

@ARTICLE{1989AJ.....97.1283L,
       author = {{Langston}, G.~I. and {Schneider}, D.~P. and {Conner}, S. and {Carilli}, C.~L. and {Lehar}, J. and {Burke}, B.~F. and {Turner}, E.~L. and {Gunn}, J.~E. and {Hewitt}, J.~N. and {Schmidt}, M.},
        title = "{MG 1654+1346: an Einstein Ring Image of a Quasar Radio Lobe}",
      journal = {\aj},
         year = 1989,
        month = may,
       volume = {97},
        pages = {1283},
          doi = {10.1086/115071},
       adsurl = {https://ui.adsabs.harvard.edu/abs/1989AJ.....97.1283L}
}

@ARTICLE{1992AJ....104..968H,
       author = {{Hewitt}, J.~N. and {Turner}, E.~L. and {Lawrence}, C.~R. and {Schneider}, D.~P. and {Brody}, J.~P.},
        title = "{A Gravitational Lens Candidate With an Unusually Red Optical Counterpart}",
      journal = {\aj},
         year = 1992,
        month = sep,
       volume = {104},
        pages = {968},
          doi = {10.1086/116290},
       adsurl = {https://ui.adsabs.harvard.edu/abs/1992AJ....104..968H}
}

@ARTICLE{1993MNRAS.261..435P,
       author = {{Patnaik}, A.~R. and {Browne}, I.~W.~A. and {King}, L.~J. and {Muxlow}, T.~W.~B. and {Walsh}, D. and {Wilkinson}, P.~N.},
        title = "{B 0218+35.7 : a gravitationally lensed system with the smallest separation.}",
      journal = {\mnras},
         year = 1993,
        month = mar,
       volume = {261},
        pages = {435},
          doi = {10.1093/mnras/261.2.435},
       adsurl = {https://ui.adsabs.harvard.edu/abs/1993MNRAS.261..435P}
}

@ARTICLE{1992MNRAS.259P...1P,
       author = {{Patnaik}, A.~R. and {Browne}, I.~W.~A. and {Walsh}, D. and {Chaffee}, F.~H. and {Foltz}, C.~B.},
        title = "{B 1422+231 : a new gravitationally lensed system at Z = 3.62.}",
      journal = {\mnras},
         year = 1992,
        month = nov,
       volume = {259},
        pages = {1P-4},
          doi = {10.1093/mnras/259.1.1P},
       adsurl = {https://ui.adsabs.harvard.edu/abs/1992MNRAS.259P...1P}
}

@ARTICLE{1993AJ....105..847L,
       author = {{Lehar}, J. and {Langston}, G.~I. and {Silber}, A. and {Lawrence}, C.~R. and {Burke}, B.~F.},
        title = "{A Gravitationally Lensed Ring in MG 1549+3047}",
      journal = {\aj},
         year = 1993,
        month = mar,
       volume = {105},
        pages = {847},
          doi = {10.1086/116476},
       adsurl = {https://ui.adsabs.harvard.edu/abs/1993AJ....105..847L}
}

@ARTICLE{1995MNRAS.274L..25J,
       author = {{Jackson}, N. and {de Bruyn}, A.~G. and {Myers}, S. and {Bremer}, M.~N. and {Miley}, G.~K. and {Schilizzi}, R.~T. and {Browne}, I.~W.~A. and {Nair}, S. and {Wilkinson}, P.~N. and {Blandford}, R.~D. and {Pearson}, T.~J. and {Readhead}, A.~C.~S.},
        title = "{1600+434: a new gravitational lens system}",
      journal = {\mnras},
         year = 1995,
        month = may,
       volume = {274},
       number = {1},
        pages = {L25-L29},
          doi = {10.1093/mnras/274.1.L25},
       adsurl = {https://ui.adsabs.harvard.edu/abs/1995MNRAS.274L..25J}
}

@ARTICLE{1995ApJ...447L...5M,
       author = {{Myers}, S.~T. and {Fassnacht}, C.~D. and {Djorgovski}, S.~G. and {Blandford}, R.~D. and {Matthews}, K. and {Neugebauer}, G. and {Pearson}, T.~J. and {Readhead}, A.~C.~S. and {Smith}, J.~D. and {Thompson}, D.~J. and {Womble}, D.~S. and {Browne}, I.~W.~A. and {Wilkinson}, P.~N. and {Nair}, S. and {Jackson}, N. and {Snellen}, I.~A.~G. and {Miley}, G.~K. and {de Bruyn}, A.~G. and {Schilizzi}, R.~T.},
        title = "{1608+656: A Quadruple-Lens System Found in the CLASS Gravitational Lens Survey}",
      journal = {\apjl},
         year = 1995,
        month = jul,
       volume = {447},
        pages = {L5},
          doi = {10.1086/309556},
       adsurl = {https://ui.adsabs.harvard.edu/abs/1995ApJ...447L...5M}
}

@ARTICLE{2013MNRAS.434.3322D,
       author = {{Deane}, R.~P. and {Rawlings}, S. and {Garrett}, M.~A. and {Heywood}, I. and {Jarvis}, M.~J. and {Kl{\"o}ckner}, H. -R. and {Marshall}, P.~J. and {McKean}, J.~P.},
        title = "{The preferentially magnified active nucleus in IRAS F10214+4724 - III. VLBI observations of the radio core}",
      journal = {\mnras},
         year = 2013,
        month = oct,
       volume = {434},
       number = {4},
        pages = {3322-3336},
          doi = {10.1093/mnras/stt1241},
archivePrefix = {arXiv},
       eprint = {1307.6566},
 primaryClass = {astro-ph.CO},
       adsurl = {https://ui.adsabs.harvard.edu/abs/2013MNRAS.434.3322D}
}

@ARTICLE{1995ApJ...449L..29G,
       author = {{Graham}, James R. and {Liu}, Michael C.},
        title = "{High-Resolution Infrared Imaging of FSC 10214+4724: Evidence for Gravitational Lensing}",
      journal = {\apjl},
         year = 1995,
        month = aug,
       volume = {449},
        pages = {L29},
          doi = {10.1086/309629},
archivePrefix = {arXiv},
       eprint = {astro-ph/9506030},
 primaryClass = {astro-ph},
       adsurl = {https://ui.adsabs.harvard.edu/abs/1995ApJ...449L..29G}
}

@ARTICLE{1997AJ....114...48L,
       author = {{Lehar}, J. and {Burke}, B.~F. and {Conner}, S.~R. and {Falco}, E.~E. and {Fletcher}, A.~B. and {Irwin}, M. and {McMahon}, R.~G. and {Muslow}, T.~W.~B. and {Schechter}, P.~L.},
        title = "{The Gravitationally Lensed Radio Source MG 0751+2716}",
      journal = {\aj},
         year = 1997,
        month = jul,
       volume = {114},
        pages = {48-53},
          doi = {10.1086/118451},
archivePrefix = {arXiv},
       eprint = {astro-ph/9702191},
 primaryClass = {astro-ph},
       adsurl = {https://ui.adsabs.harvard.edu/abs/1997AJ....114...48L}
}

@ARTICLE{1997MNRAS.289..450K,
       author = {{King}, L.~J. and {Browne}, I.~W.~A. and {Muxlow}, T.~W.~B. and {Narasimha}, D. and {Patnaik}, A.~R. and {Porcas}, R.~W. and {Wilkinson}, P.~N.},
        title = "{Multifrequency radio observations of the gravitational lens system 1938 + 666}",
      journal = {\mnras},
         year = 1997,
        month = aug,
       volume = {289},
       number = {2},
        pages = {450-456},
          doi = {10.1093/mnras/289.2.450},
       adsurl = {https://ui.adsabs.harvard.edu/abs/1997MNRAS.289..450K}
}

@ARTICLE{2015MNRAS.454..287J,
       author = {{Jackson}, Neal and {Tagore}, Amitpal S. and {Roberts}, Carl and {Sluse}, Dominique and {Stacey}, Hannah and {Vives-Arias}, Hector and {Wucknitz}, Olaf and {Volino}, Filomena},
        title = "{Observations of radio-quiet quasars at 10-mas resolution by use of gravitational lensing}",
      journal = {\mnras},
         year = 2015,
        month = nov,
       volume = {454},
       number = {1},
        pages = {287-298},
          doi = {10.1093/mnras/stv1982},
archivePrefix = {arXiv},
       eprint = {1508.05842},
 primaryClass = {astro-ph.GA},
       adsurl = {https://ui.adsabs.harvard.edu/abs/2015MNRAS.454..287J}
}

@ARTICLE{1997AaA...317L..13B,
       author = {{Bade}, N. and {Siebert}, J. and {Lopez}, S. and {Voges}, W. and {Reimers}, D.},
        title = "{RX J0911.4+0551: A new multiple QSO selected from the ROSAT All-Sky Survey.}",
      journal = {\aap},
         year = 1997,
        month = jan,
       volume = {317},
        pages = {L13-L16},
       adsurl = {https://ui.adsabs.harvard.edu/abs/1997A&A...317L..13B}
}

@ARTICLE{1998MNRAS.296..483J,
       author = {{Jackson}, N. and {Nair}, S. and {Browne}, I.~W.~A. and {Wilkinson}, P.~N. and {Muxlow}, T.~W.~B. and {de Bruyn}, A.~G. and {Koopmans}, L. and {Bremer}, M. and {Snellen}, I. and {Miley}, G.~K. and {Schilizzi}, R.~T. and {Myers}, S. and {Fassnacht}, C.~D. and {Womble}, D.~S. and {Readhead}, A.~C.~S. and {Blandford}, R.~D. and {Pearson}, T.~J.},
        title = "{B0712+472: a new radio four-image gravitational lens}",
      journal = {\mnras},
         year = 1998,
        month = may,
       volume = {296},
       number = {3},
        pages = {483-490},
          doi = {10.1046/j.1365-8711.1998.01304.x},
       adsurl = {https://ui.adsabs.harvard.edu/abs/1998MNRAS.296..483J}
}

@ARTICLE{1998AJ....115.1371S,
       author = {{Schechter}, Paul L. and {Gregg}, Michael D. and {Becker}, Robert H. and {Helfand}, David J. and {White}, Richard L.},
        title = "{The First FIRST Gravitationally Lensed Quasar: FBQ 0951+2635}",
      journal = {\aj},
         year = 1998,
        month = apr,
       volume = {115},
       number = {4},
        pages = {1371-1376},
          doi = {10.1086/300294},
archivePrefix = {arXiv},
       eprint = {astro-ph/9710120},
 primaryClass = {astro-ph},
       adsurl = {https://ui.adsabs.harvard.edu/abs/1998AJ....115.1371S}
}

@ARTICLE{1998MNRAS.301..310S,
       author = {{Sykes}, C.~M. and {Browne}, I.~W.~A. and {Jackson}, N.~J. and {Marlow}, D.~R. and {Nair}, S. and {Wilkinson}, P.~N. and {Blandford}, R.~D. and {Cohen}, J. and {Fassnacht}, C.~D. and {Hogg}, D. and {Pearson}, T.~J. and {Readhead}, A.~C.~S. and {Womble}, D.~S. and {Myers}, S.~T. and {De Bruyn}, A.~G. and {Bremer}, M. and {Miley}, G.~K. and {Schilizzi}, R.~T.},
        title = "{The complex gravitational lens system B1933+503}",
      journal = {\mnras},
         year = 1998,
        month = dec,
       volume = {301},
       number = {2},
        pages = {310-314},
          doi = {10.1046/j.1365-8711.1998.02081.x},
archivePrefix = {arXiv},
       eprint = {astro-ph/9710358},
 primaryClass = {astro-ph},
       adsurl = {https://ui.adsabs.harvard.edu/abs/1998MNRAS.301..310S}
}

@ARTICLE{1999AJ....118.1922I,
       author = {{Ibata}, Rodrigo A. and {Lewis}, Geraint F. and {Irwin}, Michael J. and {Leh{\'a}r}, Joseph and {Totten}, Edward J.},
        title = "{NICMOS and VLA Observations of the Gravitationally Lensed Ultraluminous BAL Quasar APM 08279+5255: Detection of a Third Image}",
      journal = {\aj},
         year = 1999,
        month = nov,
       volume = {118},
       number = {5},
        pages = {1922-1930},
          doi = {10.1086/301111},
archivePrefix = {arXiv},
       eprint = {astro-ph/9908052},
 primaryClass = {astro-ph},
       adsurl = {https://ui.adsabs.harvard.edu/abs/1999AJ....118.1922I}
}

@ARTICLE{1998ApJ...505..529I,
       author = {{Irwin}, Michael J. and {Ibata}, Rodrigo A. and {Lewis}, Geraint F. and {Totten}, Edward J.},
        title = "{APM 08279+5255: an Ultraluminous Broad Absorption Line Quasar at a Redshift Z = 3.87}",
      journal = {\apj},
         year = 1998,
        month = oct,
       volume = {505},
       number = {2},
        pages = {529-535},
          doi = {10.1086/306213},
archivePrefix = {arXiv},
       eprint = {astro-ph/9806171},
 primaryClass = {astro-ph},
       adsurl = {https://ui.adsabs.harvard.edu/abs/1998ApJ...505..529I}
}

@ARTICLE{1998MNRAS.300..649X,
       author = {{Xanthopoulos}, E. and {Browne}, I.~W.~A. and {King}, L.~J. and {Koopmans}, L.~V.~E. and {Jackson}, N.~J. and {Marlow}, D.~R. and {Patnaik}, A.~R. and {Porcas}, R.~W. and {Wilkinson}, P.~N.},
        title = "{The new gravitational lens system B1030+074}",
      journal = {\mnras},
         year = 1998,
        month = nov,
       volume = {300},
       number = {3},
        pages = {649-655},
          doi = {10.1046/j.1365-8711.1998.01804.x},
archivePrefix = {arXiv},
       eprint = {astro-ph/9802014},
 primaryClass = {astro-ph},
       adsurl = {https://ui.adsabs.harvard.edu/abs/1998MNRAS.300..649X}
}

@ARTICLE{1999MNRAS.303..727K,
       author = {{Koopmans}, L.~V.~E. and {de Bruyn}, A.~G. and {Marlow}, D.~R. and {Jackson}, N. and {Blandford}, R.~D. and {Browne}, I.~W.~A. and {Fassnacht}, C.~D. and {Myers}, S.~T. and {Pearson}, T.~J. and {Readhead}, A.~C.~S. and {Wilkinson}, P.~N. and {Womble}, D.},
        title = "{A new radio double lens from CLASS: B1127+385}",
      journal = {\mnras},
         year = 1999,
        month = mar,
       volume = {303},
       number = {4},
        pages = {727-735},
          doi = {10.1046/j.1365-8711.1999.02342.x},
archivePrefix = {arXiv},
       eprint = {astro-ph/9811425},
 primaryClass = {astro-ph},
       adsurl = {https://ui.adsabs.harvard.edu/abs/1999MNRAS.303..727K}
}

@ARTICLE{1999AJ....117.2565M,
       author = {{Myers}, S.~T. and {Rusin}, D. and {Fassnacht}, C.~D. and {Blandford}, R.~D. and {Pearson}, T.~J. and {Readhead}, A.~C.~S. and {Jackson}, N. and {Browne}, I.~W.~A. and {Marlow}, D.~R. and {Wilkinson}, P.~N. and {Koopmans}, L.~V.~E. and {de Bruyn}, A.~G.},
        title = "{CLASS B1152+199 and B1359+154: Two New Gravitational Lens Systems Discovered in the Cosmic Lens All-Sky Survey}",
      journal = {\aj},
         year = 1999,
        month = jun,
       volume = {117},
       number = {6},
        pages = {2565-2572},
          doi = {10.1086/300875},
archivePrefix = {arXiv},
       eprint = {astro-ph/9905043},
 primaryClass = {astro-ph},
       adsurl = {https://ui.adsabs.harvard.edu/abs/1999AJ....117.2565M}
}

@ARTICLE{1999AJ....118..654M,
       author = {{Marlow}, D.~R. and {Myers}, S.~T. and {Rusin}, D. and {Jackson}, N. and {Browne}, I.~W.~A. and {Wilkinson}, P.~N. and {Muxlow}, T. and {Fassnacht}, C.~D. and {Lubin}, L. and {Kundi{\'c}}, T. and {Blandford}, R.~D. and {Pearson}, T.~J. and {Readhead}, A.~C.~S. and {Koopmans}, L. and {de Bruyn}, A.~G.},
        title = "{CLASS B1555+375: A New Four-Image Gravitational Lens System}",
      journal = {\aj},
         year = 1999,
        month = aug,
       volume = {118},
       number = {2},
        pages = {654-658},
          doi = {10.1086/300987},
       adsurl = {https://ui.adsabs.harvard.edu/abs/1999AJ....118..654M}
}

@ARTICLE{1999AJ....117..658F,
       author = {{Fassnacht}, C.~D. and {Blandford}, R.~D. and {Cohen}, J.~G. and {Matthews}, K. and {Pearson}, T.~J. and {Readhead}, A.~C.~S. and {Womble}, D.~S. and {Myers}, S.~T. and {Browne}, I.~W.~A. and {Jackson}, N.~J. and {Marlow}, D.~R. and {Wilkinson}, P.~N. and {Koopmans}, L.~V.~E. and {de Bruyn}, A.~G. and {Schilizzi}, R.~T. and {Bremer}, M. and {Miley}, G.},
        title = "{B2045+265: A New Four-Image Gravitational Lens from CLASS}",
      journal = {\aj},
         year = 1999,
        month = feb,
       volume = {117},
       number = {2},
        pages = {658-670},
          doi = {10.1086/300724},
archivePrefix = {arXiv},
       eprint = {astro-ph/9811167},
 primaryClass = {astro-ph},
       adsurl = {https://ui.adsabs.harvard.edu/abs/1999AJ....117..658F}
}

@ARTICLE{1999MNRAS.307..225K,
       author = {{King}, Lindsay J. and {Browne}, Ian W.~A. and {Marlow}, Daniel R. and {Patnaik}, Alok R. and {Wilkinson}, Peter N.},
        title = "{Gravitationally lensed radio sources in the Jodrell Bank-VLA Astrometric Survey}",
      journal = {\mnras},
         year = 1999,
        month = aug,
       volume = {307},
       number = {2},
        pages = {225-235},
          doi = {10.1046/j.1365-8711.1999.02328.x},
       adsurl = {https://ui.adsabs.harvard.edu/abs/1999MNRAS.307..225K}
}

@ARTICLE{2000MNRAS.319L...7P,
       author = {{Phillips}, P.~M. and {Norbury}, M.~A. and {Koopmans}, L.~V.~E. and {Browne}, I.~W.~A. and {Jackson}, N.~J. and {Wilkinson}, P.~N. and {Biggs}, A.~D. and {Blandford}, R.~D. and {de Bruyn}, A.~G. and {Fassnacht}, C.~D. and {Helbig}, P. and {Mao}, S. and {Marlow}, D.~R. and {Myers}, S.~T. and {Pearson}, T.~J. and {Readhead}, A.~C.~S. and {Rusin}, D. and {Xanthopoulos}, E.},
        title = "{A new quadruple gravitational lens system: CLASS B0128+437}",
      journal = {\mnras},
         year = 2000,
        month = dec,
       volume = {319},
       number = {2},
        pages = {L7-L11},
          doi = {10.1046/j.1365-8711.2000.04033.x},
archivePrefix = {arXiv},
       eprint = {astro-ph/0009334},
 primaryClass = {astro-ph},
       adsurl = {https://ui.adsabs.harvard.edu/abs/2000MNRAS.319L...7P}
}

@ARTICLE{2000AJ....120.2868W,
       author = {{Winn}, Joshua N. and {Hewitt}, Jacqueline N. and {Schechter}, Paul L. and {Dressler}, Alan and {Falco}, E.~E. and {Impey}, C.~D. and {Kochanek}, C.~S. and {Leh{\'a}r}, J. and {Lovell}, J.~E.~J. and {McLeod}, B.~A. and {Morgan}, Nicholas D. and {Mu{\~n}oz}, J.~A. and {Rix}, H. -W. and {Ruiz}, Maria Teresa},
        title = "{PMN J1838-3427: A New Gravitationally Lensed Quasar}",
      journal = {\aj},
         year = 2000,
        month = dec,
       volume = {120},
       number = {6},
        pages = {2868-2878},
          doi = {10.1086/316874},
archivePrefix = {arXiv},
       eprint = {astro-ph/0008036},
 primaryClass = {astro-ph},
       adsurl = {https://ui.adsabs.harvard.edu/abs/2000AJ....120.2868W}
}

@ARTICLE{2001AJ....121..619M,
       author = {{Marlow}, D.~R. and {Rusin}, D. and {Norbury}, M. and {Jackson}, N. and {Browne}, I.~W.~A. and {Wilkinson}, P.~N. and {Fassnacht}, C.~D. and {Myers}, S.~T. and {Koopmans}, L.~V.~E. and {Blandford}, R.~D. and {Pearson}, T.~J. and {Readhead}, A.~C.~S. and {de Bruyn}, A.~G.},
        title = "{CLASS B0739+366: A New Two-Image Gravitational Lens System}",
      journal = {\aj},
         year = 2001,
        month = feb,
       volume = {121},
       number = {2},
        pages = {619-624},
          doi = {10.1086/318735},
archivePrefix = {arXiv},
       eprint = {astro-ph/0008037},
 primaryClass = {astro-ph},
       adsurl = {https://ui.adsabs.harvard.edu/abs/2001AJ....121..619M}
}

@ARTICLE{2001ApJ...547...60L,
       author = {{Leh{\'a}r}, J. and {Buchalter}, A. and {McMahon}, R.~G. and {Kochanek}, C.~S. and {Muxlow}, T.~W.~B.},
        title = "{An Efficient Search for Gravitationally Lensed Radio Lobes}",
      journal = {\apj},
         year = 2001,
        month = jan,
       volume = {547},
       number = {1},
        pages = {60-76},
          doi = {10.1086/318367},
archivePrefix = {arXiv},
       eprint = {astro-ph/0007458},
 primaryClass = {astro-ph},
       adsurl = {https://ui.adsabs.harvard.edu/abs/2001ApJ...547...60L}
}

@ARTICLE{2001AJ....121.1223W,
       author = {{Winn}, Joshua N. and {Hewitt}, Jacqueline N. and {Patnaik}, Alok R. and {Schechter}, Paul L. and {Schommer}, Robert A. and {L{\'o}pez}, Sebastian and {Maza}, Jos{\'e} and {Wachter}, Stefanie},
        title = "{A Nearly Symmetric Double-Image Gravitational Lens}",
      journal = {\aj},
         year = 2001,
        month = mar,
       volume = {121},
       number = {3},
        pages = {1223-1231},
          doi = {10.1086/319403},
archivePrefix = {arXiv},
       eprint = {astro-ph/0009451},
 primaryClass = {astro-ph},
       adsurl = {https://ui.adsabs.harvard.edu/abs/2001AJ....121.1223W}
}

@ARTICLE{2001AJ....122..591R,
       author = {{Rusin}, D. and {Marlow}, D.~R. and {Norbury}, M. and {Browne}, I.~W.~A. and {Jackson}, N. and {Wilkinson}, P.~N. and {Fassnacht}, C.~D. and {Myers}, S.~T. and {Koopmans}, L.~V.~E. and {Blandford}, R.~D. and {Pearson}, T.~J. and {Readhead}, A.~C.~S. and {de Bruyn}, A.~G.},
        title = "{The New Two-Image Gravitational Lens System CLASS B2319+051}",
      journal = {\aj},
         year = 2001,
        month = aug,
       volume = {122},
       number = {2},
        pages = {591-597},
          doi = {10.1086/321156},
archivePrefix = {arXiv},
       eprint = {astro-ph/0104399},
 primaryClass = {astro-ph},
       adsurl = {https://ui.adsabs.harvard.edu/abs/2001AJ....122..591R}
}

@ARTICLE{2002ApJ...564..143W,
       author = {{Winn}, Joshua N. and {Lovell}, James E.~J. and {Chen}, Hsiao-Wen and {Fletcher}, Andr{\'e} B. and {Hewitt}, Jacqueline N. and {Patnaik}, Alok R. and {Schechter}, Paul L.},
        title = "{PMN J0134-0931: A Gravitationally Lensed Quasar with Unusual Radio Morphology}",
      journal = {\apj},
         year = 2002,
        month = jan,
       volume = {564},
       number = {1},
        pages = {143-152},
          doi = {10.1086/324144},
archivePrefix = {arXiv},
       eprint = {astro-ph/0107435},
 primaryClass = {astro-ph},
       adsurl = {https://ui.adsabs.harvard.edu/abs/2002ApJ...564..143W}
}

@ARTICLE{2003MNRAS.338..957A,
       author = {{Argo}, M.~K. and {Jackson}, N.~J. and {Browne}, I.~W.~A. and {York}, T. and {McKean}, J.~P. and {Biggs}, A.~D. and {Blandford}, R.~D. and {de Bruyn}, A.~G. and {Chae}, K.~H. and {Fassnacht}, C.~D. and {Koopmans}, L.~V.~E. and {Myers}, S.~T. and {Norbury}, M. and {Pearson}, T.~J. and {Phillips}, P.~M. and {Readhead}, A.~C.~S. and {Rusin}, D. and {Wilkinson}, P.~N.},
        title = "{CLASS B0445+123: a new two-image gravitational lens system}",
      journal = {\mnras},
         year = 2003,
        month = feb,
       volume = {338},
       number = {4},
        pages = {957-961},
          doi = {10.1046/j.1365-8711.2003.06138.x},
archivePrefix = {arXiv},
       eprint = {astro-ph/0210234},
 primaryClass = {astro-ph},
       adsurl = {https://ui.adsabs.harvard.edu/abs/2003MNRAS.338..957A}
}

@ARTICLE{2002AJ....123.2925L,
       author = {{Lacy}, Mark and {Gregg}, Michael and {Becker}, Robert H. and {White}, Richard L. and {Glikman}, Eilat and {Helfand}, David and {Winn}, Joshua N.},
        title = "{The Reddest Quasars. II. A Gravitationally Lensed FeLoBAL Quasar}",
      journal = {\aj},
         year = 2002,
        month = jun,
       volume = {123},
       number = {6},
        pages = {2925-2935},
          doi = {10.1086/340568},
archivePrefix = {arXiv},
       eprint = {astro-ph/0203065},
 primaryClass = {astro-ph},
       adsurl = {https://ui.adsabs.harvard.edu/abs/2002AJ....123.2925L}
}

@ARTICLE{2002AJ....123...10W,
       author = {{Winn}, Joshua N. and {Morgan}, Nicholas D. and {Hewitt}, Jacqueline N. and {Kochanek}, Christopher S. and {Lovell}, James E.~J. and {Patnaik}, Alok R. and {Pindor}, Bart and {Schechter}, Paul L. and {Schommer}, Robert A.},
        title = "{PMN J1632-0033: A New Gravitationally Lensed Quasar}",
      journal = {\aj},
         year = 2002,
        month = jan,
       volume = {123},
       number = {1},
        pages = {10-19},
          doi = {10.1086/338094},
archivePrefix = {arXiv},
       eprint = {astro-ph/0104092},
 primaryClass = {astro-ph},
       adsurl = {https://ui.adsabs.harvard.edu/abs/2002AJ....123...10W}
}

@ARTICLE{2002AaA...382L..26R,
       author = {{Reimers}, D. and {Hagen}, H. -J. and {Baade}, R. and {Lopez}, S. and {Tytler}, D.},
        title = "{Discovery of a new quadruply lensed QSO: HS 0810+2554 - A brighter twin to PG 1115+080}",
      journal = {\aap},
         year = 2002,
        month = jan,
       volume = {382},
        pages = {L26-L28},
          doi = {10.1051/0004-6361:20011798},
       adsurl = {https://ui.adsabs.harvard.edu/abs/2002A&A...382L..26R}
}

@ARTICLE{2002AaA...395...17W,
       author = {{Wisotzki}, L. and {Schechter}, P.~L. and {Bradt}, H.~V. and {Heinm{\"u}ller}, J. and {Reimers}, D.},
        title = "{HE 0435-1223: A wide separation quadruple QSO and gravitational lens}",
      journal = {\aap},
         year = 2002,
        month = nov,
       volume = {395},
        pages = {17-23},
          doi = {10.1051/0004-6361:20021213},
archivePrefix = {arXiv},
       eprint = {astro-ph/0207062},
 primaryClass = {astro-ph},
       adsurl = {https://ui.adsabs.harvard.edu/abs/2002A&A...395...17W}
}

@ARTICLE{2003MNRAS.341...13B,
       author = {{Browne}, I.~W.~A. and {Wilkinson}, P.~N. and {Jackson}, N.~J.~F. and {Myers}, S.~T. and {Fassnacht}, C.~D. and {Koopmans}, L.~V.~E. and {Marlow}, D.~R. and {Norbury}, M. and {Rusin}, D. and {Sykes}, C.~M. and {Biggs}, A.~D. and {Blandford}, R.~D. and {de Bruyn}, A.~G. and {Chae}, K. -H. and {Helbig}, P. and {King}, L.~J. and {McKean}, J.~P. and {Pearson}, T.~J. and {Phillips}, P.~M. and {Readhead}, A.~C.~S. and {Xanthopoulos}, E. and {York}, T.},
        title = "{The Cosmic Lens All-Sky Survey - II. Gravitational lens candidate selection and follow-up}",
      journal = {\mnras},
         year = 2003,
        month = may,
       volume = {341},
       number = {1},
        pages = {13-32},
          doi = {10.1046/j.1365-8711.2003.06257.x},
archivePrefix = {arXiv},
       eprint = {astro-ph/0211069},
 primaryClass = {astro-ph},
       adsurl = {https://ui.adsabs.harvard.edu/abs/2003MNRAS.341...13B}
}

@ARTICLE{2003MNRAS.338.1084B,
       author = {{Biggs}, A.~D. and {Rusin}, D. and {Browne}, I.~W.~A. and {de Bruyn}, A.~G. and {Jackson}, N.~J. and {Koopmans}, L.~V.~E. and {McKean}, J.~P. and {Myers}, S.~T. and {Blandford}, R.~D. and {Chae}, K. -H. and {Fassnacht}, C.~D. and {Norbury}, M.~A. and {Pearson}, T.~J. and {Phillips}, P.~M. and {Readhead}, A.~C.~S. and {Wilkinson}, P.~N.},
        title = "{B0850+054: a new gravitational lens system from CLASS}",
      journal = {\mnras},
         year = 2003,
        month = feb,
       volume = {338},
       number = {4},
        pages = {1084-1088},
          doi = {10.1046/j.1365-8711.2003.06159.x},
archivePrefix = {arXiv},
       eprint = {astro-ph/0210504},
 primaryClass = {astro-ph},
       adsurl = {https://ui.adsabs.harvard.edu/abs/2003MNRAS.338.1084B}
}

@ARTICLE{2003Natur.426..810I,
       author = {{Inada}, Naohisa and {Oguri}, Masamune and {Pindor}, Bartosz and {Hennawi}, Joseph F. and {Chiu}, Kuenley and {Zheng}, Wei and {Ichikawa}, Shin-Ichi and {Gregg}, Michael D. and {Becker}, Robert H. and {Suto}, Yasushi and {Strauss}, Michael A. and {Turner}, Edwin L. and {Keeton}, Charles R. and {Annis}, James and {Castander}, Francisco J. and {Eisenstein}, Daniel J. and {Frieman}, Joshua A. and {Fukugita}, Masataka and {Gunn}, James E. and {Johnston}, David E. and {Kent}, Stephen M. and {Nichol}, Robert C. and {Richards}, Gordon T. and {Rix}, Hans-Walter and {Sheldon}, Erin Scott and {Bahcall}, Neta A. and {Brinkmann}, J. and {Ivezi{\'c}}, {\v{Z}}eljko and {Lamb}, Don Q. and {McKay}, Timothy A. and {Schneider}, Donald P. and {York}, Donald G.},
        title = "{A gravitationally lensed quasar with quadruple images separated by 14.62arcseconds}",
      journal = {\nat},
         year = 2003,
        month = dec,
       volume = {426},
       number = {6968},
        pages = {810-812},
          doi = {10.1038/nature02153},
archivePrefix = {arXiv},
       eprint = {astro-ph/0312427},
 primaryClass = {astro-ph},
       adsurl = {https://ui.adsabs.harvard.edu/abs/2003Natur.426..810I}
}

@article{Wucknitz:2009xu,
  author = "Wucknitz, Olaf",
  title = "{The gravitational lens J1131-1231 --- How to miss an opportunity and how to avoid it }",
  doi = "10.22323/1.072.0102",
  journal = "PoS",
  year = 2009,
  volume = "IX EVN Symposium",
  pages = "102"
}

@ARTICLE{2003AaA...406L..43S,
       author = {{Sluse}, D. and {Surdej}, J. and {Claeskens}, J. -F. and {Hutsem{\'e}kers}, D. and {Jean}, C. and {Courbin}, F. and {Nakos}, T. and {Billeres}, M. and {Khmil}, S.~V.},
        title = "{A quadruply imaged quasar with an optical Einstein ring candidate: 1RXS J113155.4-123155}",
      journal = {\aap},
         year = 2003,
        month = jul,
       volume = {406},
        pages = {L43-L46},
          doi = {10.1051/0004-6361:20030904},
archivePrefix = {arXiv},
       eprint = {astro-ph/0307345},
 primaryClass = {astro-ph},
       adsurl = {https://ui.adsabs.harvard.edu/abs/2003A&A...406L..43S}
}

@ARTICLE{2011ApJ...739L..28J,
       author = {{Jackson}, N.},
        title = "{The Faintest Radio Source Yet: Expanded Very Large Array Observations of the Gravitational Lens SDSS J1004+4112}",
      journal = {\apjl},
         year = 2011,
        month = sep,
       volume = {739},
       number = {1},
          eid = {L28},
        pages = {L28},
          doi = {10.1088/2041-8205/739/1/L28},
archivePrefix = {arXiv},
       eprint = {1105.2141},
 primaryClass = {astro-ph.CO},
       adsurl = {https://ui.adsabs.harvard.edu/abs/2011ApJ...739L..28J}
}

@ARTICLE{2003AJ....126..666I,
       author = {{Inada}, Naohisa and {Becker}, Robert H. and {Burles}, Scott and {Castander}, Francisco J. and {Eisenstein}, Daniel and {Hall}, Patrick B. and {Johnston}, David E. and {Pindor}, Bartosz and {Richards}, Gordon T. and {Schechter}, Paul L. and {Sekiguchi}, Maki and {White}, Richard L. and {Brinkmann}, J. and {Frieman}, Joshua A. and {Kleinman}, S.~J. and {Krzesi{\'n}ski}, Jurek and {Long}, Daniel C. and {Neilsen}, Eric H., Jr. and {Newman}, Peter R. and {Nitta}, Atsuko and {Schneider}, Donald P. and {Snedden}, S. and {York}, Donald G.},
        title = "{SDSS J092455.87+021924.9: An Interesting Gravitationally Lensed Quasar from the Sloan Digital Sky Survey}",
      journal = {\aj},
         year = 2003,
        month = aug,
       volume = {126},
       number = {2},
        pages = {666-674},
          doi = {10.1086/375906},
archivePrefix = {arXiv},
       eprint = {astro-ph/0304377},
 primaryClass = {astro-ph},
       adsurl = {https://ui.adsabs.harvard.edu/abs/2003AJ....126..666I}
}

@ARTICLE{2005AJ....130.1977H,
       author = {{Haarsma}, Deborah B. and {Winn}, Joshua N. and {Falco}, Emilio E. and {Kochanek}, Christopher S. and {Ammar}, Philip and {Boersma}, Catherine and {Fogwell}, Shannon and {Muxlow}, T.~W.~B. and {McLeod}, Brian A. and {Leh{\'a}r}, Joseph},
        title = "{The FIRST-Optical-VLA Survey for Lensed Radio Lobes}",
      journal = {\aj},
         year = 2005,
        month = nov,
       volume = {130},
       number = {5},
        pages = {1977-1995},
          doi = {10.1086/466513},
archivePrefix = {arXiv},
       eprint = {astro-ph/0507295},
 primaryClass = {astro-ph},
       adsurl = {https://ui.adsabs.harvard.edu/abs/2005AJ....130.1977H}
}

@ARTICLE{2006AJ....131....1H,
       author = {{Hennawi}, Joseph F. and {Strauss}, Michael A. and {Oguri}, Masamune and {Inada}, Naohisa and {Richards}, Gordon T. and {Pindor}, Bartosz and {Schneider}, Donald P. and {Becker}, Robert H. and {Gregg}, Michael D. and {Hall}, Patrick B. and {Johnston}, David E. and {Fan}, Xiaohui and {Burles}, Scott and {Schlegel}, David J. and {Gunn}, James E. and {Lupton}, Robert H. and {Bahcall}, Neta A. and {Brunner}, Robert J. and {Brinkmann}, Jon},
        title = "{Binary Quasars in the Sloan Digital Sky Survey: Evidence for Excess Clustering on Small Scales}",
      journal = {\aj},
         year = 2006,
        month = jan,
       volume = {131},
       number = {1},
        pages = {1-23},
          doi = {10.1086/498235},
archivePrefix = {arXiv},
       eprint = {astro-ph/0504535},
 primaryClass = {astro-ph},
       adsurl = {https://ui.adsabs.harvard.edu/abs/2006AJ....131....1H}
}

@ARTICLE{2007MNRAS.381L..55B,
       author = {{Boyce}, E.~R. and {Myers}, S.~T. and {Browne}, I.~W.~A. and {Stroman}, W.~J. and {Jackson}, N.~J.},
        title = "{J0316+4328: a probable `asymmetric double' lens}",
      journal = {\mnras},
         year = 2007,
        month = oct,
       volume = {381},
       number = {1},
        pages = {L55-L59},
          doi = {10.1111/j.1745-3933.2007.00365.x},
archivePrefix = {arXiv},
       eprint = {0707.2679},
 primaryClass = {astro-ph},
       adsurl = {https://ui.adsabs.harvard.edu/abs/2007MNRAS.381L..55B}
}

@ARTICLE{2008ApJ...686..851R,
       author = {{Riechers}, Dominik A. and {Walter}, Fabian and {Brewer}, Brendon J. and {Carilli}, Christopher L. and {Lewis}, Geraint F. and {Bertoldi}, Frank and {Cox}, Pierre},
        title = "{A Molecular Einstein Ring at z = 4.12: Imaging the Dynamics of a Quasar Host Galaxy Through a Cosmic Lens}",
      journal = {\apj},
         year = 2008,
        month = oct,
       volume = {686},
       number = {2},
        pages = {851-858},
          doi = {10.1086/591434},
archivePrefix = {arXiv},
       eprint = {0806.4616},
 primaryClass = {astro-ph},
       adsurl = {https://ui.adsabs.harvard.edu/abs/2008ApJ...686..851R}
}

@ARTICLE{2021MNRAS.508L..64M,
       author = {{Mangat}, C.~S. and {McKean}, J.~P. and {Brilenkov}, R. and {Hartley}, P. and {Stacey}, H.~R. and {Vegetti}, S. and {Wen}, D.},
        title = "{PS J1721+8842: a gravitationally lensed dual AGN system at redshift 2.37 with two radio components}",
      journal = {\mnras},
         year = 2021,
        month = nov,
       volume = {508},
       number = {1},
        pages = {L64-L68},
          doi = {10.1093/mnrasl/slab106},
archivePrefix = {arXiv},
       eprint = {2109.03253},
 primaryClass = {astro-ph.GA},
       adsurl = {https://ui.adsabs.harvard.edu/abs/2021MNRAS.508L..64M}
}

@ARTICLE{2018AaA...616L..11K,
       author = {{Krone-Martins}, A. and {Delchambre}, L. and {Wertz}, O. and {Ducourant}, C. and {Mignard}, F. and {Teixeira}, R. and {Kl{\"u}ter}, J. and {Le Campion}, J. -F. and {Galluccio}, L. and {Surdej}, J. and {Bastian}, U. and {Wambsganss}, J. and {Graham}, M.~J. and {Djorgovski}, S.~G. and {Slezak}, E.},
        title = "{Gaia GraL: Gaia DR2 gravitational lens systems. I. New quadruply imaged quasar candidates around known quasars}",
      journal = {\aap},
         year = 2018,
        month = aug,
       volume = {616},
          eid = {L11},
        pages = {L11},
          doi = {10.1051/0004-6361/201833337},
archivePrefix = {arXiv},
       eprint = {1804.11051},
 primaryClass = {astro-ph.GA},
       adsurl = {https://ui.adsabs.harvard.edu/abs/2018A&A...616L..11K}
}

@ARTICLE{2018MNRAS.479.4345A,
       author = {{Agnello}, A. and {Lin}, H. and {Kuropatkin}, N. and {Buckley-Geer}, E. and {Anguita}, T. and {Schechter}, P.~L. and {Morishita}, T. and {Motta}, V. and {Rojas}, K. and {Treu}, T. and {Amara}, A. and {Auger}, M.~W. and {Courbin}, F. and {Fassnacht}, C.~D. and {Frieman}, J. and {More}, A. and {Marshall}, P.~J. and {McMahon}, R.~G. and {Meylan}, G. and {Suyu}, S.~H. and {Glazebrook}, K. and {Morgan}, N. and {Nord}, B. and {Abbott}, T.~M.~C. and {Abdalla}, F.~B. and {Annis}, J. and {Bechtol}, K. and {Benoit-L{\'e}vy}, A. and {Bertin}, E. and {Bernstein}, R.~A. and {Brooks}, D. and {Burke}, D.~L. and {Rosell}, A. Carnero and {Carretero}, J. and {Cunha}, C.~E. and {D'Andrea}, C.~B. and {da Costa}, L.~N. and {Desai}, S. and {Drlica-Wagner}, A. and {Eifler}, T.~F. and {Flaugher}, B. and {Garc{\'\i}a-Bellido}, J. and {Gaztanaga}, E. and {Gerdes}, D.~W. and {Gruen}, D. and {Gruendl}, R.~A. and {Gschwend}, J. and {Gutierrez}, G. and {Honscheid}, K. and {James}, D.~J. and {Kuehn}, K. and {Lahav}, O. and {Lima}, M. and {Maia}, M.~A.~G. and {March}, M. and {Menanteau}, F. and {Miquel}, R. and {Ogando}, R.~L.~C. and {Plazas}, A.~A. and {Sanchez}, E. and {Scarpine}, V. and {Schindler}, R. and {Schubnell}, M. and {Sevilla-Noarbe}, I. and {Smith}, M. and {Soares-Santos}, M. and {Sobreira}, F. and {Suchyta}, E. and {Swanson}, M.~E.~C. and {Tarle}, G. and {Tucker}, D. and {Wechsler}, R.},
        title = "{DES meets Gaia: discovery of strongly lensed quasars from a multiplet search}",
      journal = {\mnras},
         year = 2018,
        month = oct,
       volume = {479},
       number = {4},
        pages = {4345-4354},
          doi = {10.1093/mnras/sty1419},
archivePrefix = {arXiv},
       eprint = {1711.03971},
 primaryClass = {astro-ph.CO},
       adsurl = {https://ui.adsabs.harvard.edu/abs/2018MNRAS.479.4345A}
}

@ARTICLE{2019MNRAS.483.2125S,
       author = {{Spingola}, C. and {McKean}, J.~P. and {Lee}, M. and {Deller}, A. and {Moldon}, J.},
        title = "{A novel search for gravitationally lensed radio sources in wide-field VLBI imaging from the mJIVE-20 survey}",
      journal = {\mnras},
         year = 2019,
        month = feb,
       volume = {483},
       number = {2},
        pages = {2125-2153},
          doi = {10.1093/mnras/sty3189},
archivePrefix = {arXiv},
       eprint = {1811.09152},
 primaryClass = {astro-ph.GA},
       adsurl = {https://ui.adsabs.harvard.edu/abs/2019MNRAS.483.2125S}
}

@ARTICLE{2020MNRAS.494.3491L,
       author = {{Lemon}, C. and {Auger}, M.~W. and {McMahon}, R. and {Anguita}, T. and {Apostolovski}, Y. and {Chen}, G.~C. -F. and {Fassnacht}, C.~D. and {Melo}, A.~D. and {Motta}, V. and {Shajib}, A. and {Treu}, T. and {Agnello}, A. and {Buckley-Geer}, E. and {Schechter}, P.~L. and {Birrer}, S. and {Collett}, T. and {Courbin}, F. and {Rusu}, C.~E. and {Abbott}, T.~M.~C. and {Allam}, S. and {Annis}, J. and {Avila}, S. and {Bertin}, E. and {Brooks}, D. and {Burke}, D.~L. and {Carnero Rosell}, A. and {Carrasco Kind}, M. and {Carretero}, J. and {Costanzi}, M. and {da Costa}, L.~N. and {De Vicente}, J. and {Desai}, S. and {Eifler}, T.~F. and {Flaugher}, B. and {Frieman}, J. and {Garc{\'\i}a-Bellido}, J. and {Gaztanaga}, E. and {Gerdes}, D.~W. and {Gruen}, D. and {Gruendl}, R.~A. and {Gschwend}, J. and {Gutierrez}, G. and {Honscheid}, K. and {James}, D.~J. and {Kim}, A. and {Krause}, E. and {Kuehn}, K. and {Kuropatkin}, N. and {Lahav}, O. and {Lima}, M. and {Lin}, H. and {Maia}, M.~A.~G. and {March}, M. and {Marshall}, J.~L. and {Menanteau}, F. and {Miquel}, R. and {Palmese}, A. and {Paz-Chinch{\'o}n}, F. and {Plazas}, A.~A. and {Roodman}, A. and {Sanchez}, E. and {Schubnell}, M. and {Serrano}, S. and {Smith}, M. and {Soares-Santos}, M. and {Suchyta}, E. and {Tarle}, G. and {Walker}, A.~R.},
        title = "{The STRong lensing Insights into the Dark Energy Survey (STRIDES) 2017/2018 follow-up campaign: discovery of 10 lensed quasars and 10 quasar pairs}",
      journal = {\mnras},
         year = 2020,
        month = may,
       volume = {494},
       number = {3},
        pages = {3491-3511},
          doi = {10.1093/mnras/staa652},
archivePrefix = {arXiv},
       eprint = {1912.09133},
 primaryClass = {astro-ph.GA},
       adsurl = {https://ui.adsabs.harvard.edu/abs/2020MNRAS.494.3491L}
}

@ARTICLE{2019arXiv191208977K,
       author = {{Krone-Martins}, A. and {Graham}, M.~J. and {Stern}, D. and {Djorgovski}, S.~G. and {Delchambre}, L. and {Ducourant}, C. and {Teixeira}, R. and {Drake}, A.~J. and {Scarano}, S., Jr. and {Surdej}, J. and {Galluccio}, L. and {Jalan}, P. and {Wertz}, O. and {Kl{\"u}ter}, J. and {Mignard}, F. and {Spindola-Duarte}, C. and {Dobie}, D. and {Slezak}, E. and {Sluse}, D. and {Murphy}, T. and {Boehm}, C. and {Nierenberg}, A.~M. and {Bastian}, U. and {Wambsganss}, J. and {LeCampion}, J. -F.},
        title = "{Gaia GraL: Gaia DR2 Gravitational Lens Systems. V. Doubly-imaged QSOs discovered from entropy and wavelets}",
      journal = {arXiv e-prints},
         year = 2019,
        month = dec,
          eid = {arXiv:1912.08977},
        pages = {arXiv:1912.08977},
          doi = {10.48550/arXiv.1912.08977},
archivePrefix = {arXiv},
       eprint = {1912.08977},
 primaryClass = {astro-ph.GA},
       adsurl = {https://ui.adsabs.harvard.edu/abs/2019arXiv191208977K}
}

@ARTICLE{2019AaA...622A.165D,
       author = {{Delchambre}, L. and {Krone-Martins}, A. and {Wertz}, O. and {Ducourant}, C. and {Galluccio}, L. and {Kl{\"u}ter}, J. and {Mignard}, F. and {Teixeira}, R. and {Djorgovski}, S.~G. and {Stern}, D. and {Graham}, M.~J. and {Surdej}, J. and {Bastian}, U. and {Wambsganss}, J. and {Le Campion}, J. -F. and {Slezak}, E.},
        title = "{Gaia GraL: Gaia DR2 Gravitational Lens Systems. III. A systematic blind search for new lensed systems}",
      journal = {\aap},
         year = 2019,
        month = feb,
       volume = {622},
          eid = {A165},
        pages = {A165},
          doi = {10.1051/0004-6361/201833802},
archivePrefix = {arXiv},
       eprint = {1807.02845},
 primaryClass = {astro-ph.CO},
       adsurl = {https://ui.adsabs.harvard.edu/abs/2019A&A...622A.165D}
}

@ARTICLE{2021ApJ...921...42S,
       author = {{Stern}, D. and {Djorgovski}, S.~G. and {Krone-Martins}, A. and {Sluse}, D. and {Delchambre}, L. and {Ducourant}, C. and {Teixeira}, R. and {Surdej}, J. and {Boehm}, C. and {den Brok}, J. and {Dobie}, D. and {Drake}, A. and {Galluccio}, L. and {Graham}, M.~J. and {Jalan}, P. and {Kl{\"u}ter}, J. and {Le Campion}, J. -F. and {Mahabal}, A. and {Mignard}, F. and {Murphy}, T. and {Nierenberg}, A. and {Scarano}, S., Jr. and {Simon}, J. and {Slezak}, E. and {Spindola-Duarte}, C. and {Wambsganss}, J.},
        title = "{Gaia GraL: Gaia DR2 Gravitational Lens Systems. VI. Spectroscopic Confirmation and Modeling of Quadruply Imaged Lensed Quasars}",
      journal = {\apj},
         year = 2021,
        month = nov,
       volume = {921},
       number = {1},
          eid = {42},
        pages = {42},
          doi = {10.3847/1538-4357/ac0f04},
archivePrefix = {arXiv},
       eprint = {2012.10051},
 primaryClass = {astro-ph.GA},
       adsurl = {https://ui.adsabs.harvard.edu/abs/2021ApJ...921...42S}
}

@ARTICLE{2023ApJ...956..117G,
       author = {{Gross}, Arran C. and {Chen}, Yu-Ching and {Foord}, Adi and {Liu}, Xin and {Shen}, Yue and {Oguri}, Masamune and {Goulding}, Andy and {Hwang}, Hsiang-Chih and {Zakamska}, Nadia L. and {Ma}, Yilun and {Nolan}, Liam},
        title = "{Varstrometry for Off-nucleus and Dual Subkiloparsec Active Galactic Nuclei (VODKA): Investigating the Nature of SDSS J0823+2418 at z = 1.81, A Likely Lensed Quasar}",
      journal = {\apj},
         year = 2023,
        month = oct,
       volume = {956},
       number = {2},
          eid = {117},
        pages = {117},
          doi = {10.3847/1538-4357/acf469},
archivePrefix = {arXiv},
       eprint = {2306.04041},
 primaryClass = {astro-ph.GA},
       adsurl = {https://ui.adsabs.harvard.edu/abs/2023ApJ...956..117G}
}

@ARTICLE{2021MNRAS.508.4625H,
       author = {{Hartley}, P. and {Jackson}, N. and {Badole}, S. and {McKean}, J.~P. and {Sluse}, D. and {Vives-Arias}, H.},
        title = "{Using strong lensing to understand the microJy radio emission in two radio quiet quasars at redshift 1.7}",
      journal = {\mnras},
         year = 2021,
        month = dec,
       volume = {508},
       number = {3},
        pages = {4625-4638},
          doi = {10.1093/mnras/stab2758},
archivePrefix = {arXiv},
       eprint = {2109.10720},
 primaryClass = {astro-ph.GA},
       adsurl = {https://ui.adsabs.harvard.edu/abs/2021MNRAS.508.4625H}
}

@INPROCEEDINGS{McKean2015,
       author = {{McKean}, J. and {Jackson}, N. and {Vegetti}, S. and {Rybak}, M. and {Serjeant}, S. and {Koopmans}, L.~V.~E. and {Metcalf}, R.~B. and {Fassnacht}, C. and {Marshall}, P.~J. and {Pandey-Pommier}, M.},
        title = "{Strong Gravitational Lensing with the SKA}",
    booktitle = {Advancing Astrophysics with the Square Kilometre Array (AASKA14)},
         year = 2015,
        month = apr,
          eid = {84},
        pages = {84},
          doi = {10.22323/1.215.0084},
archivePrefix = {arXiv},
       eprint = {1502.03362},
 primaryClass = {astro-ph.GA},
       adsurl = {https://ui.adsabs.harvard.edu/abs/2015aska.confE..84M}
}

@ARTICLE{Myers2003,
       author = {{Myers}, S.~T. and {Jackson}, N.~J. and {Browne}, I.~W.~A. and {de Bruyn}, A.~G. and {Pearson}, T.~J. and {Readhead}, A.~C.~S. and {Wilkinson}, P.~N. and {Biggs}, A.~D. and {Blandford}, R.~D. and {Fassnacht}, C.~D. and {Koopmans}, L.~V.~E. and {Marlow}, D.~R. and {McKean}, J.~P. and {Norbury}, M.~A. and {Phillips}, P.~M. and {Rusin}, D. and {Shepherd}, M.~C. and {Sykes}, C.~M.},
        title = "{The Cosmic Lens All-Sky Survey - I. Source selection and observations}",
      journal = {\mnras},
         year = 2003,
        month = may,
       volume = {341},
       number = {1},
        pages = {1-12},
          doi = {10.1046/j.1365-8711.2003.06256.x},
archivePrefix = {arXiv},
       eprint = {astro-ph/0211073},
 primaryClass = {astro-ph},
       adsurl = {https://ui.adsabs.harvard.edu/abs/2003MNRAS.341....1M}
}

@ARTICLE{Ivezic2019,
       author = {{Ivezi{\'c}}, {\v{Z}}eljko and {Kahn}, Steven M. and {Tyson}, J. Anthony and {Abel}, Bob and {Acosta}, Emily and {Allsman}, Robyn and {Alonso}, David and {AlSayyad}, Yusra and {Anderson}, Scott F. and {Andrew}, John and {Angel}, James Roger P. and {Angeli}, George Z. and {Ansari}, Reza and {Antilogus}, Pierre and {Araujo}, Constanza and {Armstrong}, Robert and {Arndt}, Kirk T. and {Astier}, Pierre and {Aubourg}, {\'E}ric and {Auza}, Nicole and {Axelrod}, Tim S. and {Bard}, Deborah J. and {Barr}, Jeff D. and {Barrau}, Aurelian and {Bartlett}, James G. and {Bauer}, Amanda E. and {Bauman}, Brian J. and {Baumont}, Sylvain and {Bechtol}, Ellen and {Bechtol}, Keith and {Becker}, Andrew C. and {Becla}, Jacek and {Beldica}, Cristina and {Bellavia}, Steve and {Bianco}, Federica B. and {Biswas}, Rahul and {Blanc}, Guillaume and {Blazek}, Jonathan and {Blandford}, Roger D. and {Bloom}, Josh S. and {Bogart}, Joanne and {Bond}, Tim W. and {Booth}, Michael T. and {Borgland}, Anders W. and {Borne}, Kirk and {Bosch}, James F. and {Boutigny}, Dominique and {Brackett}, Craig A. and {Bradshaw}, Andrew and {Brandt}, William Nielsen and {Brown}, Michael E. and {Bullock}, James S. and {Burchat}, Patricia and {Burke}, David L. and {Cagnoli}, Gianpietro and {Calabrese}, Daniel and {Callahan}, Shawn and {Callen}, Alice L. and {Carlin}, Jeffrey L. and {Carlson}, Erin L. and {Chandrasekharan}, Srinivasan and {Charles-Emerson}, Glenaver and {Chesley}, Steve and {Cheu}, Elliott C. and {Chiang}, Hsin-Fang and {Chiang}, James and {Chirino}, Carol and {Chow}, Derek and {Ciardi}, David R. and {Claver}, Charles F. and {Cohen-Tanugi}, Johann and {Cockrum}, Joseph J. and {Coles}, Rebecca and {Connolly}, Andrew J. and {Cook}, Kem H. and {Cooray}, Asantha and {Covey}, Kevin R. and {Cribbs}, Chris and {Cui}, Wei and {Cutri}, Roc and {Daly}, Philip N. and {Daniel}, Scott F. and {Daruich}, Felipe and {Daubard}, Guillaume and {Daues}, Greg and {Dawson}, William and {Delgado}, Francisco and {Dellapenna}, Alfred and {de Peyster}, Robert and {de Val-Borro}, Miguel and {Digel}, Seth W. and {Doherty}, Peter and {Dubois}, Richard and {Dubois-Felsmann}, Gregory P. and {Durech}, Josef and {Economou}, Frossie and {Eifler}, Tim and {Eracleous}, Michael and {Emmons}, Benjamin L. and {Fausti Neto}, Angelo and {Ferguson}, Henry and {Figueroa}, Enrique and {Fisher-Levine}, Merlin and {Focke}, Warren and {Foss}, Michael D. and {Frank}, James and {Freemon}, Michael D. and {Gangler}, Emmanuel and {Gawiser}, Eric and {Geary}, John C. and {Gee}, Perry and {Geha}, Marla and {Gessner}, Charles J.~B. and {Gibson}, Robert R. and {Gilmore}, D. Kirk and {Glanzman}, Thomas and {Glick}, William and {Goldina}, Tatiana and {Goldstein}, Daniel A. and {Goodenow}, Iain and {Graham}, Melissa L. and {Gressler}, William J. and {Gris}, Philippe and {Guy}, Leanne P. and {Guyonnet}, Augustin and {Haller}, Gunther and {Harris}, Ron and {Hascall}, Patrick A. and {Haupt}, Justine and {Hernandez}, Fabio and {Herrmann}, Sven and {Hileman}, Edward and {Hoblitt}, Joshua and {Hodgson}, John A. and {Hogan}, Craig and {Howard}, James D. and {Huang}, Dajun and {Huffer}, Michael E. and {Ingraham}, Patrick and {Innes}, Walter R. and {Jacoby}, Suzanne H. and {Jain}, Bhuvnesh and {Jammes}, Fabrice and {Jee}, M. James and {Jenness}, Tim and {Jernigan}, Garrett and {Jevremovi{\'c}}, Darko and {Johns}, Kenneth and {Johnson}, Anthony S. and {Johnson}, Margaret W.~G. and {Jones}, R. Lynne and {Juramy-Gilles}, Claire and {Juri{\'c}}, Mario and {Kalirai}, Jason S. and {Kallivayalil}, Nitya J. and {Kalmbach}, Bryce and {Kantor}, Jeffrey P. and {Karst}, Pierre and {Kasliwal}, Mansi M. and {Kelly}, Heather and {Kessler}, Richard and {Kinnison}, Veronica and {Kirkby}, David and {Knox}, Lloyd and {Kotov}, Ivan V. and {Krabbendam}, Victor L. and {Krughoff}, K. Simon and {Kub{\'a}nek}, Petr and {Kuczewski}, John and {Kulkarni}, Shri and {Ku}, John and {Kurita}, Nadine R. and {Lage}, Craig S. and {Lambert}, Ron and {Lange}, Travis and {Langton}, J. Brian and {Le Guillou}, Laurent and {Levine}, Deborah and {Liang}, Ming and {Lim}, Kian-Tat and {Lintott}, Chris J. and {Long}, Kevin E. and {Lopez}, Margaux and {Lotz}, Paul J. and {Lupton}, Robert H. and {Lust}, Nate B. and {MacArthur}, Lauren A. and {Mahabal}, Ashish and {Mandelbaum}, Rachel and {Markiewicz}, Thomas W. and {Marsh}, Darren S. and {Marshall}, Philip J. and {Marshall}, Stuart and {May}, Morgan and {McKercher}, Robert and {McQueen}, Michelle and {Meyers}, Joshua and {Migliore}, Myriam and {Miller}, Michelle and {Mills}, David J. and {Miraval}, Connor and {Moeyens}, Joachim and {Moolekamp}, Fred E. and {Monet}, David G. and {Moniez}, Marc and {Monkewitz}, Serge and {Montgomery}, Christopher and {Morrison}, Christopher B. and {Mueller}, Fritz and {Muller}, Gary P. and {Mu{\~n}oz Arancibia}, Freddy and {Neill}, Douglas R. and {Newbry}, Scott P. and {Nief}, Jean-Yves and {Nomerotski}, Andrei and {Nordby}, Martin and {O'Connor}, Paul and {Oliver}, John and {Olivier}, Scot S. and {Olsen}, Knut and {O'Mullane}, William and {Ortiz}, Sandra and {Osier}, Shawn and {Owen}, Russell E. and {Pain}, Reynald and {Palecek}, Paul E. and {Parejko}, John K. and {Parsons}, James B. and {Pease}, Nathan M. and {Peterson}, J. Matt and {Peterson}, John R. and {Petravick}, Donald L. and {Libby Petrick}, M.~E. and {Petry}, Cathy E. and {Pierfederici}, Francesco and {Pietrowicz}, Stephen and {Pike}, Rob and {Pinto}, Philip A. and {Plante}, Raymond and {Plate}, Stephen and {Plutchak}, Joel P. and {Price}, Paul A. and {Prouza}, Michael and {Radeka}, Veljko and {Rajagopal}, Jayadev and {Rasmussen}, Andrew P. and {Regnault}, Nicolas and {Reil}, Kevin A. and {Reiss}, David J. and {Reuter}, Michael A. and {Ridgway}, Stephen T. and {Riot}, Vincent J. and {Ritz}, Steve and {Robinson}, Sean and {Roby}, William and {Roodman}, Aaron and {Rosing}, Wayne and {Roucelle}, Cecille and {Rumore}, Matthew R. and {Russo}, Stefano and {Saha}, Abhijit and {Sassolas}, Benoit and {Schalk}, Terry L. and {Schellart}, Pim and {Schindler}, Rafe H. and {Schmidt}, Samuel and {Schneider}, Donald P. and {Schneider}, Michael D. and {Schoening}, William and {Schumacher}, German and {Schwamb}, Megan E. and {Sebag}, Jacques and {Selvy}, Brian and {Sembroski}, Glenn H. and {Seppala}, Lynn G. and {Serio}, Andrew and {Serrano}, Eduardo and {Shaw}, Richard A. and {Shipsey}, Ian and {Sick}, Jonathan and {Silvestri}, Nicole and {Slater}, Colin T. and {Smith}, J. Allyn and {Smith}, R. Chris and {Sobhani}, Shahram and {Soldahl}, Christine and {Storrie-Lombardi}, Lisa and {Stover}, Edward and {Strauss}, Michael A. and {Street}, Rachel A. and {Stubbs}, Christopher W. and {Sullivan}, Ian S. and {Sweeney}, Donald and {Swinbank}, John D. and {Szalay}, Alexander and {Takacs}, Peter and {Tether}, Stephen A. and {Thaler}, Jon J. and {Thayer}, John Gregg and {Thomas}, Sandrine and {Thornton}, Adam J. and {Thukral}, Vaikunth and {Tice}, Jeffrey and {Trilling}, David E. and {Turri}, Max and {Van Berg}, Richard and {Vanden Berk}, Daniel and {Vetter}, Kurt and {Virieux}, Francoise and {Vucina}, Tomislav and {Wahl}, William and {Walkowicz}, Lucianne and {Walsh}, Brian and {Walter}, Christopher W. and {Wang}, Daniel L. and {Wang}, Shin-Yawn and {Warner}, Michael and {Wiecha}, Oliver and {Willman}, Beth and {Winters}, Scott E. and {Wittman}, David and {Wolff}, Sidney C. and {Wood-Vasey}, W. Michael and {Wu}, Xiuqin and {Xin}, Bo and {Yoachim}, Peter and {Zhan}, Hu},
        title = "{LSST: From Science Drivers to Reference Design and Anticipated Data Products}",
      journal = {\apj},
         year = 2019,
        month = mar,
       volume = {873},
       number = {2},
          eid = {111},
        pages = {111},
          doi = {10.3847/1538-4357/ab042c},
archivePrefix = {arXiv},
       eprint = {0805.2366},
 primaryClass = {astro-ph},
       adsurl = {https://ui.adsabs.harvard.edu/abs/2019ApJ...873..111I}
}

@ARTICLE{Spergel2015,
       author = {{Spergel}, D. and {Gehrels}, N. and {Baltay}, C. and {Bennett}, D. and {Breckinridge}, J. and {Donahue}, M. and {Dressler}, A. and {Gaudi}, B.~S. and {Greene}, T. and {Guyon}, O. and {Hirata}, C. and {Kalirai}, J. and {Kasdin}, N.~J. and {Macintosh}, B. and {Moos}, W. and {Perlmutter}, S. and {Postman}, M. and {Rauscher}, B. and {Rhodes}, J. and {Wang}, Y. and {Weinberg}, D. and {Benford}, D. and {Hudson}, M. and {Jeong}, W. -S. and {Mellier}, Y. and {Traub}, W. and {Yamada}, T. and {Capak}, P. and {Colbert}, J. and {Masters}, D. and {Penny}, M. and {Savransky}, D. and {Stern}, D. and {Zimmerman}, N. and {Barry}, R. and {Bartusek}, L. and {Carpenter}, K. and {Cheng}, E. and {Content}, D. and {Dekens}, F. and {Demers}, R. and {Grady}, K. and {Jackson}, C. and {Kuan}, G. and {Kruk}, J. and {Melton}, M. and {Nemati}, B. and {Parvin}, B. and {Poberezhskiy}, I. and {Peddie}, C. and {Ruffa}, J. and {Wallace}, J.~K. and {Whipple}, A. and {Wollack}, E. and {Zhao}, F.},
        title = "{Wide-Field InfrarRed Survey Telescope-Astrophysics Focused Telescope Assets WFIRST-AFTA 2015 Report}",
      journal = {arXiv e-prints},
         year = 2015,
        month = mar,
          eid = {arXiv:1503.03757},
        pages = {arXiv:1503.03757},
          doi = {10.48550/arXiv.1503.03757},
archivePrefix = {arXiv},
       eprint = {1503.03757},
 primaryClass = {astro-ph.IM},
       adsurl = {https://ui.adsabs.harvard.edu/abs/2015arXiv150303757S}
}

@ARTICLE{Laureijs2011,
       author = {{Laureijs}, R. and {Amiaux}, J. and {Arduini}, S. and {Augu{\`e}res}, J. -L. and {Brinchmann}, J. and {Cole}, R. and {Cropper}, M. and {Dabin}, C. and {Duvet}, L. and {Ealet}, A. and {Garilli}, B. and {Gondoin}, P. and {Guzzo}, L. and {Hoar}, J. and {Hoekstra}, H. and {Holmes}, R. and {Kitching}, T. and {Maciaszek}, T. and {Mellier}, Y. and {Pasian}, F. and {Percival}, W. and {Rhodes}, J. and {Saavedra Criado}, G. and {Sauvage}, M. and {Scaramella}, R. and {Valenziano}, L. and {Warren}, S. and {Bender}, R. and {Castander}, F. and {Cimatti}, A. and {Le F{\`e}vre}, O. and {Kurki-Suonio}, H. and {Levi}, M. and {Lilje}, P. and {Meylan}, G. and {Nichol}, R. and {Pedersen}, K. and {Popa}, V. and {Rebolo Lopez}, R. and {Rix}, H. -W. and {Rottgering}, H. and {Zeilinger}, W. and {Grupp}, F. and {Hudelot}, P. and {Massey}, R. and {Meneghetti}, M. and {Miller}, L. and {Paltani}, S. and {Paulin-Henriksson}, S. and {Pires}, S. and {Saxton}, C. and {Schrabback}, T. and {Seidel}, G. and {Walsh}, J. and {Aghanim}, N. and {Amendola}, L. and {Bartlett}, J. and {Baccigalupi}, C. and {Beaulieu}, J. -P. and {Benabed}, K. and {Cuby}, J. -G. and {Elbaz}, D. and {Fosalba}, P. and {Gavazzi}, G. and {Helmi}, A. and {Hook}, I. and {Irwin}, M. and {Kneib}, J. -P. and {Kunz}, M. and {Mannucci}, F. and {Moscardini}, L. and {Tao}, C. and {Teyssier}, R. and {Weller}, J. and {Zamorani}, G. and {Zapatero Osorio}, M.~R. and {Boulade}, O. and {Foumond}, J.~J. and {Di Giorgio}, A. and {Guttridge}, P. and {James}, A. and {Kemp}, M. and {Martignac}, J. and {Spencer}, A. and {Walton}, D. and {Bl{\"u}mchen}, T. and {Bonoli}, C. and {Bortoletto}, F. and {Cerna}, C. and {Corcione}, L. and {Fabron}, C. and {Jahnke}, K. and {Ligori}, S. and {Madrid}, F. and {Martin}, L. and {Morgante}, G. and {Pamplona}, T. and {Prieto}, E. and {Riva}, M. and {Toledo}, R. and {Trifoglio}, M. and {Zerbi}, F. and {Abdalla}, F. and {Douspis}, M. and {Grenet}, C. and {Borgani}, S. and {Bouwens}, R. and {Courbin}, F. and {Delouis}, J. -M. and {Dubath}, P. and {Fontana}, A. and {Frailis}, M. and {Grazian}, A. and {Koppenh{\"o}fer}, J. and {Mansutti}, O. and {Melchior}, M. and {Mignoli}, M. and {Mohr}, J. and {Neissner}, C. and {Noddle}, K. and {Poncet}, M. and {Scodeggio}, M. and {Serrano}, S. and {Shane}, N. and {Starck}, J. -L. and {Surace}, C. and {Taylor}, A. and {Verdoes-Kleijn}, G. and {Vuerli}, C. and {Williams}, O.~R. and {Zacchei}, A. and {Altieri}, B. and {Escudero Sanz}, I. and {Kohley}, R. and {Oosterbroek}, T. and {Astier}, P. and {Bacon}, D. and {Bardelli}, S. and {Baugh}, C. and {Bellagamba}, F. and {Benoist}, C. and {Bianchi}, D. and {Biviano}, A. and {Branchini}, E. and {Carbone}, C. and {Cardone}, V. and {Clements}, D. and {Colombi}, S. and {Conselice}, C. and {Cresci}, G. and {Deacon}, N. and {Dunlop}, J. and {Fedeli}, C. and {Fontanot}, F. and {Franzetti}, P. and {Giocoli}, C. and {Garcia-Bellido}, J. and {Gow}, J. and {Heavens}, A. and {Hewett}, P. and {Heymans}, C. and {Holland}, A. and {Huang}, Z. and {Ilbert}, O. and {Joachimi}, B. and {Jennins}, E. and {Kerins}, E. and {Kiessling}, A. and {Kirk}, D. and {Kotak}, R. and {Krause}, O. and {Lahav}, O. and {van Leeuwen}, F. and {Lesgourgues}, J. and {Lombardi}, M. and {Magliocchetti}, M. and {Maguire}, K. and {Majerotto}, E. and {Maoli}, R. and {Marulli}, F. and {Maurogordato}, S. and {McCracken}, H. and {McLure}, R. and {Melchiorri}, A. and {Merson}, A. and {Moresco}, M. and {Nonino}, M. and {Norberg}, P. and {Peacock}, J. and {Pello}, R. and {Penny}, M. and {Pettorino}, V. and {Di Porto}, C. and {Pozzetti}, L. and {Quercellini}, C. and {Radovich}, M. and {Rassat}, A. and {Roche}, N. and {Ronayette}, S. and {Rossetti}, E. and {Sartoris}, B. and {Schneider}, P. and {Semboloni}, E. and {Serjeant}, S. and {Simpson}, F. and {Skordis}, C. and {Smadja}, G. and {Smartt}, S. and {Spano}, P. and {Spiro}, S. and {Sullivan}, M. and {Tilquin}, A. and {Trotta}, R. and {Verde}, L. and {Wang}, Y. and {Williger}, G. and {Zhao}, G. and {Zoubian}, J. and {Zucca}, E.},
        title = "{Euclid Definition Study Report}",
      journal = {arXiv e-prints},
         year = 2011,
        month = oct,
          eid = {arXiv:1110.3193},
        pages = {arXiv:1110.3193},
          doi = {10.48550/arXiv.1110.3193},
archivePrefix = {arXiv},
       eprint = {1110.3193},
 primaryClass = {astro-ph.CO},
       adsurl = {https://ui.adsabs.harvard.edu/abs/2011arXiv1110.3193L}
}

@ARTICLE{Yue2022,
       author = {{Yue}, Minghao and {Fan}, Xiaohui and {Yang}, Jinyi and {Wang}, Feige},
        title = "{A Mock Catalog of Gravitationally-lensed Quasars for the LSST Survey}",
      journal = {\aj},
         year = 2022,
        month = mar,
       volume = {163},
       number = {3},
          eid = {139},
        pages = {139},
          doi = {10.3847/1538-3881/ac4cb0},
archivePrefix = {arXiv},
       eprint = {2201.06761},
 primaryClass = {astro-ph.GA},
       adsurl = {https://ui.adsabs.harvard.edu/abs/2022AJ....163..139Y}
}

@MISC{Nyland2023,
        author = {{Nyland}, K. and {Alexander}, K. and {Andernach}, H. and {Callingham}, J. and {Cigan}, P. and {Clarke}, T. and {Dong}, D and {Gordon}, Y. and {Kent}, B. and {Lacy}, M. and {Law}, C. and {Myers}, S. and {Ott}, J. and {Peters}, W. and {Polisensky}, E. and {Sivakoff}, G. and {Tremblay}, C. and {Ward}, C. and {Birmingham}, S. and {Patil}, P. and {Petric}, A. and {Kooi}, J. and {The VLASS SSG}},
        title = "{VLASS Epoch 4 Science Case}",
        howpublished = {\url{https://science.nrao.edu/vlass/library/vlass-epoch-4-science-case}},
        year = 2023,
        month = Oct
}

@MISC{Carilli2015,
        author = {{Carilli}, C.~L. and {McKinnon}, M. and {Ott}, J. and {Beasley}, A. and {Isella}, A. and {Murphy}, E. and {Leroy}, A. and {Casey}, C. and {Moullet}, A. and {Lacy}, M. and {Hodge}, J. and {Bower}, G. and {Demorest}, P. and {Hull}, C. and {Hughes}, M. and {di Francesco}, J. and {Narayanan}, D. and {Kent}, B. and {Clark}, B. and {Butler}, B.},
        title = "{Next Generation Very Large Array Memo No. 5. Science Working Groups Project Overview}",
        howpublished = {\url{https://library.nrao.edu/public/memos/ngvla/NGVLA_05.pdf}},
        year = 2015,
        month = Oct
}

@ARTICLE{Jackson2024,
       author = {{Jackson}, Neal and {Badole}, Shruti and {Dugdale}, Thomas and {Stacey}, Hannah R. and {Hartley}, Philippa and {McKean}, J.~P.},
        title = "{Radio imaging of gravitationally lensed radio-quiet quasars}",
      journal = {\mnras},
         year = 2024,
        month = mar,
          doi = {10.1093/mnras/stae916},
archivePrefix = {arXiv},
       eprint = {2403.19357},
 primaryClass = {astro-ph.GA},
       adsurl = {https://ui.adsabs.harvard.edu/abs/2024MNRAS.tmp..928J}
}

@ARTICLE{Huang2020,
       author = {{Huang}, X. and {Storfer}, C. and {Ravi}, V. and {Pilon}, A. and {Domingo}, M. and {Schlegel}, D.~J. and {Bailey}, S. and {Dey}, A. and {Gupta}, R.~R. and {Herrera}, D. and {Juneau}, S. and {Landriau}, M. and {Lang}, D. and {Meisner}, A. and {Moustakas}, J. and {Myers}, A.~D. and {Schlafly}, E.~F. and {Valdes}, F. and {Weaver}, B.~A. and {Yang}, J. and {Y{\`e}che}, C.},
        title = "{Finding Strong Gravitational Lenses in the DESI DECam Legacy Survey}",
      journal = {\apj},
         year = 2020,
        month = may,
       volume = {894},
       number = {1},
          eid = {78},
        pages = {78},
          doi = {10.3847/1538-4357/ab7ffb},
archivePrefix = {arXiv},
       eprint = {1906.00970},
 primaryClass = {astro-ph.GA},
       adsurl = {https://ui.adsabs.harvard.edu/abs/2020ApJ...894...78H}
}

@ARTICLE{Lemon2024,
       author = {{Lemon}, Cameron and {Courbin}, Fr{\'e}d{\'e}ric and {More}, Anupreeta and {Schechter}, Paul and {Ca{\~n}ameras}, Raoul and {Delchambre}, Ludovic and {Leung}, Calvin and {Shu}, Yiping and {Spiniello}, Chiara and {Hezaveh}, Yashar and {Kl{\"u}ter}, Jonas and {McMahon}, Richard},
        title = "{Searching for Strong Gravitational Lenses}",
      journal = {\ssr},
         year = 2024,
        month = feb,
       volume = {220},
       number = {2},
          eid = {23},
        pages = {23},
          doi = {10.1007/s11214-024-01042-9},
archivePrefix = {arXiv},
       eprint = {2310.13466},
 primaryClass = {astro-ph.GA},
       adsurl = {https://ui.adsabs.harvard.edu/abs/2024SSRv..220...23L}
}

@ARTICLE{ODea2021,
       author = {{O'Dea}, Christopher P. and {Saikia}, D.~J.},
        title = "{Compact steep-spectrum and peaked-spectrum radio sources}",
      journal = {\aapr},
         year = 2021,
        month = dec,
       volume = {29},
       number = {1},
          eid = {3},
        pages = {3},
          doi = {10.1007/s00159-021-00131-w},
archivePrefix = {arXiv},
       eprint = {2009.02750},
 primaryClass = {astro-ph.GA},
       adsurl = {https://ui.adsabs.harvard.edu/abs/2021A&ARv..29....3O}
}

@ARTICLE{mao17,
       author = {{Mao}, S.~A. and {Carilli}, C. and {Gaensler}, B.~M. and {Wucknitz}, O. and {Keeton}, C. and {Basu}, A. and {Beck}, R. and {Kronberg}, P.~P. and {Zweibel}, E.},
        title = "{Detection of microgauss coherent magnetic fields in a galaxy five billion years ago}",
      journal = {Nature Astronomy},
         year = 2017,
        month = aug,
       volume = {1},
        pages = {621-626},
          doi = {10.1038/s41550-017-0218-x},
archivePrefix = {arXiv},
       eprint = {1708.07844},
 primaryClass = {astro-ph.GA},
       adsurl = {https://ui.adsabs.harvard.edu/abs/2017NatAs...1..621M}
}

@ARTICLE{McKean2005,
       author = {{McKean}, J.~P. and {Browne}, I.~W.~A. and {Jackson}, N.~J. and {Koopmans}, L.~V.~E. and {Norbury}, M.~A. and {Treu}, T. and {York}, T.~D. and {Biggs}, A.~D. and {Blandford}, R.~D. and {de Bruyn}, A.~G. and {Fassnacht}, C.~D. and {Mao}, S. and {Myers}, S.~T. and {Pearson}, T.~J. and {Phillips}, P.~M. and {Readhead}, A.~C.~S. and {Rusin}, D. and {Wilkinson}, P.~N.},
        title = "{CLASS B2108+213: a new wide-separation gravitational lens system}",
      journal = {\mnras},
         year = 2005,
        month = jan,
       volume = {356},
       number = {3},
        pages = {1009-1016},
          doi = {10.1111/j.1365-2966.2004.08516.x},
archivePrefix = {arXiv},
       eprint = {astro-ph/0410554},
 primaryClass = {astro-ph},
       adsurl = {https://ui.adsabs.harvard.edu/abs/2005MNRAS.356.1009M}
}

@ARTICLE{spignola2019,
       author = {{Spingola}, C. and {McKean}, J.~P. and {Massari}, D. and {Koopmans}, L.~V.~E.},
        title = "{Proper motion in lensed radio jets at redshift 3: A possible dual super-massive black hole system in the early Universe}",
      journal = {\aap},
         year = 2019,
        month = oct,
       volume = {630},
          eid = {A108},
        pages = {A108},
          doi = {10.1051/0004-6361/201935427},
archivePrefix = {arXiv},
       eprint = {1908.11756},
 primaryClass = {astro-ph.GA},
       adsurl = {https://ui.adsabs.harvard.edu/abs/2019A&A...630A.108S}
}

@ARTICLE{vegetti23,
       author = {{Vegetti}, S. and {Birrer}, S. and {Despali}, G. and {Fassnacht}, C.~D. and {Gilman}, D. and {Hezaveh}, Y. and {Perreault Levasseur}, L. and {McKean}, J.~P. and {Powell}, D.~M. and {O'Riordan}, C.~M. and {Vernardos}, G.},
        title = "{Strong gravitational lensing as a probe of dark matter}",
      journal = {arXiv e-prints},
         year = 2023,
        month = jun,
          eid = {arXiv:2306.11781},
        pages = {arXiv:2306.11781},
          doi = {10.48550/arXiv.2306.11781},
archivePrefix = {arXiv},
       eprint = {2306.11781},
 primaryClass = {astro-ph.CO},
       adsurl = {https://ui.adsabs.harvard.edu/abs/2023arXiv230611781V}
}

@ARTICLE{SKA,
       author = {{Braun}, Robert and {Bonaldi}, Anna and {Bourke}, Tyler and {Keane}, Evan and {Wagg}, Jeff},
        title = "{Anticipated Performance of the Square Kilometre Array -- Phase 1 (SKA1)}",
      journal = {arXiv e-prints},
         year = 2019,
        month = dec,
          eid = {arXiv:1912.12699},
        pages = {arXiv:1912.12699},
          doi = {10.48550/arXiv.1912.12699},
archivePrefix = {arXiv},
       eprint = {1912.12699},
 primaryClass = {astro-ph.IM},
       adsurl = {https://ui.adsabs.harvard.edu/abs/2019arXiv191212699B}
}

@ARTICLE{bechtol22,
       author = {{Bechtol}, Keith and {Birrer}, Simon and {Cyr-Racine}, Francis-Yan and {Schutz}, Katelin and {Adhikari}, Susmita and {Amin}, Mustafa and {Banerjee}, Arka and {Bird}, Simeon and {Blinov}, Nikita and {Boddy}, Kimberly K. and {Boehm}, Celine and {Bundy}, Kevin and {Buschmann}, Malte and {Chakrabarti}, Sukanya and {Curtin}, David and {Dai}, Liang and {Drlica-Wagner}, Alex and {Dvorkin}, Cora and {Erickcek}, Adrienne L. and {Gilman}, Daniel and {Heeba}, Saniya and {Kim}, Stacy and {Ir{\v{s}}i{\v{c}}}, Vid and {Leauthaud}, Alexie and {Lovell}, Mark and {Luki{\'c}}, Zarija and {Mao}, Yao-Yuan and {Mau}, Sidney and {Mitridate}, Andrea and {Mocz}, Philip and {Mu{\~n}oz}, Julian B. and {Nadler}, Ethan O. and {Peter}, Annika H.~G. and {Price-Whelan}, Adrian and {Robertson}, Andrew and {Sabti}, Nashwan and {Sehgal}, Neelima and {Shipp}, Nora and {Simon}, Joshua D. and {Singh}, Rajeev and {Van Tilburg}, Ken and {Wechsler}, Risa H. and {Widmark}, Axel and {Yu}, Hai-Bo},
        title = "{Snowmass2021 Cosmic Frontier White Paper: Dark Matter Physics from Halo Measurements}",
      journal = {arXiv e-prints},
         year = 2022,
        month = mar,
          eid = {arXiv:2203.07354},
        pages = {arXiv:2203.07354},
          doi = {10.48550/arXiv.2203.07354},
archivePrefix = {arXiv},
       eprint = {2203.07354},
 primaryClass = {hep-ph},
       adsurl = {https://ui.adsabs.harvard.edu/abs/2022arXiv220307354B}
}

@ARTICLE{powell23,
       author = {{Powell}, Devon M. and {Vegetti}, Simona and {McKean}, J.~P. and {White}, Simon D.~M. and {Ferreira}, Elisa G.~M. and {May}, Simon and {Spingola}, Cristiana},
        title = "{A lensed radio jet at milli-arcsecond resolution - II. Constraints on fuzzy dark matter from an extended gravitational arc}",
      journal = {\mnras},
         year = 2023,
        month = sep,
       volume = {524},
       number = {1},
        pages = {L84-L88},
          doi = {10.1093/mnrasl/slad074},
archivePrefix = {arXiv},
       eprint = {2302.10941},
 primaryClass = {astro-ph.CO},
       adsurl = {https://ui.adsabs.harvard.edu/abs/2023MNRAS.524L..84P}
}

@ARTICLE{lenstronomy,
       author = {{Birrer}, Simon and {Amara}, Adam},
        title = "{lenstronomy: Multi-purpose gravitational lens modelling software package}",
      journal = {Physics of the Dark Universe},
         year = 2018,
        month = dec,
       volume = {22},
        pages = {189-201},
          doi = {10.1016/j.dark.2018.11.002},
archivePrefix = {arXiv},
       eprint = {1803.09746},
 primaryClass = {astro-ph.CO},
       adsurl = {https://ui.adsabs.harvard.edu/abs/2018PDU....22..189B}
}

@ARTICLE{lenstronomy2,
       author = {{Birrer}, Simon and {Shajib}, Anowar and {Gilman}, Daniel and {Galan}, Aymeric and {Aalbers}, Jelle and {Millon}, Martin and {Morgan}, Robert and {Pagano}, Giulia and {Park}, Ji and {Teodori}, Luca and {Tessore}, Nicolas and {Ueland}, Madison and {Van de Vyvere}, Lyne and {Wagner-Carena}, Sebastian and {Wempe}, Ewoud and {Yang}, Lilan and {Ding}, Xuheng and {Schmidt}, Thomas and {Sluse}, Dominique and {Zhang}, Ming and {Amara}, Adam},
        title = "{lenstronomy II: A gravitational lensing software ecosystem}",
      journal = {The Journal of Open Source Software},
         year = 2021,
        month = jun,
       volume = {6},
       number = {62},
          eid = {3283},
        pages = {3283},
          doi = {10.21105/joss.03283},
archivePrefix = {arXiv},
       eprint = {2106.05976},
 primaryClass = {astro-ph.CO},
       adsurl = {https://ui.adsabs.harvard.edu/abs/2021JOSS....6.3283B}
}

@ARTICLE{SIE,
       author = {{Kormann}, R. and {Schneider}, P. and {Bartelmann}, M.},
        title = "{Isothermal elliptical gravitational lens models.}",
      journal = {\aap},
         year = 1994,
        month = apr,
       volume = {284},
        pages = {285-299},
       adsurl = {https://ui.adsabs.harvard.edu/abs/1994A&A...284..285K}
}

@ARTICLE{collett15,
       author = {{Collett}, Thomas E.},
        title = "{The Population of Galaxy-Galaxy Strong Lenses in Forthcoming Optical Imaging Surveys}",
      journal = {\apj},
         year = 2015,
        month = sep,
       volume = {811},
       number = {1},
          eid = {20},
        pages = {20},
          doi = {10.1088/0004-637X/811/1/20},
archivePrefix = {arXiv},
       eprint = {1507.02657},
 primaryClass = {astro-ph.CO},
       adsurl = {https://ui.adsabs.harvard.edu/abs/2015ApJ...811...20C}
}

@ARTICLE{Intema2017,
       author = {{Intema}, H.~T. and {Jagannathan}, P. and {Mooley}, K.~P. and {Frail}, D.~A.},
        title = "{The GMRT 150 MHz all-sky radio survey. First alternative data release TGSS ADR1}",
      journal = {\aap},
         year = 2017,
        month = feb,
       volume = {598},
          eid = {A78},
        pages = {A78},
          doi = {10.1051/0004-6361/201628536},
archivePrefix = {arXiv},
       eprint = {1603.04368},
 primaryClass = {astro-ph.CO},
       adsurl = {https://ui.adsabs.harvard.edu/abs/2017A&A...598A..78I}
}

@ARTICLE{Bock1999,
       author = {{Bock}, D.~C. -J. and {Large}, M.~I. and {Sadler}, Elaine M.},
        title = "{SUMSS: A Wide-Field Radio Imaging Survey of the Southern Sky. I. Science Goals, Survey Design, and Instrumentation}",
      journal = {\aj},
         year = 1999,
        month = mar,
       volume = {117},
       number = {3},
        pages = {1578-1593},
          doi = {10.1086/300786},
archivePrefix = {arXiv},
       eprint = {astro-ph/9812083},
 primaryClass = {astro-ph},
       adsurl = {https://ui.adsabs.harvard.edu/abs/1999AJ....117.1578B}
}

@ARTICLE{Morabito2022,
       author = {{Morabito}, L.~K. and {Jackson}, N.~J. and {Mooney}, S. and {Sweijen}, F. and {Badole}, S. and {Kukreti}, P. and {Venkattu}, D. and {Groeneveld}, C. and {Kappes}, A. and {Bonnassieux}, E. and {Drabent}, A. and {Iacobelli}, M. and {Croston}, J.~H. and {Best}, P.~N. and {Bondi}, M. and {Callingham}, J.~R. and {Conway}, J.~E. and {Deller}, A.~T. and {Hardcastle}, M.~J. and {McKean}, J.~P. and {Miley}, G.~K. and {Moldon}, J. and {R{\"o}ttgering}, H.~J.~A. and {Tasse}, C. and {Shimwell}, T.~W. and {van Weeren}, R.~J. and {Anderson}, J.~M. and {Asgekar}, A. and {Avruch}, I.~M. and {van Bemmel}, I.~M. and {Bentum}, M.~J. and {Bonafede}, A. and {Brouw}, W.~N. and {Butcher}, H.~R. and {Ciardi}, B. and {Corstanje}, A. and {Coolen}, A. and {Damstra}, S. and {de Gasperin}, F. and {Duscha}, S. and {Eisl{\"o}ffel}, J. and {Engels}, D. and {Falcke}, H. and {Garrett}, M.~A. and {Griessmeier}, J. and {Gunst}, A.~W. and {van Haarlem}, M.~P. and {Hoeft}, M. and {van der Horst}, A.~J. and {J{\"u}tte}, E. and {Kadler}, M. and {Koopmans}, L.~V.~E. and {Krankowski}, A. and {Mann}, G. and {Nelles}, A. and {Oonk}, J.~B.~R. and {Orru}, E. and {Paas}, H. and {Pandey}, V.~N. and {Pizzo}, R.~F. and {Pandey-Pommier}, M. and {Reich}, W. and {Rothkaehl}, H. and {Ruiter}, M. and {Schwarz}, D.~J. and {Shulevski}, A. and {Soida}, M. and {Tagger}, M. and {Vocks}, C. and {Wijers}, R.~A.~M.~J. and {Wijnholds}, S.~J. and {Wucknitz}, O. and {Zarka}, P. and {Zucca}, P.},
        title = "{Sub-arcsecond imaging with the International LOFAR Telescope. I. Foundational calibration strategy and pipeline}",
      journal = {\aap},
         year = 2022,
        month = feb,
       volume = {658},
          eid = {A1},
        pages = {A1},
          doi = {10.1051/0004-6361/202140649},
archivePrefix = {arXiv},
       eprint = {2108.07283},
 primaryClass = {astro-ph.IM},
       adsurl = {https://ui.adsabs.harvard.edu/abs/2022A&A...658A...1M}
}

@ARTICLE{Chambers2016,
       author = {{Chambers}, K.~C. and {Magnier}, E.~A. and {Metcalfe}, N. and {Flewelling}, H.~A. and {Huber}, M.~E. and {Waters}, C.~Z. and {Denneau}, L. and {Draper}, P.~W. and {Farrow}, D. and {Finkbeiner}, D.~P. and {Holmberg}, C. and {Koppenhoefer}, J. and {Price}, P.~A. and {Rest}, A. and {Saglia}, R.~P. and {Schlafly}, E.~F. and {Smartt}, S.~J. and {Sweeney}, W. and {Wainscoat}, R.~J. and {Burgett}, W.~S. and {Chastel}, S. and {Grav}, T. and {Heasley}, J.~N. and {Hodapp}, K.~W. and {Jedicke}, R. and {Kaiser}, N. and {Kudritzki}, R. -P. and {Luppino}, G.~A. and {Lupton}, R.~H. and {Monet}, D.~G. and {Morgan}, J.~S. and {Onaka}, P.~M. and {Shiao}, B. and {Stubbs}, C.~W. and {Tonry}, J.~L. and {White}, R. and {Ba{\~n}ados}, E. and {Bell}, E.~F. and {Bender}, R. and {Bernard}, E.~J. and {Boegner}, M. and {Boffi}, F. and {Botticella}, M.~T. and {Calamida}, A. and {Casertano}, S. and {Chen}, W. -P. and {Chen}, X. and {Cole}, S. and {Deacon}, N. and {Frenk}, C. and {Fitzsimmons}, A. and {Gezari}, S. and {Gibbs}, V. and {Goessl}, C. and {Goggia}, T. and {Gourgue}, R. and {Goldman}, B. and {Grant}, P. and {Grebel}, E.~K. and {Hambly}, N.~C. and {Hasinger}, G. and {Heavens}, A.~F. and {Heckman}, T.~M. and {Henderson}, R. and {Henning}, T. and {Holman}, M. and {Hopp}, U. and {Ip}, W. -H. and {Isani}, S. and {Jackson}, M. and {Keyes}, C.~D. and {Koekemoer}, A.~M. and {Kotak}, R. and {Le}, D. and {Liska}, D. and {Long}, K.~S. and {Lucey}, J.~R. and {Liu}, M. and {Martin}, N.~F. and {Masci}, G. and {McLean}, B. and {Mindel}, E. and {Misra}, P. and {Morganson}, E. and {Murphy}, D.~N.~A. and {Obaika}, A. and {Narayan}, G. and {Nieto-Santisteban}, M.~A. and {Norberg}, P. and {Peacock}, J.~A. and {Pier}, E.~A. and {Postman}, M. and {Primak}, N. and {Rae}, C. and {Rai}, A. and {Riess}, A. and {Riffeser}, A. and {Rix}, H.~W. and {R{\"o}ser}, S. and {Russel}, R. and {Rutz}, L. and {Schilbach}, E. and {Schultz}, A.~S.~B. and {Scolnic}, D. and {Strolger}, L. and {Szalay}, A. and {Seitz}, S. and {Small}, E. and {Smith}, K.~W. and {Soderblom}, D.~R. and {Taylor}, P. and {Thomson}, R. and {Taylor}, A.~N. and {Thakar}, A.~R. and {Thiel}, J. and {Thilker}, D. and {Unger}, D. and {Urata}, Y. and {Valenti}, J. and {Wagner}, J. and {Walder}, T. and {Walter}, F. and {Watters}, S.~P. and {Werner}, S. and {Wood-Vasey}, W.~M. and {Wyse}, R.},
        title = "{The Pan-STARRS1 Surveys}",
      journal = {arXiv e-prints},
         year = 2016,
        month = dec,
          eid = {arXiv:1612.05560},
        pages = {arXiv:1612.05560},
          doi = {10.48550/arXiv.1612.05560},
archivePrefix = {arXiv},
       eprint = {1612.05560},
 primaryClass = {astro-ph.IM},
       adsurl = {https://ui.adsabs.harvard.edu/abs/2016arXiv161205560C}
}

@ARTICLE{MGVLA,
       author = {{Lawrence}, C.~R. and {Bennett}, C.~L. and {Hewitt}, J.~N. and {Langston}, G.~I. and {Klotz}, S.~E. and {Burke}, B.~F. and {Turner}, K.~C.},
        title = "{5 GHz Radio Structure and Optical Identifications of Sources from the MG Survey. II. Maps and Finding Charts}",
      journal = {\apjs},
         year = 1986,
        month = may,
       volume = {61},
        pages = {105},
          doi = {10.1086/191109},
       adsurl = {https://ui.adsabs.harvard.edu/abs/1986ApJS...61..105L}
}

@ARTICLE{biggs18,
       author = {{Biggs}, A.~D. and {Browne}, I.~W.~A.},
        title = "{A revised lens time delay for JVAS B0218+357 from a reanalysis of VLA monitoring data}",
      journal = {\mnras},
         year = 2018,
        month = jun,
       volume = {476},
       number = {4},
        pages = {5393-5407},
          doi = {10.1093/mnras/sty565},
archivePrefix = {arXiv},
       eprint = {1802.10088},
 primaryClass = {astro-ph.CO},
       adsurl = {https://ui.adsabs.harvard.edu/abs/2018MNRAS.476.5393B}
}

@ARTICLE{biggs99,
       author = {{Biggs}, A.~D. and {Browne}, I.~W.~A. and {Helbig}, P. and {Koopmans}, L.~V.~E. and {Wilkinson}, P.~N. and {Perley}, R.~A.},
        title = "{Time delay for the gravitational lens system B0218+357}",
      journal = {\mnras},
         year = 1999,
        month = apr,
       volume = {304},
       number = {2},
        pages = {349-358},
          doi = {10.1046/j.1365-8711.1999.02309.x},
archivePrefix = {arXiv},
       eprint = {astro-ph/9811282},
 primaryClass = {astro-ph},
       adsurl = {https://ui.adsabs.harvard.edu/abs/1999MNRAS.304..349B}
}

@ARTICLE{koopmans2000,
       author = {{Koopmans}, L.~V.~E. and {de Bruyn}, A.~G. and {Xanthopoulos}, E. and {Fassnacht}, C.~D.},
        title = "{A time-delay determination from VLA light curves of the CLASS gravitational lens B1600+434}",
      journal = {\aap},
         year = 2000,
        month = apr,
       volume = {356},
        pages = {391-402},
          doi = {10.48550/arXiv.astro-ph/0001533},
archivePrefix = {arXiv},
       eprint = {astro-ph/0001533},
 primaryClass = {astro-ph},
       adsurl = {https://ui.adsabs.harvard.edu/abs/2000A&A...356..391K}
}

@ARTICLE{class1608A,
       author = {{Fassnacht}, C.~D. and {Pearson}, T.~J. and {Readhead}, A.~C.~S. and {Browne}, I.~W.~A. and {Koopmans}, L.~V.~E. and {Myers}, S.~T. and {Wilkinson}, P.~N.},
        title = "{A Determination of H$_{0}$ with the CLASS Gravitational Lens B1608+656. I. Time Delay Measurements with the VLA}",
      journal = {\apj},
         year = 1999,
        month = dec,
       volume = {527},
       number = {2},
        pages = {498-512},
          doi = {10.1086/308118},
archivePrefix = {arXiv},
       eprint = {astro-ph/9907257},
 primaryClass = {astro-ph},
       adsurl = {https://ui.adsabs.harvard.edu/abs/1999ApJ...527..498F}
}

@ARTICLE{class1608B,
       author = {{Koopmans}, L.~V.~E. and {Fassnacht}, C.~D.},
        title = "{A Determination of H$_{0}$ with the CLASS Gravitational Lens B1608+656. II. Mass Models and the Hubble Constant from Lensing}",
      journal = {\apj},
         year = 1999,
        month = dec,
       volume = {527},
       number = {2},
        pages = {513-524},
          doi = {10.1086/308120},
archivePrefix = {arXiv},
       eprint = {astro-ph/9907258},
 primaryClass = {astro-ph},
       adsurl = {https://ui.adsabs.harvard.edu/abs/1999ApJ...527..513K}
}

@ARTICLE{wagner19uni,
       author = {{Wagner}, Jenny},
        title = "{A Model-Independent Characterisation of Strong Gravitational Lensing by Observables}",
      journal = {Universe},
         year = 2019,
        month = jul,
       volume = {5},
       number = {7},
          eid = {177},
        pages = {177},
          doi = {10.3390/universe5070177},
archivePrefix = {arXiv},
       eprint = {1906.05285},
 primaryClass = {astro-ph.CO},
       adsurl = {https://ui.adsabs.harvard.edu/abs/2019Univ....5..177W}
}

@ARTICLE{wagner22,
       author = {{Wagner}, Jenny},
        title = "{Generalised model-independent characterisation of strong gravitational lenses. VII. Impact of source properties and higher-order lens properties on the local lens reconstruction}",
      journal = {\aap},
         year = 2022,
        month = jul,
       volume = {663},
          eid = {A157},
        pages = {A157},
          doi = {10.1051/0004-6361/202243562},
archivePrefix = {arXiv},
       eprint = {2203.06190},
 primaryClass = {astro-ph.CO},
       adsurl = {https://ui.adsabs.harvard.edu/abs/2022A&A...663A.157W}
}

@ARTICLE{wagnertessore,
       author = {{Wagner}, Jenny and {Liesenborgs}, Jori and {Tessore}, Nicolas},
        title = "{Model-independent and model-based local lensing properties of CL0024+1654 from multiply imaged galaxies}",
      journal = {\aap},
         year = 2018,
        month = apr,
       volume = {612},
          eid = {A17},
        pages = {A17},
          doi = {10.1051/0004-6361/201731932},
archivePrefix = {arXiv},
       eprint = {1709.03531},
 primaryClass = {astro-ph.CO},
       adsurl = {https://ui.adsabs.harvard.edu/abs/2018A&A...612A..17W}
}

@ARTICLE{pyhalo,
       author = {{Gilman}, Daniel and {Birrer}, Simon and {Nierenberg}, Anna and {Treu}, Tommaso and {Du}, Xiaolong and {Benson}, Andrew},
        title = "{Warm dark matter chills out: constraints on the halo mass function and the free-streaming length of dark matter with eight quadruple-image strong gravitational lenses}",
      journal = {\mnras},
         year = 2020,
        month = feb,
       volume = {491},
       number = {4},
        pages = {6077-6101},
          doi = {10.1093/mnras/stz3480},
archivePrefix = {arXiv},
       eprint = {1908.06983},
 primaryClass = {astro-ph.CO},
       adsurl = {https://ui.adsabs.harvard.edu/abs/2020MNRAS.491.6077G}
}

@ARTICLE{2016MNRAS.460.1214L,
       author = {{Ludlow}, Aaron D. and {Bose}, Sownak and {Angulo}, Ra{\'u}l E. and {Wang}, Lan and {Hellwing}, Wojciech A. and {Navarro}, Julio F. and {Cole}, Shaun and {Frenk}, Carlos S.},
        title = "{The mass-concentration-redshift relation of cold and warm dark matter haloes}",
      journal = {\mnras},
         year = 2016,
        month = aug,
       volume = {460},
       number = {2},
        pages = {1214-1232},
          doi = {10.1093/mnras/stw1046},
archivePrefix = {arXiv},
       eprint = {1601.02624},
 primaryClass = {astro-ph.CO},
       adsurl = {https://ui.adsabs.harvard.edu/abs/2016MNRAS.460.1214L}
}

@ARTICLE{2016MNRAS.455..318B,
       author = {{Bose}, Sownak and {Hellwing}, Wojciech A. and {Frenk}, Carlos S. and {Jenkins}, Adrian and {Lovell}, Mark R. and {Helly}, John C. and {Li}, Baojiu},
        title = "{The Copernicus Complexio: statistical properties of warm dark matter haloes}",
      journal = {\mnras},
         year = 2016,
        month = jan,
       volume = {455},
       number = {1},
        pages = {318-333},
          doi = {10.1093/mnras/stv2294},
archivePrefix = {arXiv},
       eprint = {1507.01998},
 primaryClass = {astro-ph.CO},
       adsurl = {https://ui.adsabs.harvard.edu/abs/2016MNRAS.455..318B}
}

@ARTICLE{gilman25,
       author = {{Gilman}, D. and {Nierenberg}, A.~M. and {Treu}, T. and {Gannon}, C. and {Du}, X. and {Paugnat}, H. and {Birrer}, S. and {Benson}, A.~J. and {Mozumdar}, P. and {Wong}, K.~C. and {Williams}, D. and {Keeley}, R.~E. and {Abazajian}, K.~N. and {Anguita}, T. and {Bennert}, V.~N. and {Djorgovski}, S.~G. and {Hoenig}, S.~H. and {Kusenko}, A. and {Malkan}, M. and {Morishita}, T. and {Motta}, V. and {Moustakas}, L.~A. and {Sheu}, W. and {Sluse}, D. and {Stern}, D. and {Stiavelli}, M.},
        title = "{JWST lensed quasar dark matter survey IV: Stringent warm dark matter constraints from the joint reconstruction of extended lensed arcs and quasar flux ratios}",
      journal = {arXiv e-prints},
         year = 2025,
        month = nov,
          eid = {arXiv:2511.07513},
        pages = {arXiv:2511.07513},
          doi = {10.48550/arXiv.2511.07513},
archivePrefix = {arXiv},
       eprint = {2511.07513},
 primaryClass = {astro-ph.CO},
       adsurl = {https://ui.adsabs.harvard.edu/abs/2025arXiv251107513G}
}

@ARTICLE{b1938vlbi,
       author = {{McKean}, J.~P. and {Spingola}, C. and {Powell}, D.~M. and {Vegetti}, S.},
        title = "{An extended and extremely thin gravitational arc from a lensed compact symmetric object at redshift of 2.059}",
      journal = {\mnras},
         year = 2025,
        month = nov,
       volume = {544},
       number = {1},
        pages = {L24-L30},
          doi = {10.1093/mnrasl/slaf039},
archivePrefix = {arXiv},
       eprint = {2510.07386},
 primaryClass = {astro-ph.GA},
       adsurl = {https://ui.adsabs.harvard.edu/abs/2025MNRAS.544L..24M}
}

@ARTICLE{2011ApJ...734..103G,
       author = {{Gralla}, Megan B. and {Gladders}, Michael D. and {Yee}, H.~K.~C. and {Barrientos}, L. Felipe},
        title = "{Constraining the Redshift Evolution of First Radio Sources in RCS1 Galaxy Clusters}",
      journal = {\apj},
         year = 2011,
        month = jun,
       volume = {734},
       number = {2},
          eid = {103},
        pages = {103},
          doi = {10.1088/0004-637X/734/2/103},
archivePrefix = {arXiv},
       eprint = {1010.6011},
 primaryClass = {astro-ph.CO},
       adsurl = {https://ui.adsabs.harvard.edu/abs/2011ApJ...734..103G}
}

@misc{SLED,
  author = {Vernardos, G. and others},
  year = {2024},
  title = {SLED: Strong Lensing Database},
  note = {Data set},
  howpublished = {GitHub repository},
  url = {https://github.com/gvernard/SLED_api}
}

@ARTICLE{2024MNRAS.530.2960N,
       author = {{Nierenberg}, A.~M. and {Keeley}, R.~E. and {Sluse}, D. and {Gilman}, D. and {Birrer}, S. and {Treu}, T. and {Abazajian}, K.~N. and {Anguita}, T. and {Benson}, A.~J. and {Bennert}, V.~N. and {Djorgovski}, S.~G. and {Du}, X. and {Fassnacht}, C.~D. and {Hoenig}, S.~F. and {Kusenko}, A. and {Lemon}, C. and {Malkan}, M. and {Motta}, V. and {Moustakas}, L.~A. and {Stern}, D. and {Wechsler}, R.~H.},
        title = "{JWST lensed quasar dark matter survey - I. Description and first results}",
      journal = {\mnras},
         year = 2024,
        month = may,
       volume = {530},
       number = {3},
        pages = {2960-2971},
          doi = {10.1093/mnras/stae499},
archivePrefix = {arXiv},
       eprint = {2309.10101},
 primaryClass = {astro-ph.CO},
       adsurl = {https://ui.adsabs.harvard.edu/abs/2024MNRAS.530.2960N}
}

@ARTICLE{1984ComAp..10...75B,
       author = {{Burke}, B.~F.},
        title = "{Gravitational lenses : observational status.}",
      journal = {Comments on Astrophysics},
         year = 1984,
        month = feb,
       volume = {10},
        pages = {75-84},
       adsurl = {https://ui.adsabs.harvard.edu/abs/1984ComAp..10...75B}
}

@ARTICLE{fleury21,
       author = {{Fleury}, Pierre and {Larena}, Julien and {Uzan}, Jean-Philippe},
        title = "{Line-of-sight effects in strong gravitational lensing}",
      journal = {\jcap},
         year = 2021,
        month = aug,
       volume = {2021},
       number = {8},
          eid = {024},
        pages = {024},
          doi = {10.1088/1475-7516/2021/08/024},
archivePrefix = {arXiv},
       eprint = {2104.08883},
 primaryClass = {astro-ph.CO},
       adsurl = {https://ui.adsabs.harvard.edu/abs/2021JCAP...08..024F}
}

@ARTICLE{jets,
       author = {{Bridle}, Alan H. and {Perley}, Richard A.},
        title = "{Extragalactic Radio Jets}",
      journal = {\araa},
         year = 1984,
        month = jan,
       volume = {22},
        pages = {319-358},
          doi = {10.1146/annurev.aa.22.090184.001535},
       adsurl = {https://ui.adsabs.harvard.edu/abs/1984ARA&A..22..319B}
}

@ARTICLE{honigagn,
       author = {{H{\"o}nig}, Sebastian F.},
        title = "{Redefining the Torus: A Unifying View of AGNs in the Infrared and Submillimeter}",
      journal = {\apj},
         year = 2019,
        month = oct,
       volume = {884},
       number = {2},
          eid = {171},
        pages = {171},
          doi = {10.3847/1538-4357/ab4591},
archivePrefix = {arXiv},
       eprint = {1909.08639},
 primaryClass = {astro-ph.GA},
       adsurl = {https://ui.adsabs.harvard.edu/abs/2019ApJ...884..171H}
}

@ARTICLE{2017NatAs...1..679R,
       author = {{Ramos Almeida}, Cristina and {Ricci}, Claudio},
        title = "{Nuclear obscuration in active galactic nuclei}",
      journal = {Nature Astronomy},
         year = 2017,
        month = oct,
       volume = {1},
        pages = {679-689},
          doi = {10.1038/s41550-017-0232-z},
archivePrefix = {arXiv},
       eprint = {1709.00019},
 primaryClass = {astro-ph.GA},
       adsurl = {https://ui.adsabs.harvard.edu/abs/2017NatAs...1..679R}
}

@ARTICLE{2011A&A...525L...8C,
       author = {{Czerny}, B. and {Hryniewicz}, K.},
        title = "{The origin of the broad line region in active galactic nuclei}",
      journal = {\aap},
         year = 2011,
        month = jan,
       volume = {525},
          eid = {L8},
        pages = {L8},
          doi = {10.1051/0004-6361/201016025},
archivePrefix = {arXiv},
       eprint = {1010.6201},
 primaryClass = {astro-ph.CO},
       adsurl = {https://ui.adsabs.harvard.edu/abs/2011A&A...525L...8C}
}

@ARTICLE{2011ApJ...739...69M,
       author = {{M{\"u}ller-S{\'a}nchez}, F. and {Prieto}, M.~A. and {Hicks}, E.~K.~S. and {Vives-Arias}, H. and {Davies}, R.~I. and {Malkan}, M. and {Tacconi}, L.~J. and {Genzel}, R.},
        title = "{Outflows from Active Galactic Nuclei: Kinematics of the Narrow-line and Coronal-line Regions in Seyfert Galaxies}",
      journal = {\apj},
         year = 2011,
        month = oct,
       volume = {739},
       number = {2},
          eid = {69},
        pages = {69},
          doi = {10.1088/0004-637X/739/2/69},
archivePrefix = {arXiv},
       eprint = {1107.3140},
 primaryClass = {astro-ph.CO},
       adsurl = {https://ui.adsabs.harvard.edu/abs/2011ApJ...739...69M}
}

@ARTICLE{2025A&A...696A.169K,
       author = {{Kim}, Jong-Seo and {M{\"u}ller}, Hendrik and {Nikonov}, Aleksei S. and {Lu}, Ru-Sen and {Knollm{\"u}ller}, Jakob and {En{\ss}lin}, Torsten A. and {Wielgus}, Maciek and {Lobanov}, Andrei P.},
        title = "{Imaging a ring-like structure and the extended jet of M87 at 86 GHz}",
      journal = {\aap},
         year = 2025,
        month = apr,
       volume = {696},
          eid = {A169},
        pages = {A169},
          doi = {10.1051/0004-6361/202452038},
archivePrefix = {arXiv},
       eprint = {2409.00540},
 primaryClass = {astro-ph.HE},
       adsurl = {https://ui.adsabs.harvard.edu/abs/2025A&A...696A.169K}
}

@ARTICLE{ehtcena,
       author = {{Janssen}, Michael and {Falcke}, Heino and {Kadler}, Matthias and {Ros}, Eduardo and {Wielgus}, Maciek and {Akiyama}, Kazunori and {Balokovi{\'c}}, Mislav and {Blackburn}, Lindy and {Bouman}, Katherine L. and {Chael}, Andrew and {Chan}, Chi-kwan and {Chatterjee}, Koushik and {Davelaar}, Jordy and {Edwards}, Philip G. and {Fromm}, Christian M. and {G{\'o}mez}, Jos{\'e} L. and {Goddi}, Ciriaco and {Issaoun}, Sara and {Johnson}, Michael D. and {Kim}, Junhan and {Koay}, Jun Yi and {Krichbaum}, Thomas P. and {Liu}, Jun and {Liuzzo}, Elisabetta and {Markoff}, Sera and {Markowitz}, Alex and {Marrone}, Daniel P. and {Mizuno}, Yosuke and {M{\"u}ller}, Cornelia and {Ni}, Chunchong and {Pesce}, Dominic W. and {Ramakrishnan}, Venkatessh and {Roelofs}, Freek and {Rygl}, Kazi L.~J. and {van Bemmel}, Ilse and {Event Horizon Telescope Collaboration} and {Alberdi}, Antxon and {Alef}, Walter and {Algaba}, Juan Carlos and {Anantua}, Richard and {Asada}, Keiichi and {Azulay}, Rebecca and {Baczko}, Anne-Kathrin and {Ball}, David and {Ball}, David and {Barrett}, John and {Benson}, Bradford A. and {Bintley}, Dan and {Bintley}, Dan and {Blundell}, Raymond and {Boland}, Wilfred and {Boland}, Wilfred and {Bower}, Geoffrey C. and {Boyce}, Hope and {Bremer}, Michael and {Brinkerink}, Christiaan D. and {Brissenden}, Roger and {Britzen}, Silke and {Broderick}, Avery E. and {Broguiere}, Dominique and {Bronzwaer}, Thomas and {Byun}, Do-Young and {Carlstrom}, John E. and {Chatterjee}, Shami and {Chen}, Ming-Tang and {Chen}, Yongjun and {Chesler}, Paul M. and {Cho}, Ilje and {Christian}, Pierre and {Conway}, John E. and {Cordes}, James M. and {Crawford}, Thomas M. and {Crew}, Geoffrey B. and {Cruz-Osorio}, Alejandro and {Cui}, Yuzhu and {Cui}, Yuzhu and {De Laurentis}, Mariafelicia and {Deane}, Roger and {Dempsey}, Jessica and {Desvignes}, Gregory and {Dexter}, Jason and {Doeleman}, Sheperd S. and {Eatough}, Ralph P. and {Farah}, Joseph and {Farah}, Joseph and {Fish}, Vincent L. and {Fomalont}, Ed and {Ford}, H. Alyson and {Fraga-Encinas}, Raquel and {Friberg}, Per and {Friberg}, Per and {Fuentes}, Antonio and {Galison}, Peter and {Gammie}, Charles F. and {Garc{\'\i}a}, Roberto and {Gelles}, Zachary and {Gentaz}, Olivier and {Georgiev}, Boris and {Georgiev}, Boris and {Gold}, Roman and {Gold}, Roman and {G{\'o}mez-Ruiz}, Arturo I. and {Gu}, Minfeng and {Gurwell}, Mark and {Hada}, Kazuhiro and {Haggard}, Daryl and {Hecht}, Michael H. and {Hesper}, Ronald and {Himwich}, Elizabeth and {Ho}, Luis C. and {Ho}, Paul and {Honma}, Mareki and {Huang}, Chih-Wei L. and {Huang}, Lei and {Hughes}, David H. and {Ikeda}, Shiro and {Inoue}, Makoto and {Inoue}, Makoto and {James}, David J. and {Jannuzi}, Buell T. and {Jeter}, Britton and {Jiang}, Wu and {Jimenez-Rosales}, Alejandra and {Jorstad}, Svetlana and {Jung}, Taehyun and {Karami}, Mansour and {Karuppusamy}, Ramesh and {Kawashima}, Tomohisa and {Keating}, Garrett K. and {Kettenis}, Mark and {Kim}, Dong-Jin and {Kim}, Jae-Young and {Kim}, Jongsoo and {Kino}, Motoki and {Kofuji}, Yutaro and {Koyama}, Shoko and {Kramer}, Michael and {Kramer}, Carsten and {Kuo}, Cheng-Yu and {Lauer}, Tod R. and {Lee}, Sang-Sung and {Levis}, Aviad and {Li}, Yan-Rong and {Li}, Zhiyuan and {Lindqvist}, Michael and {Lico}, Rocco and {Lindahl}, Greg and {Liu}, Kuo and {Lo}, Wen-Ping and {Lobanov}, Andrei P. and {Loinard}, Laurent and {Lonsdale}, Colin and {Lu}, Ru-Sen and {MacDonald}, Nicholas R. and {Mao}, Jirong and {Marchili}, Nicola and {Marscher}, Alan P. and {Mart{\'\i}-Vidal}, Iv{\'a}n and {Matsushita}, Satoki and {Matthews}, Lynn D. and {Medeiros}, Lia and {Menten}, Karl M. and {Mizuno}, Izumi and {Moran}, James M. and {Moriyama}, Kotaro and {Moscibrodzka}, Monika and {Moscibrodzka}, Monika and {Musoke}, Gibwa and {Mej{\'\i}as}, Alejandro Mus and {Nagai}, Hiroshi and {Nagar}, Neil M. and {Nakamura}, Masanori and {Narayan}, Ramesh and {Narayanan}, Gopal and {Natarajan}, Iniyan and {Nathanail}, Antonios and {Neilsen}, Joey and {Neri}, Roberto and {Noutsos}, Aristeidis and {Nowak}, Michael A. and {Okino}, Hiroki and {Olivares}, H{\'e}ctor and {Ortiz-Le{\'o}n}, Gisela N. and {Oyama}, Tomoaki and {{\"O}zel}, Feryal and {Palumbo}, Daniel C.~M. and {Park}, Jongho and {Patel}, Nimesh and {Pen}, Ue-Li and {Pi{\'e}tu}, Vincent and {Plambeck}, Richard and {PopStefanija}, Aleksandar and {Porth}, Oliver and {P{\"o}tzl}, Felix M. and {Prather}, Ben and {Preciado-L{\'o}pez}, Jorge A. and {Psaltis}, Dimitrios and {Pu}, Hung-Yi and {Pu}, Hung-Yi and {Rao}, Ramprasad},
        title = "{Event Horizon Telescope observations of the jet launching and collimation in Centaurus A}",
      journal = {Nature Astronomy},
         year = 2021,
        month = jul,
       volume = {5},
        pages = {1017-1028},
          doi = {10.1038/s41550-021-01417-w},
archivePrefix = {arXiv},
       eprint = {2111.03356},
 primaryClass = {astro-ph.GA},
       adsurl = {https://ui.adsabs.harvard.edu/abs/2021NatAs...5.1017J}
}

@ARTICLE{2023NatAs...7.1359F,
       author = {{Fuentes}, Antonio and {G{\'o}mez}, Jos{\'e} L. and {Mart{\'\i}}, Jos{\'e} M. and {Perucho}, Manel and {Zhao}, Guang-Yao and {Lico}, Rocco and {Lobanov}, Andrei P. and {Bruni}, Gabriele and {Kovalev}, Yuri Y. and {Chael}, Andrew and {Akiyama}, Kazunori and {Bouman}, Katherine L. and {Sun}, He and {Cho}, Ilje and {Traianou}, Efthalia and {Toscano}, Teresa and {Dahale}, Rohan and {Foschi}, Marianna and {Gurvits}, Leonid I. and {Jorstad}, Svetlana and {Kim}, Jae-Young and {Marscher}, Alan P. and {Mizuno}, Yosuke and {Ros}, Eduardo and {Savolainen}, Tuomas},
        title = "{Filamentary structures as the origin of blazar jet radio variability}",
      journal = {Nature Astronomy},
         year = 2023,
        month = nov,
       volume = {7},
        pages = {1359-1367},
          doi = {10.1038/s41550-023-02105-7},
archivePrefix = {arXiv},
       eprint = {2311.01861},
 primaryClass = {astro-ph.HE},
       adsurl = {https://ui.adsabs.harvard.edu/abs/2023NatAs...7.1359F}
}

@ARTICLE{walker2018,
       author = {{Walker}, R. Craig and {Hardee}, Philip E. and {Davies}, Frederick B. and {Ly}, Chun and {Junor}, William},
        title = "{The Structure and Dynamics of the Subparsec Jet in M87 Based on 50 VLBA Observations over 17 Years at 43 GHz}",
      journal = {\apj},
         year = 2018,
        month = mar,
       volume = {855},
       number = {2},
          eid = {128},
        pages = {128},
          doi = {10.3847/1538-4357/aaafcc},
archivePrefix = {arXiv},
       eprint = {1802.06166},
 primaryClass = {astro-ph.HE},
       adsurl = {https://ui.adsabs.harvard.edu/abs/2018ApJ...855..128W}
}

@ARTICLE{class2,
       author = {{Blas}, Diego and {Lesgourgues}, Julien and {Tram}, Thomas},
        title = "{The Cosmic Linear Anisotropy Solving System (CLASS). Part II: Approximation schemes}",
      journal = {\jcap},
         year = 2011,
        month = jul,
       volume = {2011},
       number = {7},
          eid = {034},
        pages = {034},
          doi = {10.1088/1475-7516/2011/07/034},
archivePrefix = {arXiv},
       eprint = {1104.2933},
 primaryClass = {astro-ph.CO},
       adsurl = {https://ui.adsabs.harvard.edu/abs/2011JCAP...07..034B}
}

@ARTICLE{class4,
       author = {{Lesgourgues}, Julien and {Tram}, Thomas},
        title = "{The Cosmic Linear Anisotropy Solving System (CLASS) IV: efficient implementation of non-cold relics}",
      journal = {\jcap},
         year = 2011,
        month = sep,
       volume = {2011},
       number = {9},
          eid = {032},
        pages = {032},
          doi = {10.1088/1475-7516/2011/09/032},
archivePrefix = {arXiv},
       eprint = {1104.2935},
 primaryClass = {astro-ph.CO},
       adsurl = {https://ui.adsabs.harvard.edu/abs/2011JCAP...09..032L}
}

@ARTICLE{classidm,
       author = {{Becker}, Niklas and {Hooper}, Deanna C. and {Kahlhoefer}, Felix and {Lesgourgues}, Julien and {Sch{\"o}neberg}, Nils},
        title = "{Cosmological constraints on multi-interacting dark matter}",
      journal = {\jcap},
         year = 2021,
        month = feb,
       volume = {2021},
       number = {2},
          eid = {019},
        pages = {019},
          doi = {10.1088/1475-7516/2021/02/019},
archivePrefix = {arXiv},
       eprint = {2010.04074},
 primaryClass = {astro-ph.CO},
       adsurl = {https://ui.adsabs.harvard.edu/abs/2021JCAP...02..019B}
}

@article{Lovell_2020,
doi = {10.3847/1538-4357/ab982a},
url = {https://doi.org/10.3847/1538-4357/ab982a},
year = {2020},
month = {jul},
publisher = {The American Astronomical Society},
volume = {897},
number = {2},
pages = {147},
author = {Lovell, Mark R.},
title = {Toward a General Parameterization of the Warm Dark Matter Halo Mass Function},
journal = {The Astrophysical Journal}
}

@article{ESTEST,
  title={An omnibus test for the two-sample problem using the empirical characteristic function},
  author={Thomas W. Epps and Kenneth J. Singleton},
  journal={Journal of Statistical Computation and Simulation},
  year={1986},
  volume={26},
  pages={177-203},
  url={https://api.semanticscholar.org/CorpusID:120867220}
}

\end{document}